\documentclass[12pt]{article}

\usepackage{cite}
\usepackage{float} 
\usepackage{graphicx}
\usepackage{amsfonts}
\usepackage{amsmath}
\usepackage{longtable}
\input{epsf}

\usepackage[colorlinks,linkcolor=blue,citecolor=red,hyperindex]{hyperref}

\def\rf#1{(\ref{eq:#1})}
\def\lab#1{\label{eq:#1}}

\def\br{\begin{eqnarray}}
\def\er{\end{eqnarray}}
\def\be{\begin{equation}}
\def\ee{\end{equation}}

\def\({\left(}
\def\){\right)}

\def\rlx{\relax\leavevmode}
\def\IR{\mathbb{R}}
\def\IC{\mathbb{C}}
\def\vp{\varphi}
\def\ve{\varepsilon}
\def\u2{\mid u\mid^2}

\relax

\newcommand{\sbr}[2]{\left\lbrack\,{#1}\, ,\,{#2}\,\right\rbrack}
\newcommand{\pbr}[2]{\{\,{#1}\, ,\,{#2}\,\}_{PB}}

\def\d{\delta}

\def\vareps{\varepsilon}

\def\tp0{\Theta_{+}^{(0)}}
\def\tm0{\Theta_{-}^{(0)}}

\def\u2{\mid u\mid^2}

\def\vp{\varphi}

\newcommand\fet[1]{\mathfrak{e}_{\tau}\left(#1\right)}
\newcommand\fez[1]{\mathfrak{e}_{\zeta}\left(#1\right)}
\newcommand\fbt[1]{\mathfrak{b}_{\tau}\left(#1\right)}
\newcommand\fbz[1]{\mathfrak{b}_{\zeta}\left(#1\right)}

\def\rlx{\relax\leavevmode}
\def\inbar{\vrule height1.5ex width.4pt depth0pt}
\def\IZ{\rlx\hbox{\sf Z\kern-.4em Z}}
\def\IN{\rlx\hbox{\rm I\kern-.18em N}}
\def\IO{\rlx\hbox{\,$\inbar\kern-.3em{\rm O}$}}
\def\IP{\rlx\hbox{\rm I\kern-.18em P}}
\def\IQ{\rlx\hbox{\,$\inbar\kern-.3em{\rm Q}$}}
\def\IF{\rlx\hbox{\rm I\kern-.18em F}}
\def\IG{\rlx\hbox{\,$\inbar\kern-.3em{\rm G}$}}
\def\IH{\rlx\hbox{\rm I\kern-.18em H}}
\def\II{\rlx\hbox{\rm I\kern-.18em I}}
\def\IK{\rlx\hbox{\rm I\kern-.18em K}}
\def\IL{\rlx\hbox{\rm I\kern-.18em L}}
\def\one{\hbox{{1}\kern-.25em\hbox{l}}}
\def\0#1{\relax\ifmmode\mathaccent''7017{#1}%
B        \else\accent23#1\relax\fi}

\begin{document}

\begin{titlepage}
\vspace*{-1cm}

\vskip 3cm

\vspace{.2in}
\begin{center}
{\large\bf  The Hidden Symmetries of Yang-Mills Theory in $\(3+1\)$-dimensions }
\end{center}

\vspace{.5cm}

\begin{center}
L. A. Ferreira$^{\dagger ,}$\footnote{laf@ifsc.usp.br}, and H. Malavazzi$^{\dagger ,}$\footnote{henrique.malavazzi@usp.br}

\vspace{.3 in}
\small

\par \vskip .2in \noindent
$^{\dagger}$Instituto de F\'\i sica de S\~ao Carlos; IFSC/USP;\\
Universidade de S\~ao Paulo, USP  \\ 
CEP 13566-590, S\~ao Carlos-SP, Brasil\\

\vskip 2cm

\begin{abstract}

We show that classical, non-supersymmetric Yang-Mills theories coupled to spin-$1/2$ and spin-$0$ elementary matter fields, in $(3+1)$-dimensional Minkowski space-time, possess exact  structures that resemble integrability, with  an infinite number of conserved charges in involution. Such structures live in the space of non-abelian electric and magnetic charges, and are based on flat connections in generalized loop spaces, presenting an $R$-matrix, and Sklyanin relation. We present two novel symmetries of Yang-Mills theories. The first one corresponds to global  transformations generated by the infinity of those conserved charges under the Poisson brackets. The gauge and matter fields, as well as Wilson lines and fluxes, have  interesting transformation laws under such a global symmetry. The second one corresponds to symmetries of the integral Yang-Mills equations, which lead to the conserved charges. They generate an infinite-dimensional group, where the elements are holonomies of connections on the loop space of functions from the circle $S^1$ to  the space-time. Our approach certainly applies to the Standard Model of the Fundamental Interactions. The conserved charges are gauge invariant, and so, in the case of QCD,  they are color singlets and perhaps are not confined. Therefore, the hadrons may carry such charges.  Our results open up the way for the construction of non-perturbative methods for Yang-Mills theories.

\end{abstract}

\normalsize
\end{center}
\end{titlepage}

\section{Introduction}
\label{sec:introduction}
\setcounter{equation}{0}

The development of exact and non-perturbative methods to solve field theories is of crucial importance for the study of non-linear and strongly coupled physical phenomena in all areas of Physics. Symmetries play a central role in most of those methods. However, it has become clear  that the   symmetries relevant for the development of those exact and non-perturbative methods are not always of the Noether type, but hidden, and sometimes non-local, symmetries that act on some special structures, like one-form flat connections.   In this paper, we discuss two novel symmetries of $(3+1)$-dimensional  Yang-Mills theories, coupled to matter fields, based on their classical integral equations \cite{ym1,ym2}. Our approach is based on methods on generalized loop spaces, and it applies, in particular, to the Standard Model of the fundamental interactions. The physically relevant non-abelian gauge theories have particle production and certainly are not integrable in the usual sense of having a factorized S-matrix.  However, we show that those same Yang-Mills theories present exact  structures, resembling integrability, with an infinite number of conserved charges in involution. Such structures  live in the space of non-abelian electric and magnetic charges, possessing a flat connection, R-matrix, and satisfying the so-called Fundamental Poisson bracket Relation (FPR), and the Sklyanin relation \cite{faddeevleshouches,babelonbook,faddeevbook,retore}. The infinity of conserved charges constitutes higher modes of the electric and magnetic charges and are obtained as the eigenvalues of a charge operator that presents an iso-spectral time evolution. Even though all those charges are in involution with the Hamiltonian,  the energy and momenta are not among those eigenvalues. The infinity of conserved charges that we construct is, in fact, associated to internal global symmetries of the Yang-Mills theories.  It is not clear if those charges may take the role of the canonical variables parameterizing the phase space, as is usual in exact integrable theories, in the Liouville sense.

The flat connection leading to the charges presents a gauge symmetry on a generalized loop space, and the corresponding infinite-dimensional gauge group plays a similar role in $(3+1)$-dimensions to that played by the Kac-Moody group in integrable theories in $(1+1)$-dimensions. Such a group, however, may not be a Lie group and may circumvent the Coleman-Mandula theorem \cite{colmand} in a novel way. The conserved charges are truly gauge invariant, and so in the case of Quantum Chromodynamics (QCD), they are color singlets. Therefore, such charges in principle are not confined and could be carried by the hadrons.  In order to contextualize  our results, we start with an  exposition of  well-known facts about integrable field theories that motivated our methods.

\centerline{\bf - Integrability in higher dimensions -} 

 Integrable field theories in $(1+1)$-dimensions constitute perhaps the best example  where exact and non-perturbative methods were developed at the classical and quantum level with great success. An important mathematical structure responsible for such developments, in two-dimensional integrable field theories, is the so-called Lax-Zakharov-Shabat equation \cite{lax,zakharov}, where the equations of motion are expressed as the zero curvature condition for a connection 
\be
\partial_0 A_1-\partial_1 A_0+\sbr{A_0}{A_1}=0
\lab{2dzc}
\ee
where $A_{\mu}$, $\mu=0,1$, is a functional of the fundamental fields of the theory and their derivatives, and takes values on a Kac-Moody algebra \cite{kacbook}. The relevant symmetries are the gauge transformations $A_{\mu}\rightarrow g\,A_{\mu}\,g^{-1} -\partial_{\mu} g\,g^{-1}$, where $g$ are special elements of the so-called Kac-Moody group (a carefully defined exponentiation of the generators of the Kac-Moody algebra in a so-called integrable representation). Such gauge transformations are not symmetries of the Lagrangian or Hamiltonian, and so are denoted hidden symmetries. They act on the connection $A_{\mu}$, which is not a physical field of the theory, preserving the zero curvature representation \rf{2dzc} of the equations of motion. Such symmetries underlie all the exact results in integrable field theories, like the inverse scattering method, dressing transformations, etc. In addition, the equation \rf{2dzc} is a conservation law as it implies that the holonomy (Wilson line) of the connection $A_{\mu}$ is independent of the path, as long as its end points are kept fixed. By imposing suitable boundary conditions, one can show that the Wilson line operator integrated on the space ($x$-axis) has an isospectral time evolution
\be
W\(\gamma_t\)=U\(t\)\,W\(\gamma_0\)\,U^{-1}\(t\)
\ee
where $\gamma_t$ and $\gamma_0$ denote the $x$-axis at the time $t=t$ and $t=0$ respectively, and $W$ is the path ordered integral $W\(\gamma\)=P\,e^{-\int_{\gamma}\, d\sigma\,A_{\mu}\,\frac{d\,x^{\mu}}{d\,\sigma}}$. Therefore, the eigenvalues of $W$, or equivalently ${\rm Tr}\, W^N$, are conserved in time. Since $A_{\mu}$ lives on an infinite-dimensional Kac-Moody algebra, one gets an infinity of conservation laws. It then follows all the impressive achievements of two-dimensional integrability, at the classical and quantum levels, like the construction of exact solutions, factorisation of the $S$-matrix, exact correlation functions, etc, \cite{faddeevbook,babelonbook}. 

It is hard to believe that physical local field theories in four dimensions can present properties like those of two-dimensional integrable theories. The reasons are manifold. The hidden (gauge) symmetries of integrable theories described above mix in a non-trivial way internal and external transformations. Good physical theories (local, unitary, Lorentz invariant, etc) in four-dimensional  Minkowski space-time are subjected   to the no-go Coleman-Mandula theorem \cite{colmand} that forbids internal and external Lie group symmetries to mix non trivially. Supersymmetry circumvents such a theorem \cite{wesszumino,haagsohnius}, and indeed supersymmetric field theories are, in general, more solvable than non-supersymmetric ones.  In this paper, we shall present novel symmetries of classical ordinary Yang-Mills theory (non-supersymmetric) that may lead to new methods suitable to solve some non-perturbative  aspects of non-abelian gauge theories. However, the relevant infinite-dimensional symmetry group  may not be of Lie type, and perhaps may circumvent the Coleman-Mandula theorem  in a novel way. Such symmetries were discovered in our attempts to generalize to higher dimensions the zero curvature \rf{2dzc},  as we now explain. 

The basic mathematical structure we need is a generalized loop space, i.e., an infinite-dimensional space of functions $f$ mapping the $(d-1)$-dimensional sphere $S^{d-1}$ into the space-time $M$, with a  base point $x_R$, i.e. 
\be
{\cal L}^{(d-1)}\equiv \{ f: \; S^{d-1}\rightarrow M\, \mid \, \mbox{\rm north pole}\rightarrow x_R\}
\lab{loopspace}
\ee
where $x_R$ is a fixed reference point in $M$, which is the image, in all mappings, of the north pole of $S^{d-1}$. Consider now a $d$-dimensional volume $\Omega$ in $M$, and take $x_R$ to lie on its border $\partial \Omega$. One can scan $\Omega$ with $(d-1)$-dimensional closed surfaces based on $x_R$, starting with an infinitesimal surface around $x_R$, and blowing it up until it reaches the border $\partial \Omega$. Every point of $\Omega$ belongs to one and only one surface used in the scanning. Such closed surfaces are, in fact,  images of the mappings $f$ of the loop space  ${\cal L}^{(d-1)}$. Therefore, $\Omega$ can be seem as a path in ${\cal L}^{(d-1)}$. We now introduce a one-form connection ${\cal A}$ on the generalized loop space ${\cal L}^{(d-1)}$, taking values on some Lie algebra ${\cal G}$, and we impose it to be flat, i.e.,  its curvature vanishes
\be
\delta {\cal A} + {\cal A}\wedge {\cal A}=0
\lab{flatconnloop}
\ee
It then follows that the holonomy of ${\cal A}$ on loop space will be path independent, as long as the  ending points of the paths are kept fixed. That is the approach proposed in \cite{afg1} to generalize to higher dimensions some of the structures of $(1+1)$-dimensional integrable field theories. The idea is to define the connection ${\cal A}$ in such a way that its zero curvature condition \rf{flatconnloop} is equivalent to the classical equations of motion of the higher-dimensional field theory of interest. Therefore, the path independency of the holonomy of ${\cal A}$ is a consequence of the dynamics of such a theory. A given path on the loop space ${\cal L}^{(d-1)}$  corresponds to a $d$-dimensional sub-manifold $\Omega$ on the space-time $M$, and the holonomy of ${\cal A}$ on that path corresponds to a volume ordered integral on $\Omega$. Deformations of the path that keep its ending points fixed are equivalent to deformations of the volume $\Omega$ that keep its border $\partial\Omega$ unchanged. By imposing appropriate boundary conditions, it is possible to obtain an isospectral  evolution of the volume ordered integral on $\Omega$ (i.e. the holonomy of ${\cal A}$), along the directions in $M$ perpendicular to $\Omega$. That corresponds to conservation  laws, since the eigenvalues of such volume-ordered integrated operator will be constant along those directions. 

If we take $M$ to be a $(d+1)$-dimensional Minkowski space-time, and $\Omega$ to be the spatial sub-manifold $\IR^d$, we get  charges that are conserved in time, in a way similar to the case of $(1+1)$-dimensional integrable field theories discussed above. In fact, for $d=1$, the relevant loop space is ${\cal L}^{(0)}$, i.e., the space of functions mapping $S^0$ into the two-dimensional space-time. But $S^0$ has only two points, and since one of them (the north pole) is always mapped into the reference point $x_R$, we get that ${\cal L}^{(0)}$ is isomorphic to the two-dimensional space-time $M$. So, the construction described above does reduce to the usual structures of $(1+1)$-dimensional integrable field theories. 

Such an approach, however, faces some difficult obstacles. First, the zero curvature condition \rf{flatconnloop} is local on the loop space ${\cal L}^{(d-1)}$, but not necessarily on the space-time $M$. Therefore, if one wants to work with local field theories in $M$, their local equations of motion will have to be equivalent to non-local conditions in $M$. Second, a given  volume $\Omega$ in  $M$ corresponds in fact to an infinity of paths on the loop space ${\cal L}^{(d-1)}$,  since the change of scanning of $\Omega$ with closed surfaces leads to different paths on ${\cal L}^{(d-1)}$. So, at the end of the day we have to show that the physics of our theory does not depend upon the loop parameterization of the volumes in $M$. 
 For a more detailed discussion of such issues, see the review \cite{afg2}. \\
 
 \centerline{\bf - Yang-Mills theories come to the rescue -}
 
 However, such an approach does overcome all those difficulties thanks to the physical properties of one of the most important theories we have, namely, gauge theories. As it is well known, the laws of electricity and magnetism were first formulated in terms of integral equations, like the Gauss and Faraday laws. The final and unified formulation of those laws   was provided  by Maxwell in terms of partial differential equations. As we know it, electrodynamics is a local field theory that admits an equivalent formulation in terms of integral equations of the form
 \be
 \mbox{\rm flux through $\partial \Omega$}\equiv \mbox{\rm charge inside $\Omega$}  
 \lab{fluxcharge}
 \ee
 where $\Omega$ is any three-dimensional volume in four-dimensional Minkowski space-time. The volume $\Omega$ can be purely spatial, like in the Gauss law, or it may extend in the  time direction,  like in the Faraday law of induction. 

Non-abelian gauge theories were formulated a la Maxwell by C. N. Yang and R. Mills \cite{ymoriginal} in terms of partial differential equations. However, as shown in \cite{ym1,ym2}, they also admit an equivalent formulation in terms of integral equations of the form \rf{fluxcharge}. Using a generalised version of the non-abelian Stokes theorem for two-form connections \cite{afg1,afg2}, it was shown in \cite{ym1,ym2} that for any three-dimensional volume $\Omega$ in the four-dimensional Minkowski space-time, the partial differential Yang-Mills equations are equivalent to the integral equations 
\be
P_2 \, e^{\int_{\partial\Omega} {\cal B} }=P_3\,e^{\int_{\Omega} {\cal A}}
\lab{ymintegralintro}
\ee
where $P_2$ denotes surface ordered integration on the border $\partial \Omega$, and $P_3$ means volume ordered integration on the volume $\Omega$. ${\cal B}$ stands for a linear combination of the Yang-Mills field tensor $F_{\mu\nu}$, and its Hodge dual ${\widetilde F}_{\mu\nu}=\frac{1}{2}\,\ve_{\mu\nu\rho\sigma}\,F^{\rho\sigma}$, conjugated with the Wilson line $W$, i.e. $W^{-1}\(\alpha\,F_{\mu\nu}+\beta\, {\widetilde F}_{\mu\nu}\)\,W$, with $\alpha$ and $\beta$ being arbitrary parameters. So, the integration on the left-hand side of \rf{ymintegralintro} gives a measure of the non-abelian electric and magnetic fluxes through $\partial \Omega$. ${\cal A}$ stands for the non-abelian electric and magnetic charges carried by the matter and gauge (gluon) fields, and so it measures the charge inside $\Omega$. Both ${\cal B}$ and ${\cal A}$ live on the Lie algebra of the gauge group $G$, and they depend upon the parameters $\alpha$ and $\beta$.  In the case of the abelian gauge group $U(1)$, \rf{ymintegralintro} reduces to the usual integral equations of classical electrodynamics. The details of the integral equation \rf{ymintegralintro} are given in Section \ref{sec:ymintegral}. 

In order to construct \rf{ymintegralintro} one needs to choose a scanning of the volume $\Omega$ with closed surfaces based on a fixed reference point $x_R$ on $\partial\Omega$, and so \rf{ymintegralintro} refers to a given path on the loop space ${\cal L}^{(2)}$ (see \rf{loopspace}), and $\partial \Omega$ corresponds to the end point of that path.  In their turn, each closed surface is scanned with loops based at $x_R$. If one changes the choice of the scanning of $\Omega$, both sides of \rf{ymintegralintro} change, but in a way that the equality still holds true. So, we can say that \rf{ymintegralintro} is covariant under the reparameterization of the volume $\Omega$. 

Since the parameters $\alpha$ and $\beta$ are arbitrary, one can expand both sides of \rf{ymintegralintro} in power series on those parameters, and so one gets an infinite number of integral Yang-Mills equations. Each one of those integral equations must be satisfied by any solution of the local partial differential Yang-Mills equations. A direct test of such equations was performed in \cite{directtest} for the case of 't Hooft-Polyakov monopoles, and they are indeed satisfied. For the case of the abelian gauge group  $U(1)$, such an expansion in powers of the parameters $\alpha$ and $\beta$ becomes trivial, and one gets the usual four integral equations of electrodynamics. 

The dynamics of the Yang-Mills theory is such that it implies that \rf{ymintegralintro} holds true on any volume $\Omega$. Therefore, if one changes the volume $\Omega \rightarrow \Omega^{\prime}$, such that the border  is not changed, i.e., $\partial\Omega=\partial\Omega^{\prime}$, one gets that 
\be
P_2 \, e^{\int_{\partial\Omega} {\cal B} }=P_3\,e^{\int_{\Omega^{\prime}} {\cal A}}\;\qquad \mbox{\rm and so}\qquad 
P_3\,e^{\int_{\Omega} {\cal A}}=P_3\,e^{\int_{\Omega^{\prime}} {\cal A}}
\lab{ymintegralintro2}
\ee
Consequently, the holonomy of the connection ${\cal A}$ on loop space is independent of the path, as long as its end points ($x_R$ and $\partial\Omega$) are kept fixed. The flatness condition \rf{flatconnloop} was proposed in \cite{afg1} in order to get that the holonomy of ${\cal A}$ was path independent. With the integral equation \rf{ymintegralintro} we get that condition directly. However, the path independency certainly leads to \rf{flatconnloop}.
Indeed, the non-abelian Stokes theorem, that leads to \rf{ymintegralintro}, implies that ${\cal A}$ is the curvature of ${\cal B}$, i.e. ${\cal A}\sim \delta {\cal B}+{\cal B}\wedge{\cal B}$, and so, it should hold on loop space what is known on ordinary manifolds, i.e. the curvature of a curvature must vanish. 

It is then very clear that the integral equations for the Yang-Mills theories, constructed in \cite{ym1,ym2}, overcome the obstacles facing the proposal put forward in \cite{afg1,afg2}, to construct integrable theories in any dimension. The partial differential Yang-Mills equations, which are local in the space-time $M$, are equivalent to the zero curvature of the connection ${\cal A}$, which is local on the loop space ${\cal L}^{(2)}$. In addition, since the connection ${\cal A}$ is path independent, it is invariant under the change of parameterization of the volumes $\Omega$. Indeed, a change of path on the loop space ${\cal L}^{(2)}$ may correspond to a change of the physical volume $\Omega$ on space-time $M$, or then to a change of parameterization of that same volume $\Omega$. Remember that a given volume $\Omega$ corresponds to an infinity of paths on ${\cal L}^{(2)}$. So, the non-locality and parameterization obstacles have been solved. 

The path independency of the holonomy of ${\cal A}$ can be used to construct conservation laws. Indeed, by imposing appropriate boundary conditions, it was shown in \cite{ym1,ym2} that such holonomy has an  iso-spectral time evolution, i.e.
\be
P_3\,e^{\int_{\IR^3_t} {\cal A}}=U\(t\)\,P_3\,e^{\int_{\IR^3_0} {\cal A}}\, U^{-1}\(t\)
\ee
where $\IR^3_t$ and $\IR^3_0$ stand for the three dimensional spatial submanifold of $M$, at time $t=t$ and $t=0$, respectively. It then follows that the eigenvalues of the operator 
\be
Q\equiv P_3\,e^{\int_{\IR^3_t} {\cal A}}=P_2 \, e^{\int_{S^2_{\infty}} {\cal B} }
\lab{chargeop}
\ee
are conserved in time, and where, in the last equality, we have used the integral equation \rf{ymintegralintro}, and where $S^2_{\infty}$ is the two-sphere at spatial infinity, i.e., the border of $\IR^3$. 

We show in Section \ref{sec:ymintegral}  that, under a general gauge transformation, the charge operator \rf{chargeop} transforms as $Q\rightarrow g_R\,Q\,g_R^{-1}$, where $g_R$ is the element of the gauge group $G$, performing the gauge transformation, evaluated at the reference point $x_R$. So, the charge operator transforms globally under local gauge transformations. Therefore, the conserved charges, which are its eigenvalues, are truly gauge invariant. That has solved a long-standing problem in the formulation of non-abelian gauge theories. The usual electric and magnetic non-abelian charges of Yang-Mills theories, discussed in the textbooks, are not really gauge invariant. 

It turns out that the eigenvalues of $Q$, namely $q_a$, $a=1,2,\ldots {\rm rank}\,G$,   are functions of the arbitrary parameters $\alpha$ and $\beta$, and therefore we have in principle an infinite number of conserved charges
\be
q_a\(\alpha\,,\,\beta\)=\sum_{m,n=0}^{\infty} \alpha^m\,\beta^n\, q_a^{(m\,,\,n)}
\ee
As we will see in this paper, the role played by the parameters $\alpha$ and $\beta$ is similar to that played by the so-called spectral parameter in integrable theories in $(1+1)$-dimensions.   

It is worth pointing out that even though the local and integral Yang-Mills equations are equivalent, the integral equations play a crucial role in understanding the global aspects of solutions. Indeed, the long-standing problem of the nature of the Wu-Yang monopole solution was solved with the help of the integral equations. The local partial differential Yang-Mills equations are not enough to settle the issue of the necessity of a source for the solution to exist. The integral Yang-Mills equations, on the other hand, lead, with an analysis based on distribution theory, to a unique form of the magnetic source needed for the Wu-Yang monopole solution to be consistent \cite{wuyangsing,mavromatos}. 

Summarizing, the Yang-Mills theory is a local field theory on the four-dimensional Minkowski space-time $M$, whose equations of motion (the local partial differential Yang-Mills equations), imply local equations on the generalised loop space ${\cal L}^{(2)}$, namely \rf{ymintegralintro} and \rf{flatconnloop}. So, Yang-Mills theories solve the problems of reparameterization and locality discussed above. In other words, the dynamics of Yang-Mills theories are equivalent to the zero curvature condition for a connection ${\cal A}$ on the loop space  ${\cal L}^{(2)}$. Therefore, the results of \cite{ym1,ym2} show that the Yang-Mills theories realize, in a quite elegant and simple way, the ideas put forward in \cite{afg1,afg2}. \\

 \centerline{\bf - The hidden symmetries of Yang- Mills theories  -}

In the present paper, we construct two types of symmetries of Yang-Mills theories based on the construction on loop space described above. The first one is the transformations of the matter and gauge fields, Wilson lines and fluxes, generated under the Poisson brackets, by the conserved charges, i.e., the eigenvalues of the operator $Q$ defined in \rf{chargeop}, or equivalently, the traces of powers of it. The second type are the symmetries of the integral equations    \rf{ymintegralintro}.

We calculate the infinitesimal transformations  generated  by the conserved charges through the Poisson brackets of the standard canonical quantization of Yang-Mills theories as described, for instance, in \cite{slavnovbook}. We work with the conserved charges expressed in terms of the traces of powers of the charge operator \rf{chargeop}, i.e.
\be
Q_N\equiv \frac{1}{N}\,{\rm Tr}\, Q^N
\lab{conservchargetrace}
\ee
The infinitesimal transformation of a given quantity $X$ is given by the equal time Poisson bracket 
\be
\delta X\equiv \ve\, \{X\,,\, Q_N\}_{PB}
\lab{symmtransf}
\ee
with $\ve$ being a constant infinitesimal parameter. 

Such global transformations are symmetries of the total  Hamiltonian $H_T=H_E+H_B+H_M+H_C$, of Yang-Mills theory, containing the usual gauge, matter, and constraints parts.  In fact, the conserved charges  weakly commute separately, i.e., when the Gauss law constraints hold true,  with each term of the total Hamiltonian, i.e. 
\be
\{H_{E/B/M/C}\,,\, Q_N\}_{PB}\cong 0
\lab{hqcommute}
\ee
with
\be
H_E= \frac{1}{2}\,\int d^3x\, {\rm Tr}\(E_i\)^2;\quad H_B= \frac{1}{2}\,\int d^3x\, {\rm Tr}\(B_i\)^2;\quad H_C= \int d^3x\, {\rm Tr}\(A_0\(e\,J_0-D_iE_i\)\)
\ee
and $H_M$ being the matter part of the Hamiltonian, with $E_i$ and $B_i$, $i=1,2,3$, being the non-abelian electric and magnetic fields, $J_0$ being the time component of the matter current, and $e$ is the gauge coupling constant. $A_0$ is the time component of the gauge field, and in the canonical formalism we use, it is a Lagrange multiplier \cite{slavnovbook}.  The symbol $\cong$ means equality when the constraints hold true. See Section \ref{sec:hamiltoniantransform} for the details. 

The striking property of the flat loop space connection ${\cal A}$ (see \rf{ymintegralintro} and \rf{ymintegralintro2}) is that it satisfies  an anomalous Fundamental Poisson bracket Relation (FPR) of the form
\br
\{{\cal A}_{\(\alpha_1\,,\,\beta_1\,,\,\zeta_1\)}\overset{\otimes}{,} {\cal A}_{\(\alpha_2\,,\,\beta_2\,,\,\zeta_2\)}\}_{PB}
&=&\delta\(\zeta_1-\zeta_2\)\left\{\sbr{{\cal R}}{{\cal A}_{\(\alpha_1\,,\,\beta_1\,,\,\zeta_1\)}\otimes \one+\one\otimes {\cal A}_{\(\alpha_2\,,\,\beta_2\,,\,\zeta_2\)}}\right.
\nonumber\\
&+&\left. \(\alpha_1-\alpha_2\)\left[ \Xi_{\rm constr.}+ \Xi_{\rm anom.}\right] \right\}
\lab{fpr}
\er
with
\be
{\cal R}=-e^2\,\vartheta\,\frac{\beta_1\,\beta_2}{\beta_1-\beta_2}\,T_a\otimes T_a
\lab{fprop}
\ee
where  $\alpha_i$ and $\beta_i$, $i=1,2$, are the parameters appearing in the integral Yang-Mills equations \rf{ymintegralintro}, $\zeta_i$ are the parameters labelling the closed surfaces, based at the reference point $x_R$, on the scanning of the volume $\Omega$, $\vartheta$ is a sign depending upon the orientation of the scanning, and $T_a$ is a basis of the Lie algebra of the compact gauge group $G$, satisfying
\be
\sbr{T_a}{T_b}=i\,f_{abc}\, T_c\;;\qquad\qquad {\rm Tr}\(T_a\,T_b\)=\delta_{ab}\;;\qquad\qquad
a,b,c=1,2,\ldots {\rm dim}\, G
\lab{liebasis}
\ee
The quantity $\Xi_{\rm constr.}$ vanishes when the constraints hold true, and $\Xi_{\rm anom.}$ is an anomalous term that will not influence the involution of the charges when the invariance under the choice of scanning (reparameterization) is guaranteed. See Section \ref{sec:fpr} for the details.

The FPR \rf{fpr} leads to a Sklyanin relation for the charge operators of the form
\br
\{Q\(\alpha_1\,,\,\beta_1\)\overset{\otimes}{,} Q\(\alpha_2\,,\,\beta_2\)\}_{PB}
&=&\sbr{{\cal R}}{Q\(\alpha_1\,,\,\beta_1\)\otimes Q\(\alpha_2\,,\,\beta_2\)}
\nonumber\\
&+&
\(\alpha_1-\alpha_2\)\left[ {\widetilde \Xi}_{\rm constr.}+ {\widetilde \Xi}_{\rm anom.}\right]
\lab{skrel}
\er
where again ${\widetilde \Xi}_{\rm constr.}$ vanishes when the constraints hold true, and ${\widetilde \Xi}_{\rm anom.}$ does not influence the involution of the charges when invariance under reparemeterization is guaranteed. See Section \ref{sec:sklyanin} for the details. 

Note that the  operator ${\cal R}$ \rf{fprop}, vanishes for either for $\beta_1=0$ or $\beta_2=0$, and the same happens for the quantities ${\widetilde \Xi}_{\rm constr.}$ and ${\widetilde \Xi}_{\rm anom.}$. Therefore, the charge operators $Q\(\alpha\,,\,\beta=0\)$, commute with all charge operators, and so they are central elements. The same is true for the loop space connections ${\cal A}_{\(\alpha\,,\,\beta=0\,,\,\zeta\)}$. 

In its turn, \rf{skrel} leads to the involution of all the conserved charges \rf{conservchargetrace}, for any power $N_i$, and any values of the spectral parameters $\alpha_i$ and $\beta_i$, $i=1,2$, and when the constraints hold true, i.e. 
\be
\{Q_{N_1}\(\alpha_1\,,\,\beta_1\)\,,\, Q_{N_2}\(\alpha_2\,,\,\beta_2\)\}_{PB}\cong 0
\lab{involution}
\ee
Therefore, we do have structures, resembling integrability, in Yang-Mills theories in the sector of the non-abelian electric and magnetic charges. Even though the Yang-Mills Hamiltonian commutes with such an infinity of charges (see \rf{hqcommute}), it is not one of them, i.e., the Hamiltonian is not a functional of the charges $Q_N$. 

Even though the conserved charges weakly commute, the transformations generated by them do not. It has to do with the fact that one can not impose the constraints and reparameterization conditions inside the Poisson brackets. Using the Jacobi identity for the Poisson bracket, one gets that the commutator of two transformations \rf{symmtransf} is 
\be
\sbr{\delta_{N_1,\alpha_1,\beta_1}}{\delta_{N_2,\alpha_2,\beta_2}}\, X=\ve_1\,\ve_2\,\(\alpha_1-\alpha_2\)\, \pbr{X}{{\rm Tr}_{RL}\(Q^{N_1}\(\alpha_1,\beta_1\)\otimes Q^{N_2}\(\alpha_2,\beta_2\)\, \Upsilon\)}
\ee
with $\ve_1$ and $\ve_2$ being the constant parameters of the two global transformations, ${\rm Tr}_{RL}$ meaning the trace on both sides of the tensor product, and $\Upsilon$ is an element of the Lie algebra of $G$, integrated over the whole $\IR^3$, depending on both sets of parameters $\alpha_i$ and $\beta_i$, $i=1,2$, and on the Wilson line, surface and volume holonomies (see \rf{upsilondef}). The quantities  ${\rm Tr}_{RL}\(Q^{N_1}\(\alpha_1,\beta_1\)\otimes Q^{N_2}\(\alpha_2,\beta_2\)\, \Upsilon\)$, Poisson commutes with the Hamiltonian, and so they are  additional conserved charges, that generates new global symmetries. However,  $\Upsilon$ vanishes when the constraints hold true, and so such charges vanish on the constrained phase space. Due to the Jacobi identity for the Poisson bracket, the transformations $\delta_{N_i,\alpha_i,\beta_i}$ satisfy the Jacobi identity, and so generate a Lie algebra. The subspaces corresponding to the same value of  $\alpha$, are abelian subalgebras, and elements of sectors for different values of $\alpha$ do not commute.   See Section \ref{sec:algebratranf} for the details. \\

\centerline{\bf - The global transformations of the local fields, Wilson lines and fluxes -}

Some objects present  simpler transformations under the  symmetry \rf{symmtransf}. We shall refer to them as {\em primary fields}, and they are the matter fields, the non-abelian electric and magnetic fluxes, and some special Wilson line operators. 

Consider a matter field multiplet transforming under a given representation $R$ of the gauge group $G$
\be
\psi_{r}\rightarrow R\(g\)_{r\,s}\,\psi_{s}\;;\qquad\qquad g\in G\;;\qquad\qquad r\,,\, s=1,2,\ldots {\rm dim}\,R
\ee
The fields $\psi$ stand for spinors (quarks) or scalars (Higgs), and the form of the transformation generated by the conserved charges is the same  for both of them, and is given by
\br
\delta\,\psi_{r}\(x\)&=&\ve\,\{\psi_{r}\(x\)\,,\, Q_N\(\alpha\,,\,\beta\)\}_{PB}
\lab{fieldtransf}\\
&=& -\ve\beta e^2\vartheta\sum_{a=1}^{{\rm dim}\,G} \left[R\(T_a\)\,\psi\(x\)\right]_{r}\,{\rm Tr}\(Q^N\(\zeta_f\)S\(x\)T_aS^{-1}\(x\)\)
\nonumber
\er
where $Q\(\zeta_f\)$ is the charge operator integrated over the whole $\IR^3_t$, $S\(x\)$ is an operator built by volume ordered integration of the same quantities leading to the charge operator $Q$, and it lies in a matrix representation of the gauge group $G$, not necessarily the same as $R$, and it is unity at the reference point $x_R$
\be
S\(x_R\)=\one
\ee
See Section \ref{sec:transflocalfields}  for details of the construction of \rf{fieldtransf}.

Note that the matrix $S$ and the charge operator $Q$, appearing in \rf{fieldtransf}, depend upon all fields (gauge and matter), but the factor ${\rm Tr}\(Q^N\, S\(x\)\, T_a\,S^{-1}\(x\)\)$ is the same for all components $\psi_r$ of the matter fields. Therefore,  it turns out  that any gauge invariant monomial of the matter fields  is also invariant under the transformations \rf{fieldtransf}. In particular, it preserves the squared modulus of the multiplets (spinors or scalars), i.e.
\be
\delta \(\psi_r^{\dagger}\,\psi_r\)=\ve\,\{\(\psi_r^{\dagger}\,\psi_r\)\,,\, Q_N\}_{PB}=0
\ee
and so the integrated (exponentiated) version of \rf{fieldtransf} should be a unitary transformation. 

In order to describe the transformations of the gauge fields and Wilson lines, we need to decompose them into special directions dictated by the scanning of volumes and surfaces. As we have said, we scan a volume $\Omega$ with closed surfaces, based at the reference point $x_R$ on its border $\partial\Omega$. The surfaces scanning $\Omega$ are labelled by a parameter $\zeta$. In their turn, each of those surfaces is scanned by loops, labelled by a parameter $\tau$, starting and ending at $x_R$. Each loop is parameterized by $\sigma$. Each point of $\Omega$ belongs to only one surface, to only one loop, and to only one point of that loop. Therefore, if we take $\Omega$ to the spatial sub-manifold $\IR^3_t$, of the Minkowski space-time, at a given fixed time $t$, we have a one-to-one correspondence  among the points of $\IR^3_t$ and those parameters, i.e. $x^i=x^i\(\sigma\,,\,\tau\,,\,\zeta\)$, $i=1,2,3$, and $x^i$ being the Cartesian coordinates of $\IR^3_t$. 

In the canonical formalism we use \cite{slavnovbook}, the time component $A_0$, of the gauge field is a Lagrange multiplier, and only the space components, $A_i$, $i=1,2,3$, correspond to physical degrees of freedom. We express the one-form gauge connection as $A=A_i\,dx^i= A_{\sigma}\, d\sigma+ A_{\tau}\,d\tau+A_{\zeta}\, d\zeta$. It turns out that the $\sigma$-component of the gauge field is invariant under the global transformations \rf{symmtransf}, i.e,  
\be
\delta A_{\sigma} \equiv \ve\, \{A_{\sigma}\,,\, Q_N\}_{PB}=0
\lab{symmtransfasigma}
\ee
The other two components $A_{\tau}$ and $A_{\zeta}$ have very precise transformation laws under \rf{symmtransf}, but they do not transform as {\em primary fields}. $\delta A_{\tau/\zeta}$ is not linear in $A_{\tau/\zeta}$, and it involves surface and volume ordered integrals of quantities appearing in the charge operator. See Section \ref{sec:gaugesector} for the details. 

The transformations of the gauge degrees of freedom are better described by the Wilson lines defined on paths in the $\sigma$, $\tau$, and $\zeta$-directions. Consider the Wilson lines $\omega_{\xi}$, defined by
\be
\frac{d\,\omega_{\xi}}{d\,\xi}+A_{i}\,\frac{d\,x^i\(\xi\)}{d\,\xi}\,\omega_{\xi}=0;\qquad\qquad \qquad \xi\equiv\sigma\,,\,\tau\,,\,\zeta
\lab{wilsonlinedefintro}
\ee
where $x^i\(\xi\)$ is a path, parameterized by $\xi$, where the other two parameters are kept fixed, i.e., $x\(\sigma\)$ is a path along the loop scanning the surfaces, with $\tau$ (loop) and $\zeta$ (surface), kept fixed. Similarly, $x^i\(\tau\)$ is a path, on a fixed surface ($\zeta$ constant), crossing the loops at fixed value of $\sigma$, and $x^i\(\zeta\)$ is a path crossing the surfaces (labelled by $\zeta$) at constant values of $\tau$ and $\zeta$. 

Due to \rf{symmtransfasigma} it turns out that the Wilson line $\omega_{\sigma}$ is invariant under the global transformations \rf{symmtransf}, i.e,  
\be
\delta \omega_{\sigma} \equiv \ve\, \{\omega_{\sigma}\,,\, Q_N\}_{PB}=0
\lab{symmtransfomegasigma}
\ee
The Wilson lines $\omega_{\tau}$ and $\omega_{\zeta}$ transform as 
\br
\delta \omega_{\tau/\zeta} &=&\ve\, \{\omega_{\tau/\zeta}\,,\, Q_N\}_{PB}
\lab{varomegataufinal2intro}\\
&=&\ve\,e^2\,\beta\,\vartheta\,\left[ \omega_{\tau/\zeta}\, T_a\,{\rm Tr}\left[Q^N\(\zeta_f\)
 S_R^{-1}T_a\,S_R\right]
- T_a\,\omega_{\tau/\zeta} {\rm Tr}\left[Q^N\(\zeta_f\)
 S^{-1}_L T_a S_L\right]\right]
 \nonumber
\er
where $Q\(\zeta_f\)$ is the charge operator integrated over the whole $\IR^3_t$, and $S_{R/L}$ are operators built by volume ordered integration of the same quantities leading to the charge operator $Q$, and they are different for the cases of $\omega_{\tau}$ and $\omega_{\zeta}$. For the case of $\omega_{\zeta}$, the second equality in \rf{varomegataufinal2intro} holds true when the constraints are imposed. 

Other quantities that transform in a simple way under the global transformations \rf{symmtransf} are the electric and magnetic fluxes defined as
\br
\mathfrak{e}_{\tau/\zeta}= \int_{\sigma_i}^{\sigma_f} d\sigma\,\omega_{\sigma}^{-1}\,E_i\,\omega_{\sigma}\,\ve_{ijk}\,
\frac{d\,x^j}{d\,\sigma}\,\frac{d\,x^k}{d\,\tau/\zeta}
\qquad\qquad 
\mathfrak{b}_{\tau/\zeta}=\int_{\sigma_i}^{\sigma_f} d\sigma\,\omega_{\sigma}^{-1}\,B_i\,\omega_{\sigma}\,\ve_{ijk}\,
\frac{d\,x^j}{d\,\sigma}\,\frac{d\,x^k}{d\,\tau/\zeta}
\lab{ebfrakdefintro}
\er
where $E_i$ and $B_i$ are the non-abelian electric and magnetic fields, and $\omega_{\sigma}$ is the Wilson line defined in \rf{wilsonlinedefintro}. The integral is over an entire loop parameterized by $\sigma$, and the Wilson line is defined on that loop, integrated from the reference point $x_R$ up to the point where the fields $E_i$ and $B_i$ are located. 

Under \rf{symmtransf} we have that 
\br
\delta\,\mathfrak{b}_{\tau/\zeta}&=&\ve\,\pbr{\mathfrak{b}_{\tau/\zeta}}{Q_N}=0
\lab{fluxtransfintro}\\
\delta\,\mathfrak{e}_{\tau/\zeta}&=&\ve\,\pbr{\mathfrak{e}_{\tau/\zeta}}{Q_N}=
-\ve\,e^2\,\beta\,\vartheta\,\sbr{T_a}{\mathfrak{e}_{\tau/\zeta}}\;{\rm Tr}\(Q\(\zeta_f\)^N\,S^{-1}_{\mathfrak{e}}\, T_a\,S_{\mathfrak{e}}\)
\nonumber
\er
where again $S_{\mathfrak{e}}$ is an operator built by volume ordered integration of the same quantities leading to the charge operator $Q$. See Sections \ref{sec:transflocalfields},  \ref{sec:transffluxes}, and \ref{sec:specialwilson}  for the details on the global transformations.

The fluxes $\mathfrak{e}_{\tau/\zeta}$ and $\mathfrak{b}_{\tau/\zeta}$ play a very important role in the properties of the Yang-Mills theory, and they are the basic constituents of the electric and magnetic charge densities of the gauge fields (gluons) on the loop space formulation of the integral equations. As we discuss in Section \ref{sec:chargeoperator}, they are perhaps the closest thing to what Polyakov used to call the rings of glue \cite{polyakovring}.

Note that the conserved charges for the $\beta=0$, i.e. $Q_N\(\alpha\,,\,\beta=0\)$, do not generate any transformation, as can be seen by setting $\beta$ to zero in \rf{fieldtransf}, \rf{varomegataufinal2intro} and \rf{fluxtransfintro}. As we commented above (see paragraph below \rf{skrel}), the corresponding charge operator leading to such charges, Poisson commutes with all other charge operators. So, they constitute central elements of the algebra of the charge operators. 

It is worth pointing out that despite the fact that the conserved charges \rf{conservchargetrace} are truly gauge invariant, and independent of the parameterizations of volumes and surfaces, the global transformations \rf{symmtransf} have to be defined on a given scanning. Indeed, suppose we consider two different parameterizations of volumes and surfaces, leading to two global transformations, and consider their difference $\delta X- \delta^{\prime} X=\ve\,\pbr{X}{Q_N-Q_N^{\prime}}$. The invariance of the charge under reparameterization, i.e. $Q_N=Q_N^{\prime}$, follows from the static integral Yang-Mills equations, which are equivalent to the Gauss constraints and Bianchi identity, $D_iB_i=0$. Since we cannot impose the constraints inside the Poisson brackets, it follows that, in principle, $\delta X\neq \delta^{\prime} X$. All that means that our global transformations \rf{symmtransf} are, in fact, defined on loop space. The transformation depends upon the path on ${\cal L}^{(2)}$ where the charge $Q_N$ is evaluated. But that is similar to Weyl's gauge principle \cite{weyl1929,dirac1931}: the gauge interaction introduces non-integrable phases on the wave functions, but the physics does not depends upon the path chosen to calculate that phase.  It is amazing that our loop space formulation of the global transformations lead to such a property that the physics does not depend upon the choice of path. That is what allows us  to define the transformations of local fields, like the matter and gauge fields. 

Note that our conserved charges are integrated over the whole spatial sub-manifold $\IR^3$ of the Minkowski space-time $M$. Therefore, in the language of the so-called generalized global symmetries \cite{ggs1,ggs2,ggs3,ggs4}, our global transformations \rf{symmtransf} should be called $0$-form symmetries and should act on point objects, like the local fields. However, our conserved charges also generate transformation of extended objects like the Wilson lines and electric and magnetic fluxes. That is a point that deserves further analysis to connect our work with those results.

Another interesting point is that our results depend crucially on the contributions of boun\-da\-ry terms. Indeed, as we show  in Section \ref{sec:symymcharges}, the global transformation of any quantity $X$, which is a function of the canonical variables, contains a volume and a boundary term (see \rf{chargepoisson5}). The symmetries that we have constructed are certainly not what is called in the li\-te\-ra\-tu\-re asymptotic symmetries \cite{strominger,barnich,henneaux}, but it would be interesting to investigate if there is any connection with those symmetries. 

In addition, the structures underlying our holonomies  may relate to what is called in the li\-te\-ra\-tu\-re higher gauge theories, gerbes, Lie two-algebras, and other categorical frameworks \cite{breen,baez1,baez2}. Our direct approach, however, dispenses with such mathematical structures, but it would be interesting to investigate further any possible connections.   \\

\centerline{\bf - The symmetries of the integral Yang-Mills equations -}

The second type of transformations  that we construct are symmetries of the Yang-Mills integral equations \rf{ymintegralintro}. We construct an infinite group of transformations on each point (function) of the loop space ${\cal L}^{(2)}$ (see \rf{loopspace}). As we have said, the construction of the integral equations \rf{ymintegralintro} requires a scanning of the volume $\Omega$ with closed surfaces based at the reference point $x_R$. So, it requires a choice of path on the loop space ${\cal L}^{(2)}$. The right-hand side of \rf{ymintegralintro} is integrated over that chosen path, and its left-hand side is defined on a point in ${\cal L}^{(2)}$, which is the end point of that path. On its turn, such a point on ${\cal L}^{(2)}$ can be seen as a closed path in ${\cal L}^{(1)}$, and so the operator on the left hand side of \rf{ymintegralintro} can be considered the holonomy of the connection ${\cal B}$ on ${\cal L}^{(1)}$.

We shall fix a closed path on ${\cal L}^{(1)}$, and so a point in ${\cal L}^{(2)}$, which we shall denote $\partial\Omega$. In addition, we parameterize the closed path in ${\cal L}^{(1)}$ with a parameter $\tau$, such that $\tau_i$ and $\tau_f$ correspond respectively to its initial and final points. We then consider all possible connections $\mathfrak{a}$ on ${\cal L}^{(1)}$, and calculate their holonomies ${\hat g}$, on the fixed path $\partial\Omega$, through the equation
\be
\frac{d\,{\hat g}\(\tau\)}{d\,\tau}-{\hat g}\(\tau\)\,\mathfrak{a}\(\tau\)=0\; \qquad\qquad \mbox{\rm such that}\qquad\qquad {\hat g}\(x_R\)=\one
\lab{holgroupeq}
\ee
Given two connections $\mathfrak{a}_1$ and $\mathfrak{a}_2$ we defined the composition law of their holonomies, ${\hat g}_1$ and ${\hat g}_2$ respectively,  as
\be
{\hat g}_3\(\tau\)= {\hat g}_1\(\tau\)\,{\hat g}_2\(\tau\)
\lab{productlawpre}
\ee
where the (matrix) product is performed on each point $\tau$ of the path $\partial\Omega$. It turns out that  ${\hat g}_3\(\tau\)$ also to satisfy \rf{holgroupeq}, and the connections must compose as
\be
\mathfrak{a}_3\(\tau\)={\hat g}_2^{-1}\(\tau\)\,\mathfrak{a}_1\(\tau\)\,{\hat g}_2\(\tau\)+ \mathfrak{a}_2\(\tau\)
\lab{compositionconn}
\ee
The set of elements of our group ${\hat G}$, however, is defined as the holonomies integrated over the whole fixed path $\partial\Omega$, and the group product law is defined as
\be
{\hat g}_3\(\tau_f\)= {\hat g}_1\(\tau_f\)\,{\hat g}_2\(\tau_f\)\;\qquad \mbox{\rm or equivalently}\qquad 
{\hat g}_3\(\partial\Omega\)= {\hat g}_1\(\partial\Omega\)\,{\hat g}_2\(\partial\Omega\)
\lab{productlawfinal}
\ee
One can easily verify that such an infinite set ${\widehat G}$ of elements ${\hat g}\(\partial\Omega\)$, with the product law \rf{productlawfinal}, satisfies all the postulates of a group. The product is associative, the product of any two elements is again an element of the set (closure), there is a unique (left and right) identity, and each element has a unique (left and right) inverse. Therefore, we can define on any point $\partial\Omega$ of ${\cal L}^{(2)}$ such an infinite dimensional group ${\hat G}$. Note, however, that  even though the  elements are holonomies, such a group is not the usual holonomy group known in the literature. The usual holonomy group has a fixed connection,
and its associated holonomy is evaluated on every closed path based on a fixed point.  The composition law gives an element, which is the holonomy of that fixed connection, calculated on the  composition of two closed loops.  Our group ${\hat G}$ is defined on a fixed  path and an infinity of connections. The composition law for the connections is given by \rf{compositionconn} on that fixed path. 

Using the non-abelian Stokes theorem, one can express each group element of ${\hat G}$ as an ordered integral on the volume $\Omega$ whose border is $\partial\Omega$, i.e., $\Omega$ is a path on ${\cal L}^{(2)}$ 
\be
{\hat g}\(\partial\Omega\)= {\hat g}\(\Omega\)
\lab{stokesgroup}
\ee
Therefore, one can view each side of the integral Yang-Mills equations \rf{ymintegralintro} as taking values on such an infinite-dimensional group ${\hat G}$. The symmetries of the integral equations \rf{ymintegralintro} are given by the left and right transformations
\be
P_2 \, e^{\int_{\partial\Omega} {\cal B} } \rightarrow {\hat g}_L\(\partial\Omega\) \,P_2 \, e^{\int_{\partial\Omega} {\cal B} } \qquad\quad  {\rm and}\qquad\quad  P_3\,e^{\int_{\Omega} {\cal A}}\rightarrow {\hat g}_L\(\Omega\)\,P_3\,e^{\int_{\Omega} {\cal A}}
\lab{lefttransfintro}
\ee
and
\be
P_2 \, e^{\int_{\partial\Omega} {\cal B} } \rightarrow P_2 \, e^{\int_{\partial\Omega} {\cal B} } \,{\hat g}_R\(\partial\Omega\) \qquad\quad  {\rm and}\qquad\quad  P_3\,e^{\int_{\Omega} {\cal A}}\rightarrow P_3\,e^{\int_{\Omega} {\cal A}}\,{\hat g}_R\(\Omega\)
\lab{righttransfintro}
\ee
It is clear from \rf{stokesgroup} that such transformations are symmetries of the integral Yang-Mills equations \rf{ymintegralintro}. In fact, the connection ${\cal B}$ will compose with the connections producing the holonomies  ${\hat g}_L\(\partial\Omega\)$ and ${\hat g}_R\(\partial\Omega\)$, in the same way as in \rf{compositionconn}. 

As we have seen, the connection ${\cal A}$ is flat due to the integral equations, i.e. 
\be
\delta {\cal A}+{\cal A}\wedge {\cal A}=0
\ee
and so it must be of the pure gauge form
\be
{\cal A}= \delta {\hat g}\(\partial\Omega\)\,{\hat g}^{-1}\(\partial\Omega\)
\ee
Under the transformations \rf{lefttransfintro} and \rf{righttransfintro} it transform, respectively, as
\be
{\cal A}\rightarrow {\hat g}_L\(\partial\Omega\)\,{\cal A}\,{\hat g}_L^{-1}\(\partial\Omega\)+ \delta {\hat g}_L\(\partial\Omega\){\hat g}_L^{-1}\(\partial\Omega\)
\ee
and
\be
{\cal A}\rightarrow {\cal A}+{\hat g}\(\partial\Omega\)\,\delta {\hat g}_R\(\partial\Omega\)\,{\hat g}_R^{-1}\(\partial\Omega\)\,{\hat g}^{-1}\(\partial\Omega\)
\ee
Consequently, one can see the infinite-dimensional group ${\hat G}$ as a group of  gauge transformations for the flat connections ${\cal A}$ on the loop space ${\cal L}^{(2)}$. 

 Note that the transformations \rf{lefttransfintro} and \rf{righttransfintro} are not symmetries of the Yang-Mills Hamiltonian and Lagrangian. Indeed, if one takes a vacuum  configuration where both ${\cal B}$ and ${\cal A}$ vanish, one gets that $P_2\,e^{\int_{\partial\Omega} {\cal B} }=P_3\,e^{\int_{\Omega} {\cal A}}=\one$. The transformations \rf{lefttransfintro} and \rf{righttransfintro} will map such vacuum configuration in non-trivial ${\cal B}^{\prime}$ and ${\cal A}^{\prime}$, such that
 \be
 P_2\,e^{\int_{\partial\Omega} {\cal B}^{\prime} }={\hat g}_{L/R}\(\partial\Omega\)\;\qquad \quad {\rm and}\qquad\quad
 P_3\,e^{\int_{\Omega} {\cal A}^{\prime}}={\hat g}_{L/R}\(\Omega\)
\ee 
So, one starts with a configuration where the energy vanishes and gets a non-trivial configuration with non-vanishing energy. The transformations \rf{lefttransfintro} and \rf{righttransfintro} must therefore act on the full space of solutions. However, we are very far from having a concrete method for constructing solutions, like we have the inverse scattering  and dressing methods in $(1+1)$-dimensional integrable field theory. The mathematical structure that makes those methods possible is the so-called  Riemann-Hilbert problem. The question of whether there exists an equivalent method for four-dimensional gauge theories is far beyond the scope of the present paper, and that is certainly an issue to be investigated further.  

One would note that such a symmetry resembles the gauge transformations of the Lax-Zakharov-Shabat equation \rf{2dzc}, discussed above. However, the infinite-dimensional group ${\hat G}$ does not seem to be a Lie group. The structure of such a group has still to be clarified, but the (infinite) dimension of the space of elements of  ${\hat G}$ at the identity element seems to be different from the dimension at other elements. So, such a space does not seem to be a manifold. Therefore, we might not have an infinite-dimensional Lie algebra associated to such a symmetry, like we have the Kac-Moody algebra in two-dimensional integrable theories. But that is an interesting point, since the fact that our symmetry is not associated to a Lie group, it might circumvent the Coleman-Mandula theorem for four-dimensional Yang-Mills theories in a novel way. See Section \ref{sec:symymintegral} for the details. 

The paper is organized as follows. In Section \ref{sec:ymintegral} we revise the non-abelian Stokes theorem for two-form connections, construct the integral equations for Yang-Mills theories, and from them the infinity of conserved charges. In Section \ref{sec:symcharges}, we present the symplectic structure we shall use to calculate the transformations generated by the charges. The properties of the charge operator, including its iso-spectral time evolution, are discussed in Section \ref{sec:chargeoperator}.  The transformations generated by the conserved charges are constructed in Section \ref{sec:symymcharges}.  The invariance of the Hamiltonian of the Yang-Mills theories under the global transformations is proved in Section \ref{sec:hamiltoniantransform}. In Section \ref{sec:transflocalfields} we show how the local matter and gauge fields transform under those global transformations, and in section \ref{sec:transffluxes}  how the electric and magnetic fluxes transform. The transformations of the Wilson lines are presented in Section  \ref{sec:specialwilson}. The integrability structures of Yang-Mills theories are presented in Sections \ref{sec:fpr} and \ref{sec:sklyanin}, where we construct respectively the Fundamental Poisson Bracket Relation for the flat connection, and the Sklyanin relation for the charge operator, and following from it the involution of the conserved charges. The infinite-dimensional group ${\hat G}$, which is a symmetry of the integral Yang-Mills equations, is constructed in Section \ref{sec:symymintegral}. Our discussions and conclusions are presented in Section \ref{sec:conclusion}. Finally, we have 10 appendices where we give the details of many calculations needed in the main parts of the paper.

\section{The Yang-Mills integral equations}
\label{sec:ymintegral}
\setcounter{equation}{0}

The symmetries of the Yang-Mills theories that we discuss in this paper rely on the integral Yang-Mills equations constructed in \cite{ym1,ym2}. In this section, we summarize the construction of those integral equations, which uses the  Stokes theorem for a two-form connection.  The proof of that theorem is quite straightforward \cite{afg1,afg2}.    

\subsection{The non-abelian Stokes theorem for two-forms}
\label{subsec:nonabelianstokes}

First of all, we have to define how integrals over volumes, surfaces, and loops are ordered\footnote{The version of the non-abelian Stokes theorem for a two-form connection that we derive here applies to volumes and surfaces with trivial topology. If one wants to consider spaces with holes and handles, one has to partition the volumes and surfaces into patches with trivial topology, and the proof becomes a lot more involved. For the case of the ordinary non-abelian Stokes theorem for one-form connections, the proof for spaces with non-trivial topology has been given in \cite{hirayama}.}.  Given a three-dimensional volume $\Omega$ in the Minkowski space-time $M$, we choose a reference point $x_R$ on its border $\partial \Omega$, and scan $\Omega$ with closed two dimensional surfaces based on $x_R$, and labelled by a parameter $\zeta$, varying from $\zeta_i$ to $\zeta_f$, such that $\zeta_i$ corresponds to the infinitesimal closed surface around $x_R$, and $\zeta_f$ corresponds to the border $\partial\Omega$. On their turn, the closed two-dimensional surfaces scanning $\Omega$ are scanned with loops, starting and ending at the reference point $x_R$, and labelled by a parameter $\tau$, varying from $\tau_i$ to $\tau_f$. We start with an infinitesimal loop around $x_R$, labelled by $\tau_i$, then vary the  loops to cover the whole surface, and end with another infinitesimal loop around $x_R$, labelled by $\tau_f$. Each closed loop is parameterised by a parameter $\sigma$, varying from $\sigma_i$ to $\sigma_f$, and both corresponding to the (initial and final) reference point $x_R$. Every point of $\Omega$ belongs to only one two-dimensional surface and to only one loop. So, loops and surfaces do not intersect each other. Therefore, the coordinates of the space-time points of the volume $\Omega$ are functions of those three parameters, i.e. $x^{\mu}\(\zeta\,,\,\tau\,,\,\sigma\)$. The choice of the scanning of $\Omega$ defines in fact a map from a $3$-sphere $S^3$, parameterised by  $\(\zeta\,,\,\tau\,,\,\sigma\)$, to the space-time $M$, such that the image is the volume $\Omega$. The reference point $x_R$ is the image of the north-pole of $S^3$, and so we have that $x^{\mu}\(\zeta\,,\,\tau\,,\,\sigma\)$ corresponds to a point in the loop space ${\cal L}^{(3)}$ (see \rf{loopspace}). Equivalently, we can consider those functions $x^{\mu}\(\zeta\,,\,\tau\,,\,\sigma\)$ as defining a path in ${\cal L}^{(2)}$, such that the parameter $\zeta$ parameterises such a path. In fact, the latter is more appropriate for our applications. 

We now consider an antisymmetric rank two tensor $B_{\mu\nu}$, and a one-form connection $C_{\mu}$,  on the Minkowski space-time $M$, and taking values on a Lie algebra, which in our applications will be the Lie algebra ${\cal G}$ of the compact gauge group $G$. We introduce a $1$-form connection on ${\cal L}^{(1)}$, through a $2$-form $B_{\mu\nu}$ and a $1$-form $C_{\mu}$ on $M$, as
\be
{\cal T}\equiv \int_{\sigma_i}^{\sigma_f}d\sigma\, W^{-1}\, B_{\mu\nu}\,W\,\frac{d\,x^{\mu}}{d\,\sigma}\, \frac{d\,x^{\nu}}{d\,\tau}
\lab{connectionl2}
\ee
where $W$ is the Wilson line operator associated to the $1$-form connection $C_{\mu}$, defined by the equation 
\be
\frac{d\,W}{d\,\sigma}+C_{\mu}\,\frac{d\,x^{\mu}}{d\,\sigma}\,W=0
\lab{wdefc}
\ee
The $1$-form connection ${\cal T}$ is defined by an integral on every loop scanning the two-dimensional surfaces, that in their turn scan $\Omega$. So, ${\cal T}$ is defined on points of ${\cal L}^{(1)}$. The Wilson line $W$ in \rf{connectionl2} is obtained by integrating \rf{wdefc} from the reference point $x_R$ ($\sigma=\sigma_i$), up to the point of the loop where $B_{\mu\nu}$ is evaluated. So, the general solution of \rf{wdefc}  is given by the series
\br
W\(\sigma\)&=&\left[\one-\int_{\sigma_i}^{\sigma}d\sigma^{\prime}\,C\(\sigma^{\prime}\)+
\int_{\sigma_i}^{\sigma}d\sigma^{\prime}\,\int_{\sigma_i}^{\sigma^{\prime}}d\sigma^{\prime\prime}\,C\(\sigma^{\prime}\)\,C\(\sigma^{\prime\prime}\)
\lab{wilsonsol}
\right.\\
&-&\left. \int_{\sigma_i}^{\sigma}d\sigma^{\prime}\,\int_{\sigma_i}^{\sigma^{\prime}}d\sigma^{\prime\prime}\,\int_{\sigma_i}^{\sigma^{\prime\prime}}d\sigma^{\prime\prime\prime}\,C\(\sigma^{\prime}\)\,C\(\sigma^{\prime\prime}\)\,C\(\sigma^{\prime\prime\prime}\)+\ldots\right]\,W_R \equiv P_1\,e^{-\int_{\sigma_i}^{\sigma}\, d\sigma^{\prime}\, C\(\sigma^{\prime}\)}\,W_R
\nonumber
\er
with $C\(\sigma\)\equiv C_{\mu}\,\frac{d\,x^{\mu}}{d\,\sigma}$, $W_R$ is a matrix integration constant, and $P_1$ denotes  path ordering. Note that $W_R$ is the value of $W$ at the reference point $x_R$, i.e. $W\(\sigma_i\)=W_R$. The calculations have to be performed in a given representation of the Lie algebra ${\cal G}$, as it involves products of the generators, and not only Lie algebra brackets of them. 

We define the holonomy of ${\cal T}$ on a given path in ${\cal L}^{(1)}$ through the equation
\be
\frac{d\,V}{d\,\tau} - V\, {\cal T}=0
\lab{holonomyl2}
\ee
Similarly, the general solution of \rf{holonomyl2} is given by the series
\br
V\(\tau\)&=&V_R\,\left[\one+\int_{\tau_i}^{\tau}d\tau^{\prime}\,{\cal T}\(\tau^{\prime}\)+
\int_{\tau_i}^{\tau}d\tau^{\prime}\,\int_{\tau_i}^{\tau^{\prime}}d\tau^{\prime\prime}\,{\cal T}\(\tau^{\prime\prime}\)\,{\cal T}\(\tau^{\prime}\)
\lab{wilsonsurfacesol}
\right.\\
&+& \left.\int_{\tau_i}^{\tau}d\tau^{\prime}\,\int_{\tau_i}^{\tau^{\prime}}d\tau^{\prime\prime}\,\int_{\tau_i}^{\tau^{\prime\prime}}d\tau^{\prime\prime\prime}\,{\cal T}\(\tau^{\prime\prime\prime}\){\cal T}\(\tau^{\prime\prime}\)\,{\cal T}\(\tau^{\prime}\)\,+\ldots\right] \equiv V_R\,P_2\,e^{\int_{\tau_i}^{\tau}\, d\tau^{\prime}\, {\cal T}\(\tau^{\prime}\)}
\nonumber
\er
with $P_2$ denoting surface ordering, and $V_R$ being a matrix integration constant.  Note that $V_R$ is the value of $V$ at the infinitesimal loop around the reference point $x_R$, i.e., $V\(\tau_i\)=V_R$. 

The question we ask now is how $V$, defined on a given  surface, varies when we change that surface infinitesimally. In other words, what is the variation $\delta V$, when we change the function $x^{\mu}\(\zeta\,,\,\tau\,,\,\sigma\)$ to $x^{\mu}\(\zeta\,,\,\tau\,,\,\sigma\)+\delta x^{\mu}\(\zeta\,,\,\tau\,,\,\sigma\)$, keeping the reference point $x_R$ fixed? Instead of performing that variation directly on the solution \rf{wilsonsurfacesol}, it is much easier to vary the defining equation \rf{holonomyl2} for $V$,  to get a differential equation for the variation $\delta V$. Integrating such an equation, up to the loop labelled by $\tau$, one gets that (see the details in \cite{afg1,afg2,ym1,ym2}, especially Appendix A.2 of \cite{ym2})
\br
\delta V\(\tau\)\,V^{-1}\(\tau\)=V\(\tau\)\,{\cal T}\(\tau\,,\,\delta\)\,V^{-1}\(\tau\)+{\cal K}\(\tau\,,\,\delta\)
\lab{deltav}
\er
with 
\be
{\cal T}\(\tau\,,\,\delta\)\equiv \int_{\sigma_i}^{\sigma_f}d\sigma\, W^{-1}\, B_{\mu\nu}\,W\,\frac{d\,x^{\mu}}{d\,\sigma}\, \delta x^{\nu}
\lab{connectionl2delta}
\ee
and
\br
{\cal K}\(\tau\,,\,\delta\)&\equiv&
\int_{\tau_i}^{\tau}d\tau^{\prime}\,\int_{\sigma_i}^{\sigma_f}d\sigma\,V\(\tau^{\prime}\)\,\left\{W^{-1}\,
\left[D_{\lambda}B_{\mu\nu}+D_{\mu}B_{\nu\lambda}+D_{\nu}B_{\lambda\mu}\right]
\,
W\frac{d\,x^{\mu}}{d\,\sigma}\,\frac{d\,x^{\nu}}{d\,\tau^{\prime}}\,
\delta x^{\lambda}\right. \nonumber\\
&-&\left. \int_{\sigma_i}^{\sigma}d\sigma^{\prime}
\sbr{B_{\kappa\rho}^W\(\sigma^{\prime}\)- H_{\kappa\rho}^W\(\sigma^{\prime}\)}
{B_{\mu\nu}^W\(\sigma\)}\frac{dx^{\kappa}}{d\sigma^{\prime}}\frac{dx^{\mu}}{d\sigma}\right.\nonumber\\
&&\left.\times
\(\frac{d\,x^{\rho}\(\sigma^{\prime}\)}{d\,\tau^{\prime}}\delta x^{\nu}\(\sigma\)
-\delta x^{\rho}\(\sigma^{\prime}\)\,\frac{d\,x^{\nu}\(\sigma\)}{d\,\tau^{\prime}}\)\right\} V^{-1}\(\tau^{\prime}\)
\lab{kdelta}
\er  
where  we have introduced  the covariant derivative $D_{\mu}*\equiv\partial_{\mu}*+\sbr{C_{\mu}}{*}$, and the curvature of the connection $C_{\mu}$,  $H_{\mu\nu}\equiv\partial_{\mu}C_{\nu}-\partial_{\nu}C_{\mu}+\sbr{C_{\mu}}{C_{\nu}}$,    and where the superscript $^W$ means conjugation by the Wilson line, i.e. $X^W\equiv W^{-1}\,X\,W$. Note that $\tau_i$ corresponds to the infinitesimal loop around the reference point $x_R$. Since we do not vary $x_R$, we have that $\delta x^{\mu}\(\tau_i\)=0$, and so when performing the integration to get \rf{deltav}, we have used the fact that ${\cal T}\(\tau_i\,,\,\delta\)=0$. 

We now consider the case where the variation of the surface corresponds to the variation of a given  closed surface in the scanning of $\Omega$, into one infinitesimally close to it, i.e. we take $\delta x^{\mu}=\frac{d\,x^{\mu}}{d\,\zeta}\,d\zeta$. First, we consider the variation of $V$, integrated up to a loop labelled by $\tau$, due to the variation of  a surface labelled by $\zeta$, to one labelled by $\zeta+d\zeta$. Replacing the variation $\delta x^{\mu}=\frac{d\,x^{\mu}}{d\,\zeta}\,d\zeta$, into \rf{deltav} and dividing both sides  by $d \zeta$, and taking the limit $d \zeta\rightarrow 0$, we get a differential equation for $V$
\br
\frac{d\,V\(\tau\)}{d\,\zeta}\,V^{-1}\(\tau\)=V\(\tau\)\,{\cal T}\(\tau\)\,V^{-1}\(\tau\)+{\cal K}\(\tau\)
\lab{deltav2}
\er
where ${\cal K\(\tau\)}$ stands for ${\cal K}\(\tau\,,\,\delta\)$, given in \rf{kdelta},  and ${\cal T}\(\tau\)$ stands for ${\cal T}\(\tau\,,\,\delta\)$, given in \rf{connectionl2delta}, when we replace $\delta x^{\mu}$ by $\delta x^{\mu}=\frac{d\,x^{\mu}}{d\,\zeta}\,d\zeta$, and divide by $d\zeta$. 

Consider now the case where $V$ is integrated over the whole surface labelled by $\zeta$, i.e., integrated up to the final loop labelled by $\tau_f$, which is also an infinitesimal loop around the reference point $x_R$, at the end of the scanning of the surface with loops. Since we do not vary the reference point, we have that $\delta x^{\mu}\(\tau_f\)=0$, and so, from \rf{connectionl2delta}, we have ${\cal T}\(\tau_f\,,\,\delta\)=0$. Therefore, in such a case \rf{deltav2} becomes 
\be
\frac{d\,V}{d\,\zeta}- {\cal K}\, V=0
\lab{holonomyl3}
\ee
where $V$ stands for $V\(\tau_f\)$, and ${\cal K}$ stands for ${\cal K}\(\tau_f\)$.

The general solution of \rf{holonomyl3} is given by the series 
\br
V\(\zeta\)&=&\left[\one+\int_{\zeta_i}^{\zeta}d\zeta^{\prime}\,{\cal K}\(\zeta^{\prime}\)+
\int_{\zeta_i}^{\zeta}d\zeta^{\prime}\,\int_{\zeta_i}^{\zeta^{\prime}}d\zeta^{\prime\prime}\,{\cal K}\(\zeta^{\prime}\)\,{\cal K}\(\zeta^{\prime\prime}\)
\lab{wilsonvolumesol}
\right.\\
&+& \left.\int_{\zeta_i}^{\zeta}d\zeta^{\prime}\,\int_{\zeta_i}^{\zeta^{\prime}}d\zeta^{\prime\prime}\,\int_{\zeta_i}^{\zeta^{\prime\prime}}d\zeta^{\prime\prime\prime}\,{\cal K}\(\zeta^{\prime}\)\,{\cal K}\(\zeta^{\prime\prime}\)\,{\cal K}\(\zeta^{\prime\prime\prime}\)+\ldots\right]\,{\widehat V}_R \equiv P_3\,e^{\int_{\zeta_i}^{\zeta}\, d\zeta^{\prime}\, {\cal A}\(\zeta^{\prime}\)}\,{\widehat V}_R
\nonumber
\er
where $P_3$ denotes volume ordering, and again ${\widehat V}_R$ is a matrix integration constant. Note that ${\widehat V}_R$ is the value of $V$ at the infinitesimal closed surface around the reference point $x_R$, i.e. $V\(\zeta_i\)={\widehat V}_R$. 

Given a volume $\Omega$ with boundary $\partial\Omega$, and scanned with given functions $x^{\mu}\(\zeta\,,\,\tau\,,\,\sigma\)$, we can determine the quantity $V$ on the surface $\partial \Omega$ in two equivalent ways. First, by integrating \rf{holonomyl2} on the whole surface $\partial \Omega$. Second, by integrating \rf{holonomyl3} on the volume $\Omega$, starting at the infinitesimally closed surface around $x_R$ ($\zeta=\zeta_i$) up to the surface corresponding to the border of $\Omega$ ($\zeta=\zeta_f$). Since those two integrals are bound to be the same, we get the non-abelian Stokes theorem \cite{afg1,afg2,ym1,ym2}
\be
V_R\,P_2\,e^{\int_{\tau_i}^{\tau_f}\, d\tau\, {\cal T}\(\tau\)}=P_3\,e^{\int_{\zeta_i}^{\zeta_f}\, d\zeta\, {\cal K}\(\zeta\)}\, V_R
\lab{stokestheo}
\ee
Note that $\zeta=\zeta_i$ corresponds to an infinitesimal closed surface around the reference point $x_R$. Therefore, the loops scanning that infinitesimal surface shrink to just one infinitesimal loop around $x_R$, corresponding to $\tau=\tau_i$. In such a case we have that $P_2\,e^{\int_{\tau_i}^{\tau_f}\, d\tau\, {\cal T}\(\tau\)}\rightarrow \one$ and $P_3\,e^{\int_{\zeta_i}^{\zeta_f}\, d\zeta\, {\cal K}\(\zeta\)}\rightarrow \one$, and so we have that the integration constants have to be the same, i.e. $V_R={\widehat V}_R$. 

\subsection{The integral Yang-Mills equations}

We shall consider a Yang-Mills theory for a compact gauge group $G$, coupled to families of spinor $\psi$ and scalar $\vp$ fields. The Lagrangian is given by
\be
{\cal L}= -\frac{1}{4}\, {\rm Tr}\(F_{\mu\nu}\,F^{\mu\nu}\)+ {\bar \psi}\(i\,\gamma^{\mu}\,D_{\mu}-m\)\psi+\(D_{\mu}\vp\)^{\dagger}D^{\mu}\vp-V\(\mid \vp\mid\)
\lab{ymlag}
\ee
where, in order to simplify the notation, we have dropped the summation over the families of spinors and scalars. In addition, we have that
\be
F_{\mu\nu}=\partial_{\mu}A_{\nu}-\partial_{\nu}A_{\mu}+i\,e\,\sbr{A_{\mu}}{A_{\nu}}
\ee
with $A_{\mu}$ being the physical gauge field, and $e$ the gauge coupling constant. The Lagrangian \rf{ymlag} is invariant under the gauge transformations
\br
A_{\mu}&\rightarrow& g\,A_{\mu}\,g^{-1}+ \frac{i}{e}\,\partial_{\mu}g\,g^{-1}\;;\qquad \quad F_{\mu\nu}\rightarrow g\,F_{\mu\nu}\,g^{-1}
\nonumber\\
\psi&\rightarrow& R^{\psi}\(g\)\,\psi\;;\qquad\qquad\qquad\qquad\;\;\; \vp\rightarrow R^{\vp}\(g\)\,\vp
\lab{gaugetransf}
\er
with $g\in G$, and where $R^{\psi}$ and $R^{\vp}$ are the representations of $G$ under which the spinors and scalar, respectively, transform. Therefore, we have that $D_{\mu}\psi=\partial_{\mu}\psi+i\,e\,R^{\psi}\(A_{\mu}\)\,\psi$, and $D_{\mu}\vp=\partial_{\mu}\vp+i\,e\,R^{\vp}\(A_{\mu}\)\,\vp$. Consequently, under the gauge transformation \rf{gaugetransf} we have that $D_{\mu}\psi\rightarrow R^{\psi}\(g\)\,D_{\mu}\psi$, and $D_{\mu}\vp\rightarrow R^{\vp}\(g\)\,D_{\mu}\vp$.

The Euler-Lagrange equations associated to the gauge fields, following from \rf{ymlag}, are the local Yang-Mills partial differential  equations, together with the Bianchi identities 
\be
D_{\mu}\,F^{\mu\nu}=e\,J^{\nu}\;;\qquad\qquad\qquad D_{\mu}\,{\widetilde F}^{\mu\nu}=0
\lab{localymeq}
\ee
where ${\widetilde F}_{\mu\nu}$ is the Hodge dual of the field tensor, i.e.
\be
{\widetilde F}_{\mu\nu}=\frac{1}{2}\,\ve_{\mu\nu\rho\lambda}\,F^{\rho\lambda}
\lab{fdual}
\ee
and $J^{\mu}$ is the matter current
\be
J^{\mu}=\left[{\bar \psi}\,\gamma^{\mu}\,R^{\psi}\(T_a\)\,\psi+\frac{i}{2}\,\(\vp^{\dagger}\,R^{\vp}\(T_a\)\,D^{\mu}\vp-\(D^{\mu}\vp\)^{\dagger}\,R^{\vp}\(T_a\)\,\vp\)\right]\,T_a
\lab{mattercurr}
\ee
Note that $J^{\mu}$ is an element of the Lie algebra of the gauge group $G$. $J^{\mu}$ transforms under the adjoint representation of $G$, irrespective of the representations under which the matter fields transform,  i.e., under \rf{gaugetransf} we have that
\be
J^{\mu}\rightarrow g\,J^{\mu}\,g^{-1}
\lab{jtransf}
\ee

The Euler-Lagrange equations associated to the matter fields, following from \rf{ymlag}, are
\be
\(i\,\gamma^{\mu}\,D_{\mu}-m\)\psi=0\;;\qquad\qquad\qquad D_{\mu}D^{\mu}\vp+\frac{\delta\,V}{\delta\,\mid \vp\mid^2}\,\vp=0
\lab{mattereq}
\ee
 
The integral Yang-Mills equations are obtained by combining the non-abelian Stokes theorem for a two-form connection \rf{stokestheo},  with the local differential Yang-Mills equations \rf{localymeq}. We take the one-form $C_{\mu}$ and two-form $B_{\mu\nu}$ to be given by
\be
C_{\mu}=i\,e\,A_{\mu}\;;\qquad\qquad\qquad B_{\mu\nu}=i\,e\,\(\alpha\,F_{\mu\nu}+\beta\,{\widetilde F}_{\mu\nu}\)
\lab{replace1}
 \ee
with $\alpha$ and $\beta$ being arbitrary (even complex) parameters.  Using the Yang-Mills equations \rf{localymeq} we get that
\be
D_{\lambda}B_{\mu\nu}+D_{\mu}B_{\nu\lambda}+D_{\nu}B_{\lambda\mu}=i\,e^2\,\beta\, {\widetilde J}_{\lambda\mu\nu}
\lab{replace2}
\ee
where ${\widetilde J}_{\mu\nu\rho}=\ve_{\mu\nu\rho\lambda}\,J^{\lambda}$, is the Hodge dual of the matter current. 
 
 Therefore, the $1$-form connection on ${\cal L}^{(1)}$ defined in \rf{connectionl2} becomes
 \be
{\cal T}\equiv i\,e\,\int_{\sigma_i}^{\sigma_f}d\sigma\, W^{-1}\, \(\alpha\,F_{\mu\nu}+\beta\,{\widetilde F}_{\mu\nu}\)\,W\,\frac{d\,x^{\mu}}{d\,\sigma}\, \frac{d\,x^{\nu}}{d\,\tau}
\lab{connectionl2ym}
\ee
From \rf{replace1} and  \rf{replace2},  and the substitution $\delta x^{\mu}=\frac{d\,x^{\mu}}{d\,\zeta}\,d\zeta$, we get that the one-form connection ${\cal K}\(\tau_f\,,\, \delta\)$ in ${\cal L}^{(2)}$,  introduced in \rf{kdelta}, which we now denote by ${\cal A}$,  becomes
\be
{\cal A}=i\,e^2\,\int_{\tau_i}^{\tau_f}d\tau\,V\(\tau\)\,{\cal J}\,V^{-1}\(\tau\)
\lab{connectionacal}
\ee
with
 \br
{\cal J}&\equiv&
\int_{\sigma_i}^{\sigma_f}d\sigma\,\left\{\beta\,
{\widetilde J}_{\mu\nu\lambda}^{W}\,\frac{d\,x^{\mu}}{d\,\sigma}\,\frac{d\,x^{\nu}}{d\,\tau}\,
\frac{d\,x^{\lambda}}{d\,\zeta}\right. \nonumber\\
&-&\left. i\,\int_{\sigma_i}^{\sigma}d\sigma^{\prime}
\sbr{\(\alpha-1\)F_{\kappa\rho}^W\(\sigma^{\prime}\)+\beta\,{\widetilde F}_{\kappa\rho}^W\(\sigma^{\prime}\)}
{\alpha\,F_{\mu\nu}^W\(\sigma\)+\beta\, {\widetilde F}_{\mu\nu}^W\(\sigma\)}\frac{dx^{\kappa}}{d\sigma^{\prime}}\frac{dx^{\mu}}{d\sigma}\right.\nonumber\\
&&\left.\times
\(\frac{d\,x^{\rho}\(\sigma^{\prime}\)}{d\,\tau}\frac{d\,x^{\nu}\(\sigma\)}{d\,\zeta}
-\frac{d\,x^{\rho}\(\sigma^{\prime}\)}{d\,\zeta}\,\frac{d\,x^{\nu}\(\sigma\)}{d\,\tau}\)\right\} 
\lab{caljdef}
\er  
where again the superscript $^W$ means $X^W\equiv W^{-1}\,X\,W$. The equation \rf{holonomyl3} then becomes 
\be
\frac{d\,V}{d\,\zeta}- {\cal A}\, V=0
\lab{holonomyl4}
\ee

 As we show in Appendix \ref{sec:proofnewj}, ${\cal J}$ defined in \rf{caljdef}, can be written as 
 \br
  {\cal J}= \beta\, {\cal J}_M+\beta\, {\cal J}_G +i\,\alpha\sbr{{\cal F}_{\tau}}{{\cal F}_{\zeta}}
  -i\,\sbr{\alpha\,{\cal F}_{\tau}+\beta\,{\widetilde{\cal F}}_{\tau}}{\alpha\,{\cal F}_{\zeta}+\beta\,{\widetilde{\cal F}}_{\zeta}}
\lab{caljdef2}
  \er
where we have defined the quantities 
\be
{\cal J}_M\equiv   \int_{\sigma_i}^{\sigma_f}d\sigma\,
W^{-1}\,{\widetilde J}_{\mu\nu\lambda}\,W\,\frac{d\,x^{\mu}}{d\,\sigma}\,\frac{d\,x^{\nu}}{d\,\tau}\,
\frac{d\,x^{\lambda}}{d\,\zeta}
\lab{jmatterdef}
 \ee
 and
 \br
 {\cal J}_G&\equiv&i\,\int_{\sigma_i}^{\sigma_f}d\sigma\,\int_{\sigma_i}^{\sigma}d\sigma^{\prime}\left\{
 \sbr{F_{\kappa\rho}^W\(\sigma^{\prime}\)\frac{dx^{\kappa}}{d\sigma^{\prime}}\frac{d\,x^{\rho}\(\sigma^{\prime}\)}{d\,\tau}}{{\widetilde F}_{\mu\nu}^W\(\sigma\)\frac{dx^{\mu}}{d\sigma}\frac{d\,x^{\nu}\(\sigma\)}{d\,\zeta}}
 \right.\nonumber\\
&-&\left. 
\sbr{F_{\kappa\rho}^W\(\sigma^{\prime}\)\frac{dx^{\kappa}}{d\sigma^{\prime}}\frac{d\,x^{\rho}\(\sigma^{\prime}\)}{d\,\zeta}}{{\widetilde F}_{\mu\nu}^W\(\sigma\)\frac{dx^{\mu}}{d\sigma}\frac{d\,x^{\nu}\(\sigma\)}{d\,\tau}}
\right\}
\lab{jgaugedef}
 \er
and 
\be
{\widetilde{\cal F}}_{\tau/\zeta}\equiv \int_{\sigma_i}^{\sigma_f}d\sigma \,W^{-1}\, {\widetilde F}_{\mu\nu}\,W\,\frac{dx^{\mu}}{d\sigma}\frac{d\,x^{\nu}}{d\,\tau/\zeta}
 \lab{fcaltildetauzetadef}
 \ee
and
 \be
 {\cal F}_{\tau/\zeta}\equiv \int_{\sigma_i}^{\sigma_f}d\sigma \,W^{-1}\, F_{\mu\nu}\,W\,\frac{dx^{\mu}}{d\sigma}\frac{d\,x^{\nu}}{d\,\tau/\zeta}=-\frac{i}{e}\,W^{-1}_c\,\frac{d\,W_c}{d\,\tau/\zeta}
 \lab{fcaltauzetadef}
 \ee
where $W_c$ is the Wilson operator obtained by integrating (see \rf{wdefc} and \rf{replace1}) 
\be
\frac{d\,W}{d\,\sigma}+i\,e\,A_{\mu}\,\frac{d\,x^{\mu}}{d\,\sigma}\,W=0
\lab{wdefa}
\ee 
on a closed loop belonging to the scanning of $\Omega$, starting and ending at the reference point $x_R$,  labelled by $\tau$, and lying on a closed surface labelled by $\zeta$. See \rf{variationw2b} for a proof of the last equality in \rf{fcaltauzetadef}, or alternatively see section 2 of \cite{afg1}. 

With such a notation we have that ${\cal T}$, defined in \rf{connectionl2ym}, can be written as
\be
{\cal T}= i\,e\, \left[ \alpha\, {\cal F}_{\tau}+\beta\, {\widetilde {\cal F}}_{\tau}\right]=\alpha\,W^{-1}_c\,\frac{d\,W_c}{d\,\tau}+ i\,e\,\beta\, {\widetilde {\cal F}}_{\tau}
\lab{tconnectionfftilde}
\ee

 Under the gauge transformations \rf{gaugetransf} we have that the Wilson line, defined in \rf{wdefa},  transforms as
 \be
 W\rightarrow g\(x\)\,W\,W^{-1}_R\,g^{-1}\(x_R\) \, W_R= g\(x\)\,W\,{\widetilde g}^{-1}\(x_R\);\qquad\qquad 
 {\widetilde g}\(x_R\)\equiv W^{-1}_R\,g\(x_R\) \, W_R
 \ee 
 where $x_R$ and $x$ are respectively the initial  and final points of the curve where $W$ is evaluated, and $W_R$ is the integration constant associated to the first order differential equation \rf{wdefa}, and it corresponds to the value of $W$ at the reference point $x_R$ (see \rf{wilsonsol}).  The integration constant $W_R$ will not play any physical role in our calculations, and in what follows, we shall denote ${\widetilde g}\(x_R\)$ by $g\(x_R\)$, as if that constant were chosen to lie in the center of the gauge group $G$.

 Therefore, for any quantity $X$ transforming under the adjoint representation of the gauge group, we have that 
 \be
 X\(x\)\rightarrow g\(x\)\,X\(x\)\,g\(x\)^{-1};\qquad \qquad
 W^{-1}\,X\(x\)\,W\rightarrow g\(x_R\)\,W^{-1}\,X\(x\)\,W\,g\(x_R\)^{-1}
  \lab{globaltransf}
  \ee
 Note that $F_{\mu\nu}$, ${\widetilde F}_{\mu\nu}$, and ${\widetilde J}_{\mu\nu\lambda}$, which transform under the adjoint representation, appear in all our formulas conjugated by the Wilson line. Consequently, the one-form connection ${\cal T}$ in ${\cal L}^{(1)}$, given in \rf{connectionl2ym}, and the quantity ${\cal J}$, defined in \rf{caljdef},  transform as 
 \be
 {\cal T}\rightarrow g\(x_R\)\,{\cal T}\,g^{-1}\(x_R\);\qquad\qquad\qquad
 {\cal J}\rightarrow g\(x_R\)\,{\cal J}\,g^{-1}\(x_R\)
 \ee
Therefore, the quantity $V\(\tau\)$ given in  \rf{wilsonsurfacesol}, transforms as
\be
V\(\tau\)\rightarrow \(V_R\,g\(x_R\)\,V_R^{-1}\)\, V\(\tau\)\, g^{-1}\(x_R\)
\ee
Consequently, the one-form connection in ${\cal L}^{(2)}$ given in \rf{connectionacal}, transform as 
\be
 {\cal A}\rightarrow \(V_R\,g\(x_R\)\,V_R^{-1}\)\,{\cal A}\,\(V_R\,g\(x_R\)\,V_R^{-1}\)^{-1}
 \ee
The left and right hand sides of   \rf{stokestheo}, transform as
\br
V_R\,P_2\,e^{\int_{\tau_i}^{\tau_f}\, d\tau\, {\cal T}\(\tau\)} \rightarrow 
\(V_R\,g\(x_R\)\,\,V_R^{-1}\)\(V_R\,P_2\,e^{\int_{\tau_i}^{\tau_f}\, d\tau\, {\cal T}\(\tau\)}\)\, g^{-1}\(x_R\)
\nonumber\\
P_3\,e^{\int_{\zeta_i}^{\zeta_f}\, d\zeta\, {\cal A}\(\zeta\)}\, V_R\rightarrow 
\(V_R\,g\(x_R\)\,\,V_R^{-1}\)\(P_3\,e^{\int_{\zeta_i}^{\zeta_f}\, d\zeta\, {\cal A}\(\zeta\)}\, V_R\)\,g^{-1}\(x_R\)
\lab{stokestransfom}
\er
Therefore, the equation \rf{stokestheo} transforms covariantly under gauge transformations.

Note that all the quantities in our formulas appear conjugated by the Wilson line operator $W$, and so, the local gauge symmetry has been wrapped up by such a conjugation. Therefore, our formulas present just a global gauge symmetry performed by the group element evaluated at the reference point, i.e. $g\(x_R\)$.  That fact  simplifies things a lot, and play a crucial role in all of our results.

The integral Yang-Mills equations are obtained from the non-abelian Stokes theorem \rf{stokestheo}, with the replacements \rf{replace1}, and the imposition, in  \rf{replace2}, of the  local Yang-Mills partial differential  equations \rf{localymeq}. So, for any three-dimensional volume $\Omega$, on the four-dimensional Minkowski space-time, we get that the classical dynamics of the Yang-Mils theory is described by the equations
 \be
V_R\,P_2\,e^{i\,e\,\int_{\partial\Omega} d\tau\, d\sigma\, W^{-1}\, \(\alpha\,F_{\mu\nu}+\beta\,{\widetilde F}_{\mu\nu}\)\,W\,\frac{d\,x^{\mu}}{d\,\sigma}\, \frac{d\,x^{\nu}}{d\,\tau}}=P_3\,e^{i\,e^2\,\int_{\Omega} d\zeta\,  d\tau\,V\,{\cal J}\,V^{-1}}\, V_R
\lab{ymintegraleqs}
\ee
That is the integral equation for the Yang-Mills theory in $(3+1)$-dimensional Minkowski space-time obtained in \cite{ym1,ym2}. 
Such an equation is equivalent to the local Yang-Mills partial differential  equations \rf{localymeq}. Indeed, if one considers the length parameter $l\equiv \(\mbox{\rm volume of $\Omega$}\)^{1/3}$, and expands both sides of \rf{ymintegraleqs} in a power series in $l$, one gets  the Yang-Mills equations \rf{localymeq} in lowest order of that expansion. 

As we have shown above, the integral equations \rf{ymintegraleqs} transform covariantly under general local gauge transformations. Note however that the tensors $F_{\mu\nu}$, ${\widetilde F}_{\mu\nu}$ and ${\widetilde J}_{\mu\nu\lambda}$, which transform under the adjoint representation of the gauge group $G$, appear in \rf{ymintegraleqs} conjugated by the Wilson line $W$. Therefore, as shown in \rf{globaltransf}, the Yang-Mills integral equations only involve quantities that transform globally under gauge transformations. 

In order to obtain the integral equations \rf{ymintegraleqs} we have to choose a scanning of the volume $\Omega$ with closed surfaces based at the reference point $x_R$. However, by the construction of the non-abelian Stokes theorem \rf{stokestheo}, it is clear that if we change the choice of scanning, both sides of \rf{ymintegraleqs} will change, but in a way that the equality between them holds true in the new scanning. So, \rf{ymintegraleqs} transforms covariantly under the change of scanning. 

The integral equation \rf{ymintegraleqs} is also independent upon the choice of the reference point $x_R$. Indeed, if one changes $x_R$ to $x^{\prime}_R$, the Wilson line will change as $W\rightarrow W\,W_{x_R\rightarrow x^{\prime}_R}$, where $W_{x_R\rightarrow x^{\prime}_R}$ is obtained by integrating \rf{wdefa} from $x_R$ to $x^{\prime}_R$ along a fixed chosen curve on $\partial\Omega$, joining the two reference points. But since everything in \rf{ymintegraleqs} is conjugated by the Wilson line, it turns out that everything gets conjugated by the constant matrix  $W_{x_R\rightarrow x^{\prime}_R}$. So, both sides of the integral Yang-Mills equations \rf{ymintegraleqs} get conjugated by $W_{x_R\rightarrow x^{\prime}_R}$, and such equations transform covariantly under the change of the reference point $x_R$. 

By expanding both sides of \rf{ymintegraleqs} in power series in the parameters $\alpha$ and $\beta$ we get, in fact, an infinite number of integral equations. For instance, the terms linear in $\alpha$ lead to the integral equation 
\be
V_R\,e\,\int_{\partial\Omega} d\tau\, d\sigma\, W^{-1}\, F_{\mu\nu}\,W\,\frac{d\,x^{\mu}}{d\,\sigma}\, \frac{d\,x^{\nu}}{d\,\tau}=i\,e^2\,\int_{\Omega} d\zeta\,  d\tau\,\sbr{{\cal F}_{\tau}}{{\cal F}_{\zeta}}\,V_R
\lab{magneticinteq}
\ee
and the terms linear in $\beta$ lead to the integral equation
\be
V_R\,e\,\int_{\partial\Omega} d\tau\, d\sigma\, W^{-1}\, {\widetilde F}_{\mu\nu}\,W\,\frac{d\,x^{\mu}}{d\,\sigma}\, \frac{d\,x^{\nu}}{d\,\tau}=e^2\,\int_{\Omega} d\zeta\,  d\tau\,\({\cal J}_M+ {\cal J}_G\)\,V_R
\lab{electricinteq}
\ee
Note that when the volume $\Omega$ is purely spatial, the right-hand sides of \rf{magneticinteq} and \rf{electricinteq} can be interpreted, respectively, as the non-abelian magnetic and electric flux through the border $\partial\Omega$. The left-hand sides of \rf{magneticinteq} and \rf{electricinteq} can be interpreted, respectively, as the non-abelian magnetic and electric charges inside the volume $\Omega$. In addition, ${\cal J}_M$ and ${\cal J}_G$ account for the charges of the matter fields and gauge fields (gluons),  respectively. Continuing the expansion, we get higher integral equations. For a check of such equations, for the case of $SU(2)$ magnetic monopoles, see \cite{directtest}. 
 
\subsection{The gauge invariant conserved charges}
\label{subsec:conservedcharges}

One of the most striking consequences of the integral Yang-Mills equation \rf{ymintegraleqs} is that the operator on its right-hand side is independent of the volume $\Omega$, as long as its border $\partial\Omega$ is kept fixed. Indeed, consider two volumes $\Omega$ and $\Omega^{\prime}$ such that their borders are the same, i.e. $\partial\Omega=\partial\Omega^{\prime}$. Therefore, the integral equations \rf{ymintegraleqs} considered on those two volumes are such that their left-hand sides are equal. Consequently, we get that
\be
V\(\Omega\)\equiv P_3\,e^{i\,e^2\,\int_{\Omega} d\zeta\,  d\tau\,V\,{\cal J}\,V^{-1}}\, V_R=P_3\,e^{i\,e^2\,\int_{\Omega^{\prime}} d\zeta\,  d\tau\,V\,{\cal J}\,V^{-1}}\, V_R\equiv V\(\Omega^{\prime}\);\qquad {\rm if} \qquad\partial\Omega=\partial\Omega^{\prime}
\lab{crucialeq}
\ee
where we have assumed that the integration constant $V_R$ is the same for both volumes. 
Since the volumes $\Omega$ and $\Omega^{\prime}$ are scanned with closed surfaces based on a given reference point $x_R$, they can be seen as  paths on the loop space ${\cal L}^{(2)}$ (see \rf{loopspace}). So, the relation \rf{crucialeq} implies that the operator $P_3\,e^{i\,e^2\,\int_{\Omega} d\zeta\,  d\tau\,V\,{\cal J}\,V^{-1}}$ is independent of the path on ${\cal L}^{(2)}$, as long as the initial and final points of the path are kept fixed. The initial point of the path is the infinitesimal closed surface around $x_R$, and the final point is the border $\partial\Omega$. Note that if we keep the volume $\Omega$ fixed and change its scanning with closed surfaces, we get that the path on ${\cal L}^{(2)}$ changes. In other words, there exists an infinite number of paths on ${\cal L}^{(2)}$ corresponding to the same physical volume $\Omega$. Therefore, the relation \rf{crucialeq} also implies that the operator $P_3\,e^{i\,e^2\,\int_{\Omega} d\zeta\,  d\tau\,V\,{\cal J}\,V^{-1}}$ is independent of the choice of scanning, i.e. it is reparameterisation invariant. 

The conserved charges were constructed in \cite{ym1,ym2} using such a path independency of the volume operator appearing on the right-hand side of \rf{ymintegraleqs}. Let us consider two paths starting and ending at the same points. The first path is made of two parts. We take the spatial sub-manifold 
$\IR^3_{0}$, at time $t=0$, and take the reference point $x_R^{0}$, at the border of it, i.e. $S^2_{\infty\,,\, t=0}$. The first part of the path is made by the scanning of  $\IR^3_{0}$ with closed two-dimensional surfaces based at $x_R^{0}$, starting with the infinitesimal surface around $x_R^{0}$, and ending at the border $S^2_{\infty\,,\, t=0}$. The second part of the path is obtained by scanning a hyper cylinder $I\times S^2_{\infty}$, with the bottom part being $S^2_{\infty\,,\, t=0}$, and the top part being $S^2_{\infty\,,\, t=t}$. The side of the hyper cylinder is in the time direction, rising from $t=0$ to $t=t$. We scan it with two-dimensional surfaces (of infinite radius) based at $x_R^{0}$. The integration of \rf{holonomyl3} with ${\cal A}$ given in \rf{connectionacal} leads to the operator
\be
V_{x_R^{0}}\(I\times S^2_{\infty}\)\, V_{x_R^{0}}\(\IR^3_{0}\)
\lab{operator1}
\ee
with
\be
V_{x_R^{0}}\(X\)\equiv P_3\,e^{i\,e^2\,\int_{X} d\zeta\,  d\tau\,V\,{\cal J}\,V^{-1}}
\ee
and the subscript $x_R^{0}$ means that the two dimensional surfaces scanning the volumes are based at the reference point $x_R^{0}$.

The second path is also made of two parts. The first part is an infinitesimally thin hyper cylinder $I\times S^2_{0}$, with the bottom part being a two sphere of vanishing radius at the reference point $x_R^{0}$ at $t=0$, and the top part being a two sphere of vanishing radius at the reference point $x_R^{t}$ at $t=t$. The second part corresponds to the spatial sub-manifold $\IR^3_{t}$, at time $t=t$. All the two-dimensional surfaces scanning such a volume are based at $x_R^{0}$, as our loop space formulation requires. The integration of \rf{holonomyl3} with ${\cal A}$ given in \rf{connectionacal} leads to the operator
\be
V_{x_R^{0}}\(\IR^3_{t}\)\, V_{x_R^{0}}\(I\times S^2_{0}\)
\lab{operator2}
\ee
From the arguments leading to \rf{crucialeq}, the operators \rf{operator1} and \rf{operator2} should be equal. 

We now impose the following boundary condition on the Yang-Mills fields. The field tensor and the matter current should fall at spatial infinity as
\be
F_{\mu\nu} \rightarrow \frac{1}{r^{\frac{3}{2}+\delta}}\qquad\qquad \qquad J_{\mu}\rightarrow \frac{1}{r^{2+\delta^{\prime}}} \qquad \qquad \qquad r\rightarrow \infty
\lab{boundcond}
\ee
with $\delta\,,\, \delta^{\prime} >0$. Such boundary conditions are sufficient for the quantity ${\cal J}$ to vanish at spatial infinity. Remember that the reference points $x_R^{0}$ and $x_R^{t}$ are at spatial infinity. Therefore, one gets that 
\be
 V_{x_R^{0}}\(I\times S^2_{\infty}\)\rightarrow\one 
 \qquad\qquad\qquad 
 V_{x_R^{0}}\(I\times S^2_{0}\)\rightarrow \one
 \lab{boundcond2}
 \ee
 When evaluating the charge operator on the spatial sub-manifold $\IR^3$, at a given time, we want the reference point to be on the border of $\IR^3$ at that same time. The shifting of the reference point amounts to the change of the initial point where the Wilson line is evaluated. Therefore, we have that
 \be
 V_{x_R^{0}}\(\IR^3_{t}\)=W^{-1}\(x_R^{t}\,,\, x_R^{0}\)\,V_{x_R^{t}}\(\IR^3_{t}\)W\(x_R^{t}\,,\, x_R^{0}\)
 \ee
 where $W\(x_R^{t}\,,\, x_R^{0}\)$ is the Wilson line obtained by integrating \rf{wdefa} along the time interval $I$, from $x_R^{0}$ to $x_R^{t}$. Therefore, we get that
 \be
 V_{x_R^{t}}\(\IR^3_{t}\)=W\(x_R^{t}\,,\, x_R^{0}\)\,  V_{x_R^{0}}\(\IR^3_{0}\) \, W^{-1}\(x_R^{t}\,,\, x_R^{0}\)
 \ee
Consequently, the operator obtained by integrating  \rf{holonomyl3}, with ${\cal A}$ given in \rf{connectionacal}, on the spatial sub-manifold $\IR^3$, at a given time, has an iso-spectral time evolution. Therefore, its eigenvalues, or equivalently, the traces of powers of it, are constant in time. Those are the conserved charges of the Yang-Mills theories. We denote the charge operator and the charges, respectively, as
\be
Q\(\alpha\,,\,\beta\)\equiv V_{x_R^{t}}\(\IR^3_{t}\)\qquad\qquad\qquad \qquad
Q_N\(\alpha\,,\,\beta\)\equiv\frac{1}{N}\, {\rm Tr}\left[Q\(\alpha\,,\,\beta\)\right]^N
\lab{chargeoperatordef}
\ee

Note that using the Yang-Mills integral equations \rf{ymintegraleqs}, we can write the charge operator either as an ordered volume integral or as an ordered surface integral, i.e., we have 

 \be
Q\(\alpha\,,\,\beta\)=V_R\,P_2\,e^{i\,e\,\int_{S^2_{\infty}} d\tau\, d\sigma\, W^{-1}\, \(\alpha\,F_{\mu\nu}+\beta\,{\widetilde F}_{\mu\nu}\)\,W\,\frac{d\,x^{\mu}}{d\,\sigma}\, \frac{d\,x^{\nu}}{d\,\tau}}=P_3\,e^{i\,e^2\,\int_{\IR^3_{t}} d\zeta\,  d\tau\,V\,{\cal J}\,V^{-1}}\, V_R
\lab{chargevolumesurface}
\ee 
where $S^2_{\infty}$ is the two-dimensional sphere at spatial infinity, i.e., the border of $\IR^3_{t}$. 

Note from \rf{stokestransfom} that, under a gauge transformation \rf{gaugetransf}, the charge operator $Q\(\alpha\,,\,\beta\)=P_3\,e^{i\,e^2\,\int_{\IR^3} d\zeta\,  d\tau\,V\,{\cal J}\,V^{-1}}\, V_R$, transforms as 
\be
Q\(\alpha\,,\,\beta\)\rightarrow \(V_R\,g\(x_R\)\,V_R^{-1}\)\, Q\(\alpha\,,\,\beta\)\, g^{-1}\(x_R\)
\ee
Therefore, in order for the conserved charges $Q_N\(\alpha\,,\,\beta\)= {\rm Tr}\(Q\(\alpha\,,\,\beta\)\)^N$, to be gauge invariant, we have to impose that the integration constant $V_R$ must satisfy 
\be
V_R\in \;\mbox{\rm center of the gauge group $G$}
\lab{intconstcenter}
\ee
That condition will play an important role in our discussion of the algebra of such conserved charges.

By expanding the charge operator in power series of the parameters $\alpha$ and $\beta$
\be
Q\(\alpha\,,\,\beta\)=\sum_{n,m=0}^{\infty}
\alpha^m\,\beta^n\, Q\(m\,,\,n\)
\lab{expandchargeoperator}
\ee
and using the fact that $W^{-1}\(x_R^{t}\,,\, x_R^{0}\)$ does not depend upon $\alpha$ and $\beta$, we get that each component of the charge operator has an iso-spectral time evolution
\be
Q\(m\,,\,n\,,\,t\)= W\(x_R^{t}\,,\, x_R^{0}\)\, Q\(m\,,\,n\,,\,0\)\, W^{-1}\(x_R^{t}\,,\, x_R^{0}\)
\lab{isospectralcomponents}
\ee
Consequently, we get an infinity of conserved charges given by
\be
Q_N\(m\,,\,n\)=\frac{1}{N}\, {\rm Tr}\left[Q\(m\,,\,n\)\right]^N
\lab{chargesexpanded}
\ee

Let us expand both sides of the  integral Yang-Mills equation \rf{ymintegraleqs}  in  power series of the parameters $\alpha$ and $\beta$
\br
V\( \partial\Omega\)&=&V_R\,P_2\,e^{i\,e\,\int_{\partial\Omega} d\tau\, d\sigma\, W^{-1}\, \(\alpha\,F_{\mu\nu}+\beta\,{\widetilde F}_{\mu\nu}\)\,W\,\frac{d\,x^{\mu}}{d\,\sigma}\, \frac{d\,x^{\nu}}{d\,\tau}}
=\sum_{n,m=0}^{\infty}\alpha^m\,\beta^n\, V\(\partial\Omega \,,\,m\,,\,n\)
\nonumber\\
V\(\Omega\)&=&P_3\,e^{i\,e^2\,\int_{\Omega} d\zeta\,  d\tau\,V\,{\cal J}\,V^{-1}}\, V_R=\sum_{n,m=0}^{\infty}\alpha^m\,\beta^n\, V\(\Omega \,,\,m\,,\,n\)
\er
and so
\be
V\(\partial\Omega \,,\,m\,,\,n\)= V\(\Omega \,,\,m\,,\,n\)
\ee
Using arguments similar to those leading to \rf{crucialeq}, we get that $V\(\Omega \,,\,m\,,\,n\)$ is independent of $\Omega$, as long as we keep its border $\partial\Omega$ fixed. That argument applies when we physically change $\Omega$, or when we just change its parameterization in the loop space ${\cal L}^{(2)}$ (remember that $\Omega$ is a path in ${\cal L}^{(2)}$). So, $V\(\Omega \,,\,m\,,\,n\)$ is independent of the scanning of $\Omega$. Consequently,  the modes $Q\(m\,,\,n\)$ of the  charge operator are independent of the scanning of the three-dimensional space  $\IR^3_{t}$. Such an argument corroborates the iso-spectral time evolution of $Q\(m\,,\,n\)$, shown in \rf{isospectralcomponents}.

Note that $Q\(\alpha\,,\,\beta\)$, and so $Q\(m\,,\,n\)$, are matrices in a given representation of the gauge group $G$. Therefore, the number of values of the integer $N$ that leads to functionally independent operators is equal to the rank of the gauge group $G$. For more details of such a construction, see \cite{ym1,ym2}.

\subsection{The invariance of the charges under reparameterization}
\label{subsec:reparameterize}

As we have argued below \rf{crucialeq}, the integral Yang-Mills equations imply that the volume ordered integral, on the right-hand side of  \rf{chargevolumesurface}, is independent of the volume as long as its border is kept fixed. That is the basic property used to show the conservation of the charges \rf{chargeoperatordef}. The scanning of the volume with closed surfaces, based at the reference point $x_R$, makes that volume correspond to a path on the loop space 
${\cal L}^{(2)}$. That path can be changed either by changing the physical volume itself or by changing the scanning of that fixed volume. So, we have an infinite number of paths in ${\cal L}^{(2)}$, corresponding to the same physical volume. 

The physical properties of the Yang-Mills theories should not depend upon such a reparameterization of the volume. The discussion below \rf{crucialeq} has shown that the integral Yang-Mills equations guarantee that the right-hand side of \rf{chargevolumesurface} is invariant under reparameterization of $\IR^3$, as long as its border is kept fixed. However, we have to keep fixed not only the physical border $S^2_{\infty}$, but also its parameterization. In other words, we have to keep fixed the point of ${\cal L}^{(2)}$ corresponding to $S^2_{\infty}$. Since what really matters is the physical surface $S^2_{\infty}$, and not its particular scanning, we have to require the invariance of conserved charges when we change the scanning of the end point $S^2_{\infty}$.   

In order to obtain how the left hand side of \rf{chargevolumesurface} varies when we change the parameterization of $S^2_{\infty}$, we consider a path in ${\cal L}^{(2)}$ from the reference point $x_R$ up to a point $\gamma$ of ${\cal L}^{(2)}$ which corresponds to a given scanning of $S^2_{\infty}$. Such a path corresponds to a given scanning of $\IR^3$ at a given fixed time. We then integrate \rf{holonomyl4} along that path to obtain $V\(\gamma\)\equiv Q\(\gamma\)$. Then we consider paths in ${\cal L}^{(2)}$ which correspond to changes of the scanning of $S^2_{\infty}$. Along those paths the $\zeta$-derivative in \rf{holonomyl4} means variations parallel to the surface $S^2_{\infty}$. Therefore, on such paths we need $\frac{d\,Q\(\gamma\)}{\d\,\zeta}=0$, and that implies that ${\cal A}=0$, along those paths.

Note that the first term of \rf{caljdef} involves the contraction of a $3$-form with  three derivatives with respect to $\sigma$, $\tau$, and $\zeta$. But since those three derivatives are now parallel to the $S^2_{\infty}$, which is two-dimensional, it follows that the first term of  \rf{caljdef} vanishes identically. Therefore, the condition for  the conserved charges \rf{chargeoperatordef} to be invariant under the parameterization of $S^2_{\infty}$ is \cite{ym1} 
\be
\delta Q_N\(\alpha\,,\,\beta\)= {\rm Tr}\left[Q^N\(\alpha\,,\,\beta\)\int_{\tau_i}^{\tau_f}d\tau V\(\tau\)\,{\cal N}\,V^{-1}\(\tau\)\right]=0
\lab{reparcondsurface0}
\ee
where
\br
&&{\cal N}=e^2\,\int_{\sigma_i}^{\sigma_f}d\sigma\, \int_{\sigma_i}^{\sigma}d\sigma^{\prime}
\sbr{\(\alpha-1\)F_{ij}^W\(\sigma^{\prime}\)+\beta\,{\widetilde F}_{ij}^W\(\sigma^{\prime}\)}
{\alpha\,F_{kl}^W\(\sigma\)+\beta\, {\widetilde F}_{kl}^W\(\sigma\)}\times 
\nonumber\\
&&\times
\frac{dx^{i}}{d\sigma^{\prime}}\frac{dx^{k}}{d\sigma}\,\(\frac{d\,x^{j}\(\sigma^{\prime}\)}{d\,\tau}\delta x^{l}\(\sigma\)
-\delta x^{j}\(\sigma^{\prime}\)\,\frac{d\,x^{l}\(\sigma\)}{d\,\tau}\)  \qquad\quad {\rm on}\quad S^2_{\infty}
\lab{reparcondsurface}
\er
with $i,j,k,l=1,2,3$, since $S^2_{\infty}$ is a spatial surface, and where we have replaced the $\zeta$-derivatives of $x^i$ by the variations $\delta x^i$, to reinforce that such variations are parallel to $S^2_{\infty}$. 

As discussed in \cite{ym1}, there are basically two sufficient conditions (but perhaps not necessary) leading  to \rf{reparcondsurface0}. Note that the parameters $\(\sigma\,,\,\tau\)$ are angles on $S^2_{\infty}$, and the variations $\delta x^i$ are parallel to $S^2_{\infty}$, and so the second line in \rf{reparcondsurface} goes as $r^4$, as the radial distance $r$ goes to infinite. Therefore, if  the components of the  tensors $F_{ij}$ and ${\widetilde F}_{ij}$ (magnetic and electric fields respectively) fall faster than $1/r^2$, as $r\rightarrow \infty$, ${\cal N}$ vanishes and the condition \rf{reparcondsurface0} is satisfied. 

The second sufficient condition satisfying \rf{reparcondsurface0} is a bit more subtle. Assume that the magnetic and electric fields fall as 
\be
F_{ij}\sim \ve_{ijk} \frac{{\hat r}_k}{r^2}\, G\({\hat r}\);\qquad\qquad \qquad {\widetilde F}_{ij}\sim \ve_{ijk} \frac{{\hat r}_k}{r^2} \,{\widetilde G}\({\hat r}\)
\lab{monopolecond}
\ee
with ${\hat r}$ being the unit radial vector, and $G\({\hat r}\)$ and ${\widetilde G}\({\hat r}\)$ being elements of the Lie algebra of the gauge group, that are covariantly constant
\be
D_k G\({\hat r}\)=0; \qquad\qquad\qquad \qquad D_k {\widetilde G}\({\hat r}\)=0
\lab{covariantconstantg}
\ee
Using \rf{wdefa} we have that \rf{covariantconstantg} implies
\be
\frac{d\;}{d\,\sigma}\(W^{-1}\, G\({\hat r}\)\,W\)=0; \qquad\qquad\qquad \qquad
\frac{d\;}{d\,\sigma}\(W^{-1}\, {\widetilde G}\({\hat r}\)\,W\)=0
\ee
Therefore, the magnetic and electric field conjugated by the Wilson line, which appear in \rf{reparcondsurface}, $F_{ij}^W$ and ${\widetilde F}_{ij}^W$ respectively, have, on the surface of $S^2_{\infty}$, a constant direction in the Lie algebra of the gauge group, i.e. 
\be
W^{-1}\,F_{ij}\,W\sim  \ve_{ijk} \frac{{\hat r}_k}{r^2}\, c;\qquad\qquad \qquad W^{-1}\,{\widetilde F}_{ij}\,W\sim  \ve_{ijk} \frac{{\hat r}_k}{r^2}\,  {\widetilde c}
\lab{fwftildewconstant}
\ee
with $c$ and ${\widetilde c}$ being the constant values of $W^{-1}\,G\,W$ and $W^{-1}\,{\widetilde G}\,W$ respectively, at the reference point $x_R$. 

If we now assume that those constant Lie algebra elements lie on the same Cartan subalgebra, then the commutator in \rf{reparcondsurface} vanishes, and so such fields configurations lead to conserved charges that are truly invariant under reparameterization. 

Note that the conditions \rf{monopolecond} and \rf{covariantconstantg} are exactly those satisfied by monopole and dyons solutions of Yang-Mills theory \cite{goddardnuytsolive,goddardolivereview,ym1,remark}. 

The question if there are other ways of satisfying \rf{reparcondsurface0} has still to be investigated.

\section{The symplectic structure of Yang-Mills theories}
\label{sec:symcharges}
\setcounter{equation}{0}

In order to find the symmetries generated by the conserved charges, constructed in the last part of section \ref{sec:ymintegral}, we need a symplectic structure for our gauge theory. We follow the standard canonical quantization of gauge theories as described, for instance, in the book by Faddeev and Slavnov \cite{slavnovbook}. We use the first-order formalism where the Lagrangian \rf{ymlag} is written as 
\br
{\cal L}&=& -\frac{1}{2}\, {\rm Tr}\left[\(\partial_{\mu}A_{\nu}-\partial_{\nu}A_{\mu}+i\,e\,\sbr{A_{\mu}}{A_{\nu}}-\frac{1}{2}\,F_{\mu\nu}\)F^{\mu\nu}\right]+ {\bar \psi}\(i\,\gamma^{\mu}\,D_{\mu}-m\)\psi
\nonumber\\
&+&\frac{1}{2}\,\left[\vp_{\mu}^{\dagger}\,D^{\mu}\vp+\(D^{\mu}\vp\)^{\dagger}\,\vp_{\mu}-V\(\mid \vp\mid\)\right]-\frac{1}{2}\,\vp_{\mu}^{\dagger}\,\vp^{\mu}
\lab{ymlagfirstorder}
\er
where $\vp_{\mu}$ and $\vp_{\mu}^{\dagger}$ are auxiliary fields. In such first order Lagrangian formalism we consider  as independent variables the fields $A_{\mu}$, $F_{\mu\nu}$, $\psi$, ${\bar \psi}$, $\vp$, $\vp^{\dagger}$, $\vp_{\mu}$ and $\vp_{\mu}^{\dagger}$. 

We use the basis $T_a$, $a=1,2,3,\dots {\rm dim}\,G$, for the Lie algebra of the compact gauge group $G$, as described in \rf{liebasis}, and use the components of the gauge fields in that basis, i.e. $A_{\mu}=A_{\mu}^a\,T_a$, and $F_{\mu\nu}=F_{\mu\nu}^a\,T_a$. In the Hamiltonian formalism, we have to split the time and space components of the fields, and so we shall denote by Latin letters the space indices, i.e., $i,j,k=1,2,3$. We define the non-abelian electric and magnetic fields as
\be
E_i=F_{0i};\qquad\qquad\qquad B_i=-\frac{1}{2}\,\ve_{ijk}\,F_{jk}
\lab{ebdef}
\ee
with $\ve_{123}=1$. The canonical momenta conjugate to the space components of the gauge fields are
\be
\pi_i^a=\frac{\delta {\cal L}}{\delta\,\partial_0\,A_i^a}=E_i^a
\ee
and for the spinor fields, we have
\be
\pi^{\psi}_{\alpha}=\frac{\delta {\cal L}}{\delta\,\partial_0\,\psi_{\alpha}}=i\psi^{\dagger}_{\alpha};\qquad\qquad \alpha=1,2,3,4
\ee
with $\alpha$ being the index for Dirac spinor components. We have dropped the group index of the representation $R^{\psi}$ under which the spinor multiplet transforms. For the boson fields, we have
\be
\pi_{\vp}=\frac{\delta {\cal L}}{\delta\,\partial_0\,\vp}= \frac{1}{2}\,\vp_0^{\dagger};\qquad\qquad
\pi_{\vp^{\dagger}}=\frac{\delta {\cal L}}{\delta\,\partial_0\,\vp^{\dagger}}= \frac{1}{2}\,\vp_0
\ee
where again we have dropped the group index of the representation $R^{\vp}$ under which the boson multiplet transforms.

In terms of the canonical variables, the Lagrangian density \rf{ymlagfirstorder} becomes
\be
{\cal L}= \pi_i^a\,\partial_0A_i^a+\pi^{\psi}_{\alpha} \,\partial_0\psi_{\alpha}+\pi_{\vp}\,\partial_0\vp+\pi_{\vp^{\dagger}}\,\partial_0\vp^{\dagger} -{\cal H}+A_0^a\,{\cal C}_a
\ee
where the density of the Hamiltonian is
\br
{\cal H}&=&\frac{1}{2}\left[\(E_i^a\)^2+\(B _i^a\)^2\right]+i{\bar\psi}\gamma_i\,D_i\psi+m{\bar\psi}\,\psi
+\frac{1}{2}\,\left[\vp_i^{\dagger}\,D_i\vp+\(D_i\vp\)^{\dagger}\,\vp_i\right]
\nonumber\\
&+&V\(\mid \vp\mid\)+2\,\pi_{\vp}\,\pi_{\vp^{\dagger}}-\frac{1}{2}\,\vp_i^{\dagger}\,\vp_i
\lab{hamiltonian}
\er
The constraints ${\cal C}_a$, $a=1,2,3,\ldots {\rm dim}\,G$, are given by
\be
{\cal C}_a=\(D_iE_i\)_a-e\,\psi^{\dagger}\,R^{\psi}\(T_a\)\,\psi+i\,e\,\left[\pi_{\vp}\,R^{\vp}\(T_a\)\,\vp-
\vp^{\dagger}\,R^{\vp}\(T_a\)\,\pi_{\vp^{\dagger}}\right]
\lab{constraint}
\ee
The time components of the gauge fields $A_0^a$ play the role of Lagrange multipliers.  

Note that the Euler-Lagrangian equations associated to $\vp_{\mu}$, $\vp_{\mu}^{\dagger}$ and $F_{ij}$, following from \rf{ymlagfirstorder}, are
\be
\vp_{\mu}=D_{\mu}\vp;\qquad \vp_{\mu}^{\dagger}=\(D_{\mu}\vp\)^{\dagger};\qquad 
F_{ij}=\partial_iA_j-\partial_jA_i+i\,e\,\sbr{A_i}{A_j}
\lab{auxiliareqom}
\ee
Therefore, we can use such equations to eliminate the variables $\vp_{i}$, $\vp_{i}^{\dagger}$ and $F_{ij}$, and the independent pairs of canonical variables become $\(A_i\,,\, E_i\)$, $\(\psi\,,\,\pi^{\psi}\)$,  $\(\vp\,,\, \pi_{\vp}\)$, and $\(\vp^{\dagger}\,,\, \pi_{\vp^{\dagger}}\)$. 

The complete Hamiltonian is then given by
\be
H_T=\int d^3x\, \( {\cal H}-A_0^a\,{\cal C}_a\)
\lab{completeham}
\ee

The equal time canonical Poisson brackets are given by
\br
\pbr{A_i^a\(x\)}{A_j^b\(y\)}&=&0
\nonumber\\
\pbr{\pi_i^a\(x\)}{\pi_j^b\(y\)}&=&0
\nonumber\\
\pbr{A_i^a\(x\)}{\pi_j^b\(y\)}&=&\delta^{ab}\,\delta_{ij}\,\delta^{(3)}\(x-y\)
\lab{pbrel}\\
\pbr{\psi_{\alpha}\(x\)}{\pi^{\psi}_{\beta}\(y\)}&=&\delta^{\alpha\beta}\,\delta^{(3)}\(x-y\)
\nonumber\\
\pbr{\vp\(x\)}{\pi_{\vp}\(y\)}&=&\delta^{(3)}\(x-y\)
\nonumber\\
\pbr{\vp^{\dagger}\(x\)}{\pi_{\vp^{\dagger}}\(y\)}&=&\delta^{(3)}\(x-y\)
\nonumber
\er
where we have dropped the indices associated to the representations $R^{\psi}$ and $R^{\vp}$, under which the spinor and boson  multiplets, respectively, transform. They account for a Kronecker delta on those indices on the right-hand side of the Poisson brackets for those fields. From \rf{pbrel} we have that
\be
\pbr{E_i^a\(x\)}{B_j^b\(y\)}= -\ve_{ijk}\left[e\,f_{abc}\,A^c_k\(x\)\,\delta^{(3)}\(x-y\)-\delta^{ab}\,\frac{\partial\,\delta^{(3)}\(x-y\)}{\partial\,y^k}\right]
\lab{ebpbrel}
\ee

The time component of the matter current \rf{mattercurr}, in the first-order formalism, is given by
\be
J_0= J_0^a\,T_a=\left[\rho_{a}^{\psi}+\rho_{a}^{\vp}\right]\,T_a
\lab{jomatterdef}
\ee
with
\be
\rho_{a}^{\psi}=-i\,\pi^{\psi}\,R^{\psi}\(T_a\)\,\psi\qquad\qquad
\rho_{a}^{\vp}=-i\left[\pi_{\vp}\,R^{\vp}\(T_a\)\,\vp-\vp^{\dagger}\,R^{\vp}\(T_a\)\,\pi_{\vp^{\dagger}}\right]
\lab{matterdensities}
\ee
Therefore, the constraint \rf{constraint} can be written as
\be
{\cal C}={\cal C}_a\,T_a= D_iE_i-e\,J_0
\lab{constraint2}
\ee

The constraints \rf{constraint2} are the generators of gauge transformations under the Poisson bracket, of the canonical variables. Indeed, we have that
\br
\pbr{{\cal C}_a\(x\)}{\psi\(y\)}&=&-i\,e\,R^{\psi}\(T_a\)\,\psi\(x\)\,\delta^{(3)}\(x-y\)
\nonumber\\
\pbr{{\cal C}_a\(x\)}{\pi^{\psi}\(y\)}&=&i\,e\,\pi^{\psi}\(x\)\,R^{\psi}\(T_a\)\,\delta^{(3)}\(x-y\)
\nonumber\\
\pbr{{\cal C}_a\(x\)}{\vp\(y\)}&=&-i\,e\,R^{\vp}\(T_a\)\,\vp\(x\)\,\delta^{(3)}\(x-y\)
\nonumber\\
\pbr{{\cal C}_a\(x\)}{\pi_{\vp}\(y\)}&=&i\,e\,\pi_{\vp}\(x\)\,R^{\vp}\(T_a\)\,\delta^{(3)}\(x-y\)
\lab{constraintgengauge}\\
\pbr{{\cal C}_a\(x\)}{E_i^b\(y\)}&=& -e\,f_{abc}\,E_i^c\(x\)\,\delta^{(3)}\(x-y\)
\nonumber\\
\pbr{{\cal C}_a\(x\)}{A_i^b\(y\)}&=& -e\,f_{abc}\,A^c_i\(x\)\,\delta^{(3)}\(x-y\)-\delta^{ab}\,\frac{\partial\,\delta^{(3)}\(x-y\)}{\partial\,x^i}
\nonumber
\er

\section{The charge operator}
\label{sec:chargeoperator}
\setcounter{equation}{0}

The charge operator $Q\(\alpha\,,\,\beta\)$, defined in \rf{chargeoperatordef}, is the holonomy of the loop space connection ${\cal A}$, defined in  \rf{connectionacal}, over a path on the loop space ${\cal L}^{(2)}$, corresponding to the whole three dimensional spatial sub-manifold $\IR^3_{t}$, at a given fixed time $t$. As explained at the beginning of Section \ref{sec:ymintegral}, in order to perform such a volume ordered integral, we scan  $\IR^3_{t}$ with closed two dimensional surfaces, labelled by $\zeta$, based at a reference point $x_R$ on the border $S^2_{\infty , t}$ of $\IR^3_{t}$. Each closed surface is scanned with loops, labelled by $\tau$, starting and ending at the reference point $x_R$. The loops are parameterized by $\sigma$. Therefore, the Cartesian coordinates of $\IR^3_{t}$, $x^i$, $i=1,2,3$,  become functions of those three parameters, i.e. $x^i=x^i\(\zeta\,,\,\tau\,,\,\sigma\)$. Note that the scanning of $\IR^3_{t}$ is such that each of its points belongs to one and only one closed surface, and to one and only one non-self-intersecting loop, scanning such a surface. Consequently, there is a one-to-one correspondence between the triples $\(x^1\,,\,x^2\,,\,x^3\)$ and $\(\zeta\,,\,\tau\,,\,\sigma\)$. In addition, as shown in \rf{crucialeq}, the charge operator $Q\(\alpha\,,\,\beta\)$ is independent of the path on loop space, as long as its end points are kept fixed, and so it is invariant under the change of  scanning of $\IR^3_{t}$, i.e. it is reparameterization invariant.  

According to \rf{holonomyl3}, the charge operator is obtained by integrating the holonomy equation 
\be
\frac{d\,Q\(\alpha\,,\,\beta\)}{d\,\zeta}-{\cal A}\(\alpha\,,\,\beta\)\,Q\(\alpha\,,\,\beta\)=0
\lab{chargeophol}
\ee
with ${\cal A}\(\alpha\,,\,\beta\)$ given by \rf{connectionacal}. As we are restricted to the three dimensional space $\IR^3_{t}$, it is convenient to introduce the quantities
\br
\mathfrak{e}_{\tau/\zeta}\(\sigma\)&\equiv& \int_{\sigma_i}^{\sigma} d\sigma^{\prime}\,W^{-1}\,E_i\,W\,\ve_{ijk}\,
\frac{d\,x^j}{d\,\sigma^{\prime}}\,\frac{d\,x^k}{d\,\tau/\zeta}
\nonumber\\
\mathfrak{b}_{\tau/\zeta}\(\sigma\)&\equiv &\int_{\sigma_i}^{\sigma} d\sigma^{\prime}\,W^{-1}\,B_i\,W\,\ve_{ijk}\,
\frac{d\,x^j}{d\,\sigma^{\prime}}\,\frac{d\,x^k}{d\,\tau/\zeta}
\lab{ebfrakdef}
\er
 where $E_i$ and $B_i$ are respectively the non-abelian electric and magnetic fields defined in \rf{ebdef}. Note that such quantities correspond to electric and magnetic fluxes through  strips, attached to the loop up to the point $\sigma$, which are parallel to the surface labelled by $\zeta$, in the case of $\mathfrak{e}_{\tau}$ and $\mathfrak{b}_{\tau}$, and perpendicular to it in the case of $\mathfrak{e}_{\zeta}$ and $\mathfrak{b}_{\zeta}$. Note that $\mathfrak{e}_{\tau/\zeta}\(\sigma_f\)$ and $\mathfrak{b}_{\tau/\zeta}\(\sigma_f\)$ correspond (up to a minus sign)  respectively to  ${\widetilde{\cal F}}_{\tau/\zeta}$ and ${{\cal F}}_{\tau/\zeta}$,  given in \rf{fcaltildetauzetadef} and \rf{fcaltauzetadef}, when the loops are purely spatial.

 As discussed in \rf{globaltransf}, under local gauge transformations \rf{gaugetransf}, such quantities transform globally
 \be
 \mathfrak{e}_{\tau/\zeta}\(\sigma\)\rightarrow g\(x_R\)\,\mathfrak{e}_{\tau/\zeta}\(\sigma\)\,g^{-1}\(x_R\);\qquad\qquad\qquad
 \mathfrak{b}_{\tau/\zeta}\(\sigma\)\rightarrow g\(x_R\)\,\mathfrak{b}_{\tau/\zeta}\(\sigma\)\,g^{-1}\(x_R\)
 \lab{globalfefbtransf}
 \ee
 with $g\(x_R\)$ being the gauge group element, performing the gauge transformation, evaluated at the reference point $x_R$.
 
The surface holonomy equation, defined in \rf{holonomyl2} and with ${\cal T}$ given in \rf{tconnectionfftilde}, becomes, in the case of the spatial sub-manifold $\IR^3_{t}$, 
 \be
\frac{d\,V}{d\,\tau} - V\, {\cal T}_{\tau}=0;\qquad\qquad \qquad {\cal T}_{\tau}=-i\,e\left[\alpha\,\mathfrak{b}_{\tau}\(\sigma_f\)+\beta\,\mathfrak{e}_{\tau}\(\sigma_f\)\right]
\lab{holonomyl2fluxes}
\ee
where we have used \rf{fdual} and \rf{ebdef} to write ($\ve_{0123}=\ve_{123}=1$)
\be
F_{ij}= - \ve_{ijk}\, B_k;\qquad\qquad\qquad {\widetilde F}_{ij}= - \ve_{ijk}\, E_k
\lab{ebdefdual}
\ee
We have added a subscript $\tau$ to ${\cal T}$, i.e. ${\cal T}_{\tau}$, for a convenience that will become clear later on. 

In the case of purely spatial loops the quantity ${\cal J}$, given in \rf{caljdef2}, becomes
 \br
  {\cal J}_{\rm spatial}= \beta\( \rho_M+ \rho_G\) +\alpha\,\rho_{\rm mag.}
  -i\,\sbr{\alpha\,\fbt{\sigma_f}+\beta\,\fet{\sigma_f}}{\alpha\,\fbz{\sigma_f}+\beta\,\fez{\sigma_f}}
\lab{caljdef2static}
  \er
with $\rho_M$ and $\rho_G$ corresponding respectively to ${\cal J}_M$ and ${\cal J}_G$, defined in \rf{jmatterdef} and \rf{jgaugedef},  restricted to space loops, and given by
\be
\rho_M\(\tau\,,\,\zeta\)=  -\int_{\sigma_i}^{\sigma_f}d\sigma\,
W^{-1}\,J_0\,W\,\ve_{ijk}\,\frac{d\,x^{i}}{d\,\sigma}\,\frac{d\,x^{j}}{d\,\tau}\,
\frac{d\,x^{k}}{d\,\zeta}
\lab{jmatterdefstatic}
 \ee
with $J_0$ given in \rf{jomatterdef}, and 
\br
\rho_G\(\tau\,,\,\zeta\)= i\,\int_{\sigma_i}^{\sigma_f}d\sigma\,\left\{ \sbr{\fbt{\sigma}}{\frac{d\,\fez{\sigma}}{d\,\sigma}}-
\sbr{\fbz{\sigma}}{\frac{d\,\fet{\sigma}}{d\,\sigma}}
\right\}
\lab{caljgdefstatic}
\er
In addition, we have denoted
\be
\rho_{\rm mag.}=i\,\sbr{\fbt{\sigma_f}}{\fbz{\sigma_f}}
\lab{caljmagdefstatic}
\ee
Note that the first order (in $\alpha$ and $\beta$) integral Yang-Mills equations \rf{magneticinteq} and \rf{electricinteq}, applied to  a volume $\Omega$ being the three-dimensional  space sub-manifold $\IR^3_{t}$, become respectively 
\be
\int_{S^2_{\infty , t}} d\tau\, d\sigma\, W^{-1}\, B_i\,W\,\ve_{ijk}\,\frac{d\,x^{j}}{d\,\sigma}\, \frac{d\,x^{k}}{d\,\tau}=-e\,\int_{\IR^3_{t}} d\zeta\,  d\tau\,\rho_{\rm mag.}
\lab{magneticinteqstatic}
\ee
and 
\be
\int_{S^2_{\infty , t}} d\tau\, d\sigma\, W^{-1}\, E_i\,W\,\ve_{ijk}\,\frac{d\,x^{j}}{d\,\sigma}\, \frac{d\,x^{k}}{d\,\tau}
=-e\,\int_{\IR^3_{t}} d\zeta\,  d\tau\,\(\rho_M+ \rho_G\)
\lab{electricinteqstatic}
\ee
where we have chosen the integration constant $V_R$ equal to unity.

We have seen in \rf{expandchargeoperator} that we can expand the charge operator in power series in $\alpha$ and $\beta$, and each component has an iso-spectral time evolution, with the charges being given by \rf{chargesexpanded}. The right-hand sides of \rf{magneticinteqstatic} and \rf{electricinteqstatic} correspond in fact to the first two operators in that expansion, i.e.   
\be
Q\( 1\,,\, 0\)= i\,e^2\, \int_{\IR^3_{t}} d\zeta\,  d\tau\,\rho_{\rm mag.}
\lab{chargeop10}
\ee
with $\rho_{\rm mag.}$ given in \rf{caljmagdefstatic}, and 
\be
Q\( 0\,,\, 1\)= i\,e^2\, \int_{\IR^3_{t}} d\zeta\,  d\tau\,\(\rho_M+ \rho_G\)
\lab{chargeop01}
\ee
with $\rho_M$ and $\rho_G$ given by \rf{jmatterdefstatic} and  \rf{caljgdefstatic} respectively. The conserved magnetic and electric charges are respectively the eigenvalues of the operators \rf{chargeop10} and \rf{chargeop01}, or equivalently traces of powers of them (see \rf{chargesexpanded}). Therefore, the number of magnetic and electric charges, in lowest order in $\alpha$ and $\beta$, is equal to the rank of the gauge group $G$, since the operators \rf{chargeop10} and \rf{chargeop01} are elements of the Lie algebra of $G$. The higher charge operators live on the enveloping algebra associated to a given representation of $G$. Note that in the case of an abelian gauge group, like in Maxwell theory, the electric \rf{caljgdefstatic} and magnetic \rf{caljmagdefstatic} charge densities vanish, and the only source of the gauge field is the matter charge density \rf{jmatterdefstatic}. 

Clearly, \rf{magneticinteqstatic} and \rf{electricinteqstatic} are respectively the integral non-abelian magnetic and electric Gauss laws for Yang-Mills theory. On the left-hand side, we have, respectively, the non-abelian magnetic and electric fluxes through the two-dimensional sphere $S^2_{\infty , t}$ at spatial infinity. On the right-hand side, we have respectively the volume integral of the densities of magnetic and electric charges over the whole space $\IR^3_{t}$. 

Therefore, $\rho_{\rm mag.}$ and $\rho_G$ are respectively the magnetic and electric charge densities associated to the Yang-Mills gauge fields (gluons). Note, however, that they are not local in space-time, but instead defined on the loops parameterized by $\sigma$. Indeed, $\rho_{\rm mag.}$ and $\rho_G$  are built out of the fluxes $\mathfrak{e}_{\tau/\zeta}\(\sigma\)$ and  $\mathfrak{b}_{\tau/\zeta}\(\sigma\)$, defined in \rf{ebfrakdef}, and those are non-local quantities on space-time. So, $\rho_{\rm mag.}$ and $\rho_G$  are local densities on the loop space ${\cal L}^{(1)}$ \rf{loopspace}, of maps from the circle $S^1$ to the space-time $M$, based at the reference point $x_R$. So, we are inclined to interpret the sources of the non-abelian magnetic and electric fields as arranged on loops. That resembles  the rings of glue of \cite{polyakovring}. However, in our formulation, the rings are not the Wilson loops, but instead the quantities $\rho_{\rm mag.}$ and $\rho_G$, which are defined on the loops.  Note in addition that all the quantities are conjugated by the Wilson line, and so they transform globally under local gauge transformations (see \rf{globalfefbtransf}).

\subsection{The algebra of the charge densities}

Using the results of Appendix \ref{sec:pbchrgedensities}, we get that the electric, magnetic, and matter charge densities satisfy the following algebra under the Poisson bracket
\br
\pbr{\rho_G^a\(\tau\,,\,\zeta\)}{\rho_G^b\(\tau^{\prime}\,,\,\zeta^{\prime}\)}&=&-\vartheta\,f_{abc}\,\rho_G^c\(\tau\,,\,\zeta\)\,\delta\(\zeta-\zeta^{\prime}\)\,\delta\(\tau-\tau^{\prime}\)
\nonumber\\
\pbr{\rho_M^a\(\tau\,,\,\zeta\)}{\rho_M^b\(\tau^{\prime}\,,\,\zeta^{\prime}\)}&=&-\vartheta\,f_{abc}\,\rho_M^c\(\tau\,,\,\zeta\)\,\delta\(\zeta-\zeta^{\prime}\)\,\delta\(\tau-\tau^{\prime}\)
\\
\pbr{\rho_{\rm mag.}^a\(\tau\,,\,\zeta\)}{\rho_{\rm mag.}^b\(\tau^{\prime}\,,\,\zeta^{\prime}\)}&=&0
\nonumber
\er
where we have written them in terms of the basis of Lie algebra given in \rf{liebasis}, i.e. 
  $\rho_{G/M/{\rm mag.}}=\rho_{G/M/{\rm mag.}}^a\, T_a$. In addition, they commute among themselves 
\br
\pbr{\rho_G^a\(\tau\,,\,\zeta\)}{\rho_M^b\(\tau^{\prime}\,,\,\zeta^{\prime}\)}&=&0
\nonumber\\
\pbr{\rho_G^a\(\tau\,,\,\zeta\)}{\rho_{\rm mag.}^b\(\tau^{\prime}\,,\,\zeta^{\prime}\)}&=&0
\nonumber\\
\pbr{\rho_M^a\(\tau\,,\,\zeta\)}{\rho_{\rm mag.}^b\(\tau^{\prime}\,,\,\zeta^{\prime}\)}&=&0
\er

\section{The symmetries generated by the conserved charges} 
\label{sec:symymcharges}
\setcounter{equation}{0}

The first type of symmetries of Yang-Mills theories, which we discuss in this paper,  are the canonical transformations generated by the conserved charges, defined in \rf{chargeoperatordef}, under the Poisson brackets, i.e., the transformation of a given quantity $X$ is given by
\be
\delta X\equiv \ve\,\pbr{X}{Q_N\(\alpha\,,\,\beta\)}=\frac{\ve}{N}\,\pbr{X}{{\rm Tr}\left[Q\(\alpha\,,\,\beta\)\right]^N}
\lab{symmetrycharges}
\ee
with $\ve$ being an infinitesimal constant parameter, and $Q\(\alpha\,,\,\beta\)$ being the charge operator obtained by integrating \rf{chargeophol}.

As we have discussed on the first paragraph of Section \ref{sec:chargeoperator}, the points of the spatial sub-manifold $\IR^3_{t}$ are in one-to-one correspondence with the parameters $\(\zeta\,,\,\tau\,,\,\sigma\)$ of the scanning of $\IR^3_{t}$. Therefore, derivatives with respect to those parameters are spatial derivatives, and consequently commute with the Poisson bracket. Therefore, from \rf{chargeophol} we have that
\be
\frac{d\,\pbr{X}{Q\(\alpha\,,\,\beta\)}}{d\,\zeta}-\pbr{X}{{\cal A}\(\alpha\,,\,\beta\)}\,Q\(\alpha\,,\,\beta\)-
{\cal A}\(\alpha\,,\,\beta\)\,\pbr{X}{Q\(\alpha\,,\,\beta\)}=0
\lab{chargepoisson}
\ee
where we have assumed that the quantity $X$ is independent of the parameters $\(\zeta\,,\,\tau\,,\,\sigma\)$ of the scanning of $\IR^3_{t}$, used to obtain the charge operator through \rf{chargeophol}. The quantity $X$ can itself be obtained by an ordered integral, but it involves scanning parameters independent of $\(\zeta\,,\,\tau\,,\,\sigma\)$. Note that in the case where $X$ is a matrix, we are assuming that $X$ in \rf{chargepoisson}, and in the equations that follow below, stands for any given entry of that matrix. 

From \rf{chargeophol} we have that the inverse charge operator satisfies
\be
\frac{d\,Q^{-1}\(\alpha\,,\,\beta\)}{d\,\zeta}+Q^{-1}\(\alpha\,,\,\beta\)\,{\cal A}\(\alpha\,,\,\beta\)\,=0
\lab{invchargeophol}
\ee
Therefore, multiplying \rf{chargepoisson} from the left by $Q^{-1}\(\alpha\,,\,\beta\)$, and multiplying \rf{invchargeophol} from the right by $\pbr{X}{Q\(\alpha\,,\,\beta\)}$, and adding them up, we get that
\be
\frac{d\,\left[Q^{-1}\(\alpha\,,\,\beta\)\,\pbr{X}{Q\(\alpha\,,\,\beta\)}\right]}{d\,\zeta}-Q^{-1}\(\alpha\,,\,\beta\)\pbr{X}{{\cal A}\(\alpha\,,\,\beta\)}\,Q\(\alpha\,,\,\beta\)=0
\lab{chargepoisson2}
\ee
Integrating \rf{chargepoisson2} we get
\be
\pbr{X}{Q\(\zeta\)}=Q\(\zeta\)\int_{\zeta_i}^{\zeta} d\zeta^{\prime}\;Q^{-1}\(\zeta^{\prime}\)\,\pbr{X}{{\cal A}\(\zeta^{\prime}\)}\,Q\(\zeta^{\prime}\)
\lab{chargepoisson3}
\ee
where, to simplify the notation, we have dropped the explicit dependence on $\alpha$ and $\beta$, and have used the fact that $Q\(\zeta_i\)$ is an integration constant, corresponding to the value of the charge operator at the infinitesimal closed surface around the reference point $x_R$, labelled by $\zeta_i$. Being an integration constant, it does not depend upon the canonical variables (fields), and so Poisson commutes with any $X$. 

The same reasoning can be applied to the holonomy $V$ defined in \rf{holonomyl2fluxes}. Indeed, since the derivative with respect to $\tau$ is a spatial derivative, and so it commutes with the Poisson bracket, we get  from \rf{holonomyl2fluxes} that
\be
\frac{d\,\pbr{X}{V}}{d\,\tau} - \pbr{X}{V}\, {\cal T}_{\tau}- V\, \pbr{X}{{\cal T}_{\tau}}=0
\lab{holonomyl2fluxespoisson}
\ee
In addition, it follows from \rf{holonomyl2fluxes} that the inverse operator $V^{-1}$ satisfy
 \be
\frac{d\,V^{-1}}{d\,\tau} +  {\cal T}_{\tau}\,V^{-1}=0
\lab{holonomyl2fluxesinv}
\ee
Combining \rf{holonomyl2fluxespoisson} with \rf{holonomyl2fluxesinv} we get
\be
\frac{d\,\left[\pbr{X}{V}\,V^{-1}\right]}{d\,\tau} = V\, \pbr{X}{{\cal T}_{\tau}}\,V^{-1}
\lab{holonomyl2fluxespoisson2}
\ee
and so
\be
\pbr{X}{V\(\tau\)}\,V^{-1}\(\tau\)= \int_{\tau_i}^{\tau}d\tau^{\prime}\;  V\(\tau^{\prime}\)\, \pbr{X}{{\cal T}_{\tau}\(\tau^{\prime}\)}\,V^{-1}\(\tau^{\prime}\)
\lab{holonomyl2fluxespoisson3}
\ee
where we have used the fact that $V\(\tau_i\)$ is an integration constant, and so independent of the canonical variables. So, it Poisson commutes with any $X$. Using the fact that the Poisson bracket is a derivation, i.e., it satisfies Leibniz rule, we get from \rf{holonomyl2fluxespoisson3} that 
\be
\pbr{X}{V\(\tau\)\,T_a\,V^{-1}\(\tau\)}=\sbr{\int_{\tau_i}^{\tau}d\tau^{\prime}\;  V\(\tau^{\prime}\)\, \pbr{X}{{\cal T}_{\tau}\(\tau^{\prime}\)}\,V^{-1}\(\tau^{\prime}\)}{V\(\tau\)\,T_a\,V^{-1}\(\tau\)}
\lab{holonomyl2fluxespoisson4}
\ee

Using \rf{connectionacal} and \rf{holonomyl2fluxespoisson4}, we get that the Poisson bracket appearing on the right-hand side of \rf{chargepoisson3} becomes
\br
\pbr{X}{{\cal A}\(\zeta\)}&=&i\,e^2\,\int_{\tau_i}^{\tau_f} d\tau\; \left[ V\(\tau\)\,\pbr{X}{ {\cal J}_{\rm spatial}\(\tau\)}\,V^{-1}\(\tau\)
\right.
\lab{pbxcala}
\\
&+& \left. \sbr{\int_{\tau_i}^{\tau}d\tau^{\prime}\;  V\(\tau^{\prime}\)\, \pbr{X}{{\cal T}_{\tau}\(\tau^{\prime}\)}\,V^{-1}\(\tau^{\prime}\)}{V\(\tau\)\,{\cal J}_{\rm spatial}\(\tau\)\,V^{-1}\(\tau\)}\right]
\nonumber
\er
where ${\cal J}_{\rm spatial}$ is defined in \rf{caljdef2static}. There are two important manipulations in the evaluation of \rf{pbxcala}. The first one is to write $V\(\tau\)\,{\cal J}_{\rm spatial}\(\tau\)\,V^{-1}\(\tau\)=\frac{d\;}{d\,\tau}\int_{\tau_i}^{\tau}d\tau^{\prime}\,V\(\tau^{\prime}\)\,{\cal J}_{\rm spatial}\(\tau^{\prime}\)\,V^{-1}\(\tau^{\prime}\)$, and then integrate by parts the second term on the right hand side of \rf{pbxcala} to get
\br
&&i\,e^2\,\int_{\tau_i}^{\tau_f} d\tau\; \sbr{\int_{\tau_i}^{\tau}d\tau^{\prime}\;  V\(\tau^{\prime}\)\, \pbr{X}{{\cal T}_{\tau}\(\tau^{\prime}\)}\,V^{-1}\(\tau^{\prime}\)}{V\(\tau\)\,{\cal J}_{\rm spatial}\(\tau\)\,V^{-1}\(\tau\)}
\nonumber\\
&=&\sbr{\int_{\tau_i}^{\tau_f}d\tau\;  V\(\tau\)\, \pbr{X}{{\cal T}_{\tau}\(\tau\)}\,V^{-1}\(\tau\)}{\frac{d\,Q\(\zeta\)}{d\,\zeta}\,Q^{-1}\(\zeta\)}
\nonumber\\
&-&\int_{\tau_i}^{\tau_f} d\tau\;\sbr{ V\(\tau\)\, \pbr{X}{{\cal T}_{\tau}\(\tau\)}\,V^{-1}\(\tau\)}{{\cal A}\(\tau\)}
\lab{secondtermpbxa}
\er
where we have used the fact that ${\cal T}_{\tau}\(\tau_i\)={\cal J}_{\rm spatial}\(\tau_i\)=0$, since $\tau_i$ corresponds to the infinitesimal loop around the reference point $x_R$, and so the $\sigma$-integrals in those quantities vanish. In addition, following \rf{connectionacal}, we have denoted
\be
{\cal A}\(\tau\)=i\,e^2\,\int_{\tau_i}^{\tau}d\tau^{\prime}\,V\(\tau^{\prime}\)\,{\cal J}_{\rm spatial}\(\tau^{\prime}\)\,V^{-1}\(\tau^{\prime}\)
\lab{connectionacaltau}
\ee
and, from \rf{chargeophol}, we have used the fact that ${\cal A}\(\tau_f\)=\frac{d\,Q\(\zeta\)}{d\,\zeta}\,Q^{-1}\(\zeta\)$. 

The second important manipulation involves the use of equations \rf{deltav2} and \rf{holonomyl2fluxes} for the derivatives of $V$ with respect to $\zeta$ and $\tau$, respectively. Using the replacement \rf{replace1} and the notation introduced in \rf{ebfrakdef}, we get that equation \rf{deltav2}, for the case of the volume $\Omega$ being the three dimensional sub-manifold $\IR^3_{t}$, can be written as
\be
V\(\tau\)\,{\cal T}_{\zeta}\(\tau\)\,V^{-1}\(\tau\)=\frac{d\,V\(\tau\)}{d\,\zeta}\,V^{-1}\(\tau\)-{\cal K}\(\tau\)
\lab{deltav3}
\ee
with
\be
{\cal T}_{\zeta}=-i\,e\left[\alpha\,\mathfrak{b}_{\zeta}\(\sigma_f\)+\beta\,\mathfrak{e}_{\zeta}\(\sigma_f\)\right]
\lab{tzetadef}
\ee
and 
\br
{\cal K}\(\tau\)&=&\int_{\tau_i}^{\tau}d\tau^{\prime}\,V\(\tau^{\prime}\)\left\{-ie\int_{\sigma_i}^{\sigma_f}d\sigma W^{-1}\,
\left[\alpha\, D_{i}B_{i}+\beta\, D_{i}E_{i}\right]
W\ve_{jkl}\frac{d\,x^{j}}{d\,\sigma}\,\frac{d\,x^{k}}{d\,\tau^{\prime}}\,\frac{d\,x^{l}}{d\,\zeta}
\right.
\lab{ktaustatic}\\
&+&\left. i\,e^2\left(\beta\, \rho_G +\alpha\,\rho_{\rm mag.}
  -i\,\sbr{\alpha\,\fbt{\sigma_f}+\beta\,\fet{\sigma_f}}{\alpha\,\fbz{\sigma_f}+\beta\,\fez{\sigma_f}}\right)
\right\}\,V^{-1}\(\tau^{\prime}\)
\nonumber
\er
where we have used manipulations similar to those leading \rf{caljdef} to \rf{caljdef2} and then to  \rf{caljdef2static}. Note that in obtaining \rf{deltav3} we did not use the Yang-Mills differential equations. Indeed, as explained in \rf{deltav},  \rf{deltav3} is an identity obtained by the variation of the equation \rf{holonomyl2fluxes}, which defines $V$ in the static case. 

Note that if we take $\tau=\tau_f$, we get from the arguments leading to \rf{holonomyl3}, that \rf{deltav3} becomes
\be
{\cal K}\(\tau_f\)= \frac{d\,V\(\tau_f\)}{d\,\zeta} V^{-1}\(\tau_f\)
\lab{deltav4}
\ee

Using \rf{holonomyl2fluxes}, \rf{deltav3}, and \rf{tzetadef}, we can write the first term on the right-hand side of \rf{pbxcala}  as
\br
&&i\,e^2\,\int_{\tau_i}^{\tau_f} d\tau\; V\,\pbr{X}{ {\cal J}_{\rm spatial}}\,V^{-1}
\nonumber\\
&=&
i\,e^2\,\int_{\tau_i}^{\tau_f} d\tau V\left[\pbr{X}{ \beta\( \rho_M+ \rho_G\) +\alpha\,\rho_{\rm mag.}}
+\frac{i}{e^2}\pbr{X}{\sbr{{\cal T}_{\tau}}{{\cal T}_{\zeta}}}
\right]\,V^{-1}
\nonumber\\
&=&\int_{\tau_i}^{\tau_f} d\tau V\left[\pbr{X}{i e^2\left[ \beta\( \rho_M+ \rho_G\) +\alpha\,\rho_{\rm mag.}\right]+ \frac{d\,{\cal T}_{\zeta}}{d\,\tau}-\frac{d\,{\cal T}_{\tau}}{d\,\zeta}}\right]V^{-1}
\nonumber\\
&+&\int_{\tau_i}^{\tau_f} d\tau\left[\frac{d\;}{d\,\zeta}\(V\pbr{X}{{\cal T}_{\tau}}V^{-1}\)
-\frac{d\;}{d\,\tau}\(V\pbr{X}{{\cal T}_{\zeta}}V^{-1}\)\right]
\nonumber\\
&+&\int_{\tau_i}^{\tau_f} d\tau\,\sbr{V\(\tau\)\pbr{X}{{\cal T}_{\tau}}\,V^{-1}\(\tau\)}{{\cal K}\(\tau\)}
\lab{firsttermpbxa}
\er
where we have used the fact that $X$ does not depend upon the parameters $\(\zeta\,,\,\tau\,,\,\sigma\)$. 

Using \rf{secondtermpbxa} and \rf{firsttermpbxa} we get that  \rf{pbxcala} becomes 
\br
\pbr{X}{{\cal A}\(\zeta\)}&=&\int_{\tau_i}^{\tau_f} d\tau\left\{ V\(\tau\)\,\pbr{X}{{\cal M}}\,V^{-1}\(\tau\)
-\frac{d\;}{d\,\tau}\(V\(\tau\)\pbr{X}{{\cal T}_{\zeta}\(\tau\)}\,V^{-1}\(\tau\)\)
\right.
\nonumber\\
&+&\left. \sbr{V\(\tau\)\pbr{X}{{\cal T}_{\tau}\(\tau\)}\,V^{-1}\(\tau\)}{{\cal K}\(\tau\)- {\cal A}\(\tau\)}
\right\}
\nonumber\\
&+& \frac{d\;}{d\,\zeta}\int_{\tau_i}^{\tau_f} d\tau\(V\(\tau\)\pbr{X}{{\cal T}_{\tau}\(\tau\)}V^{-1}\(\tau\)\)
\nonumber\\
&+&\sbr{\int_{\tau_i}^{\tau_f}d\tau\;  V\(\tau\)\, \pbr{X}{{\cal T}_{\tau}\(\tau\)}\,V^{-1}\(\tau\)}{\frac{d\,Q\(\zeta\)}{d\,\zeta}\,Q^{-1}\(\zeta\)}
\lab{pbxcala2}
\er
where we have defined 
\be
{\cal M}\equiv i e^2\left[ \beta\( \rho_M+ \rho_G\) +\alpha\,\rho_{\rm mag.}\right]+ \frac{d\,{\cal T}_{\zeta}}{d\,\tau}-\frac{d\,{\cal T}_{\tau}}{d\,\zeta}
\lab{calmdef}
\ee

In the scanning of a given volume $\Omega$, the reference point $x_R$ lies on the border $\partial \Omega$, and it is kept fixed. The values $\tau_i$ and $\tau_f$ of the parameter $\tau$ correspond to infinitesimal loops around the reference point $x_R$. Therefore, we must have (see Appendix \ref{sec:scanning} for an example of scanning of $\IR^3_{t}$)
\be
\frac{d\,x^{\mu}}{d\,\tau}= \frac{d\,x^{\mu}}{d\,\zeta}=0\qquad\qquad {\rm at}\qquad \tau=\tau_i\quad {\rm and} \quad \tau=\tau_f
\ee
Therefore, from \rf{ebfrakdef} we conclude that 
\be
\mathfrak{e}_{\tau/\zeta}=\mathfrak{b}_{\tau/\zeta}=0 \qquad\qquad {\rm at}\qquad \tau=\tau_i\quad {\rm and} \quad \tau=\tau_f
\ee
Consequently, from \rf{tzetadef} we get that
\be
\int_{\tau_i}^{\tau_f} d\tau\,\frac{d\;}{d\,\tau}\(V\(\tau\)\pbr{X}{{\cal T}_{\zeta}\(\tau\)}\,V^{-1}\(\tau\)\)=0
\ee

From \rf{caljdef2static}, \rf{jmatterdefstatic}, \rf{connectionacaltau} and \rf{ktaustatic} we get that
\br
&&{\cal K}\(\tau\)- {\cal A}\(\tau\)=
\lab{kminusa}\\
&&=-ie\,\int_{\tau_i}^{\tau}d\tau^{\prime}\,V\(\tau^{\prime}\)\int_{\sigma_i}^{\sigma_f}d\sigma W^{-1}\,
\left[\alpha\, D_{i}B_{i}+\beta\, \(D_{i}E_{i}-e\,J_0\)\right]
W\ve_{jkl}\frac{d\,x^{j}}{d\,\sigma}\,\frac{d\,x^{k}}{d\,\tau^{\prime}}\,\frac{d\,x^{l}}{d\,\zeta}\,V^{-1}\(\tau^{\prime}\)
\nonumber
\er
As we discussed in Section \ref{sec:symcharges}, $A_i$ and $E_i$ are independent canonical variables, and $B_i$ is expressed in terms of $A_i$. Therefore, $D_{i}B_{i}=0$, as it is the static Bianchi identity. On the other hand, we have from  \rf{constraint2} that $\(D_{i}E_{i}-e\,J_0\)$ are the constraints. Therefore, \rf{kminusa} vanishes when the constraints hold true. 

In Appendix \ref{sec:proofcalm} we show that the quantity ${\cal M}$, defined in \rf{calmdef}, is given by
\br
{\cal M}= i\,e\,\int_{\sigma_i}^{\sigma_f}d\sigma\; W^{-1}\left[\alpha\, D_{i}B_{i}+\beta\, \(D_{i}E_{i}-e\,J_0\)\right]
W\ve_{jkl}\frac{d\,x^{j}}{d\,\sigma}\,\frac{d\,x^{k}}{d\,\tau}\,\frac{d\,x^{l}}{d\,\zeta}
\lab{calmdef2}
\er
and so, it is homogeneous on the static Bianchi identity and the constraints. However, contrary to ${\cal K}\(\tau\)- {\cal A}\(\tau\)$, the quantity ${\cal M}$ appears in \rf{pbxcala2} inside the Poisson bracket, and therefore we can not set the constraints straight to zero. The static Bianchi identity, however, can be set to zero inside the Poisson bracket. 

Consequently \rf{pbxcala2} becomes 
\br
\pbr{X}{{\cal A}\(\zeta\)}&=&ie\beta\vartheta \int_{\tau_i}^{\tau_f} d\tau V\(\tau\)\int_{\sigma_i}^{\sigma_f}d\sigma \pbr{X}{{\cal C}_a}\,\,W^{-1}\,T_a\,W V^{-1}\(\tau\) \Delta\(\sigma\,,\,\tau\,,\,\zeta\)
\nonumber\\
&+&Q\(\zeta\)\, \frac{d\;}{d\,\zeta}\left[Q^{-1}\(\zeta\)\,\int_{\tau_i}^{\tau_f} d\tau\(V\(\tau\)\pbr{X}{{\cal T}_{\tau}\(\tau\)}V^{-1}\(\tau\)\)\,Q\(\zeta\)\right]\,Q^{-1}\(\zeta\)
\nonumber\\
&+& ie\beta\vartheta\, {\cal X}
\lab{pbxcala3}
\er
where  ${\cal C}_a$ are the constraints defined in \rf{constraint2}, and where we have defined
\br
{\cal X}&\equiv& \int_{\tau_i}^{\tau_f} d\tau \left\{V\(\tau\)\int_{\sigma_i}^{\sigma_f}d\sigma \pbr{X}{W^{-1}\,T_a\,W} V^{-1}\(\tau\) \,{\cal C}_a\,\Delta\(\sigma\,,\,\tau\,,\,\zeta\)
\right. 
\lab{calxdef}\\
&-&\left. \sbr{V\(\tau\)\pbr{X}{{\cal T}_{\tau}\(\tau\)}V^{-1}\(\tau\)}{\int_{\tau_i}^{\tau}d\tau^{\prime}V\(\tau^{\prime}\)\int_{\sigma_i}^{\sigma_f}d\sigma W^{-1}\,{\cal C}\,WV^{-1}\(\tau^{\prime}\)\Delta\(\sigma ,\tau^{\prime} ,\zeta\)}
\right\}
\nonumber
\er
We have also denoted the Jacobian of the transformation $\(x^1\,,\,x^2\,,\,x^3\)\rightarrow \(\sigma\,,\,\tau\,,\,\zeta\)$ as
\be
\Delta\(\sigma\,,\,\tau\,,\,\zeta\) \equiv \vartheta\; \ve_{ijk}\frac{d\,x^{i}}{d\,\sigma}\,\frac{d\,x^{j}}{d\,\tau}\,\frac{d\,x^{k}}{d\,\zeta};\qquad\qquad\qquad \vartheta=\pm 1
\lab{jacobiandef}
\ee
The sign $\vartheta$ is due to the orientation of the scanning of $\IR^3_{t}$. If we take the vector $\frac{d\,x^i}{d\,\zeta}$, to point outward to the closed surfaces scanning $\IR^3_{t}$, we get that $\vartheta=1$, when the cross product $\ve_{ijk}\,\frac{d\,x^i}{d\,\sigma}\,\frac{d\,x^j}{d\,\tau}$ is also a vector  pointing outwards the closed surface. When that cross product points inwards to the surface, we get $\vartheta=-1$. Note that $\frac{d\,x^i}{d\,\sigma}$ and $\frac{d\,x^i}{d\,\tau}$ are tangent vectors to the closed surfaces.  Therefore, $\Delta\(\sigma\,,\,\tau\,,\,\zeta\)$ is always positive. See Appendix \ref{sec:scanning} for an example of a  scanning of $\IR^3_{t}$. 

Therefore, the relation \rf{chargepoisson3} becomes 
\br
&&\pbr{X}{Q\(\zeta\)}=\left[\int_{\tau_i}^{\tau_f} d\tau\(V\(\tau\)\pbr{X}{{\cal T}_{\tau}\(\tau\)}V^{-1}\(\tau\)\)\right]_{\zeta=\zeta}
\,Q\(\zeta\)
\nonumber
\\
&+&ie\beta\,\vartheta\, Q\(\zeta\)\,\int_{\zeta_i}^{\zeta} d\zeta^{\prime}\;Q^{-1}\(\zeta^{\prime}\)\,\int_{\tau_i}^{\tau_f} d\tau\; V\(\tau\)
\int_{\sigma_i}^{\sigma_f}d\sigma \pbr{X}{{\cal C}_a}\,W^{-1}T_aW\times
\nonumber\\
&\times&V^{-1}\(\tau\)Q\(\zeta^{\prime}\)\Delta\(\sigma\,,\,\tau\,,\,\zeta^{\prime}\)
\lab{chargepoisson4}\\
&+& ie\beta\vartheta\, Q\(\zeta\)\int_{\zeta_i}^{\zeta} d\zeta^{\prime}\;Q^{-1}\(\zeta^{\prime}\)\,{\cal X}\,Q\(\zeta^{\prime}\)
\nonumber
\er
where we have used the fact that (see Appendix \ref{sec:scanning} for an example of scanning of $\IR^3_{t}$)
\be
\frac{d\,x^{\mu}}{d\,\tau}=0\qquad\qquad {\rm at}\qquad \zeta=\zeta_i
\ee
since $\zeta_i$ corresponds to the infinitesimal closed surface around the reference point $x_R$, which is kept fixed in the scanning. Therefore, from \rf{ebfrakdef} we have that 
\be
\mathfrak{e}_{\tau}=\mathfrak{b}_{\tau}=0 \qquad\qquad {\rm at}\qquad \zeta=\zeta_i
\ee
and so, from \rf{holonomyl2fluxes}, we have that  ${\cal T}_{\tau}\(\tau\)\mid_{\zeta=\zeta_i}=0$. 

From \rf{symmetrycharges} and \rf{chargepoisson4}, we get that the transformations generated by the conserved charges are given by
\br
&&\delta X=\ve\,\pbr{X}{Q_N\(\alpha\,,\,\beta\)}
\nonumber\\
&&=\ve\, {\rm Tr}\left[Q^N\(\zeta_f\)\left\{\left[\int_{\tau_i}^{\tau_f} d\tau\(V\(\tau\)\pbr{X}{{\cal T}_{\tau}\(\tau\)}V^{-1}\(\tau\)\)\right]_{\zeta=\zeta_f}
\right. \right.
\lab{chargepoisson5}\\
&&+\left. \left. ie\beta\vartheta 
\int_{\zeta_i}^{\zeta_f} d\zeta Q^{-1}\(\zeta\)\int_{\tau_i}^{\tau_f} d\tau V\(\tau\)
\int_{\sigma_i}^{\sigma_f}d\sigma \pbr{X}{{\cal C}_a} W^{-1}T_aWV^{-1}\(\tau\)Q\(\zeta\)\Delta\(\sigma ,\tau ,\zeta\)
\right.\right.\nonumber\\
&&+\left.\left. ie\beta\vartheta\,\int_{\zeta_i}^{\zeta_f} d\zeta\;Q^{-1}\(\zeta\)\,{\cal X}\,Q\(\zeta\)
\right\}\right]
\nonumber
\er

Note that the transformations generated by the purely magnetic charges, corresponding to $\beta=0$,  act only at spatial infinity, and so, they are like asymptotic symmetries. Indeed, we have that
 \br
\delta_{(\beta=0)} X&=&\ve\,\pbr{X}{Q_N\(\alpha\,,\,0\)}
\lab{chargepoissonpuremagnetic}\\
&=&-i\,e\,\alpha\,\ve\, {\rm Tr}\left[Q^N\(\zeta_f, \alpha , 0\)\left[\int_{\tau_i}^{\tau_f} d\tau\(V_{\beta=0}\(\tau\)\pbr{X}{\mathfrak{b}_{\tau}\(\sigma_f\)}V_{\beta=0}^{-1}\(\tau\)\)\right]_{\zeta=\zeta_f}
\right]
\nonumber
\er

 By expanding both sides of \rf{chargepoisson4} in power series in $\alpha$ and $\beta$, we obtain the transformations generated by the conserved charges associated to each component in the expansion \rf{expandchargeoperator}  of the charge operator. The transformations generated by the charges, defined in \rf{chargesexpanded},  associated to the operators in lowest order  in that expansion, given in \rf{chargeop10} and \rf{chargeop01}, are  
 \br
\delta^{(N,1,0)} X&=&\ve\,\pbr{X}{Q_N\(1\,,\,0\)}
\nonumber\\
&=&-i\,e\,\ve\, {\rm Tr}\left[Q^{N-1}\(1\,,\,0\)\,\left[\int_{\tau_i}^{\tau_f} d\tau\pbr{X}{\mathfrak{b}_{\tau}\(\sigma_f\)}\right]_{\zeta=\zeta_f}\right]
\lab{transformcharge10}
\er
and
\br
\delta^{(N,0,1)} X&=&\ve\,\pbr{X}{Q_N\(0\,,\,1\)}
\nonumber\\
&=&\ve\, {\rm Tr}\left\{Q^{N-1}\(0\,,\,1\)\left[-i\,e\,\left[\int_{\tau_i}^{\tau_f} d\tau\pbr{X}{\mathfrak{e}_{\tau}\(\sigma_f\)}\right]_{\zeta=\zeta_f}
\right.\right.
\nonumber
\\
&+&\left.\left. ie\,\vartheta\, \int_{\zeta_i}^{\zeta_f} d\zeta\;\int_{\tau_i}^{\tau_f} d\tau\; 
\int_{\sigma_i}^{\sigma_f}d\sigma \pbr{X}{{\cal C}_a}\,W^{-1}T_aW\,\Delta\(\sigma\,,\,\tau\,,\,\zeta\)
\right.\right.
\lab{transformcharge01}
\\
&+&\left.\left. ie\,\vartheta\, \int_{\zeta_i}^{\zeta_f} d\zeta\;\int_{\tau_i}^{\tau_f} d\tau \int_{\sigma_i}^{\sigma_f}d\sigma \pbr{X}{W^{-1}\,T_a\,W}  \Delta\(\sigma\,,\,\tau\,,\,\zeta\)\,{\cal C}_a\right]\right\}
\nonumber
\er
As we have commented on the paragraph below \rf{chargeop10} and \rf{chargeop01}, we have rank $G$ magnetic and rank $G$ electric conserved charges associated to the operators  \rf{chargeop10} and \rf{chargeop01}, respectively. Consequently, the number of independent transformations in \rf{transformcharge10}, and similarly in \rf{transformcharge01}, is rank $G$. Note that such counting is true irrespective of the fact that in \rf{transformcharge01} we have the term  $\pbr{X}{{\cal C}_a}$ that generates  the usual local gauge transformations,  whose number of independent transformations equals the dimension of $G$. Note in addition that \rf{transformcharge10} and \rf{transformcharge01} (and also \rf{chargepoisson5}) are global transformations.

\subsection{The structure of the transformations generated by the charges}
\label{subsec:structuretransf}
 
We have shown in Section \ref{subsec:conservedcharges} that the charge operator $Q\(\alpha\,,\,\beta\)$, as well as it modes $Q\(m\,,\,n\)$, are independent of the scanning of $\IR^3_t$. Therefore, the charges $Q_N\(\alpha\,,\,\beta\)$ and $Q_N\(m\,,\,n\)$, defined respectively in  \rf{chargeoperatordef} and \rf{chargesexpanded}, are also independent of the scanning of $\IR^3_t$. Such an independency is a consequence of the integral Yang-Mills equations \rf{ymintegraleqs}, and so of the equations of motion of the Yang-Mills theories. If we denote by  $Q_N\(\alpha\,,\,\beta\)$ and $Q_N^{\prime}\(\alpha\,,\,\beta\)$ the conserved charges evaluated for two different scannings of $\IR^3_t$, we get from \rf{symmetrycharges} that
\be
\delta X- \delta^{\prime} X=\ve\,\pbr{X}{Q_N\(\alpha\,,\,\beta\)-Q_N^{\prime}\(\alpha\,,\,\beta\)}
\lab{differentscannintransf}
\ee
Since we can not use the equations of motion inside the Poisson brackets, we can not equate $Q_N\(\alpha\,,\,\beta\)$ to $Q_N^{\prime}\(\alpha\,,\,\beta\)$, in \rf{differentscannintransf}. 
 Consequently, even though the  conserved charges are independent of the scanning of $\IR^3_t$, the transformations they generate may depend upon it. Such a dependency of the transformation is encoded in the choice of paths, on the loops spaces ${\cal L}^{(2)}$, ${\cal L}^{(1)}$ and space-time itself, of the operators $Q\(\zeta\)$, $V\(\tau\)$ and $W\(\sigma\)$, respectively. But that dependency upon the choice of paths is similar to the  Weyl's non-integrable phases of the wave functions in the presence of electromagnetic or any non-abelian gauge interactions \cite{weyl1929,dirac1931}. At the end of the day, the physics does not depend upon the choice of path on which such non-integrable phases are evaluated and, as we will see, the same happens for the transformations \rf{symmetrycharges} generated by the conserved charges.

 We observe that the transformations \rf{chargepoisson5}, generated by the conserved charges, have an interesting structure. The third term on the right-hand side of \rf{chargepoisson5} vanishes when the constraints \rf{constraint2} hold true, but it is important in the  calculation of the algebra of the transformations, as we will see in the next sections. The second term in \rf{chargepoisson5} is closely related to global gauge transformations. Indeed, let us write it as
 \br
{\cal O}\(X\)&\equiv& \int_{\zeta_i}^{\zeta_f} d\zeta \int_{\tau_i}^{\tau_f} d\tau 
\int_{\sigma_i}^{\sigma_f}d\sigma \pbr{X}{{\cal C}_a} \otimes Q^{-1}\(\zeta\)V\(\tau\)W^{-1}T_aWV^{-1}\(\tau\)Q\(\zeta\)\Delta\(\sigma ,\tau ,\zeta\)
\nonumber\\
&=& \int \, d^3 x\, d_{ba}\(x\)  \, \pbr{X}{{\cal C}_a}\otimes T_b \equiv {\cal O}_b\(X\)\otimes T_b
\lab{ocaldef}
\er
where we have used the definition of the Jacobian  \rf{jacobiandef}, and the definition of the adjoint representation of the gauge group $G$
\be
g\,T_a\,g^{-1}=T_b\,d_{ba}\(g\)
\lab{adjointrepdef} 
\ee 
to denote
\br
 Q^{-1}\(\zeta\)\, V\(\tau\)\, W^{-1}\(\sigma\)\,T_a\,W\(\sigma\)\,V^{-1}\(\tau\)\,Q\(\zeta\)&=&T_b\,d_{ba}\(Q^{-1}\(\zeta\)\, V\(\tau\)\, W^{-1}\(\sigma\)\)
 \nonumber\\
&\equiv& T_b\, d_{ba}\(x\) 
 \er
Note that $ d_{ba}\(x\) $ is a non-integrable factor in the sense that  it depends upon the choice of the scanning of $\IR^3$, for the evaluation of the operators $Q$, $V$, and $W$. In order to simplify the notation, we have denoted $d_{ba}\(Q^{-1}\(\zeta\)\, V\(\tau\)\, W^{-1}\(\sigma\)\)=d_{ba}\(x\)$, meaning that we integrate $Q$ up to the surface labeled by $\zeta$, $V$ is integrated along that surface $\zeta$, up to the loop labeled by $\tau$, and $W$ is integrated along the loop labeled by $\tau$ up to the point $\sigma$. In other words, $x$ corresponds, in the scanning of $\IR^3$, to the point $\(x^1\,,\,x^2\,,\,x^3\)=\(\sigma\,,\,\tau\,,\,\zeta\)$.

Note in addition that ${\cal O}\(X\)$ is a linear operator, i.e. ${\cal O}\(X_1+X_2\)={\cal O}\(X_1\)+{\cal O}\(X_2\)$, ${\cal O}\(\gamma\, X\)=\gamma\, {\cal O}\(X\)$, for $\gamma$ not dependent upon the canonical variables, and it satisfies Leibniz rule ${\cal O}\(X_1\,X_2\)=X_1\,{\cal O}\(X_2\)+{\cal O}\(X_1\)\,X_2$, and commutes with the space derivatives ${\cal O}\(\frac{\partial\,X}{\partial\,y^i}\)=\frac{\partial\;}{\partial\,y^i}\,{\cal O}\(X\)$.
 
Using \rf{constraintgengauge} we get that the action of ${\cal O}$ on the canonical variables is given by
\br
{\cal O}_b\(\psi\(y\)\)&=&i\,e\,d_{ba}\(y\)\,R^{\psi}\(T_a\)\,\psi\(y\);\qquad\quad 
{\cal O}_b\(\pi^{\psi}\(y\)\)=-i\,e\,d_{ba}\(y\)\,\pi^{\psi}\(y\)\,R^{\psi}\(T_a\)
\nonumber\\
{\cal O}_b\(\vp\(y\)\)&=&i\,e\,d_{ba}\(y\)\,R^{\vp}\(T_a\)\,\vp\(y\);\qquad\quad 
{\cal O}_b\(\pi^{\vp}\(y\)\)=-i\,e\,d_{ba}\(y\)\,\pi^{\vp}\(y\)\,R^{\vp}\(T_a\)
\nonumber\\
{\cal O}_b\(E_i\(y\)\)&=&i\,e\,d_{ba}\(y\)\,\sbr{T_a}{E_i\(y\)}
\lab{operatorgauge1}
\er
  In addition, we have that
\br
{\cal O}_b\(A_i\(y\)\)&=&i\,e\,d_{ba}\(y\)\,\sbr{T_a}{A_i\(y\)}-\frac{\partial\, d_{ba}\(y\)}{\partial\,y^i}\,T_a
+\int d^3x\;\frac{\partial\;}{\partial\,x^i}\(d_{ba}\(x\)\,\delta^{(3)}\(x-y\)\)\,T_a
\nonumber\\
&=& - D_i \(d_{ba}\(y\)\,T_a\)+\int d^3x\;\frac{\partial\;}{\partial\,x^i}\(d_{ba}\(x\)\,\delta^{(3)}\(x-y\)\)\,T_a
\lab{operatorgauge2}
\er
Note that the surface term can not be neglected because $d_{ba}\(Q^{-1}\(\zeta\)\, V\(\tau\)\, W^{-1}\(\sigma\)\)$ does not have to fall off rapidly at spatial infinity. 
 
The first term on the right-hand side of \rf{chargepoisson5} is a surface term and will act on the fields at spatial infinity only. Note that $\zeta_f$ corresponds, in the scanning of $\IR^3_t$, to the closed surface on the border of $\IR^3_t$, i.e., the two-dimensional sphere at spatial infinity  $S^2_{\infty}$. Therefore, the integral Yang-Mills equation \rf{ymintegraleqs}, for $\Omega\equiv \IR^3_t$, implies that $Q\(\zeta_f\)\equiv Q\(\IR^3_t\)=V\(S^2_{\infty}\)$. Note  that the integral Yang-Mills equation  $Q\(\IR^3_t\)=V\(S^2_{\infty}\)$, is equivalent to the local static Yang-Mills equations $D_iB_i=0$ and $D_iE_i=e\,J_0$, i.e the static Bianchi identity and the constraints  \rf{constraint2}. Therefore
\be
{\cal C}_a=0\quad {\rm and}\quad D_iB_i=0 \quad \rightarrow \quad Q\(\zeta_f\)\equiv Q\(\IR^3_t\)=V\(S^2_{\infty}\)
\lab{surfacetransf1}
\ee
where $V\(S^2_{\infty}\)$ is obtained by integrating \rf{holonomyl2fluxes} on the two sphere at spatial infinity, i.e. $S^2_{\infty}$. Using \rf{holonomyl2fluxespoisson3} we get that
\br
\left[\int_{\tau_i}^{\tau_f} d\tau\(V\(\tau\)\pbr{X}{{\cal T}_{\tau}\(\tau\)}V^{-1}\(\tau\)\)\right]_{\zeta=\zeta_f}=
\pbr{X}{V\(S^2_{\infty}\)}\,V^{-1}\(S^2_{\infty}\)
\lab{surfacetransf2}
\er
Therefore, from \rf{ocaldef}, \rf{surfacetransf1} and \rf{surfacetransf2}, we can write \rf{chargepoisson5}  as
\br
\delta X&=&\ve\,\pbr{X}{Q_N\(\alpha\,,\,\beta\)}
\nonumber\\
&\cong&\frac{\ve}{N}\pbr{X}{{\rm Tr}\left[V^N\(S^2_{\infty}\)\right]} 
+ i\,e\,\ve\, \beta\,\vartheta\,{\rm Tr}\left[Q^N\(\zeta_f\)\,{\cal O}\(X\)\right]
\lab{chargetransformconstraint}
\er
where the symbol $\cong$ means equality when the constraints hold true. Note that the third term on the right-hand side of \rf{chargepoisson5} has been dropped in \rf{chargetransformconstraint}  because we have imposed the constraint in \rf{surfacetransf1}. However, we have not used the constraint inside the Poisson bracket. Indeed, we have that $\pbr{X}{{\rm Tr}\left[Q^N\(\IR^3_t\)\right]}\neq \pbr{X}{{\rm Tr}\left[V^N\(S^2_{\infty}\)\right]}$. 

From \rf{chargetransformconstraint} we then see that the transformations generated by the conserved charges \rf{chargeoperatordef} are made of two parts. The first one, $\pbr{X}{{\rm Tr}\left[V^N\(S^2_{\infty}\)\right]}$,  is the transformation generated, under the Poisson bracket, by the conserved charges written as surface ordered integrals, i.e., the non-abelian magnetic and electric fluxes at spatial infinity. The second part, ${\rm Tr}\left[Q^N\(\zeta_f\)\,{\cal O}\(X\)\right]$, is integrated over the whole three dimensional volume $\IR^3$, and it is related to a kind of global gauge transformation (global because $\ve$ is constant), involving  the non-integrable factor $d_{ba}\(Q^{-1}\(\zeta\)\, V\(\tau\)\, W^{-1}\(\sigma\)\)$. As we will see in the next sections, depending upon the nature of the quantity $X$, the second part of the transformation \rf{chargetransformconstraint} may also generate surface terms, which in some special cases may cancel out the first part.

As we show in the Section \ref{sec:hamiltoniantransform}, the Hamiltonian \rf{completeham} is invariant under the transformations \rf{chargepoisson5}, and so under \rf{chargetransformconstraint}, and consequently they are global symmetries of the Yang-Mills theories.

\section{The invariance of the Hamiltonian under the charge transformations}
\label{sec:hamiltoniantransform}
\setcounter{equation}{0}

We now show that the transformations \rf{chargepoisson5}, generated by the conserved charges $Q_N\(\alpha\,,\,\beta\)$, leave the Hamiltonian invariant and so they are global symmetries of the Yang-Mills theories. 
We write the complete Hamiltonian \rf{completeham} as 
\be
H_T= H_E+H_B+H_{\psi}+H_{\vp}-H_C
\lab{completehamiltonian}
\ee
with
\br
H_E &=&\frac{1}{2}\,\int d^3x\,\(E_i^a\)^2\qquad\qquad\qquad\;
H_B=\frac{1}{2}\,\int d^3x\,\(B _i^a\)^2
\nonumber\\
H_C &=&\int d^3x\, A_0^a\,{\cal C}_a\qquad\qquad\qquad\qquad
H_{\psi}=\int d^3x\left[i{\bar\psi}\gamma_i\,D_i\psi+m{\bar\psi}\,\psi\right]
\nonumber\\
H_{\vp}&=&\int d^3x\,\left[2\,\pi_{\vp}\,\pi_{\vp^{\dagger}}+\frac{1}{2}\,\(D_i\vp\)^{\dagger} D_i\vp
+V\(\mid \vp\mid\)\right]
\lab{hamiltonianparts}
\er
where we have used \rf{auxiliareqom} to eliminate $\vp_i$ and $\vp_i^{\dagger}$ from the expression of $H_{\vp}$. We shall show the invariance of each term of the Hamiltonian \rf{completehamiltonian} under the transformation \rf{chargepoisson5}. 

\subsection{The transformations of $H_{\psi}$ and $H_{\vp}$}

Using \rf{constraintgengauge} we get that the following parts of the densities of $H_{\psi}$ and $H_{\vp}$, commute with the constraints, i.e.
\br
 \pbr{{\bar \psi}\psi\(x\)}{{\cal C}_a\(y\)}=\pbr{\pi_{\vp}\pi_{\vp^{\dagger}}\(x\)}{{\cal C}_a\(y\)}
 =\pbr{V\(\mid \vp\mid\)}{{\cal C}_a\(y\)}=0
\lab{vanishingdensityPB}
\er
The  terms in the densities of $H_{\psi}$ and $H_{\vp}$ containing covariant derivatives  do not Poisson commute with the constraints because of the appearance of derivatives of the Dirac delta function, which lead to boundary terms when the integration in $\IR^3$ is performed. Indeed, from \rf{operatorgauge1} and \rf{operatorgauge2} we have that
\br
&&{\cal O}_b\({\bar\psi}\(y\)\gamma_i\,D_i\psi\(y\)\)=i\,e\,{\bar\psi}\(y\)\left[\int d^3x\frac{\partial\;}{\partial\,x^i}\(\gamma_i\,d_{ba}\(x\)\,\delta^{(3)}\(x-y\)\)\right]\, R^{\psi}\(T_a\)\,\psi\(y\)
\lab{operatorpsicurrent1}\\
&&=\left[
\vartheta\, \int_{\tau_i}^{\tau_f}d\tau\,\int_{\sigma_i}^{\sigma_f}d\sigma\,d_{ba}\(x\)\,\delta^{(3)}\(x-y\)\,\ve_{ijk}\,\frac{d\,x^j}{d\,\sigma}\,\frac{d\,x^k}{d\,\tau}\right]_{\zeta=\zeta_f}\,i\,e\, {\bar\psi}\(y\)\,\gamma_i\,R^{\psi}\(T_a\)\,\psi\(y\)
\nonumber
\er
where we have used the Abelian Gauss theorem \rf{abeliangauss}. Therefore
\br
&&{\cal O}\(i\,{\bar\psi}\(y\)\gamma_i\,D_i\psi\(y\)\)=-e\,\vartheta\,Q^{-1}\(\zeta_f\)\times
\lab{operatorpsicurrent2}\\
&&\times\left[
 \int_{\tau_i}^{\tau_f}d\tau\,\int_{\sigma_i}^{\sigma_f}d\sigma\, V\(\tau\)\,W^{-1}\(\sigma\)\,J_i^{\psi}\(y\)\,W\(\sigma\)\,V^{-1}\(\tau\)\delta^{(3)}\(x-y\)\,\ve_{ijk}\,\frac{d\,x^j}{d\,\sigma}\,\frac{d\,x^k}{d\,\tau}\right]_{\zeta=\zeta_f}\, Q\(\zeta_f\)
 \nonumber
\er
with $J_i^{\psi}$ being the space components of the spinor  current, given in \rf{mattercurr}, i.e. 
\be
J_i^{\psi}= {\bar \psi}\,\gamma_i\,R^{\psi}\(T_a\)\,\psi\, T_a
\lab{spinorcurrent}
\ee
Similarly, one can show that
\br
&&{\cal O}\(\frac{1}{2}\,\(D_i\vp\(y\)\)^{\dagger}\,D_i\vp\(y\)\)=-e\,\vartheta\,Q^{-1}\(\zeta_f\)\times
\lab{operatorpsicurrent3}\\
&&\times\left[
 \int_{\tau_i}^{\tau_f}d\tau\,\int_{\sigma_i}^{\sigma_f}d\sigma\, V\(\tau\)\,W^{-1}\(\sigma\)\,J_i^{\vp}\(y\)\,W\(\sigma\)\,V^{-1}\(\tau\)\delta^{(3)}\(x-y\)\,\ve_{ijk}\,\frac{d\,x^j}{d\,\sigma}\,\frac{d\,x^k}{d\,\tau}\right]_{\zeta=\zeta_f}\, Q\(\zeta_f\)
 \nonumber
\er
with $J_i^{\vp}$ being the space components of the  scalar current, given in \rf{mattercurr}, i.e. 
\be
J_i^{\vp}=\frac{i}{2}\left[\vp^{\dagger}\,R^{\vp}\(T_a\)\,D_{i}\vp-\(D_{i}\vp\)^{\dagger}\,R^{\vp}\(T_a\)\,\vp\right]\,T_a
\lab{scalarcurrent}
\ee

Note that all terms in $H_{\psi}$ and $H_{\vp}$ Poisson commute with the quantity ${\cal T}_{\tau}$ given in  \rf{holonomyl2fluxes}, except for the gauge field $A_i$ contained in the covariant derivatives of the matter fields.  Using \rf{pbafrakb} and \rf{pbafrake}, one can check that
\be
\pbr{H_{\psi/\vp}}{{\cal T}_{\tau}}=i\,e^2\,\beta\,\int_{\sigma_i}^{\sigma_f}d\sigma\, W^{-1}\(\sigma\) J_i^{\psi/\vp}\(y\)\,W\(\sigma\)\, \ve_{ijk}\,\frac{d\,y^j}{d\,\sigma}\,\frac{d\,y^k}{d\,\tau}
\lab{pbhpsivpttau}
\ee
with $J_i^{\psi}$ and $J_i^{\vp}$ given in \rf{spinorcurrent} and \rf{scalarcurrent}, respectively.   

Consequently we get, from \rf{vanishingdensityPB}, \rf{operatorpsicurrent2}, \rf{operatorpsicurrent3} and \rf{pbhpsivpttau}, that the first and second terms in \rf{chargepoisson5} vanish when we take $X$ to be  $H_{\psi}$ or $H_{\vp}$. In addition, the third term in \rf{chargepoisson5} vanishes when the constraints hold true. Therefore 
\be
\delta H_{\psi/\vp}=\ve\,\pbr{H_{\psi/\vp}}{Q_N\(\alpha\,,\,\beta\)}\cong 0
\lab{hpsiveinv}
\ee
where the symbol $\cong$ means equality when the constraints hold true. 

\subsection{The transformation of $H_E$}

Using \rf{constraintgengauge} we get that the density of $H_E$ commutes with the constraints, i.e.
\be
\pbr{\(E_i^b\(x\)\)^2}{{\cal C}_a\(y\)}=0
\lab{vanishingdensityPB2}
\ee
Using \rf{pbefrake} and \rf{pbefrakb2}  we get that
\br
\pbr{H_{E}}{{\cal T}_{\tau}}&=&e^2\,\beta\,
\int_{\sigma_i}^{\sigma_f}d\sigma^{\prime}\,
\sbr{\frac{d\,\mathfrak{e}_{\tau}\(\sigma^{\prime}\)}{d\,\sigma^{\prime}}}{\int_{\sigma_i}^{\sigma^{\prime}}d\sigma^{\prime\prime}\,W^{-1}\(\sigma^{\prime\prime}\)\,E_i\(z\)\,W\(\sigma^{\prime\prime}\)\,\frac{d\,z^i}{d\,\sigma^{\prime\prime}}}
\nonumber\\
&+&e^2\,\alpha\,\,\sbr{\mathfrak{b}_{\tau}\(\sigma_f\)}{\int_{\sigma_i}^{\sigma_f}d\sigma^{\prime}\,W^{-1}\(\sigma^{\prime}\)\,E_i\(y\)\,W\(\sigma^{\prime}\)\,\frac{d\,y^i}{d\sigma^{\prime}}}
\nonumber\\
&+&i\,e\,\alpha\,\frac{d\;}{d\,\tau}\,\int_{\sigma_i}^{\sigma_f}d\sigma^{\prime}\,W^{-1}\(\sigma^{\prime}\)\,E_i\(y\)\,W\(\sigma^{\prime}\)\,\frac{d\,y^i}{d\sigma^{\prime}}
\er
The surface, in the scanning of $\IR^3$, corresponding to $\zeta=\zeta_f$ is the two sphere $S^2_{\infty}$ at spatial infinity. The parameters $\sigma$ and $\tau$ on such a sphere are angles, and so the tangent vectors behave as
\be
\frac{d\,y^j}{d\,\sigma/\tau}\rightarrow r  \qquad \qquad {\rm as} \qquad \qquad r\rightarrow \infty
\lab{tangentvectorsinfinity}
\ee
From the boundary conditions \rf{boundcond} and \rf{tangentvectorsinfinity}, we have from \rf{ebfrakdef} that
\be
\mathfrak{e}_{\tau} \rightarrow \frac{1}{r^{\delta-1/2}};\qquad \mathfrak{b}_{\tau} \rightarrow \frac{1}{r^{\delta-1/2}} \qquad \qquad {\rm as} \qquad \qquad r\rightarrow \infty
\lab{ebfrakbehaviour}
\ee
and
\be
E_i\,\frac{d\,y^i}{d\sigma}\rightarrow \frac{1}{r^{\delta+1/2}}\qquad \qquad {\rm as} \qquad \qquad r\rightarrow \infty
\ee
Therefore, the first term on the right-hand side of \rf{chargepoisson5} vanishes as 
\be
\left[\int_{\tau_i}^{\tau_f} d\tau\(V\(\tau\)\pbr{H_{E}}{{\cal T}_{\tau}\(\tau\)}V^{-1}\(\tau\)\)\right]_{\zeta=\zeta_f}
\rightarrow \frac{s_1}{r^{2\,\delta}}+ \frac{s_2}{r^{\delta+1/2}}\qquad \quad {\rm as} \qquad \quad r\rightarrow \infty
\ee
where $s_1$ and $s_2$ are finite as $r\rightarrow \infty$. From \rf{vanishingdensityPB2} we have that the second term in \rf{chargepoisson5} vanishes when we take $X$ to be $H_E$. The third term in \rf{chargepoisson5} vanishes when the constraints hold true. Consequently $H_{E}$ is  invariant under the transformations \rf{chargepoisson5}, i.e. 
\be
\delta H_{E}=\ve\,\pbr{H_{E}}{Q_N\(\alpha\,,\,\beta\)}\cong 0
\lab{heinv}
\ee
where again the symbol $\cong$ means equality when the constraints hold true. 

\subsection{The transformation of $H_C$}

Using \rf{constraint2} and \rf{operatorgauge1} we get that
\br
{\cal O}_b\({\cal C}\(y\)\)&=&i\,e\,d_{ba}\(y\)\,\sbr{T_a}{{\cal C}\(y\)}+i\,e\,\sbr{T_a}{E_i\(y\)}\,\int d^3x\, \frac{\partial\;}{\partial\,x^i}\left[d_{ba}\(x\)\,\delta^{(3)}\(x-y\)\right]
\nonumber\\
&\cong&  i\,e\,\vartheta\,\sbr{T_a}{E_i\(y\)}\int_{\tau_i}^{\tau_f}d\tau\,\int_{\sigma_i}^{\sigma_f}d\sigma\,\ve_{ijk}\frac{d\,x^j}{d\,\sigma}\,\frac{d\,x^k}{d\,\tau}\,d_{ba}\(x\)\,\delta^{(3)}\(x-y\)\mid_{\zeta=\zeta_f}
\er
where the last equality is a weak equality ($\cong$), because we have used the constraints to get rid of the term $\sbr{T_a}{{\cal C}\(y\)}$. In addition, we have used the abelian Gauss theorem \rf{abeliangauss}. Therefore
\be
{\cal O}_b\(H_C\)\cong - i\,e\,\vartheta\,\int_{\tau_i}^{\tau_f}d\tau\,\int_{\sigma_i}^{\sigma_f}d\sigma\,\ve_{ijk}\frac{d\,x^j}{d\,\sigma}\,\frac{d\,x^k}{d\,\tau}\,d_{ba}\(x\)\,{\rm Tr}\(T_a\sbr{A_0\(x\)}{E_i\(x\)}\)\mid_{\zeta=\zeta_f}
\ee
Note that $A_0$ is a Lagrange multiplier, and so,  it drops out of the Poisson bracket. 

In order for the electric field  to satisfy the boundary condition \rf{boundcond}, i.e. $E_i=F_{0i}\rightarrow \frac{1}{r^{\frac{3}{2}+\delta}}$,  we need (if $A_i$ is time dependent)
\be
A_{0}\rightarrow \frac{1}{r^{1/2+\delta}}; \qquad A_{i}\rightarrow \frac{1}{r^{3/2+\delta}}\qquad {\rm as} \qquad  r\rightarrow \infty \qquad {\rm for} \qquad i=1,2,3
\lab{bcaoai}
\ee
If $A_i$ is time independent, it may fall to zero more slowly than that. 
The parameters $\sigma$ and $\tau$ on the sphere at spatial infinity, i.e. $\zeta=\zeta_f$, are angles, and so the tangent vectors behave as $\frac{d\,x^j}{d\,\sigma/\tau}\rightarrow r$, as $r\rightarrow \infty$. Therefore, we have that
\be
{\cal O}_b\(H_C\) \rightarrow \frac{1}{r^{2\,\delta}}\qquad {\rm as} \qquad  r\rightarrow \infty 
\ee
Consequently, the second term on the right-hand side of \rf{chargepoisson5} vanishes when we take $X$ to be $H_C$. 

Using \rf{holonomyl2fluxes}, \rf{pbhcmathfracke} and \rf{pbhcmathfrackb} we get that
\be
\pbr{H_{C}}{{\cal T}_{\tau}\(\tau\)}=i\,e\,\sbr{{\cal T}_{\tau}\(\tau\)}{W_R^{-1}\,A_0\(x_R\)\,W_R}
\lab{pbhccalt}
\ee
Since the reference point $x_R$ lies  on the border of $\IR^3$, and so at spatial infinity, we get from \rf{pbhccalt}, \rf{bcaoai} and \rf{ebfrakbehaviour} that
\be
\left[\int_{\tau_i}^{\tau_f} d\tau\(V\(\tau\)\pbr{H_{C}}{{\cal T}_{\tau}\(\tau\)}V^{-1}\(\tau\)\)\right]_{\zeta=\zeta_f}
\rightarrow \frac{1}{r^{2\,\delta}}\qquad \quad {\rm as} \qquad \quad r\rightarrow \infty
\ee
Therefore, the first term on the right-hand side of \rf{chargepoisson5} also vanishes when we take $X$ to be $H_C$. The third term on the right-hand side of \rf{chargepoisson5} vanishes by the imposition of the constraints. Consequently, we get that $H_C$ is invariant under the transformations generated by the conserved charges \rf{chargeoperatordef}, i.e. 
\be
\delta H_{C}=\ve\,\pbr{H_{C}}{Q_N\(\alpha\,,\,\beta\)}\cong 0
\lab{hcinv}
\ee
where again the symbol $\cong$ means equality when the constraints hold true.

\subsection{The transformation of $H_B$}

From the properties of the operator ${\cal O}$, defined in \rf{ocaldef}, we get from \rf{ebdef} that
\be
{\cal O}_b\(B_i\(y\)\)=-\ve_{ijk}\, D_j {\cal O}_b\(A_k\(y\)\)
\ee
Therefore, from \rf{hamiltonianparts} and \rf{operatorgauge2} we get that
\br
{\cal O}_b\(H_B\)&=&\int d^3y\; {\rm Tr}\left[ B_i\(y\)\,{\cal O}_b\(B_i\(y\)\)\right]
\nonumber\\
&=& \ve_{ijk}\int d^3y\,{\rm Tr}\left[B_i\(y\)\,D_j\,D_k\(d_{ba}\(y\)\,T_a\)\right]
\lab{ocalbhb1}\\
&-&
\int d^3y\,\left[{\rm Tr}\(B_i\(y\)\,T_a\)\,\frac{\partial\,{\cal S}_{ij}^{ba}\(y\)}{\partial\,y^j}-i\,e\,{\rm Tr}\(\sbr{A_j\(y\)}{B_i\(y\)}\,T_a\)\,{\cal S}_{ij}^{ba}\(y\)\right]
\nonumber
\er
where we have defined
\br
{\cal S}_{ij}^{ba}\(y\)&\equiv&  \ve_{ijk}\int d^3x\;\frac{\partial\;}{\partial\,x^k}\left[d_{ba}\(x\)\,\delta^{(3)}\(x-y\)\right]
\nonumber\\
&=&\vartheta\,\int_{\tau_i}^{\tau_f}d\tau\,\int_{\sigma_i}^{\sigma_f}d\sigma\,d_{ba}\(x\)\,\delta^{(3)}\(x-y\)
\(\frac{d\,x^i}{d\,\sigma}\,\frac{d\,x^j}{d\,\tau}-\frac{d\,x^j}{d\,\sigma}\,\frac{d\,x^i}{d\,\tau}\)\mid_{\zeta=\zeta_f}
\lab{calsabijdef}
\er
and where we have used \rf{abeliangauss}. Performing integration by parts in \rf{ocalbhb1} we get
\br
{\cal O}_b\(H_B\)&=&\ve_{ijk}\int d^3y\,\frac{\partial\;}{\partial y^j}\,\frac{\partial\;}{\partial y^k}\left[d_{ba}\(y\){\rm Tr}\(B_i\(y\)\,T_a\)\right]
+\int d^3y\,{\rm Tr}\left[D_j\,B_i\(y\)\,T_a\right]\,{\cal S}_{ij}^{ba}\(y\)
\nonumber\\
&-&\int d^3y\,\frac{\partial\;}{\partial y^j}\,\left[{\rm Tr}\(B_i\(y\)\,T_a\)\,{\cal S}_{ij}^{ba}\(y\) \right]
\lab{ocalbhb2}
\er
where we have used the fact that $\ve_{ijk}\,D_jD_k\,B_i=i\,e\,\sbr{B_i}{B_i}=0$. 
Note that the first term on the right-hand side of \rf{ocalbhb2} does not vanish because the derivatives do not commute when acting on the non-integrable factor $d_{ba}\(y\)=d_{ba}\(Q^{-1}\,V\,W^{-1}\)$. However, we show in \rf{decayratefunnyterm} that such a term decays as $\frac{1}{r^{\frac{1}{2}+\delta}}$ as   $r\rightarrow \infty$. In addition, we show in \rf{decayratefunnyterm2} that the second term in \rf{ocalbhb2} decays as $\frac{1}{r^{\frac{1}{4}+\frac{3}{2}\delta}}$ as   $r\rightarrow \infty$, and in \rf{decayratefunnyterm3} we show that the third term in \rf{ocalbhb2} vanishes identically. Therefore, ${\cal O}_b\(H_B\)$ is a surface term, and we have that it behaves as 
\be
{\cal O}_b\(H_B\) \rightarrow \frac{1}{r^{\frac{1}{4}+\frac{3}{2}\delta}}\qquad {\rm as}\qquad   r\rightarrow \infty
\lab{calohbdecayrate}
\ee

From \rf{pbbfrakb} we have that
\be
\pbr{H_B}{\mathfrak{b}_{\tau}\(\sigma\)}=0
\ee
and from \rf{pbbfrakb2} that 
\br
\pbr{H_B}{\mathfrak{e}_{\tau}\(\sigma_f\)}=\int_{\sigma_i}^{\sigma_f} d\sigma^{\prime}\(\frac{d\,y^i}{d\,\sigma^{\prime}}\, \frac{d\,y^j}{d\,\tau}-\frac{d\,y^j}{d\,\sigma^{\prime}}\,\frac{d\,y^i}{d\,\tau}\)\, W^{-1}\(\sigma^{\prime}\)\,D_jB_i\,W\(\sigma^{\prime}\)
\nonumber
\er
Therefore, using \rf{decayratefunnyterm4} and \rf{tangentvectorsinfinity} we get that 
\be
\left[\int_{\tau_i}^{\tau_f} d\tau\(V\(\tau\)\pbr{H_{B}}{{\cal T}_{\tau}\(\tau\)}V^{-1}\(\tau\)\)\right]_{\zeta=\zeta_f}
\rightarrow \frac{1}{r^{\frac{1}{4}+\frac{3}{2}\delta}} \qquad \quad {\rm as} \qquad \quad r\rightarrow \infty
\lab{surfacehbdecayrate}
\ee
Consequently, due to \rf{surfacehbdecayrate} and \rf{calohbdecayrate}, we observe that the first and second terms in \rf{chargepoisson5} vanish when we take $X$ to be $H_B$. In addition,  the third term in \rf{chargepoisson5} vanishes when we impose the constraints. Therefore, we have that $H_{B}$ is  invariant under the transformations \rf{chargepoisson5}, i.e. 
\be
\delta H_{B}=\ve\,\pbr{H_{B}}{Q_N\(\alpha\,,\,\beta\)}\cong 0
\lab{hbinv}
\ee

Therefore, we have from \rf{hpsiveinv} ,\rf{heinv}, \rf{hcinv} and \rf{hbinv}, that the complete Hamiltonian $H_T$, given in \rf{completehamiltonian}, is invariant under the transformations  \rf{chargepoisson5} generated by the conserved charges, i.e. 
\be
\delta H_{T}=\ve\,\pbr{H_{T}}{Q_N\(\alpha\,,\,\beta\)}\cong 0
\lab{hbinv2}
\ee
where again the symbol $\cong$ means equality when the constraints hold true. 

We have then shown that the complete Hamiltonian \rf{completehamiltonian} is invariant under \rf{chargepoisson5}, and so the transformations generated by the conserved charges  $Q_N\(\alpha\,,\,\beta\)$ are global symmetries of the Yang-Mills theories.

\section{The transformations of the local fields}
\label{sec:transflocalfields}
\setcounter{equation}{0}

\subsection{The matter sector}

From the canonical Poisson bracket relations \rf{pbrel} we have that the spinor   and scalar fields, $\psi$ and $\vp$,  commute with the gauge fields. Therefore, the matter fields commute with the Wilson line operators as well as with ${\cal T}_{\tau}$, given in \rf{holonomyl2fluxes}. Consequently, the first and third terms on the right-hand side of \rf{chargepoisson5} vanish when $X$ is a matter field. Indeed, ${\cal X}$, given in \rf{calxdef}, vanishes for $X$ being a matter field, and so we do not have to impose the constraints \rf{constraint2} to get rid of the third term in \rf{chargepoisson5}. Therefore, using \rf{constraintgengauge}, we get that the transformations \rf{chargepoisson5} for the spinor fields are given by
\br
\delta\, \psi\(x\,,\,t\)&=&-\ve\,e^2\,\beta\,\vartheta\,R^{\psi}\(T_a\)\,\psi\(x\,,\, t\)\times
\nonumber\\
&\times&{\rm Tr}\left[Q^N\(\zeta_f\)\,Q^{-1}\(\zeta_x\)\, V\(\tau_x\)\,W^{-1}\(\sigma_x\)\,T_a\,W\(\sigma_x\)\,V^{-1}\(\tau_x\)\,Q\(\zeta_x\)\right]
\lab{spinortransf}
\er
where we have used the fact that
\be
\delta^{(3)}\(x-y\)\,\Delta\(\sigma\,,\,\tau\,,\,\zeta\)=\delta\(\zeta-\zeta^{\prime}\)\,\delta\(\tau-\tau^{\prime}\)\,\delta\(\sigma-\sigma^{\prime}\)
\lab{jacobiandeltarel}
\ee
and where we have that, in the scanning of $\IR^3_{t}$,  $\(x^1\,,\,x^2\,,\,x^3\)$ corresponds to $\(\zeta\,,\,\tau\,,\,\sigma\)$, and $\(y^1\,,\,y^2\,,\,y^3\)$  to $\(\zeta^{\prime}\,,\,\tau^{\prime}\,,\,\sigma^{\prime}\)$. In addition, we have that $\psi\(x\,,\,t\)$ sits on a fixed point of $\IR^3_{t}$, corresponding to the parameters $\(\zeta_x\,,\,\tau_x\,,\,\sigma_x\)$. Then, $Q\(\zeta_x\)$ is the operator obtained by integrating \rf{chargeophol} up to the closed surfaced labelled by $\zeta_x$, in the scanning of $\IR^3_{t}$. Similarly, $V\(\tau_x\)$ is the operator obtained by integrating \rf{holonomyl2fluxes} up to the loop labelled by $\tau_x$, on the surface $\zeta_x$, and $W\(\sigma_x\)$ is the Wilson line operator obtained by integrating \rf{wdefa}, along the loop $\tau_x$, up to the point  labelled by $\sigma_x$. 

The transformations \rf{chargepoisson5} for the scalar fields are similar to \rf{spinortransf}, i.e.
\br
\delta\, \vp\(x\,,\,t\)&=&-\ve\,e^2\,\beta\,\vartheta\,R^{\vp}\(T_a\)\,\vp\(x\,,\, t\)\times
\nonumber\\
&\times&{\rm Tr}\left[Q^N\(\zeta_f\)\,Q^{-1}\(\zeta_x\)\, V\(\tau_x\)\,W^{-1}\(\sigma_x\)\,T_a\,W\(\sigma_x\)\,V^{-1}\(\tau_x\)\,Q\(\zeta_x\)\right]
\lab{scalartransf}
\er 

Note that the matter fields do not transform under the purely magnetic charges. Indeed, \rf{chargepoissonpuremagnetic}  and \rf{transformcharge10} become 
\be
\delta_{(\beta=0)} \psi=0;\qquad\qquad\qquad \delta^{(N,1,0)} \psi=0
\ee
and the same for the scalars $\vp$. The transformation \rf{transformcharge01} becomes
\br 
\delta^{(N,0,1)}\psi\(x\,,\, t\)&=&-\ve\,e^2\,\vartheta\,R^{\psi}\(T_a\)\,\psi\(x\,,\, t\){\rm Tr}\left[Q^{N-1}\(0\,,\,1\)\,W^{-1}\(\sigma_x\)\,T_a\,W\(\sigma_x\)\right]
\lab{spinortransf01}\\
&=&-\ve\,e^2\,\vartheta\,R^{\psi}\(W_x\)\,R^{\psi}\(T_a\)\,R^{\psi}\(W^{-1}_x\)\psi\(x\,,\, t\){\rm Tr}\left[Q^{N-1}\(0\,,\,1\)T_a\right]
\nonumber
\er
where in the last equality we have used the fact that the adjoint representation of a simple compact Lie group is real and unitary, and so orthogonal, i.e. $T_a\otimes g\,T_a\,g^{-1}=T_a\otimes T_b\,d_{ba}\(g\)=T_a\,d_{ab}\(g^{-1}\)\otimes T_b= g^{-1}\,T_b\,g\otimes T_b$. Note that $\psi$ picks the inverse of the Weyl non-integrable phase, i.e. $R^{\psi}\(W^{-1}_x\)$ \cite{weyl1929,dirac1931}. Therefore, we can interpret it as if $\psi$ is parallel transported to the reference point $x_R$, $\psi\rightarrow R^{\psi}\(W^{-1}_x\)\psi$, it  is then rotated by $R^{\psi}\(T_a\)$ at $x_R$, and it is  parallel transported back to $x$, i.e.  $\psi\rightarrow R^{\psi}\(W_x\)\,R^{\psi}\(T_a\)\,R^{\psi}\(W^{-1}_x\)\psi$. That same picture applies to the general transformation \rf{spinortransf}.

For a simple Lie group the transformation \rf{spinortransf01} vanishes for $N=1$, since ${\rm Tr}\(T_a\)=0$. The relevant values of $N$ are the orders of the Casimir operators of the gauge group $G$, and from  \rf{chargeop01} we observe that  the transformation \rf{spinortransf01} is proportional to $e^{2\,N}$. Every $G$ has a quadratic Casimir, and for $N=2$, \rf{spinortransf01} becomes
\be
\delta^{(2,0,1)}\psi\(x\,,\, t\)=-i\,\ve\,e^4\,\vartheta\,R^{\psi}\(W_x\left[\int_{\IR^3_{t}} d\zeta\,  d\tau\,\(\rho_M+ \rho_G\)\right]\,W^{-1}_x\)\,\psi\(x\,,\, t\)
\lab{spinortransf201}
\ee
where we have used \rf{chargeop01}. Note that the factor picked by the spinor field $\psi$ in \rf{spinortransf201} involves the  total electric charge operator $Q\(0\,,\,1\)$, given in \rf{chargeop01}, since its eigenvalues are conserved. The electric charge associated to $\psi$ alone is not conserved. Note in addition that \rf{spinortransf201} is a global transformation, even though the phase factor is field dependent. The same reasoning applies to the scalar fields $\vp$. 

\subsection{The gauge sector}
\label{sec:gaugesector}

We  first call the attention to the fact that the transformations \rf{chargepoissonpuremagnetic}  generated by pure magnetic conserved charges leave the gauge field $A_i$ and magnetic field $B_i$ invariant. Indeed, from \rf{pbafrakb} and  \rf{pbbfrakb} we get  they are invariant under the transformations \rf{chargepoissonpuremagnetic} 
\be
\delta_{(\beta=0)}\, A_i=0;\qquad\qquad\qquad \delta_{(\beta=0)} \, B_i=0
\ee
In order to calculate the transformations of the gauge field,  under  \rf{chargepoisson5}, we shall not 
work with Cartesian components of the one-form gauge field, i.e. $A=A_i\,dx^i$, but instead we shall use the fact that, in the scanning of $\IR^3$,  $\(x^1\,,\,x^2\,,\,x^3\)$ corresponds to $\(\zeta_x\,,\,\tau_x\,,\, \sigma_x\)$, and work with $A= A_{\sigma_x} d\sigma_x+A_{\tau_x} d\tau_x+A_{\zeta_x}d\zeta_x$, where  $A_{\sigma_x}= A_i\(x\)\,\frac{d\,x^i}{d\,\sigma_x}$, $A_{\tau_x}=A_i\(x\)\,\frac{d\,x^i}{d\,\tau_x}$, and $A_{\zeta_x}=A_i\(x\)\,\frac{d\,x^i}{d\,\zeta_x}$.

\subsubsection{The transformation of $A_i^a\(x\)\,\frac{d\,x^i}{d\,\sigma_x}$}

From \rf{holonomyl2fluxes}, \rf{pbafrakb} and \rf{pbafrake2} we conclude that the first and third terms on the left hand side of \rf{chargepoisson5} vanish when we take $X\equiv A_i^a\(x\)\,\frac{d\,x^i}{d\,\sigma_x}$. Using \rf{constraintgengauge} we get from \rf{chargepoisson5} that
\be
\pbr{A_i^a\(x\)\,\frac{d\,x^i}{d\,\sigma_x}}{Q_N\(\alpha\,,\,\beta\)}=
 ie\beta\vartheta\,{\rm Tr}\left[Q^N\(\zeta_f\) \left( I^{(1)}_{i,a,\sigma_x}+ I^{(2)}_{i,a,\sigma_x}\right)\right]
 \ee
 with
 \br
 I^{(1)}_{i,a,\sigma_x}&=&i\,e\,\int_{\zeta_i}^{\zeta_f} d\zeta Q^{-1}\(\zeta\)\int_{\tau_i}^{\tau_f} d\tau V\(\tau\)
\int_{\sigma_i}^{\sigma_f}d\sigma  W^{-1}\sbr{A_i\(x\)}{T_a}WV^{-1}\(\tau\)Q\(\zeta\)\times 
\nonumber\\
&\times& \delta^{(3)}\(x-y\)\Delta\(\sigma ,\tau ,\zeta\)\,\frac{d\,x^i}{d\,\sigma_x}
\lab{termI}
\er
and
\be
I^{(2)}_{i,a,\sigma_x}=\int_{\zeta_i}^{\zeta_f} d\zeta Q^{-1}\(\zeta\)\int_{\tau_i}^{\tau_f} d\tau V\(\tau\)
\int_{\sigma_i}^{\sigma_f}d\sigma  W^{-1}T_aWV^{-1}\(\tau\)Q\(\zeta\)\frac{\partial\,\delta^{(3)}\(x-y\)}{\partial\,y^i}\frac{d\,x^i}{d\,\sigma_x}\Delta\(\sigma ,\tau ,\zeta\)
\lab{termII0}
\ee
where, in the scanning of $\IR^3$, $\(y^1\,,\,y^2\,,\,y^3\)$ corresponds to $\(\zeta\,,\,\tau\,,\, \sigma\)$. Since $\Delta$ is the Jacobian, given in \rf{jacobiandef}, we integrate $I^{(2)}_{i,a,\sigma_x}$ by parts in the coordinate $y^i$, and use Gauss theorem on the surface term to get
\br
&&I^{(2)}_{i,a,\sigma_x}=\frac{d\,x^i}{d\,\sigma_x}\,\left[- \frac{\partial\;}{\partial\,x^i}\left[Q^{-1}\(\zeta_x\) V\(\tau_x\)
  W^{-1}\(\sigma_x\)T_aW\(\sigma_x\)V^{-1}\(\tau_x\)Q\(\zeta_x\)\right]
  \right.
  \lab{termII}\\
 &+& \left.
 Q^{-1}\(\zeta_f\)\left[
\int_{S^2_{\infty}}d\tau d\sigma V\(\tau\)  W^{-1}\(\sigma\)T_aW\(\sigma\)V^{-1}\(\tau\)\vartheta\ve_{ijk}\frac{d\,y^j}{d\,\sigma}\frac{d\,y^k}{d\,\tau}\delta^{(3)}\(x-y\)\right]_{\zeta=\zeta_f}Q\(\zeta_f\)
\right]
\nonumber
\er
since the border $S^2_{\infty}$ of $\IR^3$ corresponds to the surface $\zeta=\zeta_f$. However, the second term on the right-hand side of \rf{termII} vanishes since
\be
\ve_{ijk}\frac{d\,x^i}{d\,\sigma_x}\frac{d\,y^j}{d\,\sigma}\frac{d\,y^k}{d\,\tau}\delta^{(3)}\(x-y\)=0
\ee
as the delta function imply that $\frac{d\,x^i}{d\,\sigma_x}$ and $\frac{d\,y^j}{d\,\sigma}$, are parallel. Consequently, \rf{termII} becomes
\br
I^{(2)}_{i,a,\sigma_x}&=& -Q^{-1}\(\zeta_x\) V\(\tau_x\)\frac{d\;}{d\,\sigma_x}\left[
  W^{-1}\(\sigma_x\)T_aW\(\sigma_x\)\right]V^{-1}\(\tau_x\)Q\(\zeta_x\)
  \nonumber\\
  &=&-i\,e\,Q^{-1}\(\zeta_x\) V\(\tau_x\)
  W^{-1}\(\sigma_x\)\sbr{A_i}{T_a}W\(\sigma_x\)V^{-1}\(\tau_x\)Q\(\zeta_x\)\frac{d\,x^i}{d\,\sigma_x}
 \lab{termII2} 
\er
where we have used \rf{wdefastatic}. But \rf{termII2} cancels \rf{termI}, and so
\be
\delta\(A_i^a\(x\)\,\frac{d\,x^i}{d\,\sigma_x}\)=\ve\,\pbr{A_i^a\(x\)\,\frac{d\,x^i}{d\,\sigma_x}}{Q_N\(\alpha\,,\,\beta\)}=0
\lab{transfgaugesigma}
\ee
Therefore, the component of the gauge field $A_i^a\(x\)\,\frac{d\,x^i}{d\,\sigma_x}$, is invariant under the global transformations \rf{chargepoisson5}, generated by the conserved charges $Q_N\(\alpha\,,\,\beta\)$.

\subsubsection{The transformation of $A_i^a\(x\)\,\frac{d\,x^i}{d\,\tau_x}$}

Again, using  \rf{holonomyl2fluxes}, \rf{pbafrakb} and \rf{pbafrake2} we get that the first and third terms on the left hand side of \rf{chargepoisson5} vanish when we take $X\equiv A_i^a\(x\)\,\frac{d\,x^i}{d\,\tau_x}$. Then, using \rf{constraintgengauge} we get from \rf{chargepoisson5} that
\be
\pbr{A_i^a\(x\)\,\frac{d\,x^i}{d\,\tau_x}}{Q_N\(\alpha\,,\,\beta\)}=
 ie\beta\vartheta\,{\rm Tr}\left[Q^N\(\zeta_f\) \left( I^{(1)}_{i,a,\tau_x}+ I^{(2)}_{i,a,\tau_x}\right)\right]
  \lab{qgaugetau}
 \ee
where $I^{(s)}_{i,a,\tau_x}$ have the same expressions as $I^{(s)}_{i,a,\sigma_x}$, $s=1,2$, given in \rf{termI} and \rf{termII0}, replacing $\frac{d\,x^i}{d\,\sigma_x}$ by  $\frac{d\,x^i}{d\,\tau_x}$. Then integrating $I^{(2)}_{i,a,\tau_x}$ by parts and using Gauss theorem like we did in \rf{termII}, and using the fact that
\be
\ve_{ijk}\frac{d\,x^i}{d\,\tau_x}\frac{d\,y^j}{d\,\sigma}\frac{d\,y^k}{d\,\tau}\delta^{(3)}\(x-y\)=0
\ee
we  get that
\br
I^{(2)}_{i,a,\tau_x}&=& -Q^{-1}\(\zeta_x\) \frac{d\;}{d\,\tau_x}\left[V\(\tau_x\)
  W^{-1}\(\sigma_x\)T_aW\(\sigma_x\)V^{-1}\(\tau_x\)\right]Q\(\zeta_x\)
  \nonumber\\
  &=&-i\,e\,Q^{-1}\(\zeta_x\) V\(\tau_x\)
  W^{-1}\(\sigma_x\)\sbr{A_i}{T_a}W\(\sigma_x\)V^{-1}\(\tau_x\)Q\(\zeta_x\)\frac{d\,x^i}{d\,\tau_x}
 \lab{termII3} \\
&-& Q^{-1}\(\zeta_x\) V\(\tau_x\)\sbr{{\cal T}_{\tau}\(\tau_x\)+i\,e\, \mathfrak{b}_{\tau}\(\sigma_x\)}{W^{-1}\(\sigma_x\)T_aW\(\sigma_x\)}V^{-1}\(\tau_x\)Q\(\zeta_x\)
\nonumber
\er
where we have used \rf{holonomyl2fluxes} and \rf{bfrakdefsimple}. The first term on the right hand side of \rf{termII3} cancels with $I^{(1)}_{i,a,\tau_x}$, and so we get that
\br
&&\delta\(A_i^a\(x\)\,\frac{d\,x^i}{d\,\tau_x}\)=\ve\,\pbr{A_i^a\(x\)\,\frac{d\,x^i}{d\,\tau_x}}{Q_N\(\alpha\,,\,\beta\)}
=-ie\beta\vartheta\,\ve\,{\rm Tr}\left[Q^N\(\zeta_f\)\times
\right.
\lab{transfgaugetau}\\
 &&\left. \times\, Q^{-1}\(\zeta_x\) V\(\tau_x\)\sbr{{\cal T}_{\tau}\(\tau_x\)+i\,e\, \mathfrak{b}_{\tau}\(\sigma_x\)}{W^{-1}\(\sigma_x\)T_aW\(\sigma_x\)}V^{-1}\(\tau_x\)Q\(\zeta_x\)\right]
 \nonumber
\er

\subsubsection{The transformation of $A_i^a\(x\)\,\frac{d\,x^i}{d\,\zeta_x}$}

In such a case, all the three terms of \rf{chargepoisson5} contribute. Using  \rf{holonomyl2fluxes}, \rf{pbafrakb} and \rf{pbafrake3}, we get 
\br
&&\pbr{A_i^a\(x\)\,\frac{d\,x^i}{d\,\zeta_x}}{Q_N\(\alpha\,,\,\beta\)}=
 ie\beta\vartheta\,{\rm Tr}\left[Q^N\(\zeta_f\) \left[ I^{(1)}_{i,a,\zeta_x}+ I^{(2)}_{i,a,\zeta_x}
 \right.\right. \nonumber\\
 &&-\left.\left. \left[V\(\tau_x\)\,W^{-1}\(\sigma_x\)\, T_a\,W\(\sigma_x\)\,V^{-1}\(\tau_x\)\right]_{\zeta=\zeta_f}
  \right.\right. 
 \lab{qgaugezeta} \\
 &&+ ie\beta\vartheta\,Q^{-1}\(\zeta_x\) \left.\left.
 \left[V\(\tau_x\)W^{-1}\(\sigma_x\)\, T_a\,W\(\sigma_x\) V^{-1}\(\tau_x\)\,,\,
 \right.\right.\right.\nonumber\\
&&\left.\left.\left. \,\,\int_{\tau_i}^{\tau_x}d\tau^{\prime}V\(\tau^{\prime}\)\int_{\sigma_i}^{\sigma_f}d\sigma W^{-1}\,{\cal C}\,WV^{-1}\(\tau^{\prime}\)\Delta\(\sigma ,\tau^{\prime} ,\zeta\)\right]\,Q\(\zeta_x\)
 \right]\right]
 \nonumber
 \er
where $I^{(s)}_{i,a,\zeta_x}$ have the same expressions as $I^{(s)}_{i,a,\sigma_x}$, $s=1,2$, given in \rf{termI} and \rf{termII0}, replacing $\frac{d\,x^i}{d\,\sigma_x}$ by  $\frac{d\,x^i}{d\,\zeta_x}$. Integrating $I^{(2)}_{i,a,\zeta_x}$ by parts and using Gauss theorem like we did in \rf{termII}, we get

\br
&&I^{(2)}_{i,a,\zeta_x}=-\frac{d\;}{d\,\zeta_x}\,\left[Q^{-1}\(\zeta_x\) V\(\tau_x\)
  W^{-1}\(\sigma_x\)T_aW\(\sigma_x\)V^{-1}\(\tau_x\)Q\(\zeta_x\)\right]
  \lab{termIIagain}\\
 &+& 
 Q^{-1}\(\zeta_f\)\left[
\int_{S^2_{\infty}}d\tau d\sigma V\(\tau\)  W^{-1}\(\sigma\)T_aW\(\sigma\)V^{-1}\(\tau\)\vartheta\ve_{ijk}\frac{d\,y^j}{d\,\sigma}\frac{d\,y^k}{d\,\tau}\frac{d\,x^i}{d\,\zeta_x}\delta^{(3)}\(x-y\)\right]_{\zeta=\zeta_f}Q\(\zeta_f\)
\nonumber
\er
Using \rf{jacobiandef}, \rf{jacobiandeltarel}, \rf{chargeophol}, \rf{invchargeophol}, \rf{deltav3} and \rf{bfrakdefsimple}, we get 
\br
&&I^{(2)}_{i,a,\zeta_x}=Q^{-1}\(\zeta_f\)\left[
V\(\tau_x\)  W^{-1}\(\sigma_x\)T_aW\(\sigma_x\)V^{-1}\(\tau_x\)\right]_{\zeta=\zeta_f}Q\(\zeta_f\)
\nonumber\\
&&+Q^{-1}\(\zeta_x\)\sbr{{\cal A}\(\tau_f,\zeta_x\)}{V\(\tau_x\)
  W^{-1}\(\sigma_x\)T_aW\(\sigma_x\)V^{-1}\(\tau_x\)}Q\(\zeta_x\)
  \nonumber\\
&&-Q^{-1}\(\zeta_x\)\sbr{V\(\tau_x\)\,{\cal T}_{\zeta}\(\tau_x\)\,V^{-1}\(\tau_x\)+{\cal K}\(\tau_x\)}{V\(\tau_x\)
  W^{-1}\(\sigma_x\)T_aW\(\sigma_x\)V^{-1}\(\tau_x\)}Q\(\zeta_x\)
  \nonumber\\
&&-i\,e\,Q^{-1}\(\zeta_x\)V\(\tau_x\)\,\sbr{\mathfrak{b}_{\zeta}\(\sigma_x\)+W^{-1}\(\sigma_x\)A_i\,\frac{d\,x^i}{d\,\zeta_x}W\(\sigma_x\)}{
  W^{-1}\(\sigma_x\)T_aW\(\sigma_x\)}V^{-1}\(\tau_x\)Q\(\zeta_x\)
  \nonumber
\er
From  \rf{jacobiandef} we then get
\br
&&I^{(1)}_{i,a,\zeta_x}+ I^{(2)}_{i,a,\zeta_x}=Q^{-1}\(\zeta_f\)\left[
V\(\tau_x\)  W^{-1}\(\sigma_x\)T_aW\(\sigma_x\)V^{-1}\(\tau_x\)\right]_{\zeta=\zeta_f}Q\(\zeta_f\)
\nonumber\\
&&-Q^{-1}\(\zeta_x\)V\(\tau_x\)\,\sbr{{\cal T}_{\zeta}\(\tau_x\)+i\,e\,\mathfrak{b}_{\zeta}\(\sigma_x\)}{
  W^{-1}\(\sigma_x\)T_aW\(\sigma_x\)}V^{-1}\(\tau_x\)Q\(\zeta_x\)
  \nonumber\\
&&+Q^{-1}\(\zeta_x\)\sbr{{\cal A}\(\tau_f,\zeta_x\)- {\cal K}\(\tau_x,\zeta_x\)}{V\(\tau_x\)
  W^{-1}\(\sigma_x\)T_aW\(\sigma_x\)V^{-1}\(\tau_x\)}Q\(\zeta_x\)  
  \lab{i1plusi2zetax}
  \er
Therefore, using \rf{i1plusi2zetax}, \rf{kminusa}, \rf{constraint2}, and  the static Bianchi identity, i.e. $D_iB_i=0$, we get that \rf{qgaugezeta} becomes 
\br
&&\delta\(A_i^a\(x\)\,\frac{d\,x^i}{d\,\zeta_x}\)=\ve\,\pbr{A_i^a\(x\)\,\frac{d\,x^i}{d\,\zeta_x}}{Q_N\(\alpha\,,\,\beta\)}
=-ie\beta\vartheta\,\ve\,{\rm Tr}\left[Q^N\(\zeta_f\)Q^{-1}\(\zeta_x\)\times
\right.
 \nonumber\\
 &&\left. \times\,  \left\{V\(\tau_x\)\sbr{{\cal T}_{\zeta}\(\tau_x\)+i\,e\, \mathfrak{b}_{\zeta}\(\sigma_x\)}{W^{-1}\(\sigma_x\)T_aW\(\sigma_x\)}V^{-1}\(\tau_x\)\right.\right.
\lab{transfgaugezeta}\\
&& - \left. \left.\sbr{{\cal A}\(\tau_f,\zeta_x\)- {\cal A}\(\tau_x,\zeta_x\)}{V\(\tau_x\)
  W^{-1}\(\sigma_x\)T_aW\(\sigma_x\)V^{-1}\(\tau_x\)}\right\}
 Q\(\zeta_x\)\right]
 \nonumber
\er

\section{The transformations  of the fluxes $\mathfrak{b}_{\tau/\zeta}$ and $\mathfrak{e}_{\tau/\zeta}$}
\label{sec:transffluxes}
\setcounter{equation}{0}

Using \rf{nicepbw} and \rf{transfgaugesigma}, we get that the Wilson line is invariant under the transformations generated by the charges 
\be
\delta\,W\(\sigma\)=\ve\,\pbr{W\(\sigma\)}{Q_N\(\alpha\,,\,\beta\)}=0
\lab{transfwilsonline}
\ee
Therefore, from \rf{bfrakdefsimple2}, we have that the flux $\mathfrak{b}_{\tau/\zeta}\(\sigma_f\)$, integrated over an entire loop, is invariant too, i.e. 
\be
\delta\,\mathfrak{b}_{\tau/\zeta}\(\sigma_f\)=\ve\,\pbr{\mathfrak{b}_{\tau/\zeta}\(\sigma_f\)}{Q_N\(\alpha\,,\,\beta\)}=0
\lab{transfmagnetixfluxfull}
\ee
On the other hand, if we do not integrate over the entire loop, we get from \rf{bfrakdefsimple}, \rf{transfgaugetau}, and \rf{transfgaugezeta} 
\br
\delta\,\mathfrak{b}_{\tau/\zeta}\(\sigma\)&=&\ve\,\pbr{\mathfrak{b}_{\tau/\zeta}\(\sigma\)}{Q_N\(\alpha\,,\,\beta\)}\qquad\qquad\qquad\qquad \qquad\qquad{\rm for}\quad \sigma<\sigma_f
\\
 &=&i\,e\,\beta\,\vartheta\,\ve\; T_a\;{\rm Tr}\left[Q^N\(\zeta_f\)\, Q^{-1}\(\zeta\) V\(\tau\)\sbr{{\cal T}_{\tau/\zeta}\(\tau\)+i\,e\, \mathfrak{b}_{\tau/\zeta}\(\sigma\)}{T_a}V^{-1}\(\tau\)Q\(\zeta\)\right]
 \nonumber
\er
where the final point of the loop, over which $\mathfrak{b}_{\tau/\zeta}\(\sigma\)$ is integrated, is $\(x^1\,,\,x^2\,,\,x^2\)=\(\zeta\,,\,\tau\,,\,\sigma\)$, and where we have used \rf{adjointrepproperty}.

From \rf{chargepoisson3}, \rf{pbxcala} and \rf{pbmathfracketauzetacalt} we get that
\br
&&\{\mathfrak{e}_{\tau/\zeta}\(\sigma\)\,\overset{\otimes}{,}\,Q\(\zeta_f\)\}_{PB}=
i\,e^2\,\one\otimes Q\(\zeta_f\)\int_{\zeta_i}^{\zeta_f} d\zeta^{\prime}\int_{\tau_i}^{\tau_f} d\tau^{\prime}\;
\one \otimes \(Q^{-1}\(\zeta^{\prime}\)\,V\(\tau^{\prime}\)\)\times
\nonumber\\
&&\times \,
\{\mathfrak{e}_{\tau/\zeta}\(\sigma\)\,\overset{\otimes}{,}\, {\cal J}_{\rm spatial}\(\tau^{\prime}\)\}_{PB}\,\one \otimes \(V^{-1}\(\tau^{\prime}\)\,Q\(\zeta^{\prime}\)\)
\lab{pbemathfrackchargeop}
\er
with ${\cal J}_{\rm spatial}$  defined in \rf{caljdef2static}, and where the points of the loop where $\mathfrak{e}_{\tau/\zeta}\(\sigma\)$ sits are $x^i\(\sigma\,,\,\tau\,,\,\zeta\)$, and the integral are over the points $y^i\(\sigma^{\prime}\,,\,\tau^{\prime}\,,\,\zeta^{\prime}\)$.

From \rf{pbemathfrackchargeop}, \rf{emathfrackjspatial} and \rf{symmetrycharges} we get that
\br
\delta\,\mathfrak{e}_{\tau/\zeta}\(\sigma\)&=&\ve\,\pbr{\mathfrak{e}_{\tau/\zeta}\(\sigma\)}{Q_N\(\alpha\,,\,\beta\)}=
\\
&=&-\ve\,e^2\,\beta\,\vartheta\,\sbr{T_a}{\mathfrak{e}_{\tau/\zeta}\(\sigma\)}\;{\rm Tr}\(Q\(\zeta_f\)^N\,Q^{-1}\(\zeta\)\,V\(\tau\)\,T_a\,V^{-1}\(\tau\)\,Q\(\zeta\)\)
\nonumber
\er

\section{The transformations of the Wilson line operators}
\label{sec:specialwilson}
\setcounter{equation}{0}

We have seen in \rf{transfwilsonline} that the Wilson line $W(\sigma)$, defined in \rf{wdefa}, is invariant under the transformations  generated by the charges. However,  Wilson lines defined on different paths  are not necessarily invariant under those transformations, and we shall now consider them.  

As explained before, we scan $\mathbb{R}^3_t$, at a given fixed time $t$, with closed surfaces labelled by $\zeta$, based at the reference point $x_R$. Each such surface is scanned by loops, starting and ending at $x_R$, and labelled by $\tau$. Every loop is parameterized by $\sigma$. Therefore, any point of $\mathbb{R}^3_t$ belongs to only one surface, to only one loop at that surface, and to only one point of that loop. Consequently,  we have that the points of $\mathbb{R}^3_t$ are functions of those three parameters, i.e. $x^i=x^i\(\sigma ,\tau ,\zeta\)$. The reference point $x_R$ is the initial point of every loop, corresponding to $\sigma=\sigma_i$. In addition, $\tau_i$ corresponds to the infinitesimal loop around $x_R$, and $\zeta_i$ to the infinitesimal closed surface around $x_R$. The Wilson lines $W(\sigma)$, defined in \rf{wdefa}, are associated to  paths in the $\sigma$-direction, where $\tau$ and $\zeta$ are kept fixed.  

We shall consider Wilson lines, $\omega_\tau$  and $\omega_\zeta$, associated with the same gauge connection $A_i$, but defined on paths in the $\tau$-direction and $\zeta$-direction, respectively, in the scanning of $\mathbb{R}^3_t$.  In other words, $\omega_\tau$ is the holonomy of $A_i$, on a path $x^i\(\tau\)$, where $\sigma$ and $\zeta$ are kept fixed, and starting at the reference point $x_R\equiv x^i\(\tau_i\)$. In its turn, $\omega_\zeta$  is the holonomy of $A_i$, on a path $x^i\(\zeta\)$, where $\sigma$ and $\tau$ are kept fixed, and starting at the reference point $x_R\equiv x^i\(\zeta_i\)$. Such Wilson line operators are obtained by integrating the holonomy equations
\be
\frac{d \omega_{\tau/\zeta}}{d\tau/\zeta} + i\,e\,A_i \,\frac{dx^i}{d\tau/\zeta}\,\omega_{\tau/\zeta} = 0;\qquad \qquad \qquad  \omega_{\tau}(\tau_i)=\omega_{\zeta}(\zeta_i) = W_R
\lab{holeqwtwz}
\ee
where we have chosen the integration constant to be  $W_R$, which is the same as the value of the Wilson line $W\(\sigma\)$, defined in \rf{wdefa}, at the reference point $x_R$. 

Since the holonomy equation \rf{holeqwtwz} is similar to the one satisfied by  $W(\sigma)$, i.e. \rf{wdefa},  we can use the same reasonings leading to \rf{nicepbw},  to write the variations of the holonomies  $\omega_\tau$ and $\omega_\zeta$, under the  transformations \rf{symmetrycharges},  generated by the charges,   as
\br
\delta \omega_{\xi}(\xi) &=& \vareps\pbr{\omega_{\xi}(\xi)}{Q_N(\alpha,\,\beta)} 
\nonumber\\
&=&- i\,e\,\vareps\,\omega_{\xi}(\xi)\int_{\xi_i}^{\xi}d\xi_x\,\omega^{-1}_{\xi}(\xi_x)\,\pbr{A_i\(x\)}{Q_N\(\alpha,\beta\)}\,\omega_{\xi}(\xi_x)\,\frac{d\,x^i}{d\,\xi_x}
\lab{transfwilsontauzeta}
\er
where $\xi$ stands for either $\tau$ or $\zeta$, and where the point $x^i$ corresponds to $x^i=x^i\(\sigma,\tau_x,\zeta\)$ (with $\sigma$ and $\zeta$ fixed), for the case $\xi\equiv \tau$, and to $x^i=x^i\(\sigma,\tau,\zeta_x\)$ (with $\sigma$ and $\tau$ fixed), for the case $\xi\equiv \zeta$. 

We shall need some formulas, which take the same form for both cases, $\xi\equiv \tau$ and $\xi\equiv \zeta$. Using \rf{bfrakdefsimple} and \rf{holeqwtwz} we get
\be
i\,e\,\mathfrak{b}_{\xi}\(\sigma\)=\frac{d\,\(W^{-1}\(\sigma\)\,\omega_{\xi}\(\xi_x\)\)}{d\,\xi_x}\,\(W^{-1}\(\sigma\)\,\omega_{\xi}\(\xi_x\)\)^{-1}
\lab{mathfrakbniceder}
\ee
From \rf{adjointrepdef} and \rf{adjointrepproperty} we have that 
\be
\omega^{-1}_{\xi}(\xi_x)\, T_a\,\omega_{\xi}(\xi_x)\otimes W^{-1}\(\sigma\)T_aW\(\sigma\)=
T_a\otimes W^{-1}\(\sigma\)\,\omega_{\xi}(\xi_x)\,T_a\,\omega^{-1}_{\xi}(\xi_x)\,W\(\sigma\)
\lab{adjointomegaw}
\ee
and so
\be
\sbr{i\,e\, \mathfrak{b}_{\xi}\(\sigma\)}{W^{-1}\(\sigma\)\,\omega_{\xi}(\xi_x)\,T_a\,\omega^{-1}_{\xi}(\xi_x)\,W\(\sigma\)}=\frac{d\;}{d\,\xi_x}\left[W^{-1}\(\sigma\)\,\omega_{\xi}(\xi_x)\,T_a\,\omega^{-1}_{\xi}(\xi_x)\,W\(\sigma_x\)\right]
\lab{totaldermatfrakb}
\ee
From \rf {transfwilsontauzeta} and \rf{transfgaugetau}  we get, for the case $\xi\equiv \tau$,  
\br
&&\delta \omega_{\tau}(\tau) =
-\ve\,e^2\,\beta\,\vartheta\,\,\omega_{\tau}(\tau)\int_{\tau_i}^{\tau}d\tau_x\,\omega^{-1}_{\tau}(\tau_x)\, T_a\,\omega_{\tau}(\tau_x) \times
\lab{varomegaxi}\\
&&\times {\rm Tr}\left[Q^N\(\zeta_f\)
 Q^{-1}\(\zeta\) V\(\tau_x\)\sbr{{\cal T}_{\tau}\(\tau_x\)+i\,e\, \mathfrak{b}_{\tau}\(\sigma\)}{W^{-1}\(\sigma\)T_aW\(\sigma\)}V^{-1}\(\tau_x\)Q\(\zeta\)\right]
 \nonumber
\er
Using \rf{holonomyl2fluxes}, \rf{adjointomegaw} and \rf{totaldermatfrakb} we get from \rf{varomegaxi} that 
\br
&&\delta \omega_{\tau}(\tau) =
-\ve\,e^2\,\beta\,\vartheta\,\,\omega_{\tau}(\tau)\, T_a\,\int_{\tau_i}^{\tau}d\tau_x\,\times
\lab{varomegatau}\\
&&\times {\rm Tr}\left[Q^N\(\zeta_f\)
 Q^{-1}\(\zeta\) \frac{d\;}{d\,\tau_x}\left[V\(\tau_x\)W^{-1}\(\sigma\)\,\omega_{\tau}(\tau_x)\,T_a\,\omega^{-1}_{\tau}(\tau_x)\,W\(\sigma\)\,V^{-1}\(\tau_x\)\right]Q\(\zeta\)\right]
 \nonumber
\er
 Using the boundary condition given in \rf{holeqwtwz}, and the fact that $V\(\tau_i\)=V_R$ must lie in the center of the gauge group (see \rf{intconstcenter}), we get 
\br
&&\delta \omega_{\tau}(\tau) =
-\ve\,e^2\,\beta\,\vartheta\,\,\omega_{\tau}(\tau)\, T_a\,\times
\lab{varomegataufinal}\\
&&\times {\rm Tr}\left[Q^N\(\zeta_f\)
 Q^{-1}\(\zeta\) \left[V\(\tau\)W^{-1}\(\sigma\)\,\omega_{\tau}(\tau)\,T_a\,\omega^{-1}_{\tau}(\tau)\,W\(\sigma\)\,V^{-1}\(\tau\) - T_a\right]Q\(\zeta\)\right]
 \nonumber
\er
Note that, using  \rf{adjointrepdef} and \rf{adjointrepproperty}, the transformation \rf{varomegataufinal} can be written as
\br
&&\delta \omega_{\tau}(\tau) =
\ve\,e^2\,\beta\,\vartheta\,\left\{ \omega_{\tau}(\tau)\, T_a\,{\rm Tr}\left[Q^N\(\zeta_f\)
 Q^{-1}\(\zeta\)\,T_a\,Q\(\zeta\)\right]\right.
\lab{varomegataufinal2}\\
&&-\left. T_a\,\omega_{\tau}(\tau)\, {\rm Tr}\left[Q^N\(\zeta_f\)
 Q^{-1}\(\zeta\) \,V\(\tau\)W^{-1}\(\sigma\)\,T_a\,W\(\sigma\)\,V^{-1}\(\tau\) Q\(\zeta\)\right]\right\}
 \nonumber
\er
where, as explained above,  $\sigma$ and $\zeta$ are kept fixed along the path defining the holonomy $\omega_{\tau}$.

From \rf {transfwilsontauzeta} and \rf{transfgaugezeta}  we get, for the case $\xi\equiv \zeta$,  
\br
&&\delta \omega_{\zeta}(\zeta) =
-\ve\,e^2\,\beta\,\vartheta\,\,\omega_{\zeta}(\zeta)\int_{\zeta_i}^{\zeta}d\zeta_x\,\omega^{-1}_{\zeta}(\zeta_x)\, T_a\,\omega_{\zeta}(\zeta_x) {\rm Tr}\left[Q^N\(\zeta_f\)Q^{-1}\(\zeta_x\)\times
\right.
 \nonumber\\
 &&\left. \times\,  \left\{V\(\tau\)\sbr{{\cal T}_{\zeta}\(\tau\)+i\,e\, \mathfrak{b}_{\zeta}\(\sigma\)}{W^{-1}\(\sigma\)T_aW\(\sigma\)}V^{-1}\(\tau\)\right.\right.
\lab{varomegazeta1}\\
&& - \left. \left.\sbr{{\cal A}\(\tau_f,\zeta_x\)- {\cal A}\(\tau,\zeta_x\)}{V\(\tau\)
  W^{-1}\(\sigma\)T_aW\(\sigma\)V^{-1}\(\tau\)}\right\}
 Q\(\zeta_x\)\right]
 \nonumber
\er
Using \rf{deltav3}, \rf{adjointomegaw} and \rf{totaldermatfrakb}, we get from \rf{varomegazeta1}
\br
&&\delta \omega_{\zeta}(\zeta) =
-\ve\,e^2\,\beta\,\vartheta\,\,\omega_{\zeta}(\zeta)\, T_a\,\int_{\zeta_i}^{\zeta}d\zeta_x\, {\rm Tr}\left[Q^N\(\zeta_f\) \,Q^{-1}\(\zeta_x\)\times\right.
\nonumber\\
&&\times \left. \left\{\sbr{{\cal A}\(\tau,\zeta_x\)-{\cal K}\(\tau,\zeta_x\)-{\cal A}\(\tau_f,\zeta_x\)}{V\(\tau\)\,W^{-1}\(\sigma\)\,\omega_{\zeta}(\zeta_x)\,T_a\,\omega^{-1}_{\zeta}(\zeta_x)\,W\(\sigma\)\,V^{-1}\(\tau\)}
\right.\right.
\nonumber\\
&&+\left.\left. \frac{d\;}{d\,\zeta_x}\left[ V\(\tau\)\,W^{-1}\(\sigma\)\,\omega_{\zeta}(\zeta_x)\,T_a\,\omega^{-1}_{\zeta}(\zeta_x)\,W\(\sigma\)\,V^{-1}\(\tau\)\right]\right\} Q\(\zeta_x\)\right]
\lab{varomegazeta2} 
\er
Using \rf{kminusa} we see that, when the constraints \rf{constraint2}, and the Bianchi identity, $D_iB_i=0$, hold true, we have that ${\cal A}\(\tau,\zeta_x\)-{\cal K}\(\tau,\zeta_x\)$, vanishes. From \rf{chargeophol} we have that ${\cal A}\(\tau_f,\zeta_x\)=\frac{d\,Q\(\zeta_x\)}{d\,\zeta_x}\,Q^{-1}\(\zeta_x\)$. Therefore 
\br
&&\delta \omega_{\zeta}(\zeta) \cong 
-\ve\,e^2\,\beta\,\vartheta\,\,\omega_{\zeta}(\zeta)\, T_a\,\int_{\zeta_i}^{\zeta}d\zeta_x\, {\rm Tr}\left[Q^N\(\zeta_f\) \,\times\right.
\nonumber\\
&&\left.\left. \frac{d\;}{d\,\zeta_x}\left[ Q^{-1}\(\zeta_x\)\,V\(\tau\)\,W^{-1}\(\sigma\)\,\omega_{\zeta}(\zeta_x)\,T_a\,\omega^{-1}_{\zeta}(\zeta_x)\,W\(\sigma\)\,V^{-1}\(\tau\)\,Q\(\zeta_x\)\right]\right\} \right]
\lab{varomegazeta3} 
\er
where the symbol $\cong$ means equality when the constraints \rf{constraint2} hold true. 

 Therefore, using the fact that $\zeta_i$ corresponds to the infinitesimal surface around the reference point $x_R$,  that $Q\(\zeta_i\)=V_R$, lies in the center of the gauge group,  from the boundary conditions given in \rf{holeqwtwz}, and the relations \rf{adjointrepdef} and \rf{adjointrepproperty},  we get
\br
&&\delta \omega_{\zeta}(\zeta) \cong 
\ve\,e^2\,\beta\,\vartheta\,\,\left\{\omega_{\zeta}(\zeta)\, T_a\, {\rm Tr}\left[Q^N\(\zeta_f\) \,T_a\right]\right.
\lab{varomegazetafinal}\\
&&- \left. T_a\, \omega_{\zeta}(\zeta)\,  {\rm Tr}\left[Q^N\(\zeta_f\) \,
Q^{-1}\(\zeta\)\,V\(\tau\)\,W^{-1}\(\sigma\)\,T_a\,W\(\sigma\)\,V^{-1}\(\tau\)\,Q\(\zeta\)\right]\right\}
\nonumber
\er
where, as explained above, $\tau$ and $\sigma$ are kept fixed along the path defining the holonomy $\omega_{\zeta}$.
 
Note, from \rf{varomegataufinal2} and \rf{varomegazetafinal}, that under the global transformations generated by the conserved charges,  the infinitesimal variations of $\omega_{\tau/\zeta}$ are in linear in the products $T_a\,\omega_{\tau/\zeta}$ and $\omega_{\tau/\zeta}\,T_a$. Remember that,  in the definition of the Wilson lines $ \omega_{\tau/\zeta}$, given in\rf{holeqwtwz}, the gauge field $A_i$ lies in any  chosen  faithful representation of the gauge group. Therefore, the  generators $T_a$, appearing multiplying $ \omega_{\tau/\zeta}$, in \rf{varomegataufinal2} and \rf{varomegazetafinal}, lies in that same representation.  

Therefore, the Wilson lines $ \omega_{\tau/\zeta}$ transform under the global symmetries generated by the conserved charges like ''primary fields'', in the sense that they are rotated by the matrix $T_a$, and multiplied by  non-integrable phases, containing the operators $W$, $V$, and $Q$.

\section{The Fundamental Poisson Bracket Relation}
\label{sec:fpr}
\setcounter{equation}{0}

The algebra of the conserved charges under the Poisson bracket can be calculated using the so-called Fundamental Poisson Bracket Relation (FPR) and the Sklyanin relation. Those are well-known structures in the theory of two-dimensional integrable field theories \cite{faddeevleshouches,babelonbook,retore}, but we now show that Yang-Mills theories in four dimensions possess similar structures on the loop space ${\cal L}^{(2)}$. We will show that the one-form connection ${\cal A}$ on the loop space ${\cal L}^{(2)}$, defined in \rf{connectionacal}, satisfies a FPR. We start by evaluating the Poisson bracket among the entries of the matrix ${\cal A}$ when it is evaluated on a purely spatial surface, i.e., the connection given in \rf{connectionacaltau} for $\tau=\tau_f$. We shall consider one connection on a purely spatial surface labelled by $\zeta_1$, and with its points parameterized by $x^i\equiv x^i=\(\sigma_1\,,\,\tau_1\,,\, \zeta_1\)$, and for the arbitrary parameters $\alpha_1$ and $\beta_1$. Similarly, the second connection is defined on a purely spatial surface labelled by $\zeta_2$, with points parameterized by $y^i\equiv y^i=\(\sigma_2\,,\,\tau_2\,,\, \zeta_2\)$, and  parameters $\alpha_2$ and $\beta_2$.

Using \rf{pbxcala3}, \rf{pbcaltcala} and \rf{pbwcala} we have that
\br
&&\{{\cal A}\(\zeta_1\,,\,\alpha_1\,,\,\beta_1\)\,\overset{\otimes}{,}\,{\cal A}\(\zeta_2\,,\,\alpha_2\,,\,\beta_2\)\}_{PB}=i\,e\,\beta_1\,\vartheta\,T_b\otimes \one \times
\nonumber\\
&&\times \int_{\tau_i}^{\tau_f} d\tau_1\int_{\sigma_i}^{\sigma_f}d\sigma_1\,
\Delta\(\sigma_1\,,\,\tau_1\,,\,\zeta_1\)\, d_{ba}\(V_{(1)}\(\tau_1\)\,W^{-1}\(\sigma_1\)\)\,
\{{\cal C}_a\(x\)\,\overset{\otimes}{,}\,{\cal A}\(\zeta_2\,,\,\alpha_2\,,\,\beta_2\)\}_{PB}
\nonumber\\
&&+i\,e^3\,\beta_1\,\beta_2\,\vartheta\,\left[\frac{d\;}{d\,\zeta_1}\left[{\cal Y}\(\zeta_1\,,\,\tau_f\)\,\delta\(\zeta_1-\zeta_2\)\right]- \delta\(\zeta_1-\zeta_2\)\,\sbr{{\cal A}\(\zeta_1\,,\,\alpha_1\,,\,\beta_1\)\otimes \one}{{\cal Y}\(\zeta_1\,,\,\tau_f\)}\right]
\nonumber\\
&&+e^4\,\beta_1^2\,\beta_2\,\delta\(\zeta_1-\zeta_2\)\,\int_{\tau_i}^{\tau_f} d\tau_1\,
\sbr{\frac{d\,{\cal Y}\(\zeta_1\,,\,\tau_1\)}{d\,\tau_1}}{T_b\otimes \one}\,\times
\nonumber\\
&&\times\, \int_{\tau_i}^{\tau_1}d\tau_1^{\prime}\,\int_{\sigma_i}^{\sigma_f}d\sigma_1\,\Delta\(\sigma_1\,,\,\tau_1^{\prime}\,,\,\zeta_1\)\, d_{ba}\(V_{(1)}\(\tau_1^{\prime}\)\,W^{-1}\(\sigma_1\)\)\, {\cal C}_a\(\sigma_1\,,\,\tau_1^{\prime}\,,\,\zeta_1\)
\lab{prepbacalacal}
\er
where we have used \rf{chargeophol} and \rf{adjointrepdef}, and have defined
\br
&&{\cal Y}\(\zeta_1\,,\,\tau\)\equiv \int_{\tau_i}^{\tau} d\tau_1\,V_{(1)}\(\tau_1\)\otimes V_{(2)}\(\tau_1\)\sbr{\IC}{
\,\mathfrak{e}_{\tau}\(\sigma_f\,,\,\tau_1\,,\,\zeta_1\)\,\otimes\one}\,V_{(1)}^{-1}\(\tau_1\)\otimes V_{(2)}^{-1}\(\tau_1\)
\nonumber\\
&&=-\int_{\tau_i}^{\tau} d\tau_1\,V_{(1)}\(\tau_1\)\otimes V_{(2)}\(\tau_1\)\sbr{\IC}{
\one \otimes \mathfrak{e}_{\tau}\(\sigma_f\,,\,\tau_1\,,\,\zeta_1\)}\,V_{(1)}^{-1}\(\tau_1\)\otimes V_{(2)}^{-1}\(\tau_1\)
\lab{calrdef}
\er
where we have used \rf{casimirzero}, and where the subscript $(s)$ in $V_{(s)}$, $s=1,2$, means they depend upon the parameters $\alpha_s$ and $\beta_s$. 

We now apply \rf{pbxcala3} on the first term on the right hand side of \rf{prepbacalacal},  and use \rf{calowzero}, \rf{calocalttau} and \rf{calicdtensor} to get
\br
&&T_b\otimes \one \
 \int_{\tau_i}^{\tau_f} d\tau_1\int_{\sigma_i}^{\sigma_f}d\sigma_1\,
\Delta\(\sigma_1\,,\,\tau_1\,,\,\zeta_1\)\, d_{ba}\(x\)\,
\{{\cal C}_a\(x\)\,\overset{\otimes}{,}\,{\cal A}\(\zeta_2\,,\,\alpha_2\,,\,\beta_2\)\}_{PB}=
\nonumber\\
&&=-e^2\,\beta_2\,\left\{
\vartheta\,\delta\(\zeta_1-\zeta_2\)\,\int_{\tau_i}^{\tau_f} d\tau_2\int_{\sigma_i}^{\sigma_f}d\sigma_2\,
\Delta\(\sigma_2\,,\,\tau_2\,,\,\zeta_2\)\times
\right. 
\nonumber\\
&&\times\, V_{(1)}\(\tau_2\)\otimes V_{(2)}\(\tau_2\)\,\sbr{\IC}{W^{-1}\(\sigma_2\)\,{\cal C}\,W\(\sigma_2\)\otimes \one}\,V_{(1)}^{-1}\(\tau_2\)\otimes V_{(2)}^{-1}\(\tau_2\)
\nonumber\\
&&\left.  -\delta\(\zeta_1-\zeta_2\)\,\frac{d\, {\cal Y}\(\zeta_2\,,\,\tau_f\)}{d\,\zeta_2}\,
+ \frac{d\, \({\cal Y}\(\zeta_1\,,\,\tau_f\)\,\delta\(\zeta_1-\zeta_2\)\)}{d\,\zeta_1}
\right.
\nonumber\\
&&
+\left. \delta\(\zeta_1-\zeta_2\)\,\sbr{\one\otimes {\cal A}\(\zeta_2\,,\,\alpha_2\,,\,\beta_2\)}{{\cal Y}\(\zeta_2\,,\,\tau_f\)}
\right.
\nonumber\\
&&+\left. i\,e\,\beta_2\,\vartheta\,\delta\(\zeta_1-\zeta_2\)\,\int_{\tau_i}^{\tau_f}d\tau_2
\sbr{\frac{d\, {\cal Y}\(\zeta_2\,,\,\tau_2\)}{d\,\tau_2}}{\one\otimes T_b}\times \right.
\nonumber\\
&& \times \left.\int_{\tau_i}^{\tau_2}d\tau_2^{\prime}\,\int_{\sigma_i}^{\sigma_f}d\sigma_2\,\Delta\(\sigma_2\,,\,\tau_2^{\prime}\,,\,\zeta_2\)\, d_{ba}\(V_{(2)}\(\tau_2^{\prime}\)\,W^{-1}\(\sigma_2\)\)\, {\cal C}_a\(\sigma_2\,,\,\tau_2^{\prime}\,,\,\zeta_2\)
\right\}
\nonumber
\er
where we have used \rf{adjointrepproperty} 

We then get
\br
&&\{{\cal A}\(\zeta_1\,,\,\alpha_1\,,\,\beta_1\)\,\overset{\otimes}{,}\,{\cal A}\(\zeta_2\,,\,\alpha_2\,,\,\beta_2\)\}_{PB}=-i e^3 \beta_1\beta_2\vartheta\delta\(\zeta_1-\zeta_2\)\left\{
\int_{\tau_i}^{\tau_f} d\tau_2\frac{\(-i\)}{e\(\beta_1-\beta_2\)}
\times
\right. 
\nonumber\\
&&\times\,\left.  \sbr{V_{(1)}\(\tau_2\)\otimes V_{(2)}\(\tau_2\)\,\IC\,V_{(1)}^{-1}\(\tau_2\)\otimes V_{(2)}^{-1}\(\tau_2\)}{\frac{d\,{\widetilde{\cal M}}_{(1)}\(\tau_2\)}{d\,\tau_2}\otimes \one+\one\otimes \frac{d\,{\widetilde{\cal M}}_{(2)}\(\tau_2\)}{d\,\tau_2}}\right.
\nonumber\\
&&+\left.\sbr{{\cal A}\(\zeta_1\,,\,\alpha_1\,,\,\beta_1\)\otimes \one+ \one\otimes{\cal A}\(\zeta_1\,,\,\alpha_2\,,\,\beta_2\)}{{\cal Y}\(\zeta_1\,,\,\tau_f\)} -\frac{d\, {\cal Y}\(\zeta_2\,,\,\tau_f\)}{d\,\zeta_2}\right.
\nonumber\\
&&+\left. \int_{\tau_i}^{\tau_f} d\tau_1\,
\sbr{\frac{d\,{\cal Y}\(\zeta_1\,,\,\tau_1\)}{d\,\tau_1}}{{\widetilde{\cal M}}_{(1)}\(\tau_1\)\otimes \one+\one\otimes {\widetilde{\cal M}}_{(2)}\(\tau_1\)}
\right\}
\lab{prepbawitha}
\er
where we have introduced 
\be
{\widetilde{\cal M}}_{(s)}\(\tau_1\)\equiv i\,e\,\beta_s\,\vartheta
\int_{\tau_i}^{\tau_1}d\tau_1^{\prime}\,\int_{\sigma_i}^{\sigma_f}d\sigma_1\,\Delta\(z\)\, 
V_{(s)}\(\tau_1^{\prime}\)\,W^{-1}\(\sigma_1\)\,{\cal C}\(z\)\,W\(\sigma_1\)\,V_{(s)}^{-1}\(\tau_1^{\prime}\)
\lab{calmtildedef}
\ee
where $s=1,2$, and where we have denoted $z^i=z^i\(\sigma_1\,,\,\tau_1^{\prime}\,,\,\zeta_1\)$.

We now use \rf{holonomyl2fluxes}, \rf{casimirzero} and \rf{calrdef} to write
\br
&&-i\,e\,\(\beta_1-\beta_2\)\,{\cal Y}\(\zeta_1\,,\,\tau\)=\int_{\tau_i}^{\tau} d\tau_1\,V_{(1)}\(\tau_1\)\otimes V_{(2)}\(\tau_1\)\,\times
\nonumber\\
&&\times \sbr{\IC}{V_{(1)}^{-1}\(\tau_1,\zeta_1\) \frac{d V_{(1)}\(\tau_1,\zeta_1\)}{d\tau_1}
\,\otimes\one+\one\otimes V_{(2)}^{-1}\(\tau_1,\zeta_1\) \frac{d V_{(2)}\(\tau_1,\zeta_1\)}{d \tau_1}}\,V_{(1)}^{-1}\(\tau_1\)\otimes V_{(2)}^{-1}\(\tau_1\)
\nonumber\\
&&+\(\alpha_1-\alpha_2\)\,{\widetilde {\cal Z}}\(\tau\,,\,\zeta_1\)
\nonumber\\
&&=\IC-V_{(1)}\(\tau\)\otimes V_{(2)}\(\tau\)\,\IC\,V_{(1)}^{-1}\(\tau\)\otimes V_{(2)}^{-1}\(\tau\)
+\(\alpha_1-\alpha_2\)\,{\widetilde {\cal Z}}\(\tau\,,\,\zeta_1\)
\lab{preanomalyfpr}
\er
where we have used the fact that $V_{(s)}$ at the reference point $x_R$ lies in the center of $G$ (see \rf{intconstcenter}), and 
where we have defined
\be
{\widetilde {\cal Z}}\(\tau\,,\,\zeta_1\)\equiv i\,e\,\int_{\tau_i}^{\tau} d\tau_1\,V_{(1)}\(\tau_1\)\otimes V_{(2)}\(\tau_1\)\sbr{\IC}{
\mathfrak{b}_{\tau}\(\sigma_f\,,\,\tau_1\,,\,\zeta_1\)\,\otimes\one}\,V_{(1)}^{-1}\(\tau_1\)\otimes V_{(2)}^{-1}\(\tau_1\)
\lab{calztauzetadef}
\ee

Note that if we use the Bianchi identity $D_iB_i=0$, we get from \rf{calmtildedef}, \rf{kminusa},  \rf{constraint2} and \rf{jacobiandef}, that 
\be
{\widetilde{\cal M}}_{(s)}\(\tau_1\)={\cal A}\(\zeta_1\,,\,\tau_1\,\,\alpha_s\,,\,\beta_s\)-{\cal K}\(\zeta_1\,,\,\tau_1\,,\,\alpha_s\,,\,\beta_s\)
\lab{mtildecalcalkrelation}
\ee

Replacing the expression for ${\cal Y}$, given in \rf{preanomalyfpr}, in the last term on the right hand side of \rf{prepbawitha}, and integrating by parts the terms not containing ${\widetilde {\cal Z}}$, and  using  \rf{mtildecalcalkrelation}, we get
\br
&&\{{\cal A}\(\zeta_1\,,\,\alpha_1\,,\,\beta_1\)\,\overset{\otimes}{,}\,{\cal A}\(\zeta_2\,,\,\alpha_2\,,\,\beta_2\)\}_{PB}=-i e^3 \beta_1\beta_2\vartheta\delta\(\zeta_1-\zeta_2\)\left\{-\frac{d\, {\cal Y}\(\zeta_1\,,\,\tau_f\)}{d\,\zeta_1}
\right.
\nonumber\\
&&+\left.\frac{i}{e\(\beta_1-\beta_2\)}\left\{ -\sbr{\IC+\(\alpha_1-\alpha_2\)\,{\widetilde {\cal Z}}\(\tau_f\,,\,\zeta_1\)}{{\cal A}\(\zeta_1\,,\,\alpha_1\,,\,\beta_1\)\otimes \one+ \one\otimes{\cal A}\(\zeta_1\,,\,\alpha_2\,,\,\beta_2\)}
\right.\right.
\nonumber\\
&&+\left.\left. \sbr{V_{(1)}\(\tau_f\)\otimes V_{(2)}\(\tau_f\)\,\IC\,V_{(1)}^{-1}\(\tau_f\)\otimes V_{(2)}^{-1}\(\tau_f\)}{{\cal K}\(\zeta_1\,,\,\alpha_1\,,\,\beta_1\)\otimes \one+ \one\otimes{\cal K}\(\zeta_1\,,\,\alpha_2\,,\,\beta_2\)}
\right.\right.
\nonumber\\
&&+\(\alpha_1-\alpha_2\)\,\left.\left. \int_{\tau_i}^{\tau_f} d\tau_1\,
\sbr{\frac{d\,{\widetilde {\cal Z}}\(\zeta_1\,,\,\tau_1\)}{d\,\tau_1}}{{\widetilde{\cal M}}_{(1)}\(\tau_1\)\otimes \one+\one\otimes {\widetilde{\cal M}}_{(2)}\(\tau_1\)}
\right\} \right\}
\lab{prepbawitha2}
\er
where we have denoted ${\cal K}\(\zeta_1\,,\,\tau_f\,,\,\alpha_s\,,\,\beta_s\)\equiv {\cal K}\(\zeta_1\,,\,\alpha_s\,,\,\beta_s\)$.

In the $\zeta_1$-derivative of ${\cal Y}$ in \rf{prepbawitha2}, we use \rf{preanomalyfpr}  and \rf{deltav4}, and in the commutator of ${\widetilde {\cal Z}}$ with ${\cal A}$, we use \rf{chargeophol} to express ${\cal A}$ as the $\zeta_1$-derivative of $Q$. We then get
\br
&&\{{\cal A}_{(1)}\(\zeta_1\)\,\overset{\otimes}{,}\,{\cal A}_{(2)}\(\zeta_2\)\}_{PB}=\delta\(\zeta_1-\zeta_2\)\sbr{{\cal R}\(\beta_1\,,\,\beta_2\)}{{\cal A}_{(1)}\(\zeta_1\)\otimes \one+ \one\otimes{\cal A}_{(2)}\(\zeta_1\)}
\nonumber\\
&&
- \(\alpha_1-\alpha_2\)\,\delta\(\zeta_1-\zeta_2\)\left\{
\int_{\tau_i}^{\tau_f} d\tau_1\,
\sbr{\frac{d\,{\cal Z}\(\zeta_1\,,\,\tau_1\)}{d\,\tau_1}}{{\widetilde{\cal M}}_{(1)}\(\tau_1\)\otimes \one+\one\otimes {\widetilde{\cal M}}_{(2)}\(\tau_1\)}\right.
\nonumber\\
&&-\left. Q_{\otimes}\(\zeta_1\)\frac{d\;}{d\,\zeta_1}\(Q_{\otimes}^{-1}\(\zeta_1\)\,{\cal Z}\(\zeta_1\,,\,\tau_f\)\,Q_{\otimes}\(\zeta_1\)\)\, Q_{\otimes}^{-1}\(\zeta_1\)
\right\}
\lab{fprfinal}
\er
where we have denoted ${\cal A}_{(s)}\(\zeta_1\)\equiv{\cal A}\(\zeta_1\,,\,\alpha_s\,,\,\beta_s\)$, and 
\be
Q_{\otimes}\(\zeta_1\)\equiv 
Q\(\zeta_1\,,\,\alpha_1\,,\,\beta_1\)\otimes Q\(\zeta_1\,,\,\alpha_2\,,\,\beta_2\)
\lab{qotimesdef}
\ee
In addition, we have defined 
\be
{\cal R}\(\beta_1\,,\,\beta_2\)\equiv - e^2 \,\vartheta\,\frac{\beta_1\beta_2}{\(\beta_1-\beta_2\)}\,\IC
\ee
and have rescaled ${\widetilde {\cal Z}}$, defined in \rf{calztauzetadef}, as
\br
&&{\cal Z}\(\zeta_1\,,\,\tau\)\equiv  - e^2 \,\vartheta\,\frac{\beta_1\beta_2}{\(\beta_1-\beta_2\)} {\widetilde {\cal Z}}\(\zeta_1\,,\,\tau\)=ie\int_{\tau_i}^{\tau} d\tau_1\,V_{(1)}\(\zeta_1\,,\,\tau_1\)\otimes V_{(2)}\(\zeta_1\,,\,\tau_1\)\times
\nonumber\\
&& \times \sbr{{\cal R}\(\beta_1,\beta_2\)}{
\mathfrak{b}_{\tau}\(\sigma_f\,,\,\tau_1,\zeta_1\)\,\otimes\one}\,V_{(1)}^{-1}\(\zeta_1\,,\,\tau_1\)\otimes V_{(2)}^{-1}\(\zeta_1\,,\,\tau_1\)
\lab{calztauzetadefscaled}
\er

Note from \rf{calmtildedef}, that ${\cal M}_{(s)}$ vanishes when the constraints \rf{constraint2} hold true. Therefore, the second term on the right-hand side \rf{fprfinal} drops when we impose the constraints. However, our FPR has an anomaly due to the third term in \rf{fprfinal}. As we will see in Section \ref{sec:sklyanin}, such an anomaly does not prevent the infinity of conserved charges  to Poisson commute. Of course, we get a pure FPR, without anomalies and without imposing the constraints,  when the connections ${\cal A}$ in \rf{fprfinal} correspond to the same value of the magnetic parameter $\alpha$.

\section{The  Sklyanin Relation}
\label{sec:sklyanin}
\setcounter{equation}{0}

The Fundamental Poisson Bracket relation \rf{fprfinal} leads to the so-called Sklyanin relation, which relates the Poisson brackets of the entries of the charge operator $Q\(\zeta\,,\,\alpha\,,\,\beta\)$ to commutators of an $r$-matrix with products of those entries. As we will see, the Sklyanin relation leads to the involution of the infinity of conserved charges. 

Using \rf{chargepoisson3} and \rf{fprfinal},  and adopting the notation \rf{qotimesdef},  we get
\br
&&\left\{Q\(\zeta_f\,,\,\alpha_1\,,\,\beta_1\)\,\overset{\otimes}{,}\,Q\(\zeta_f\,,\,\alpha_2\,,\,\beta_2\)\right\}_{PB}=Q_{\otimes}\(\zeta_f\)\int_{\zeta_i}^{\zeta_f}d\zeta\,
Q_{\otimes}^{-1}\(\zeta\)\times
\nonumber\\
&&\times \left\{\sbr{{\cal R}\(\beta_1\,,\,\beta_2\)}{{\cal A}_{(1)}\(\zeta\)\otimes \one+ \one\otimes{\cal A}_{(2)}\(\zeta\)}
\right.
\nonumber\\
&&
- \(\alpha_1-\alpha_2\)\,\left\{
\int_{\tau_i}^{\tau_f} d\tau\,
\sbr{\frac{d\,{\cal Z}\(\zeta\,,\,\tau\)}{d\,\tau}}{{\widetilde{\cal M}}_{(1)}\(\tau\)\otimes \one+\one\otimes {\widetilde{\cal M}}_{(2)}\(\tau\)}\right.
\nonumber\\
&&-\left. \left. Q_{\otimes}\(\zeta\)\frac{d\;}{d\,\zeta}\(Q_{\otimes}^{-1}\(\zeta\)\,{\cal Z}\(\zeta\,,\,\tau_f\)\,Q_{\otimes}\(\zeta\)\)\, Q_{\otimes}^{-1}\(\zeta\)
\right\}Q_{\otimes}\(\zeta\)\right\}
\er
We now use \rf{chargeophol} to obtain
\br
&&\left\{Q\(\zeta_f\,,\,\alpha_1\,,\,\beta_1\)\,\overset{\otimes}{,}\,Q\(\zeta_f\,,\,\alpha_2\,,\,\beta_2\)\right\}_{PB}=Q_{\otimes}\(\zeta_f\)\int_{\zeta_i}^{\zeta_f}d\zeta\,
\times
\nonumber\\
&&\times \left\{\frac{d\;}{d\,\zeta}\left\{Q_{\otimes}^{-1}\(\zeta\)\left[{\cal R}\(\beta_1\,,\,\beta_2\)+\(\alpha_1-\alpha_2\)\,{\cal Z}\(\zeta\,,\,\tau_f\)\right]\,Q_{\otimes}\(\zeta\)\right\}
\right.
\\
&&
- \left.\(\alpha_1-\alpha_2\)\,Q_{\otimes}^{-1}\(\zeta\)
\int_{\tau_i}^{\tau_f} d\tau\,
\sbr{\frac{d\,{\cal Z}\(\zeta\,,\,\tau\)}{d\,\tau}}{{\widetilde{\cal M}}_{(1)}\(\tau\)\otimes \one+\one\otimes {\widetilde{\cal M}}_{(2)}\(\tau\)}Q_{\otimes}\(\zeta\)\right\}
\nonumber
\er
In order for the conserved charges to be gauge invariant, the charge operator at $\zeta=\zeta_i$, i.e. $Q\(\zeta_i\)=V_R$, has to lie in the center of the gauge group (see \rf{intconstcenter}). In addition, we have from \rf{calztauzetadefscaled} that ${\cal Z}\(\zeta_i\,,\,\tau\)=0$, since $\mathfrak{b}_{\tau}\(\sigma_f\,,\,\tau,\zeta_i\)=0$, as the magnetic flux  goes to zero. 

Consequently
\br
&&\left\{Q\(\zeta_f\,,\,\alpha_1\,,\,\beta_1\)\,\overset{\otimes}{,}\,Q\(\zeta_f\,,\,\alpha_2\,,\,\beta_2\)\right\}_{PB}=\sbr{{\cal R}\(\beta_1\,,\,\beta_2\)}{Q\(\zeta_f\,,\,\alpha_1\,,\,\beta_1\)\otimes Q\(\zeta_f\,,\,\alpha_2\,,\,\beta_2\)}
\nonumber\\
&&+\(\alpha_1-\alpha_2\)\,{\cal Z}\(\zeta_f\,,\,\tau_f\)\,Q\(\zeta_f\,,\,\alpha_1\,,\,\beta_1\)\otimes Q\(\zeta_f\,,\,\alpha_2\,,\,\beta_2\)
\lab{skrelfinal}\\
&&-\(\alpha_1-\alpha_2\)Q_{\otimes}\(\zeta_f\)\int_{\zeta_i}^{\zeta_f}d\zeta\,Q_{\otimes}^{-1}\(\zeta\)
\int_{\tau_i}^{\tau_f} d\tau
\sbr{\frac{d\,{\cal Z}\(\zeta,\tau\)}{d\,\tau}}{{\widetilde{\cal M}}_{(1)}\(\tau\)\otimes \one+\one\otimes {\widetilde{\cal M}}_{(2)}\(\tau\)}Q_{\otimes}\(\zeta\)
\nonumber
\er
Similar to the case of the FPR \rf{fprfinal}, the Sklyanin relation \rf{skrelfinal} has a term that vanishes when the constraints \rf{constraint2} are imposed (the third term on the right-hand side of \rf{skrelfinal}), and an anomalous term (the second one). However, the anomalous term does not prevent the involution of the conserved charges, as we now explain. 

The reason behind the cancellation of the anomaly in the Poisson brackets among the conserved charges is the requirement of the invariance of those charges under reparameterization of the two-sphere $S^2_{\infty}$, the boundary of $\IR^3$. In Section \ref{subsec:reparameterize}, we have discussed two sufficient conditions for that invariance under reparameterization. 

The first one is that the electric and magnetic fields should fall at spatial infinity faster than $1/r^2$, where $r$ is the radial distance. But that condition implies that the flux $\mathfrak{b}_{\tau}\(\sigma_f\,,\,\tau_1,\zeta_f\)$, appearing in ${\cal Z}\(\zeta_f,\tau_f\)$, on the second line of  \rf{skrelfinal} (see  definition of ${\cal Z}$ in \rf{calztauzetadefscaled}), vanishes, since it is evaluated on a loop at spatial infinity ($\zeta=\zeta_f$). Therefore, under such a boundary condition, the Sklyanin relation \rf{skrelfinal} ceases to be anomalous.

The second sufficient condition  discussed in Section \ref{subsec:reparameterize}, namely \rf{fwftildewconstant}, implies the cancellation of the anomaly, not on the Sklyanin relation \rf{skrelfinal}, but on  the Poisson bracket of the conserved charge \rf{chargeoperatordef} with the charge operator. From  \rf{chargeoperatordef} and \rf{skrelfinal} we get 
\br
&&\pbr{Q_N\(\alpha_1,\beta_1\)}{Q\(\zeta_f\,,\,\alpha_2\,,\,\beta_2\)}= {\rm Tr}_L\(Q^{N-1}\(\zeta_f,\alpha_1,\beta_1\)\otimes \one \sbr{{\cal R}}{\one\otimes Q\(\zeta_f,\alpha_2,\beta_2\)}\)
\nonumber\\
&&-  \frac{i e^3\vartheta\beta_1\beta_2\(\alpha_1-\alpha_2\)}{\(\beta_1-\beta_2\)}
\int_{\tau_i}^{\tau_f} d\tau {\rm Tr}\left[\sbr{\mathfrak{b}_{\tau}\(\sigma_f,\tau,\zeta_f\)}{V_{(1)}^{-1}\(\zeta_f,\tau\)Q^N\(\zeta_f,\alpha_1,\beta_1\)V_{(1)}\(\zeta_f,\tau\)}T_a\right]\times
\nonumber\\
&&\times \,V_{(2)}\(\zeta_f,\tau\)\,T_a\,V_{(2)}^{-1}\(\zeta_f,\tau\)\,Q\(\zeta_f,\alpha_2,\beta_2\)
\lab{precommuting}\\
&&-\(\alpha_1-\alpha_2\)\,{\rm Tr}_L\(Q^{N}\(\zeta_f\,,\,\alpha_1\,,\,\beta_1\)\otimes Q\(\zeta_f\,,\,\alpha_2\,,\,\beta_2\)\, \Upsilon\)
\nonumber
\er
where we have denoted
\br
\Upsilon\equiv \int_{\zeta_i}^{\zeta_f}d\zeta\,Q_{\otimes}^{-1}\(\zeta\)
\int_{\tau_i}^{\tau_f} d\tau
\sbr{\frac{d\,{\cal Z}\(\zeta,\tau\)}{d\,\tau}}{{\widetilde{\cal M}}_{(1)}\(\tau\)\otimes \one+\one\otimes {\widetilde{\cal M}}_{(2)}\(\tau\)}Q_{\otimes}\(\zeta\)
\lab{upsilondef}
\er
Note from \rf{ebfrakdef} that $\mathfrak{b}_{\tau}\(\sigma_f\,,\,\tau,\zeta_f\)$ is defined on a loop at spatial infinity ($\zeta=\zeta_f$). From \rf{fwftildewconstant} and \rf{ebfrakdef}, we have that
\be
\mathfrak{b}_{\tau}\(\sigma_f\,,\,\tau,\zeta_f\)\sim c\, \int_{\sigma_i}^{\sigma_f}d\sigma\, \ve_{ijk}\,\frac{{\hat r}_i}{r^2}\,\frac{d\,x^j}{d\,\sigma}\, \frac{d\,x^k}{d\,\tau};\qquad \qquad 
\mathfrak{e}_{\tau}\(\sigma_f\,,\,\tau,\zeta_f\)\sim {\widetilde c}\, \int_{\sigma_i}^{\sigma_f}d\sigma\, \ve_{ijk}\,\frac{{\hat r}_i}{r^2}\,\frac{d\,x^j}{d\,\sigma}\, \frac{d\,x^k}{d\,\tau}
\lab{mathfrakebinfinity}
\ee
with ${\hat r}$ being the unit radial vector, and $c$ and ${\widetilde c}$ being constant elements of a given Cartan subalgebra of the Lie algebra of the gauge group $G$. 

 But $V_{(i)}\(\zeta_f,\tau\)$, $i=1,2$, is defined on the surface $S^2_{\infty}$, the border of $\IR^3$, and so from \rf{holonomyl2fluxes} and \rf{mathfrakebinfinity}  we see that it is an exponentiation of the same Cartan subalgebra generators, $c$ and ${\widetilde c}$. If we now impose the constraints \rf{constraint2} and the static Bianchi identity $D_iB_i=0$, which are equivalent to the static Yang-Mills differential equations, it follows that we are imposing the static integral Yang-Mills equation. Consequently, we can express the charge operator $Q\(\zeta_f\,,\,\alpha_1\,,\,\beta_1\)$, as a surface ordered integral on $S^2_{\infty}$ through \rf{chargevolumesurface}. Therefore, from \rf{mathfrakebinfinity}  we conclude that such a charge operator is also an exponentiation of those same Cartan subalgebra generators, $c$ and ${\widetilde c}$. Consequently, we have that 
 \be
 V_{(i)}^{-1}\(\tau\)Q^N\(\zeta_f\,,\,\alpha_1\,,\,\beta_1\)V_{(i)}\equiv \mbox{\rm exponentiation of  $c$ and ${\widetilde c}$};\qquad\qquad i=1,2
 \lab{exponentiationcc}
 \ee
 
 We then observe that the commutator inside the trace, on the second line of \rf{precommuting}, vanishes. But since such a result was obtained by   imposing the constraints to hold, we get that  ${\widetilde{\cal M}}_{(1)}$, given in \rf{calmtildedef}, vanishes, and so does $\Upsilon$, given in \rf{upsilondef}. Consequently, we get that the Sklyanin relation is in fact non-anomalous, i.e. 
\be
\pbr{Q_N\(\alpha_1,\beta_1\)}{Q\(\zeta_f\,,\,\alpha_2\,,\,\beta_2\)}\cong {\rm Tr}_L\(Q^{N-1}\(\zeta_f\,,\,\alpha_1\,,\,\beta_1\)\otimes \one \sbr{{\cal R}}{\one\otimes Q\(\zeta_f\,,\,\alpha_2\,,\,\beta_2\)}\)
\lab{sklyaninlax}
\ee
where the symbol $\cong$ means equality when the constraints \rf{constraint2} hold true. 

We now get, from \rf{sklyaninlax}, that the conserved charges \rf{chargeoperatordef}  are involution
\be
\pbr{Q_N\(\alpha_1,\beta_1\)}{Q_M\(\alpha_2\,,\,\beta_2\)}\cong 0
\lab{involutionfinal}
\ee
Such a result is true for all values of the parameters $\alpha_i$ and $\beta_i$, $i=1,2$. Therefore, expanding the charge operators in power series on those parameters, as in \rf{expandchargeoperator}, we get that the infinity of charges \rf{chargesexpanded} are in involution. Consequently, we have an exact integrability structure in Yang-Mills theories on the sector of the non-abelian magnetic and electric charges and their higher modes.

\subsection{The algebra of the transformations generated by the charges}
\label{sec:algebratranf}

We now show that, despite the fact that the conserved charges are in involution, the transformations generated by them do not quite commute. That is a consequence of the anomalies of the Sklyanin relation \rf{skrelfinal}, and the fact that we   can not impose the constraints inside the Poisson bracket.

Consider the commutator of two transformations of the type \rf{symmetrycharges}. Using the Jacobi identity for the Poisson bracket, we get
\be
\sbr{\delta_{N_1,\alpha_1,\beta_1}}{\delta_{N_2,\alpha_2,\beta_2}}\, X=-\ve_1\,\ve_2 \,\pbr{X}{\pbr{Q_{N_1}\(\alpha_1\,,\,\beta_1\)}{Q_{N_2}\(\alpha_2\,,\,\beta_2\)}}
\lab{commutatortransf}
\ee
Using \rf{precommuting} we get
\br
&&\pbr{Q_{N_1}\(\alpha_1,\beta_1\)}{Q_{N_2}\(\alpha_2\,,\,\beta_2\)}= -\(\alpha_1-\alpha_2\)\left\{
{\rm Tr}_{RL}\(Q^{N_1}_{(1)}\(\zeta_f\)\otimes Q^{N_2}_{(2)}\(\zeta_f\)\, \Upsilon\) \right. 
\nonumber\\
&&+\left.i e^3 \,\vartheta\,\frac{\beta_1\beta_2}{\(\beta_1-\beta_2\)}\,
\int_{\tau_i}^{\tau_f} d\tau \,{\rm Tr}\(\sbr{\mathfrak{b}_{\tau}\(\sigma_f\,,\,\tau,\zeta_f\)}{V_{(1)}^{-1}\(\zeta_f,\tau\)Q^{N_1}_{(1)}\(\zeta_f\)\,V_{(1)}\(\zeta_f,\tau\)}\,T_a\)\times \right.
\nonumber\\
&&\times \left. \,{\rm Tr}\(T_a\,V_{(2)}^{-1}\(\zeta_f,\tau\)\,Q^{N_2}_{(2)}\(\zeta_f\)V_{(2)}\(\zeta_f,\tau\)\)\right\}
\lab{pbqn1qn2}
\er
where the subscript $(s)$, means the operator depends upon $\(\alpha_s\,,\,\beta_s\)$.

As explained above, the right-hand side of \rf{pbqn1qn2} vanishes when we impose the constraints \rf{constraint2}, and the static Bianchi identity $D_iB_i=0$. However, those quantities are inside the Poisson bracket in \rf{commutatortransf}, and we can not impose the constraints inside the Poisson bracket.

The Poisson bracket of $X$ with the second term on the right-hand side of \rf{pbqn1qn2} leads to
\br
&&\int_{\tau_i}^{\tau_f} d\tau \,{\rm Tr}\(\pbr{X}{\mathfrak{b}_{\tau}\(\sigma_f\,,\,\tau,\zeta_f\)}\,T_a\)\times 
\nonumber\\
&&\times  \,{\rm Tr}\(\sbr{V_{(1)}^{-1}\(\zeta_f,\tau\)Q^{N_1}_{(1)}\(\zeta_f\)\,V_{(1)}\(\zeta_f,\tau\)}{V_{(2)}^{-1}\(\zeta_f,\tau\)\,Q^{N_2}_{(2)}\(\zeta_f\)V_{(2)}\(\zeta_f,\tau\)}\,T_a\)
\nonumber\\
&&-\int_{\tau_i}^{\tau_f} d\tau \,{\rm Tr}\(\pbr{X}{V_{(1)}^{-1}\(\zeta_f,\tau\)Q^{N_1}_{(1)}\(\zeta_f\)\,V_{(1)}\(\zeta_f,\tau\)}\,T_a\)\times 
\nonumber\\
&&\times  \,{\rm Tr}\(\sbr{\mathfrak{b}_{\tau}\(\sigma_f\,,\,\tau,\zeta_f\)}{V_{(2)}^{-1}\(\zeta_f,\tau\)\,Q^{N_2}_{(2)}\(\zeta_f\)V_{(2)}\(\zeta_f,\tau\)}\,T_a\)
\nonumber\\
&&+\int_{\tau_i}^{\tau_f} d\tau \,{\rm Tr}\(\sbr{\mathfrak{b}_{\tau}\(\sigma_f\,,\,\tau,\zeta_f\)}{V_{(1)}^{-1}\(\zeta_f,\tau\)Q^{N_1}_{(1)}\(\zeta_f\)\,V_{(1)}\(\zeta_f,\tau\)}\,T_a\)\times 
\nonumber\\
&&\times  \,{\rm Tr}\(T_a\,\pbr{X}{V_{(2)}^{-1}\(\zeta_f,\tau\)\,Q^{N_2}_{(2)}\(\zeta_f\)V_{(2)}\(\zeta_f,\tau\)}\)
\lab{wierdpoissonbracket}
\er
where we have used the cyclic property of the trace, and on the first two terms in \rf{wierdpoissonbracket}, we have used \rf{casimirzero}. Using \rf{mathfrakebinfinity} and \rf{exponentiationcc}, we conclude that all the three terms of \rf{wierdpoissonbracket} vanish, since $c$ and ${\widetilde c}$ commute. 

Consequently, we get that 
\be
\sbr{\delta_{N_1,\alpha_1,\beta_1}}{\delta_{N_2,\alpha_2,\beta_2}}\, X\cong \ve_1\,\ve_2 \,\(\alpha_1-\alpha_2\)\pbr{X}{{\rm Tr}_{RL}\(Q^{N_1}_{(1)}\(\zeta_f\)\otimes Q^{N_2}_{(2)}\(\zeta_f\)\, \Upsilon\)}
\lab{comutatortwotransf}
\ee
Therefore, the commutator of two transformations generated by the conserved charges \rf{chargeoperatordef} does not vanish, since the Poisson brackets on the right-hand side of \rf{comutatortwotransf} do not vanish for arbitrary $X$.  

Note from \rf{calmtildedef} and \rf{upsilondef}, that $\Upsilon$ is linear in the constraints ${\cal C}={\cal C}_a\,T_a$. Therefore, only the Poisson bracket $\pbr{X}{{\cal C}_a}$ matters in \rf{comutatortwotransf}, since the other terms have the constraint ${\cal C}_a$ outside the Poisson bracket, and so they vanish when the constraints are imposed. 

In Appendix \ref{sec:htwocom} we show that ${\rm Tr}_{RL}\(Q^{N_1}_{(1)}\(\zeta_f\)\otimes Q^{N_2}_{(2)}\(\zeta_f\)\, \Upsilon\)$ Poisson commutes with all terms in the total Hamiltonian $H_T$, given in \rf{completehamiltonian}. Therefore, it also generates symmetries of the Yang-Mills theories, i.e. 
\be
\delta_{\Upsilon} H_T\equiv \ve\,\pbr{H_T}{{\rm Tr}_{RL}\(Q^{N_1}_{(1)}\(\zeta_f\)\otimes Q^{N_2}_{(2)}\(\zeta_f\)\, \Upsilon\)}\cong 0
\ee
However, the conserved charge ${\rm Tr}_{RL}\(Q^{N_1}_{(1)}\(\zeta_f\)\otimes Q^{N_2}_{(2)}\(\zeta_f\)\, \Upsilon\)$, vanishes on the constrained phase space, defined by the constraints \rf{constraint2}.

\section{The symmetries of  Yang-Mills integral equations}
\label{sec:symymintegral}
\setcounter{equation}{0}

The second type of hidden symmetry of the Yang-Mills theories is a symmetry of the integral equations \rf{ymintegraleqs}. Those equations are defined on any three-dimensional volume $\Omega$, with border $\partial \Omega$, and their construction requires a scanning of $\Omega$ with closed surfaces based at the reference point $x_R$. So, given a scanning, the border $\partial \Omega$ is a point in the loop space
\be
{\cal L}^{(2)}\equiv \{ f: \; S^{2}\rightarrow M\, \mid \, \mbox{\rm north pole}\rightarrow x_R\}
\lab{loopspacecall2}
\ee
and the volume $\Omega$  is a path on that loop space. 

We will now construct on every point of ${\cal L}^{(2)}$ an infinite-dimensional group, which is a symmetry group of the integral equations \rf{ymintegraleqs}. In fact, as we will see, both sides of \rf{ymintegraleqs} are elements of that group. 

Consider a one-form connection $\mathfrak{a}$  on the loop space 
\be
{\cal L}^{(1)}\equiv \{ f: \; S^{1}\rightarrow M\, \mid \, \mbox{\rm north pole}\rightarrow x_R\}
\lab{loopspacecall1}
\ee
which take values on the Lie algebra of the gauge group $G$. The loops in ${\cal L}^{(1)}$ are functions $x^{\mu}\(\sigma\)$, from the circle $S^1$, parameterized by $\sigma$, to the space-time $M$, with coordinates $x^{\mu}$, such that the north pole of $S^1$ is always mapped into the reference point $x_R$. We then consider a path ${\cal L}^{(1)}$, parameterized by $\tau$, i.e., each loop (point) in the path is labelled by a given value of $\tau$.

Given $\mathfrak{a}\(\tau\)$, we define two holonomies ${\hat g}\(\tau\)$ and ${\widetilde g}\(\tau\)$,  on a given path in ${\cal L}^{(1)}$ parameterized by $\tau$, by the  equations
\be
\frac{d\,{\hat g}\(\tau\)}{d\,\tau}-{\hat g}\(\tau\)\,\mathfrak{a}\(\tau\)=0
\lab{hatgholonomy}
\ee
and
\be
\frac{d\,{\widetilde g}\(\tau\)}{d\,\tau}+ \mathfrak{a}\(\tau\)\,{\widetilde g}\(\tau\)\,=0
\lab{tildegholonomy}
\ee
On the infinitesimal loop around the reference point $x_R$, they satisfy the boundary condition
\be
{\hat g}\(x_R\)=\one;\qquad\qquad \qquad {\widetilde g}\(x_R\)=\one
\lab{hattildebdcond}
\ee
In fact, \rf{hattildebdcond} are the integration constants for the first-order differential equations \rf{hatgholonomy} and \rf{tildegholonomy}. 

Note that \rf{hatgholonomy} and \rf{hattildebdcond} involve the product of $\mathfrak{a}\(\tau\)$ with ${\hat g}\(\tau\)$ and ${\widetilde g}\(\tau\)$. Since $\mathfrak{a}\(\tau\)$ lies on the Lie algebra of $G$, and the holonomies must take values on $G$, the equations \rf{hatgholonomy} and \rf{hattildebdcond} have to be defined on a given representation of $G$. Therefore, $\one$ appearing in \rf{hattildebdcond} must be the unit matrix on that representation. The choice of representation of $G$, however, is arbitrary, as long as it is a faithful representation.

The motivation to introduce two holonomies is that each one is the inverse of the other, as we now explain. From \rf{hatgholonomy}, \rf{tildegholonomy} and \rf{hattildebdcond} we have that
\be
\frac{d\,\({\hat g}\,{\widetilde  g}\)}{d\,\tau}={\hat g}\,\mathfrak{a}\,{\widetilde  g}-{\hat g}\,\mathfrak{a}\,{\widetilde  g}=0 \qquad\rightarrow \qquad {\hat g}\,{\widetilde  g}={\rm constant}={\hat g}\(x_R\)\,{\widetilde  g}\(x_R\)=\one
\ee
So, ${\widetilde  g}$ is a right inverse of ${\hat g}$. Now
\be
\frac{d\,\({\widetilde  g}\,{\hat g}\)}{d\,\tau}=-\mathfrak{a}\,{\widetilde  g}\,{\hat g}+{\widetilde  g}\,{\hat g}\,\mathfrak{a}=\sbr{{\widetilde  g}\,{\hat g}}{\mathfrak{a}}
\ee
Introduce a matrix $\mathfrak{a}^{\prime}$ as $\mathfrak{a}={\widetilde  g}\,\mathfrak{a}^{\prime}\,{\hat g}$. Then, using the fact that ${\hat g}\,{\widetilde  g}=\one$, we have
\be
\frac{d\,\({\widetilde  g}\,{\hat g}\)}{d\,\tau}=\sbr{{\widetilde  g}\,{\hat g}}{{\widetilde  g}\,\mathfrak{a}^{\prime}\,{\hat g}}={\widetilde  g}\sbr{\one}{\mathfrak{a}^{\prime}}{\hat g}=0\qquad\rightarrow \qquad {\widetilde  g}\,{\hat g}\,={\rm constant}={\widetilde  g}\(x_R\){\hat g}\(x_R\)=\one
\ee
Therefore, ${\widetilde  g}$ is also a left inverse of ${\hat g}$.

Given two one-forms in ${\cal L}^{(1)}$, $\mathfrak{a}_1$ and $\mathfrak{a}_2$, we construct, through \rf{hatgholonomy}, the respectives holonomies ${\hat g}_1\(\tau\)$ and ${\hat g}_2\(\tau\)$. Using \rf{hatgholonomy} we have that  the $\tau$-derivative of their matrix product satisfy
\be
\frac{d\,\({\hat g}_1\,{\hat g}_2\)}{d\,\tau}={\hat g}_1\,\mathfrak{a}_1\,{\hat g}_2+ {\hat g}_1\,{\hat g}_2\,\mathfrak{a}_2={\hat g}_1\,{\hat g}_2\,\left[\mathfrak{a}_2+{\hat g}_2^{-1}\,\mathfrak{a}_1\,{\hat g}_2\right]
\ee
We now define the one-form $\mathfrak{a}_3\(\tau\)$ by
\be
\mathfrak{a}_3\(\tau\)\equiv \mathfrak{a}_2\(\tau\)+{\hat g}_2^{-1}\(\tau\)\,\mathfrak{a}_1\(\tau\)\,{\hat g}_2\(\tau\)
\lab{compositionconnection}
\ee
and therefore the matrix product 
\be
{\hat g}_3\(\tau\)\equiv {\hat g}_1\(\tau\)\,{\hat g}_2\(\tau\)
\lab{compositiongroup}
\ee
satisfy the holonomy equation
\be
\frac{d\,{\hat g}_3\(\tau\)}{d\,\tau}-{\hat g}_3\(\tau\)\,\mathfrak{a}_3\(\tau\)=0
\lab{hatgholonomy3}
\ee
We then have defined compositions of holonomies ${\hat g}$ and one-form connections $\mathfrak{a}$.  

Certainly, the composition of holonomies is associative as it is defined by the matrix product on a given representation of the gauge group $G$. Indeed, given three holonomies, we have that
\be
\({\hat g}_1\,{\hat g}_2\)\,{\hat g}_3={\hat g}_1\,\({\hat g}_2\,{\hat g}_3\)
\ee
In addition, the composition of connections is also associative. Indeed, we have
\be
\mathfrak{a}_{\(1\,2\)\,3}\equiv \mathfrak{a}_3+{\hat g}_3^{-1}\left[\mathfrak{a}_2+{\hat g}_2^{-1}\,\mathfrak{a}_1\,{\hat g}_2\right]{\hat g}_3=\left[\mathfrak{a}_3+ {\hat g}_3^{-1}\mathfrak{a}_2{\hat g}_3\right]+\({\hat g}_2\,{\hat g}_3\)^{-1}\,\mathfrak{a}_1\, {\hat g}_2\,{\hat g}_3\equiv \mathfrak{a}_{1\,\(2\,3\)}
\ee

The identity element ${\hat g}_{\rm id}$, of such a composition of holonomies is the holonomy associated to the vanishing connection, i.e. $\mathfrak{a}_{{\hat g}_{\rm id}}=0$. Indeed, from \rf{hatgholonomy} and \rf{hattildebdcond}, we have that for $\mathfrak{a}=0$
\be
\frac{d\,{\hat g}_{\rm id}\(\tau\)}{d\,\tau}=0\qquad \rightarrow \qquad {\hat g}_{\rm id}=\one
\ee
Similarly, from \rf{tildegholonomy} for $\mathfrak{a}=0$, and \rf{hattildebdcond}, we get that ${\widetilde g}_{\rm id}=\one$, and so ${\widetilde g}_{\rm id}={\hat g}_{\rm id}$. 

From the composition of connections \rf{compositionconnection}, we have  for any ${\hat g}$ that
\be
\mathfrak{a}_{{\hat g}{\hat g}_{\rm ig}}\(\tau\)= \mathfrak{a}_{{\hat g}_{\rm id}}\(\tau\)+{\hat g}_{\rm id}^{-1}\(\tau\)\,\mathfrak{a}\(\tau\)\,{\hat g}_{\rm id}\(\tau\)=\mathfrak{a}\(\tau\)=\mathfrak{a}\(\tau\)+{\hat g}^{-1}\(\tau\)\, \mathfrak{a}_{{\hat g}_{\rm id}}\(\tau\)\,{\hat g}\(\tau\)=\mathfrak{a}_{{\hat g}_{\rm id}{\hat g}}\(\tau\)
\ee
Therefore, ${\hat g}_{\rm id}$ is indeed a left and right identity. 

Note that any right and left identity is unique independently of the fact of the product
being associative or not. Suppose there exist two identities ${\hat g}_{\rm id}$ and ${\hat g}_{\rm id}^{\prime}$ such that ${\hat g}{\hat g}_{\rm id}= {\hat g}_{\rm id}{\hat g}= {\hat g}_{\rm id}^{\prime}{\hat g}= {\hat g}{\hat g}_{\rm id}^{\prime}={\hat g}$, for any ${\hat g}$. 
 Then for ${\hat g}=  {\hat g}_{\rm id}$ we have ${\hat g}_{\rm id}{\hat g}_{\rm id}^{\prime}={\hat g}_{\rm id}$, 
 and
for ${\hat g}= {\hat g}_{\rm id}^{\prime}$ we have ${\hat g}_{\rm id}{\hat g}_{\rm id}^{\prime}= {\hat g}_{\rm id}^{\prime}$. Therefore ${\hat g}_{\rm id}={\hat g}_{\rm id}^{\prime}$, and the identity is unique.

Similarly, suppose that ${\hat g}$ has two right inverses ${\widetilde g}_1$  and ${\widetilde g}_2$ such that ${\hat g}{\widetilde g}_1 = {\hat g}{\widetilde g}_2 = {\hat g}_{\rm id}$, 
and suppose ${\widetilde g}_3$ is a left inverse of ${\hat g}$, i.e. ${\widetilde g}_3{\hat g}={\hat g}_{\rm id}$. Then ${\widetilde g}_3\({\hat g}{\widetilde g}_1\) = {\widetilde g}_3\({\hat g}{\widetilde g}_2\)$ and
using associativity we get $\({\widetilde g}_3{\hat g}\){\widetilde g}_1 = \({\widetilde g}_3{\hat g}\){\widetilde g}_2$ and so ${\hat g}_{\rm id}{\widetilde g}_1 = {\hat g}_{\rm id}{\widetilde g}_2$ and then ${\widetilde g}_1 = {\widetilde g}_2$. 
Therefore, the right inverse is unique. A similar argument can be used to
show the uniqueness of the left inverse. Now if ${\widetilde g}_3$ and ${\widetilde g}_1$ are respectively the left and right inverses of ${\hat g}$, we have ${\widetilde g}_3{\hat g}= {\hat g}_{\rm id}= {\hat g}{\widetilde g}_1$ and then using associativity
we get $\({\widetilde g}_3{\hat g}\){\widetilde g}_1 = {\hat g}_{\rm id}{\widetilde g}_1 = {\widetilde g}_1 = {\widetilde g}_3\({\hat g}{\widetilde g}_1\) = {\widetilde g}_3{\hat g}_{\rm id}= {\widetilde g}_3$. So the left and right inverses
are the same. Consequently, ${\widetilde  g}$ is the unique right and left inverse of ${\hat g}$.

We now consider all possible one-forms $\mathfrak{a}$ in ${\cal L}^{(1)}$, and construct all possible holonomies ${\hat g}$ and ${\widetilde g}$, through \rf{hatgholonomy} and \rf{tildegholonomy} respectively. Therefore, by construction, the set of holonomies close under the compositions \rf{compositionconnection} and \rf{compositiongroup}. Consequently, from the discussion above, the set of all holonomies ${\hat g}\(\tau_f\)$ and ${\widetilde g}\(\tau_f\)$ forms a group which we shall denote by ${\hat G}$. Note that, even tough we have defined the composition of holonomies in \rf{compositiongroup}, on each point $\tau$ of the path, the elements of our group ${\hat G}$ are holonomies integrated along the entire loop, ${\hat g}\(\tau_f\)$ and ${\widetilde g}\(\tau_f\)$, i.e. integrated up to the final point  $\tau=\tau_f$. 

Note that even though the elements of ${\hat G}$ are holonomies, ${\hat G}$ is not a holonomy group in the usual sense defined in the literature. The usual holonomy group is constructed for a fixed connection, and the composition is given through the composition of loops. Our group ${\hat G}$ is defined on a fixed loop in ${\cal L}^{(2)}$, and we compose the connections as in \rf{compositionconnection}. Therefore, we define a group ${\hat G}$ on each point of ${\cal L}^{(2)}$, and the product law is pointwise defined. That is similar to the usual gauge group $G$, which is defined on each point of the space-time $M$, and the product is pointwise.

 We can define connections on ${\cal L}^{(1)}$ through local quantities on the space-time $M$. Indeed, consider a one-form $a_{\mu}$ and a two-form $b_{\mu\nu}$, on the four dimensional Minkowski space-time $M$, with both taking values on the Lie algebra of the gauge group $G$, i.e. $a_{\mu}=a_{\mu}^a\,T_a$, and $b_{\mu\nu}=b_{\mu\nu}^a\,T_a$, with $T_a$ being a basis of the gauge Lie algebra, see \rf{liebasis}. A one-form in ${\cal L}^{(1)}$ can be constructed as
\be
\mathfrak{a}\(\tau\)=\int_{\sigma_i}^{\sigma_f}d\sigma\, \omega^{-1}\,b_{\mu\nu}\,\omega\, \frac{d\,x^{\mu}}{d\,\sigma}\,\frac{d\,x^{\nu}}{d\,\tau}
\lab{mathfrakadef}
\ee
where $\omega$ is the holonomy on $M$ of the one-form connection $a_{\mu}$, i.e. 
\be
\frac{d\,\omega}{d\,\sigma}+a_{\mu}\,\frac{d\,x^{\mu}}{d\,\sigma}\, \omega =0
\lab{omegaholonomy}
\ee
Note that the composition \rf{compositionconnection} of two one-form connections in ${\cal L}{(1)}$, $\mathfrak{a}_1$ and $\mathfrak{a}_2$, of the form \rf{mathfrakadef}, may not be of the form \rf{mathfrakadef}, since that composition is not local in $M$. Therefore, we may not be able to write the composed one-form $\mathfrak{a}_3$, as in \rf{mathfrakadef}, in terms of local one-form $a_{\mu}^3$,  and two-form $b_{\mu\nu}^3$, in $M$.  

We now want to use a Stokes-like theorem to express the elements of the infinite-dimensional group ${\hat G}$, as holonomies in the loop space ${\cal L}^{(2)}$, defined in \rf{loopspacecall2}. In order to do that we consider variations $\delta x^{\mu}\(\sigma\)$, of the loops (points) in ${\cal L}^{(1)}$, and calculate how the holonomies ${\hat g}$, defined in \rf{hatgholonomy}, vary. Performing such variation on the equation \rf{hatgholonomy}, we get 
\be
\frac{d\,\delta{\hat g}\(\tau\)}{d\,\tau}-\delta {\hat g}\(\tau\)\,\mathfrak{a}\(\tau\) - {\hat g}\(\tau\)\,\delta\mathfrak{a}\(\tau\)=0
\lab{hatgholonomyvary}
\ee
Note that the variation of the one-form $\mathfrak{a}\(\tau\)$ is induced by  the variation of the point in ${\cal L}^{(1)}$ where it sits, i.e. $\delta\mathfrak{a}\(\tau\)=\mathfrak{a}\(x^{\mu}\(\sigma\)+\delta x^{\mu}\(\sigma\)\)-\mathfrak{a}\(x^{\mu}\(\sigma\)\)$, where $x^{\mu}\(\sigma\)$ is the loop (point) in ${\cal L}^{(1)}$, labelled by $\tau$. It is not a variation of the parameters of that one-form, which may take it to another one-form in  ${\cal L}^{(1)}$. 

We now multiply \rf{hatgholonomyvary}, from the right by ${\widetilde g}\(\tau\)$, defined in \rf{tildegholonomy}, for the same one-form $\mathfrak{a}\(\tau\)$ that leads to ${\hat g}\(\tau\)$, i.e. ${\widetilde g}\(\tau\)$ is the inverse of ${\hat g}\(\tau\)$. Then we multiply \rf{tildegholonomy} from the left, by $\delta{\hat g}\(\tau\)$, and add  them up to get
\be
\frac{d\,\(\delta{\hat g}\(\tau\)\,{\widetilde g}\(\tau\)\)}{d\,\tau} - {\hat g}\(\tau\)\,\delta\mathfrak{a}\(\tau\)\,{\widetilde g}\(\tau\)=0
\lab{hatgholonomyvary2}
\ee
Therefore
\be
\delta{\hat g}\(\tau_f\)\,{\widetilde g}\(\tau_f\)= \int_{\tau_i}^{\tau_f}d\tau\, {\hat g}\(\tau\)\,\delta\mathfrak{a}\(\tau\)\,{\widetilde g}\(\tau\)
\lab{hatgholonomyvary3}
\ee
where we have assumed that the initial point of the loop is not varied, i.e. $\delta{\hat g}\(\tau_i\)=0$. Note that despite the fact that the initial and final points of the loop coincide, the variation $\delta{\hat g}\(\tau_f\)$ does not vanish, because ${\hat g}\(\tau_f\)$ is the result of the integration along the loop, and so depends upon the variation on each point of the loop. 

We can look at \rf{hatgholonomyvary3} as a differential equation on ${\cal L}^{(2)}$. Indeed, let us take a path on ${\cal L}^{(2)}$ parameterized by $\zeta$, and define the holonomy in ${\cal L}^{(2)}$, as 
\be
\frac{d\,{\hat g}\(\zeta,\tau_f\)}{d\,\zeta} = \left[\int_{\tau_i}^{\tau_f}d\tau\, {\hat g}\(\tau\)\,\frac{d\,\mathfrak{a}\(\tau\)}{d\,\zeta}\,{\widetilde g}\(\tau\)\right]\,{\hat g}\(\zeta,\tau_f\)
\lab{hatgholonomyvary4}
\ee
as ${\widetilde g}$ is the inverse of ${\hat g}$. If we use local coordinates in ${\cal L}^{(1)}$, one can verify that the integral on the right-hand side of \rf{hatgholonomyvary4} will involve the curvature of the one-form $\mathfrak{a}$. 

We can now calculate the group element ${\hat g}$ in two ways. First through \rf{hatgholonomy}, as an holonomy integrated on a loop $\partial \Omega$, in ${\cal L}^{(1)}$. Second by integrating \rf{hatgholonomyvary4} from an infinitesimal loop in ${\cal L}^{(1)}$, around the reference point $x_R$, up to  the loop $\partial \Omega$. Since, from our considerations, those two quantities 
must be the same, we get that ${\hat g}$ can be defined on the loop  $\partial \Omega$, in ${\cal L}^{(1)}$,  (a surface in $M$), or by a path in ${\cal L}^{(2)}$ (a volume $\Omega$ in $M$), with final point $\partial \Omega$, i.e. 
\be
{\hat g}\(\partial \Omega\)={\hat g}\(\Omega\)
\lab{stokesghat}
\ee
That is a Stokes-like theorem for ${\hat g}$, on the same lines as we derived that theorem in Section \ref{subsec:nonabelianstokes} for a two-form connection on $M$.

We then observe that both sides of the integral Yang-Mills equations \rf{ymintegraleqs} are  elements of our infinite-dimensional group ${\hat G}$. In fact, they correspond to holonomies of a connection $\mathfrak{a}$ of the form \rf{mathfrakadef}, where we take 
\be
b_{\mu\nu}=i\,e\,\(\alpha\,F_{\mu\nu}+\beta\,{\widetilde F}_{\mu\nu}\);\qquad\qquad a_{\mu}=i\,e\, A_{\mu}
\ee

We are then inclined to try to define left and right symmetries of the integral Yang-Mills equation \rf{ymintegraleqs} as follows. In order to simplify the notation, we introduce  
\br
V\(\partial\Omega\)&\equiv& P_2\,e^{i\,e\,\int_{\partial\Omega} d\tau\, d\sigma\, W^{-1}\, \(\alpha\,F_{\mu\nu}+\beta\,{\widetilde F}_{\mu\nu}\)\,W\,\frac{d\,x^{\mu}}{d\,\sigma}\, \frac{d\,x^{\nu}}{d\,\tau}}
\nonumber\\
V\(\Omega\)&\equiv&P_3\,e^{i\,e^2\,\int_{\Omega} d\zeta\,  d\tau\,V\,{\cal J}\,V^{-1}}
\lab{twosideofintymeqs}
\er
where we have dropped the integration constant $V_R$, as it has to lie in the center of the gauge group $G$ (see \rf{intconstcenter}). Then, \rf{ymintegraleqs} becomes
\be
V\(\partial\Omega\)= V\(\Omega\)
\lab{intymvsimple}
\ee
From \rf{stokesghat} we then observe that the left and right transformations
\br
V\(\partial\Omega\)\rightarrow {\hat g}_L\(\partial \Omega\)\, V\(\partial\Omega\); \qquad\qquad
V\(\Omega\)\rightarrow {\hat g}_L\( \Omega\)\, V\(\Omega\)
\lab{lefttransf}
\er
and
\br
V\(\partial\Omega\)\rightarrow  V\(\partial\Omega\)\,{\hat g}_R\(\partial \Omega\); \qquad\qquad
V\(\Omega\)\rightarrow  V\(\Omega\)\,{\hat g}_R\( \Omega\)
\lab{righttransf}
\er
with ${\hat g}_{L/R}$ being elements of ${\hat G}$, leave the integral Yang-Mills equation \rf{intymvsimple} invariant. 

We have shown in Section \ref{subsec:nonabelianstokes} how the surface holonomy $V\(\partial\Omega\)$ varies when we vary $\partial\Omega$. In fact, we have shown in Section \ref{subsec:conservedcharges} that such a variation leads to the definition  of the connection ${\cal A}$, given  in \rf{connectionacal}, and it relates to $V$ as 
\be
{\cal A}=\frac{d\,V}{d\,\zeta}\,V^{-1} 
\lab{acaloneform}
\ee
Since the variation of the parameter $\zeta$ accounts for the variation of $\partial\Omega$, we conclude that  ${\cal A}$ is a one-form in ${\cal L}^{(2)}$ given by 
\be
{\cal A}\(\partial\Omega\)=\delta\,V\(\partial\Omega\)\,V^{-1}\(\partial\Omega\) 
\lab{acaloneform2}
\ee
where $\delta$ can be seen as exterior derivative in ${\cal L}^{(2)}$, and $\partial\Omega$ is any point  in ${\cal L}^{(2)}$. 

From \rf{acaloneform2} one can calculate how ${\cal A}$ transforms under \rf{lefttransf} and \rf{righttransf}, since we know how to evaluate $\delta {\hat g}\(\partial\Omega\)$. It is
\br
{\cal A}\rightarrow {\hat g}_L\(\partial\Omega\)\,{\cal A}\,\, {\hat g}_L^{-1}\(\partial\Omega\) +\delta {\hat g}_L\(\partial\Omega\)\,{\hat g}_L^{-1}\(\partial\Omega\)
\lab{connectiontransf1}
\er
and
\br
{\cal A}\rightarrow  {\cal A} +
V\(\partial\Omega\)\,\delta {\hat g}_R\(\partial \Omega\)\,{\hat g}_R^{-1}\(\partial \Omega\)\,V^{-1}\(\partial\Omega\) 
\lab{connectiontransf2}
\er
Note that \rf{connectiontransf1} is like a gauge transformation of the pure gauge connection \rf{acaloneform2}, associated to the infinite-dimensional group ${\hat G}$.

There is a very important point to emphasize here. Even though the transformations  \rf{lefttransf} and \rf{righttransf} leave the integral Yang-Mills equations \rf{intymvsimple} invariant, we can not claim yet that they are symmetries of the Yang-Mills theories. The reason is that given a field configuration of Yang-Mills theory ($A_{\mu}$, $F_{\mu\nu}$, etc),  we can evaluate both sides of \rf{intymvsimple}. However, when we transform both sides of \rf{intymvsimple}, under \rf{lefttransf} and \rf{righttransf}, we should be able to evaluate the new Yang-Mills field configuration that leads to the new both sides of \rf{intymvsimple}. 

That problem appears already in integrable field theories in $(1+1)$-dimensions, where the field equations admit a zero curvature condition
\be
\partial_t A_x- \partial_x A_t+\sbr{A_t}{A_x}=0
\lab{laxeq}
\ee
with $A_t$ and $A_x$, the Lax pair, being elements of a Kac-Moody algebra (or loop algebra) which are functionals of the physical fields. The hidden symmetries of those theories are the gauge transformations 
\be
A_{\mu}\rightarrow {\widetilde A}_{\mu}=g\,A_{\mu}\,g^{-1}-\partial_{\mu}g\,g^{-1} \qquad\qquad \qquad \mu=t\,,\,x
\lab{1+1transf}
\ee
with $g$ being elements of the loop group (or Kac-Moody group when one has a so-called integrable representation of it). Even though \rf{1+1transf} leaves \rf{laxeq}, it can not be considered yet as a symmetry of the integrable field theory, because one has to express the transformed ${\widetilde A}_{\mu}$ in terms of the physical fields.  What makes that possible is the so-called Riemann-Hilbert problem that has a complex analysis version leading to the inverse scattering method, and an algebraic version that leads to the so-called dressing method. Those techniques allow the mapping from one solution (of the physical fields) to another solution. 

The great challenge that our approach faces is the construction of a method that allows us to express the transformations \rf{lefttransf} and \rf{righttransf}, which leave the integral equation \rf{intymvsimple} invariant, into symmetries of Yang-Mills theories, that map solutions into solutions. In other words, one has to find the equivalent of the Riemann-Hilbert problem (if it exists) for non-abelian gauge theories. That for sure is far beyond the scope of the present paper, but we believe that the approach described in this section is on the right direction. 

It may happen that the mapping among solutions will not come in terms of the local Yang-Mills fields. The electric and magnetic fluxes \rf{ebfrakdef}, and perhaps their extension in the time direction, may be objects more suitable for that task, as they are defined on points of the loop space ${\cal L}^{(1)}$. In fact, the integral Yang-Mills equations \rf{intymvsimple} can be expressed in terms of them.

The other challenge of our approach is to get a clear understanding of the structure of the infinite-dimensional group ${\hat G}$. We suspect that the space formed by the elements of that group may not be a manifold, and therefore it may not be a Lie group. It seems to us that the infinite dimension of that manifold is different at different points of it. If that proves to be the case, we might have a very interesting situation. If it is not a Lie group, it may circumvent the Coleman-Mandula theorem in a novel way. It might be an explanation of why a non-supersymmetric theory like the  Yang-Mills theory has an exact integrability structure, as the one we have proved it has.

\section{Conclusions}
\label{sec:conclusion}
\setcounter{equation}{0}

We have shown in this paper that classical, non-supersymmetric Yang-Mills theories in $(3+1)$-dimensional Minkowski space-time, coupled to spin-$1/2$ and spin-$0$ matter fields, present  exact  structures, resembling integrability,  involving the  non-abelian electric and magnetic conserved charges, and their higher modes. In addition, we have shown that those theories possess two types of novel hidden symmetries. Our results apply to the Standard Model of the Fundamental Interactions, in particular to Quantum Chromodynamics (QCD), and it opens the way to the development of new methods to study non-perturbative aspects of non-abelian gauge theories. 

Our results were made possible thanks to the developments of references \cite{afg1,afg2}, that proposed an approach to construct integrable theories in higher dimensions using flat connections on generalized loop spaces, and to references \cite{ym1,ym2}, that implemented those ideas in the context of gauge theories through the use of integral equations, which guaranteed the path independency of the holonomies of those connections, and so their flatness. In order for those ideas to work, one had to show that the local partial differential equations of motion of a field theory had to be equivalent to the flatness condition of a connection on loop space, which is local on loop space but highly non-local on space-time. It is an extremely amazing fact that the theories that satisfy those conditions are exactly the most relevant ones  to describe the fundamental interactions, namely non-abelian gauge theories. The particular property relevant for such an achievement is that the laws of electrodynamics  and Yang-Mills theories can be formulated in terms of integral equations, which state the equality of the charge inside a given three-volume in space-time, to the flux of  the fields associated to them, at the border of that volume. Those integral Yang-Mills equations, proposed in \cite{ym1,ym2}, lead to the construction of an infinite number of gauge-invariant conserved charges, namely the non-abelian electric and magnetic charges and their higher modes. 

Those charges are constructed as the eigenvalues of a charge operator that depends on two arbitrary parameters $\alpha$ and $\beta$. They are not charges of the Noether type. Their conservation comes from the path independency of the holonomy of the loop space connection, and not from  symmetries of the Yang-Mills Lagrangian. Contrary to the Yang-Mills charges discussed in the literature and in textbooks, those charges are truly gauge invariant. If one expands the charge operator in power series in the parameters $\alpha$ and $\beta$, one gets a number of charges equal to the rank of the gauge group, for each term of the series, and so an infinite number of them.  

We have shown that the loop space connection ${\cal A}$, when evaluated on a spatial surface at a given fixed time, satisfies a Fundamental Poisson Bracket Relation (FPR), in a way that reminds the FPR encountered in integrable field theories in $(1+1)$-dimensions. In fact, the associated $R$-matrix is similar to that one appearing in the rational FPR for current algebras \cite{faddeevleshouches}. The FPR for the loop space connection  ${\cal A}$, however, is anomalous in the sense that it has two terms that become irrelevant when the constraints are imposed, and the conditions for reparameterization invariance are applied. 

The FPR for the loop space connection ${\cal A}$ leads in a straightforward way to the Sklyanin relation for the charge operator, which is also anomalous. The involution of the infinity of magnetic and electric charges follows from the Sklyanin relation, once the anomalous terms disappear by the imposition of the constraints and reparameterization conditions. 

Therefore, non-abelian gauge theories possess exact  structures, resembling integrability, with an infinite number of conserved charges in involution. Such a fact, however, does not mean that Yang-Mills  theories are integrable in the usual sense encountered in the literature. The integrability structures that we have constructed do not lead, for instance, to the factorization of the $S$-matrix. Indeed,  it is well known that Yang-Mills theories do not possess such a property, as particle production is a common place in their processes. The infinity of conserved charges involves the non-abelian electric and magnetic charges and their higher modes. The usual concept of integrability involves the energy and momentum and their higher modes instead.  Even though our conserved charges Poisson commute with the Hamiltonian, the latter, as well as the momenta, do not appear as  eigenvalues of our charge operator.  

The novel hidden symmetries that we have constructed are of two types. The first ones are the transformations generated by the conserved charges under the Poisson brackets, and so, they are canonical transformations. Since the  Hamiltonian Poisson commutes with the charges, those transformations are global symmetries of the Yang-Mills theories. We have evaluated how the matter fields, gauge fields, the Wilson lines, and the magnetic and electric fluxes transform under those symmetries. Those transformations are very interesting, and they present  non-integrable factors involving the Wilson line, surface, and volume ordered integrals (holonomies) of the connections in the loop spaces ${\cal L}^{(1)}$ and ${\cal L}^{(2)}$. Even though the charges commute among themselves, the transformations generated by them do not. That is due to the fact that the involution of the charges holds true when the constraints are imposed. The commutator of two transformations leads to a transformation generated by new charges that vanish when the constraints are imposed. Those new charges, however, Poisson commute with the Hamiltonian and so are symmetries of Yang-Mills theories. 

The second type of hidden symmetries that we have discovered corresponds to the transformations that leave the integral Yang-Mills equations invariant. They are generated by an infinite-dimensional group, whose elements are holonomies of one-form connections on the loop space ${\cal L}^{(1)}$. For every point of ${\cal L}^{(1)}$ we defined a composition of any two given connections, which in its turn lead to the composition of the corresponding holonomies, that are the group elements. Such a group differs from the usual holonomy group defined in the literature, whose elements are the holonomies of a  fixed connection evaluated on every closed path based on a fixed point. The    composition law of such a group is defined through the composition of the loops. We instead fix the loop in  ${\cal L}^{(1)}$ and compose the connections on that fixed loop. 

Even though our infinite group leaves the integral Yang-Mills equations invariant,  it is not a symmetry of the Hamiltonian. So, in principle, it can define maps between two quite different configurations. With a suitable choice of connection, one can, for instance,  map a vacuum configuration into a non-trivial solution like  a magnetic monopole. Therefore, such a group may play a role, in non-abelian gauge theories, similar to that played by the Kac-Moody group in integrable field theories in $(1+1)$-dimensions. However, the powerful methods applicable for such theories, like the inverse scattering method, dressing method, etc, rely, among other things,  on the so-called Riemann-Hilbert problem.  It is not known yet if the group we have constructed may lead to similar  powerful structures applicable to gauge theories.  That is one of the challenges that our approach puts forward, and the answer to such questions is well beyond the scope of the present paper. 

The other interesting aspect of such a group is that it may not be a Lie group. We have to investigate that further, but it seems that the space formed by the group elements may not be a manifold. Its (infinite) dimension may vary from point to point. If that turns out to be true, the symmetries of the integral Yang-Mills equations, defined by such a group, may not be subjected to the Coleman-Mandula no-go theorem \cite{colmand}. 

 The $(2+1)$-dimensional Yang-Mills theories also present integral equations and conserved charges similar to the ones discussed in this paper \cite{ym2}. We have shown that such theories also have exact  structures, resembling integrability, with an infinite number of charges in involution, with an FPR for the loop space connection, and a Sklyanin relation for the charge operator. They also possess two types of hidden symmetries, one generated by the charges under the Poisson bracket, and another one by an infinite-dimensional group of holonomies, which are symmetries of the integral equations.  Such results are presented in a separate paper \cite{threeym}, and they will certainly  serve as a laboratory to study the structures of Yang-Mills, discussed here,  in a simpler scenario.

Our results raise some important and interesting questions related to the nature of the conserved charges. As we have shown, they are truly gauge invariant, and so, in principle, such charges are physical observables. In the case of QCD, they are color singlets (i.e., transform under the scalar representation of $SU(3)$ color) and so may not be confined. Therefore, the hadrons could carry such charges, even though the quarks may not. In order to analyze  such an issue  it has been considered classical Yang-Mills in $(1+1)$-dimensions and the equivalent of our three-dimensional charges has been constructed  \cite{twoclassical}.  Then it has been considered the quantum  version of that two-dimensional gauge theory on a lattice, and it has been evaluated the expectation values of the charges in some particle configurations, at the strong coupling limit, where the plaquette action is dominated by the hopping term. The results are quite interesting \cite{twolattice}. The mesons and baryons do present non-vanishing expectation values of the charges. The quarks, however, not being  color singlets, present  vanishing expectation values for the charges. One certainly has to study such an issue in a more complete and realistic scenario, since the charges may have the potential of being order parameters of some phases of non-abelian gauge theories. 

Clearly, the  most important issue still to be investigated is the  role of such  structures, which resemble integrability, on the quantum Yang-Mills theories. One can approach such a problem from different perspectives and methods. It would be interesting to check if the hidden symmetries we presented in this paper might be broken or not at the quantum level. In addition, the conservation of the charges may present quantum anomalies, and they may introduce selection rules in many processes, which will be important in the properties of quantum Yang-Mills. Certainly, the  results presented in this paper open up several ways to investigate the non-perturbative aspects of non-abelian gauge theories.\\

\vspace{2 cm}

\noindent {\bf Acknowledgements} We are grateful to D. Melnikov and A. Pinzul for providing us with the first motivations for this work, and to A. A. Sharapov for very helpful discussions on the transformations generated by the charges. We are grateful to G. Luchini for the collaboration in the early stage of this work, and to P. A. Faria da Veiga and R. Mistry for invaluable discussions of the properties of the charges.  We are also indebted to  H. N. S\'a Earp, I. Mencattini, and M. Pedroni for discussions on the structure of the infinite-dimensional group. We are grateful to the participants of the conferences InTropea2023, especially to the organizer R. Tateo, and  The Fourth Patricio Letelier School on Mathematical Physics - 2024, where preliminary versions of this work were presented. We benefited from discussions with G. Dito, V. Kupriyanov, J. S\'anchez Guill\'en, and P. Vieira. LAF acknowledges the financial support of Fapesp
(Funda\c c\~ao de Amparo \`a Pesquisa do Estado de S\~ao Paulo) grant 2022/00808-7, and CNPq
(Conselho Nacional de Desenvolvimento Cient\'ifico e Tecnol\'ogico) grant 307833/2022-4. HM acknowledges the financial support of Fapesp  grant 2021/10141-7.

\newpage

\appendix

\section{A particular scanning of $\IR^3_{t}$} 
\label{sec:scanning}
\setcounter{equation}{0}

An example of a scanning of $\IR^3_{t}$ is
\br
x^1&=&\zeta\,\cos^2\tau\(1-\cos\sigma\)-L;\qquad\qquad 0\leq \sigma\leq 2\,\pi;\qquad\quad L\rightarrow \infty
\nonumber\\
x^2&=&-\zeta\,\cos\tau\,\sin\sigma;\qquad\qquad\qquad\quad -\frac{\pi}{2}\leq \tau \leq \frac{\pi}{2}
\lab{nicescan}\\
x^3&=&\zeta\,\cos\tau\,\sin\tau\(1-\cos\sigma\);\qquad\qquad 0\leq \zeta < \infty
\nonumber
\er
The reference point is $x_R=\(-L\,,\,0\,,\,0\)$, the surfaces of constant $\zeta$ are spheres of radius $\zeta$, with center at $\(-L+\zeta\,,\,0\,,\, 0\)$. Indeed $\left[\(x^1+L-\zeta\)^2+\(x^2\)^2+\(x^3\)^2\right]=\zeta^2$. The Jacobian, defined in \rf{jacobiandef},  becomes
\be
\ve_{ijk}\,\frac{d\,x^i}{d\,\sigma}\,\frac{d\,x^j}{d\,\tau}\,\frac{d\,x^k}{d\,\zeta}=4\,\zeta^2\,\cos^3\tau\,\sin^4\(\frac{\sigma}{2}\)
\ee
and so it is positive, and $\vartheta=1$. Note that the Jacobian only vanishes at the reference point $x_R$, since it corresponds to $\sigma=0\,{\rm or}\,2\,\pi$, or to $\zeta=0$ (the infinitesimally small sphere around $x_R$, or to $\tau=-\frac{\pi}{2}\,{\rm or}\,\frac{\pi}{2}$ (the infinitesimally small loops around $x_R$).

\section{Proof of relation \rf{caljdef2}}
\label{sec:proofnewj}
\setcounter{equation}{0}

Note that, in \rf{caljdef}, whenever we have the commutator of two matrices of the same type, we can use the following manipulations to eliminate the intertwined integrals
\br
&&\int_{\sigma_i}^{\sigma_f}d\sigma\,\int_{\sigma_i}^{\sigma}d\sigma^{\prime}\,\sbr{M_{\kappa\rho}\(\sigma^{\prime}\)}{M_{\mu\nu}\(\sigma\)}\, \frac{dx^{\kappa}}{d\sigma^{\prime}}\frac{dx^{\mu}}{d\sigma}\,
 \(\frac{d\,x^{\rho}\(\sigma^{\prime}\)}{d\,\tau}\frac{d\,x^{\nu}\(\sigma\)}{d\,\zeta}
-\frac{d\,x^{\rho}\(\sigma^{\prime}\)}{d\,\zeta}\,\frac{d\,x^{\nu}\(\sigma\)}{d\,\tau}\) 
\nonumber\\
&=&\int_{\sigma_i}^{\sigma_f}d\sigma\,\int_{\sigma_i}^{\sigma}d\sigma^{\prime}\left[
\sbr{M_{\kappa\rho}\(\sigma^{\prime}\)\frac{dx^{\kappa}}{d\sigma^{\prime}}\frac{d\,x^{\rho}\(\sigma^{\prime}\)}{d\,\tau}}{M_{\mu\nu}\(\sigma\)\frac{dx^{\mu}}{d\sigma}\frac{d\,x^{\nu}\(\sigma\)}{d\,\zeta}}
\right.\nonumber\\
&+&\left.
\sbr{M_{\mu\nu}\(\sigma\)\frac{dx^{\mu}}{d\sigma}\frac{d\,x^{\nu}\(\sigma\)}{d\,\tau}}{M_{\kappa\rho}\(\sigma^{\prime}\)\frac{dx^{\kappa}}{d\sigma^{\prime}}\frac{d\,x^{\rho}\(\sigma^{\prime}\)}{d\,\zeta}}
\right]
\nonumber\\
&=&\left[\int_{\sigma_i}^{\sigma_f}d\sigma\,\int_{\sigma_i}^{\sigma}d\sigma^{\prime}+
\int_{\sigma_i}^{\sigma_f}d\sigma^{\prime}\,\int_{\sigma_i}^{\sigma^{\prime}}d\sigma\right]
\sbr{M_{\kappa\rho}\(\sigma^{\prime}\)\frac{dx^{\kappa}}{d\sigma^{\prime}}\frac{d\,x^{\rho}\(\sigma^{\prime}\)}{d\,\tau}}{M_{\mu\nu}\(\sigma\)\frac{dx^{\mu}}{d\sigma}\frac{d\,x^{\nu}\(\sigma\)}{d\,\zeta}}
\nonumber\\
&=&\sbr{\int_{\sigma_i}^{\sigma_f}d\sigma^{\prime}\,M_{\kappa\rho}\(\sigma^{\prime}\)\frac{dx^{\kappa}}{d\sigma^{\prime}}\frac{d\,x^{\rho}\(\sigma^{\prime}\)}{d\,\tau}}{\int_{\sigma_i}^{\sigma_f}d\sigma\,M_{\mu\nu}\(\sigma\)\frac{dx^{\mu}}{d\sigma}\frac{d\,x^{\nu}\(\sigma\)}{d\,\zeta}}
\er

\section{The quantity ${\cal M}$ given in \rf{calmdef}}
\label{sec:proofcalm}
\setcounter{equation}{0}

The Wilson line operator $W$ is defined on a given curve $x^{\mu}\(\sigma\)$, parameterized by $\sigma$ through the equation \rf{wdefa}, such that $x^{\mu}\(\sigma_i\)$ corresponds to the reference point $x_R$. If we vary such a curve, $x^{\mu}\(\sigma\)\rightarrow x^{\mu}\(\sigma\)+ \delta x^{\mu}\(\sigma\)$, keeping the reference point $x_R$ fixed we get that $W$, integrated up to a given point $\sigma$, changes as (for the details of such a calculation see \cite{afg1,afg2,ym1,ym2}, specially Section 2 of \cite{afg1})
\be
W^{-1}\(\sigma\)\delta W\(\sigma\)=-i\,e\,W^{-1}\(\sigma\)\,A_{\mu}\(\sigma\)\,W\(\sigma\)\,\delta x^{\mu}\(\sigma\)+i\,e\,\int_{\sigma_i}^{\sigma}d\sigma^{\prime}\,W^{-1}\,F_{\mu\nu}\,W\,\frac{d\,x^{\mu}}{d\,\sigma^{\prime}}\,\delta x^{\nu}
\lab{variationw}
\ee 
In the scanning of a given volume $\Omega$, we have the loops labelled by $\tau$ and the closed surfaces labelled by $\zeta$. So, we can consider the variations of the Wilson line when we vary a given  loop in the scanning to one infinitesimally close to it on the same surface, and in such a case we have $\delta x^{\mu}= \frac{d\,x^{\mu}}{d\,\tau}\,d\tau$. Similarly, we can consider the variation of $W$ when we vary a given loop on a given surface to one lying on a surface infinitesimally close to it, and in such a case we have $\delta x^{\mu}= \frac{d\,x^{\mu}}{d\,\zeta}\,d\zeta$. Therefore, from \rf{variationw} we get differential equations for $W$ associated to such variations
\be
W^{-1}\(\sigma\)\frac{d\, W}{d\,\tau/\zeta}\(\sigma\)=-i\,e\,W^{-1}\(\sigma\)\,A_{\mu}\(\sigma\)\,W\(\sigma\)\,\frac{d\,x^{\mu}}{d\,\tau/\zeta}\(\sigma\)+i\,e\,\int_{\sigma_i}^{\sigma}d\sigma^{\prime}\,W^{-1}\,F_{\mu\nu}\,W\,\frac{d\,x^{\mu}}{d\,\sigma^{\prime}}\,\frac{d\,x^{\nu}}{d\,\tau/\zeta}
\lab{variationw2}
\ee
If we close the loop, integrating up to $\sigma_f$, we get
\be
W^{-1}\(\sigma_f\)\frac{d\, W}{d\,\tau/\zeta}\(\sigma_f\)=i\,e\,\int_{\sigma_i}^{\sigma_f}d\sigma^{\prime}\,W^{-1}\,F_{\mu\nu}\,W\,\frac{d\,x^{\mu}}{d\,\sigma^{\prime}}\,\frac{d\,x^{\nu}}{d\,\tau/\zeta}
\lab{variationw2b}
\ee
That is so, because we always keep the reference point fixed, and so (see Appendix \ref{sec:scanning} for an example of scanning of $\IR^3_{t}$)
 \be
\frac{d\,x^{\mu}}{d\,\tau}= \frac{d\,x^{\mu}}{d\,\zeta}=0\qquad\qquad {\rm at}\qquad \sigma=\sigma_i\quad {\rm and} \quad \sigma=\sigma_f
\lab{vanishsigmader}
\ee

For the case where the volume $\Omega$ corresponds to the spatial sub-manifold $\IR^3_t$, we can use \rf{variationw2} to express the quantity $\mathfrak{b}_{\tau/\zeta}\(\sigma\)$, defined in \rf{ebfrakdef}, as 
 \be
 \mathfrak{b}_{\tau/\zeta}\(\sigma\)=- W^{-1}\(\sigma\)\,A_i\,W\(\sigma\)\,\frac{d\,x^i}{d\,\tau/\zeta}+\frac{i}{e}\,W^{-1}\(\sigma\)\,\frac{d\,W\(\sigma\)}{d\,\tau/\zeta}; \qquad\qquad {\rm for}\qquad \sigma< \sigma_f
 \lab{bfrakdefsimple}
 \ee
 When we close the loop, i.e., take the integral in \rf{ebfrakdef} up to $\sigma_f$, it turns out, due to \rf{vanishsigmader}, that 
 \be
 \mathfrak{b}_{\tau/\zeta}\(\sigma_f\)=\frac{i}{e}\,W^{-1}\(\sigma_f\)\,\frac{d\,W\(\sigma_f\)}{d\,\tau/\zeta}
 \lab{bfrakdefsimple2}
 \ee
 
 For any given Lie algebra valued quantity $L$ conjugated with the Wilson line,  we have from \rf{bfrakdefsimple}, that
 \br
 \frac{d\;}{d\,\tau/\zeta}\(W^{-1}\(\sigma\)\,L\(\sigma\) \,W\(\sigma\)\)&=&W^{-1}\(\sigma\)\,D_i L\(\sigma\) \,W\(\sigma\)\,  \frac{d\,x^i}{d\,\tau/\zeta}
 \nonumber\\
 &-&i\,e\,\sbr{W^{-1}\(\sigma\)\,L\(\sigma\) \,W\(\sigma\)}{\mathfrak{b}_{\tau/\zeta}\(\sigma\) }
 \lab{identity1}
 \er
 where $D_i$ is the covariant derivative in the adjoint representation, i.e. 
 $D_i*=\partial_i*+i\,e\,\sbr{A_i}{*}$. In addition, from the defining equation of the Wilson line \rf{wdefa}, we have that
 \be
 \frac{d\;}{d\,\sigma}\(W^{-1}\(\sigma\)\,L\(\sigma\) \,W\(\sigma\)\)=W^{-1}\(\sigma\)\,D_i L\(\sigma\) \,W\(\sigma\)\,  \frac{d\,x^i}{d\,\sigma} 
 \lab{identity2}
\ee
We shall need the identity
\be
\frac{d\;}{d\,\tau}\(\ve_{ijk}\frac{d\,x^j}{d\,\sigma}\frac{d\,x^k}{d\,\zeta}\)-\frac{d\;}{d\,\zeta}\(\ve_{ijk}\frac{d\,x^j}{d\,\sigma}\frac{d\,x^k}{d\,\tau}\)=\frac{d\;}{d\,\sigma}\(\ve_{ijk}\frac{d\,x^j}{d\,\tau}\frac{d\,x^k}{d\,\zeta}\)
\lab{identity3}
\ee
Therefore, using \rf{identity1}, \rf{identity2} and \rf{identity3} we have from \rf{ebfrakdef}, \rf{holonomyl2fluxes} and \rf{tzetadef} that
\br
&&\frac{d\,{\cal T}_{\tau}}{d\,\zeta}-\frac{d\,{\cal T}_{\zeta}}{d\,\tau}=-i\,e\,\int_{\sigma_i}^{\sigma_f}d\sigma\left\{
-\frac{d\;}{d\,\sigma}\(W^{-1}\(\alpha\,B_i+\beta\,E_i\)W\,\ve_{ijk}\frac{d\,x^j}{d\,\tau}\frac{d\,x^k}{d\,\zeta}\)
\right. \nonumber\\
&+&\left. W^{-1}D_l\(\alpha\,B_i+\beta\,E_i\)W\,\ve_{ijk}\(\frac{d\,x^j}{d\,\sigma}\frac{d\,x^k}{d\,\tau}\frac{d\,x^l}{d\,\zeta}-\frac{d\,x^j}{d\,\sigma}\frac{d\,x^k}{d\,\zeta}\frac{d\,x^l}{d\,\tau}+\frac{d\,x^j}{d\,\tau}\frac{d\,x^k}{d\,\zeta}\frac{d\,x^l}{d\,\sigma}\)
\right. \nonumber\\
&+&\left. i\,e\,\sbr{W^{-1}\(\alpha\,B_i+\beta\,E_i\)W}{\mathfrak{b}_{\tau}\(\sigma\)}\,\ve_{ijk}\frac{d\,x^j}{d\,\sigma}\frac{d\,x^k}{d\,\zeta}
\right. \nonumber\\
&-&\left. i\,e\,\sbr{W^{-1}\(\alpha\,B_i+\beta\,E_i\)W}{\mathfrak{b}_{\zeta}\(\sigma\)}\,\ve_{ijk}\frac{d\,x^j}{d\,\sigma}\frac{d\,x^k}{d\,\tau}
\right\}
\lab{proofcurlt}
\er
From \rf{vanishsigmader} we observe that the first term on the right-hand side of \rf{proofcurlt} vanishes.  In addition, we have that
\be
\ve_{ijk}\(\frac{d\,x^j}{d\,\sigma}\frac{d\,x^k}{d\,\tau}\frac{d\,x^l}{d\,\zeta}-\frac{d\,x^j}{d\,\sigma}\frac{d\,x^k}{d\,\zeta}\frac{d\,x^l}{d\,\tau}+\frac{d\,x^j}{d\,\tau}\frac{d\,x^k}{d\,\zeta}\frac{d\,x^l}{d\,\sigma}\)=
\delta_{il}\, \ve_{jkm}\frac{d\,x^j}{d\,\sigma}\frac{d\,x^k}{d\,\tau}\frac{d\,x^m}{d\,\zeta}
\ee
Using \rf{ebfrakdef}, \rf{caljgdefstatic} and \rf{caljmagdefstatic}, we have that
\br
&&\int_{\sigma_i}^{\sigma_f}d\sigma\left\{\sbr{W^{-1}\(\alpha\,B_i+\beta\,E_i\)W}{\mathfrak{b}_{\tau}\(\sigma\)}\,\ve_{ijk}\frac{d\,x^j}{d\,\sigma}\frac{d\,x^k}{d\,\zeta}
\right.\nonumber\\
&&\left. -\sbr{W^{-1}\(\alpha\,B_i+\beta\,E_i\)W}{\mathfrak{b}_{\zeta}\(\sigma\)}\,\ve_{ijk}\frac{d\,x^j}{d\,\sigma}\frac{d\,x^k}{d\,\tau}\right\}
\nonumber\\
&=& \int_{\sigma_i}^{\sigma_f}d\sigma\left\{\sbr{\alpha\,\frac{d\,\mathfrak{b}_{\zeta}\(\sigma\)}{d\,\sigma}+\beta\, \frac{d\,\mathfrak{e}_{\zeta}\(\sigma\)}{d\,\sigma}}{\mathfrak{b}_{\tau}\(\sigma\)} 
-\sbr{\alpha\,\frac{d\,\mathfrak{b}_{\tau}\(\sigma\)}{d\,\sigma}+\beta\, \frac{d\,\mathfrak{e}_{\tau}\(\sigma\)}{d\,\sigma}}{\mathfrak{b}_{\zeta}\(\sigma\)}\right\}
\nonumber\\
&=&i\(\beta\,\rho_G+\alpha\,\rho_{\rm mag.}\)
\er
where we have used the fact that, due to \rf{ebfrakdef}, we have $\mathfrak{b}_{\tau/\zeta}\(\sigma_i\)=0$.

Therefore, \rf{proofcurlt} becomes 
\be
\frac{d\,{\cal T}_{\tau}}{d\,\zeta}-\frac{d\,{\cal T}_{\zeta}}{d\,\tau}-ie^2\(\beta\,\rho_G+\alpha\,\rho_{\rm mag.}\) =-ie
\int_{\sigma_i}^{\sigma_f}d\sigma W^{-1}\(\alpha\,D_l B_l+\beta\,D_l E_l\)W\,\ve_{ijk}\frac{d\,x^i}{d\,\sigma}\frac{d\,x^j}{d\,\tau}\frac{d\,x^k}{d\,\zeta}
\lab{proofcurlt2}
\ee
Adding $\(-i\,e^2\, \beta\,\rho_M\)$ to both sides of \rf{proofcurlt2} (with $\rho_M$ given in \rf{jmatterdefstatic}), we get that \rf{calmdef} is indeed equal to  \rf{calmdef2}.

\section{Poisson brackets for the gauge fields} 
\label{sec:pbgauge}
\setcounter{equation}{0}

In order to evaluate the transformations of the gauge fields,  the electric and magnetic fields, we need the Poisson brackets involving the Wilson line. Note that, since we are in $\IR^3_{t}$, the derivatives with respect to $\sigma$, $\tau$, and $\zeta$ are space derivatives, and so they commute with the equal time Poisson bracket. Therefore, the Poisson bracket of a given quantity $X$ with a spatial Wilson line can be extracted directly from the  holonomy equations restricted to space loops (see \rf{wdefa})
\be
\frac{d\,W}{d\,\sigma}+i\,e\,A_{i}\,\frac{d\,x^{i}}{d\,\sigma}\,W=0; \qquad\qquad 
\frac{d\,W^{-1}}{d\,\sigma}-i\,e\,W^{-1}\,A_{i}\,\frac{d\,x^{i}}{d\,\sigma}=0
\lab{wdefastatic}
\ee 
Indeed, we get that
\be
\frac{d\,\(W^{-1}\,\pbr{X}{W}\)}{d\,\sigma}+i\,e\,W^{-1}\,\pbr{X}{A_i}\,W\, \frac{d\,x^{i}}{d\,\sigma}=0
\ee
where we have used the fact that $X$ does not depend upon $\sigma$. Therefore, since $W\(\sigma_i\)$ is a constant of integration independent of the fields, we get that
\be
\pbr{X}{W\(\sigma\)}=-i\,e\,W\(\sigma\)\,\int_{\sigma_i}^{\sigma} d\sigma^{\prime}\,W^{-1}\(\sigma^{\prime}\)\,\pbr{X}{A_i\(\sigma^{\prime}\)}\,W\(\sigma^{\prime}\)\, \frac{d\,x^{i}}{d\,\sigma^{\prime}}
\lab{nicepbw}
\ee
In addition, from the properties of the Poisson bracket, we  get that
\be
\pbr{X}{W^{-1}\,T_a\,W}=\sbr{W^{-1}\,T_a\,W}{W^{-1}\,\pbr{X}{W}}
\lab{nicepbw2}
\ee

From the Poisson bracket relations \rf{pbrel} we observe that the gauge field $A_i$ and magnetic field $B_i$ commute. In addition, from \rf{nicepbw} we have that $A_i$ commutes with the Wilson line $W$. Therefore,  from \rf{ebfrakdef} we have that
\be
\pbr{A_i^a\(x\,,\,t\)}{\mathfrak{b}_{\tau/\zeta}\(\sigma\)}=0
\lab{pbafrakb}
\ee
From \rf{pbrel}, \rf{ebfrakdef}, \rf{nicepbw} and \rf{nicepbw2} we have that
\br
\pbr{A_i^a\(x\,,\,t\)}{\mathfrak{e}_{\tau}\(\sigma\)}=\int_{\sigma_i}^{\sigma} d\sigma^{\prime}\,\delta^{(3)}\(x-y\)\,\ve_{ijk}\,\frac{d\,y^j}{d\,\sigma^{\prime}}\, \frac{d\,y^k}{d\,\tau}\,W^{-1}\(\sigma^{\prime}\)\, T_a\,W\(\sigma^{\prime}\)
\lab{pbafrake}
\er
where we have used, in the scanning of $\IR^3_t$, the correspondence
\be
\(x^1\,,\,x^2\,,\,x^3\)\equiv \(\zeta_x\,,\,\tau_x\,,\,\sigma_x\);\qquad\qquad\qquad
\(y^1\,,\,y^2\,,\,y^3\)\equiv \(\zeta\,,\,\tau\,,\,\sigma^{\prime}\)
\lab{correspondxparameter}
\ee
We observe that 
\be
\delta^{(3)}\(x-y\)\,\ve_{ijk}\,\frac{d\,x^i}{d\,\sigma_x/\tau_x}\, \frac{d\,y^j}{d\,\sigma^{\prime}}\, \frac{d\,y^k}{d\,\tau}=0
\lab{deltaidentity}
\ee
since the three dimensional Dirac delta function impose the points $x^i$ and $y^i$ to coincide, and when that happens the vectors $\frac{d\,x^i}{d\,\sigma_x}$ and  $\frac{d\,y^j}{d\,\sigma^{\prime}}$ are parallel, as well as  $\frac{d\,x^i}{d\,\tau_x}$ and  $\frac{d\,y^k}{d\,\tau}$. Therefore
\br
\pbr{A_i^a\(x\,,\,t\)\,\frac{d\,x^i}{d\,\sigma_x}}{\mathfrak{e}_{\tau}\(\sigma\)}&=&0
\lab{pbafrake2}\\
\pbr{A_i^a\(x\,,\,t\)\,\frac{d\,x^i}{d\,\tau_x}}{\mathfrak{e}_{\tau}\(\sigma\)}&=&0
\nonumber
\er
On the other hand, using \rf{jacobiandef} and \rf{jacobiandeltarel}, we get
\be
\pbr{A_i^a\(x\,,\,t\)\,\frac{d\,x^i}{d\,\zeta_x}}{\mathfrak{e}_{\tau}\(\sigma\)}=\vartheta\, \delta\(\zeta-\zeta_x\)\,\delta\(\tau-\tau_x\)\, \theta\(\sigma-\sigma_x\)\,W^{-1}\(\sigma_x\)\, T_a\,W\(\sigma_x\)
\lab{pbafrake3}
\ee
where $\theta\(\sigma-\sigma^{\prime}\)$ is the step function, i.e. $\theta\(\sigma-\sigma^{\prime}\)=1$ for $\sigma \geq \sigma^{\prime}$ and zero otherwise. 

From \rf{pbrel} and \rf{nicepbw} we have that the magnetic field $B_i$ commutes with the Wilson line $W$, and so, from \rf{ebfrakdef}, we have that 
\be
\pbr{B_i^a\(x\,,\,t\)}{\mathfrak{b}_{\tau/\zeta}\(\sigma\)}=0
\lab{pbbfrakb}
\ee
Using \rf{ebpbrel}, we have from \rf{ebfrakdef} that
\br
&&\pbr{B_i^a\(x\,,\,t\)}{\mathfrak{e}_{\tau/\zeta}\(\sigma\)}=\int_{\sigma_i}^{\sigma} d\sigma^{\prime}\(\frac{d\,y^i}{d\,\sigma^{\prime}}\, \frac{d\,y^j}{d\,\tau/\zeta}-\frac{d\,y^j}{d\,\sigma^{\prime}}\,\frac{d\,y^i}{d\,\tau/\zeta}\)\,\times
\lab{pbbfrakb2}\\
&&\times W^{-1}\(\sigma^{\prime}\)\(-T_a\,\frac{\partial\,\delta^{(3)}\(x-y\)}{\partial\,x^j}+i\,e\,\sbr{A_j\(x\)}{T_a}\,\delta^{(3)}\(x-y\)\)W\(\sigma^{\prime}\)
\nonumber
\er

Using \rf{nicepbw}, \rf{nicepbw2} and \rf{pbrel} we get
\be
\pbr{E_i^a\(x\,,\,t\)}{\mathfrak{e}_{\tau/\zeta}\(\sigma\)}=
i\,e\,\int_{\sigma_i}^{\sigma}d\sigma^{\prime}\,
\sbr{\frac{d\,\mathfrak{e}_{\tau/\zeta}\(\sigma^{\prime}\)}{d\,\sigma^{\prime}}}{\int_{\sigma_i}^{\sigma^{\prime}}d\sigma^{\prime\prime}\,W^{-1}\(\sigma^{\prime\prime}\)\,T_a\,W\(\sigma^{\prime\prime}\)\,\frac{d\,z^i}{d\,\sigma^{\prime\prime}}\,\delta\(x-z\)}
\lab{pbefrake}
\ee
where we have used, in the scanning of $\IR^3_t$, besides \rf{correspondxparameter}, the correspondence
\be
\(z^1\,,\,z^2\,,\,z^3\)\equiv \(\zeta\,,\,\tau\,,\,\sigma^{\prime\prime}\)
\lab{correspondxparameter2}
\ee
Note that we have
\be
\ve_{ijk}\,\frac{d\,z^i}{d\,\sigma^{\prime\prime}}\,\frac{d\,x^j}{d\,\sigma_x} \,\frac{d\,x^k}{d\,\tau_x/\zeta_x} \,\delta\(x-z\)=0
\lab{sigmaprimeprimesigmaxratuzetaxvainh}
\ee
as the delta function imposes $\frac{d\,z^i}{d\,\sigma^{\prime\prime}}$ and $\frac{d\,x^j}{d\,\sigma_x}$ to be parallel. Therefore
\be
\pbr{E_i^a\(x\,,\,t\)\,\ve_{ijk}\,\frac{d\,x^j}{d\,\sigma_x} \,\frac{d\,x^k}{d\,\tau_x/\zeta_x} }{\mathfrak{e}_{\tau/\zeta}\(\sigma\)}=0
\lab{pbetauzetamathfracketauzeta}
\ee

Using \rf{bfrakdefsimple}, \rf{nicepbw}, \rf{nicepbw2}, and \rf{pbrel} we get that
\br
\pbr{E_i^a\(x\,,\,t\)}{\mathfrak{b}_{\tau/\zeta}\(\sigma\)}&=&W^{-1}\(\sigma\)\,T_a\,W\(\sigma\)\,\frac{d\,w^i}{d\tau/\zeta}\,\,\delta^{(3)}\(x-w\)
\nonumber\\
&+&i\,e\,\sbr{\mathfrak{b}_{\tau/\zeta}\(\sigma\)}{\int_{\sigma_i}^{\sigma}d\sigma^{\prime}\,W^{-1}\(\sigma^{\prime}\)\,T_a\,W\(\sigma^{\prime}\)\,\frac{d\,y^i}{d\sigma^{\prime}}\,\delta^{(3)}\(x-y\)}
\nonumber\\
&-&\frac{d\;}{d\,\tau/\zeta}\,\int_{\sigma_i}^{\sigma}d\sigma^{\prime}\,W^{-1}\(\sigma^{\prime}\)\,T_a\,W\(\sigma^{\prime}\)\,\frac{d\,y^i}{d\sigma^{\prime}}\,\delta^{(3)}\(x-y\)
\lab{pbefrakb}
\er
where we have used, in the scanning of $\IR^3_t$, besides \rf{correspondxparameter} and \rf{correspondxparameter2}, the correspondence
\be
\(w^1\,,\,w^2\,,\,w^3\)\equiv \(\zeta\,,\,\tau\,,\,\sigma\)
\lab{correspondxparameter3}
\ee
If we take $\sigma=\sigma_f$ and use \rf{bfrakdefsimple2} instead of  \rf{bfrakdefsimple}, we get
\br
\pbr{E_i^a\(x\,,\,t\)}{\mathfrak{b}_{\tau/\zeta}\(\sigma_f\)}&=&i\,e\,\sbr{\mathfrak{b}_{\tau/\zeta}\(\sigma_f\)}{\int_{\sigma_i}^{\sigma_f}d\sigma^{\prime}\,W^{-1}\(\sigma^{\prime}\)\,T_a\,W\(\sigma^{\prime}\)\,\frac{d\,y^i}{d\sigma^{\prime}}\,\delta^{(3)}\(x-y\)}
\nonumber\\
&-&\frac{d\;}{d\,\tau/\zeta}\,\int_{\sigma_i}^{\sigma_f}d\sigma^{\prime}\,W^{-1}\(\sigma^{\prime}\)\,T_a\,W\(\sigma^{\prime}\)\,\frac{d\,y^i}{d\sigma^{\prime}}\,\delta^{(3)}\(x-y\)
\lab{pbefrakb2}
\er
Using relations similar to \rf{sigmaprimeprimesigmaxratuzetaxvainh} we get, from \rf{pbefrakb2}, that
\be
\pbr{E_i^a\(x\,,\,t\)\,\ve_{ijk}\,\frac{d\,x^j}{d\,\sigma_x} \,\frac{d\,x^k}{d\,\tau_x/\zeta_x} }{\mathfrak{b}_{\tau/\zeta}\(\sigma_f\)}=0
\lab{pbbtauzetamathfracketauzeta}
\ee

\section{Poisson brackets involving $H_C$ and $H_B$} 
\label{sec:pbhchb}
\setcounter{equation}{0}

Using \rf{constraintgengauge} we get that $H_C$, defined in \rf{hamiltonianparts}, satifies
\be
\pbr{H_C}{A_i\(y\)}=D_i A_0\(y\)-\int d^3x\, \frac{\partial\;}{\partial\,x^i}\left[A_0\(x\)\,\delta^{(3)}\(x-y\)\right]
\ee
as $A_0$ is a Lagrange multiplier, and so, drops out of the Poisson bracket, and where $D_i A_0=\partial_iA_0+i\,e\,\sbr{A_i}{A_0}$. Using \rf{wdefastatic}, we get that for any quantity $X$,  
\be
\frac{d\;}{d\,\sigma}\(W^{-1}\(\sigma\)\,X\,W\(\sigma\)\)=W^{-1}\(\sigma\)\,\left(\partial_i X+i\,e\,\sbr{A_i}{X}\right)\,W\(\sigma\)\, \frac{d\,x^i}{d\,\sigma}
\ee
Therefore, from \rf{nicepbw} we get
\br
\pbr{H_C}{W\(\sigma\)}&=&-i\,e\,W\(\sigma\)\left[\int_{\sigma_i}^{\sigma} d\sigma^{\prime}\,\frac{d\;}{d\,\sigma^{\prime}}\left[W^{-1}\(\sigma^{\prime}\)\,A_0\(\sigma^{\prime}\)\,W\(\sigma^{\prime}\)\right]
\right.
\\
&-&\left. \int_{\sigma_i}^{\sigma} d\sigma^{\prime}\,W^{-1}\(\sigma^{\prime}\)\,T_b\,W\(\sigma^{\prime}\)\,\frac{d\,y^{i}}{d\,\sigma^{\prime}}\,\int d^3x\, \frac{\partial\;}{\partial\,x^i}\left[A^b_0\(x\)\,\delta^{(3)}\(x-y\)\right] \right]
\nonumber
\er
Using \rf{abeliangauss} we get
\br
&&\frac{d\,y^{i}}{d\,\sigma^{\prime}}\,\int d^3x\, \frac{\partial\;}{\partial\,x^i}\left[A^b_0\(x\)\,\delta^{(3)}\(x-y\)\right] =
\nonumber\\
&&=
\vartheta\, \int_{\tau_i}^{\tau_f}d\tau\,\int_{\sigma_i}^{\sigma_f}d\sigma\,A^b_0\(x\)\,\delta^{(3)}\(x-y\)\,\ve_{ijk}\,\frac{d\,y^{i}}{d\,\sigma^{\prime}}\,\frac{d\,x^j}{d\,\sigma}\,\frac{d\,x^k}{d\,\tau}\mid_{\zeta=\zeta_f}=0
\er
since the vectors $\frac{d\,y^{i}}{d\,\sigma^{\prime}}$ and $\frac{d\,x^j}{d\,\sigma}$ become parallel due to the delta function. Therefore
\be
\pbr{H_C}{W\(\sigma\)}=-i\,e\left[A_0\(x\(\sigma\)\)\,W\(\sigma\)-W\(\sigma\)\,W^{-1}_R\,A_0\(x_R\)\,W_R\right]
\lab{pbhcw}
\ee
where $W_R=W\(\sigma_i\)$, and $x_R$ is the reference point sitting on the border of $\IR^3$, at the initial and final points of every loop. 

Using \rf{constraintgengauge} and \rf{pbhcw}, we get from \rf{ebfrakdef} that
\be
\pbr{H_C}{\mathfrak{e}_{\tau/\zeta}\(\sigma_f\)}=i\,e\,\sbr{\mathfrak{e}_{\tau/\zeta}\(\sigma_f\)}{W^{-1}_R\,A_0\(x_R\)\,W_R}
\lab{pbhcmathfracke}
\ee
Similarly, from \rf{bfrakdefsimple2} and \rf{pbhcw} we get that
\be
\pbr{H_C}{\mathfrak{b}_{\tau/\zeta}\(\sigma_f\)}=i\,e\,\sbr{\mathfrak{b}_{\tau/\zeta}\(\sigma_f\)}{W^{-1}_R\,A_0\(x_R\)\,W_R}
\lab{pbhcmathfrackb}
\ee

We now evaluate some quantities important for the transformation of $H_B$. Note that in the derivation of the abelian Gauss theorem \rf{abeliangauss} we never use the fact that the derivatives commute when acting on the quantity $b_l$. Therefore, we can use \rf{abeliangauss} to get
\br
&&\ve_{ijk}\int d^3y\,\frac{\partial\;}{\partial y^j}\,\frac{\partial\;}{\partial y^k}\left[d_{ba}\(y\){\rm Tr}\(B_i\(y\)\,T_a\)\right]=
\nonumber\\
&&=\vartheta\,\int_{\tau_i}^{\tau_f}d\tau\,\int_{\sigma_i}^{\sigma_f}d\sigma\,\frac{\partial\;}{\partial y^k}\left[d_{ba}\(y\){\rm Tr}\(B_i\(y\)\,T_a\)\right]\,\(\frac{d\,y^k}{d\,\sigma}\,\frac{d\,y^i}{d\,\tau}-\frac{d\,y^i}{d\,\sigma}\,\frac{d\,y^k}{d\,\tau}\)\mid_{\zeta=\zeta_f}
\nonumber\\
&&\rightarrow \frac{1}{r^{\frac{1}{2}+\delta}}\qquad {\rm as} \qquad r\rightarrow \infty
\lab{decayratefunnyterm}
\er
where we have used the boundary conditions \rf{boundcond}  and \rf{tangentvectorsinfinity}. Now, using \rf{calsabijdef}, we have
\br
&&\int d^3y\,{\rm Tr}\left[D_j\,B_i\(y\)\,T_a\right]\,{\cal S}_{ij}^{ba}\(y\)=
\nonumber\\
&&=\vartheta\,\int_{\tau_i}^{\tau_f}d\tau\,\int_{\sigma_i}^{\sigma_f}d\sigma\,d_{ba}\(x\)\,{\rm Tr}\left[D_j\,B_i\(x\)\,T_a\right]
\(\frac{d\,x^i}{d\,\sigma}\,\frac{d\,x^j}{d\,\tau}-\frac{d\,x^j}{d\,\sigma}\,\frac{d\,x^i}{d\,\tau}\)\mid_{\zeta=\zeta_f}
\er
In order for the magnetic field to satisfy the boundary condition \rf{boundcond}, we need 
\be
A_{i}\rightarrow \frac{1}{r^{1/2+\delta}} \qquad {\rm as} \qquad  r\rightarrow \infty \qquad {\rm for} \qquad \delta\geq 1/2; \quad i=1,2,3
\ee
and
\be
A_{i}\rightarrow \frac{1}{r^{3/4+\delta/2}} \qquad {\rm as} \qquad  r\rightarrow \infty \qquad {\rm for} \qquad 0<\delta\leq 1/2; \quad i=1,2,3
\ee
Therefore, the slowest possible decay rate for the covariant derivative of the magnetic field is
\be
D_j\,B_i \rightarrow \frac{1}{r^{\frac{9}{4}+\frac{3}{2}\delta}}\qquad {\rm as} \qquad  r\rightarrow \infty\qquad {\rm for} \qquad 0<\delta\leq 1/2; \quad i,j=1,2,3
\lab{decayratefunnyterm4}
\ee
Consequently, using \rf{tangentvectorsinfinity}, we get 
\be
\int d^3y\,{\rm Tr}\left[D_j\,B_i\(y\)\,T_a\right]\,{\cal S}_{ij}^{ba}\(y\)= \frac{1}{r^{\frac{1}{4}+\frac{3}{2}\delta}}\qquad {\rm as} \qquad  r\rightarrow \infty\qquad {\rm for} \qquad 0<\delta\leq 1/2
\lab{decayratefunnyterm2}
\ee

Using \rf{abeliangauss} and \rf{calsabijdef} we get
\br
&&\int d^3y\,\frac{\partial\;}{\partial y^j}\,\left[{\rm Tr}\(B_i\(y\)\,T_a\)\,{\cal S}_{ij}^{ba}\(y\) \right]=
\nonumber\\
&&=\int_{\tau_i}^{\tau_f}d\tau^{\prime}\,\int_{\sigma_i}^{\sigma_f}d\sigma^{\prime}\,\ve_{jmn}\frac{d\,y^m}{d\,\sigma^{\prime}}\,\frac{d\,y^n}{d\,\tau^{\prime}}\,{\rm Tr}\(B_i\(y\)\,T_a\)\times
\nonumber\\
&&\times\;
\int_{\tau_i}^{\tau_f}d\tau\,\int_{\sigma_i}^{\sigma_f}d\sigma\,d_{ba}\(x\)\,\delta^{(3)}\(x-y\)
\(\frac{d\,x^i}{d\,\sigma}\,\frac{d\,x^j}{d\,\tau}-\frac{d\,x^j}{d\,\sigma}\,\frac{d\,x^i}{d\,\tau}\)\mid_{\zeta=\zeta_f}=0
\lab{decayratefunnyterm3}
\er
It vanishes because
\be
\ve_{jmn}\frac{d\,y^m}{d\,\sigma^{\prime}}\,\frac{d\,y^n}{d\,\tau^{\prime}}\,\frac{d\,x^j}{d\,\tau}\,\delta^{(3)}\(x-y\)=0;\qquad\qquad
\ve_{jmn}\frac{d\,y^m}{d\,\sigma^{\prime}}\,\frac{d\,y^n}{d\,\tau^{\prime}}\,\frac{d\,x^j}{d\,\sigma}\,\delta^{(3)}\(x-y\)=0
\ee

\section{The algebra of the fluxes $\mathfrak{e}_{\tau/\zeta}$ and $\mathfrak{b}_{\tau/\zeta}$} 
\label{sec:algebrafluxes}
\setcounter{equation}{0}

Let us  evaluate the Poisson brackets among the fluxes $\mathfrak{e}_{\tau/\zeta}\(\sigma\)$ and $\mathfrak{b}_{\tau/\zeta}\(\sigma\)$, defined in \rf{ebfrakdef}.   Since the Wilson line only involves the gauge fields $A_i$, we get from \rf{pbrel} and \rf{nicepbw} that
\be
\{W\(\sigma\)\,\overset{\otimes}{,}\,W\(\sigma^{\prime}\)\}_{PB}=0
\lab{nicepbw3}
\ee
For the same reasons, we get from \rf{bfrakdefsimple} that 
\be
\{\mathfrak{b}_{\tau/\zeta}\(\sigma\)\,\overset{\otimes}{,}\,W\(\sigma^{\prime}\)\}_{PB}=0
\lab{nicepbfbw}
\ee
and also 
\be
\{\mathfrak{b}_{\tau/\zeta}\(\sigma\)\,\overset{\otimes}{,}\,\mathfrak{b}_{\tau^{\prime}/\zeta^{\prime}}\(\sigma^{\prime}\)\}_{PB}=0
\lab{nicepbfb}
\ee
Note that in \rf{nicepbw3}, \rf{nicepbfbw} and \rf{nicepbfb} the Wilson lines and the magnetic fluxes do not have to sit on the same loop of the scanning of $\IR^3_{t}$. They can be on different loops of different surfaces. 

Consider now the quantity $E_i^a\,\ve_{ijk}\,
\frac{d\,x^j}{d\,\sigma}\,\frac{d\,x^k}{d\,\tau/\zeta}$ evaluated on a point in $\IR^3_{t}$, which under the scanning gets  labelled by $\(\sigma\,,\,\tau\,,\,\zeta\)$, i.e.  $x^i\(\sigma\,,\,\tau\,,\,\zeta\)$. Take now a Wilson line evaluated from the reference point $x_R$ to a point $\sigma^{\prime}$, along a loop labelled by $\tau^{\prime}$ and sitting on a surface labelled by $\zeta^{\prime}$. So, the points along that Wilson line are given by $y^i\(\sigma^{\prime}\,,\,\tau^{\prime}\,,\,\zeta^{\prime}\)$. From \rf{pbrel} and \rf{nicepbw} we get that
\br
&&\pbr{E_i^a\,\ve_{ijk}\,
\frac{d\,x^j}{d\,\sigma}\,\frac{d\,x^k}{d\,\tau/\zeta}}{W\(\sigma^{\prime}\)}=
\nonumber\\
&&=i\,e\,W\(\sigma^{\prime}\)\,\int_{\sigma_i}^{\sigma^{\prime}} d\sigma^{\prime\prime}\,W^{-1}\(\sigma^{\prime\prime}\)\,T_a\,W\(\sigma^{\prime\prime}\)\,\ve_{ijk}\,\frac{d\,y^i}{d\,\sigma^{\prime\prime}}\,\frac{d\,x^j}{d\,\sigma}\,\frac{d\,x^k}{d\,\tau/\zeta}\,\delta^{(3)}\(x-y\)
\lab{prepbew}
\er
Note that the three dimensional Dirac delta function imply that \rf{prepbew} does not vanish only if the points $x^i\(\sigma\,,\,\tau\,,\,\zeta\)$ and $y^i\(\sigma^{\prime}\,,\,\tau^{\prime}\,,\,\zeta^{\prime}\)$ coincide. But when that happens the cross product of tangent vectors $\ve_{ijk}\,\frac{d\,y^i}{d\,\sigma^{\prime\prime}}\,\frac{d\,x^j}{d\,\sigma}$, vanishes since $\frac{d\,y^i}{d\,\sigma^{\prime\prime}}$ and $\frac{d\,x^j}{d\,\sigma}$ will be parallel. Consequently,  \rf{prepbew} vanishes, and from \rf{ebfrakdef} we get that 
\be
\{\mathfrak{e}_{\tau/\zeta}\(\sigma\)\,\overset{\otimes}{,}\,W\(\sigma^{\prime}\)\}_{PB}=0
\lab{nicepbfew}
\ee
and consequently the electric fluxes  Poisson commute, i.e. 
\be
\{\mathfrak{e}_{\tau/\zeta}\(\sigma\)\,\overset{\otimes}{,}\,\mathfrak{e}_{\tau^{\prime}/\zeta^{\prime}}\(\sigma^{\prime}\)\}_{PB}=0
\lab{nicepbfe}
\ee
From \rf{bfrakdefsimple2} we observe that $\mathfrak{b}_{\tau/\zeta}\(\sigma_f\)$ only involves the Wilson loop $W\(\sigma_f\)$, and so from \rf{nicepbfew}  we get that
\be
\{\mathfrak{e}_{\tau/\zeta}\(\sigma\)\,\overset{\otimes}{,}\,\mathfrak{b}_{\tau^{\prime}/\zeta^{\prime}}\(\sigma_f\)\}_{PB}=0
\lab{nicepbfbloop}
\ee

However, for $\sigma < \sigma_f$, we have from \rf{bfrakdefsimple} that $\mathfrak{b}_{\tau/\zeta}\(\sigma\)$ involves, besides $W\(\sigma\)$, the gauge field $A_i$, and that does not Poisson commute with the electric field. Therefore, from  the fact that \rf{prepbew} vanishes, and using \rf{pbrel}, \rf{ebfrakdef}, \rf{bfrakdefsimple}, \rf{nicepbw} and \rf{nicepbw2}, we get that 
\br
\{\mathfrak{e}_{\tau/\zeta}\(\sigma\)\,\overset{\otimes}{,}\,\mathfrak{b}_{\tau^{\prime}/\zeta^{\prime}}\(\sigma^{\prime}\)\}_{PB}&=&
\int_{\sigma_i}^{\sigma} d\sigma^{\prime\prime}\,
W^{-1}\(\sigma^{\prime\prime}\)\,T_a\,W\(\sigma^{\prime\prime}\)\otimes 
W^{-1}\(\sigma^{\prime}\)\,T_a\,W\(\sigma^{\prime}\)\times
\nonumber\\
&\times&\ve_{ijk}\,\frac{d\,x^j}{d\,\sigma^{\prime\prime}}\,\frac{d\,x^k}{d\,\tau/\zeta}\,\frac{d\,y^i}{d\,\tau^{\prime}/\zeta^{\prime}}\,
\delta^{(3)}\(x-y\)
\lab{prepbeb}
\er
where we have used the same notation as above, i.e. the points of the loop where $\mathfrak{e}_{\tau/\zeta}\(\sigma\)$ sits are $x^i\(\sigma\,,\,\tau\,,\,\zeta\)$, and the points of the loop where $\mathfrak{b}_{\tau/\zeta}\(\sigma^{\prime}\)$ sits are 
$y^i\(\sigma^{\prime}\,,\,\tau^{\prime}\,,\,\zeta^{\prime}\)$. The integration in $\sigma^{\prime\prime}$ is along the loop where $\mathfrak{e}_{\tau/\zeta}\(\sigma\)$  sits. Again the Dirac delta function implies that \rf{prepbeb} is only non-vanishing when the points $x^i\(\sigma\,,\,\tau\,,\,\zeta\)$ and $y^i\(\sigma^{\prime}\,,\,\tau^{\prime}\,,\,\zeta^{\prime}\)$ coincide. But the cross product of tangent vectors $\ve_{ijk}\,\frac{d\,x^k}{d\,\tau/\zeta}\,\frac{d\,y^i}{d\,\tau^{\prime}/\zeta^{\prime}}$ is only non-vanishing when the tangent vectors do not get parallel in the limit $x\rightarrow y$. But that can not happen only for the cases  $\ve_{ijk}\,\frac{d\,x^k}{d\,\tau}\,\frac{d\,y^i}{d\,\zeta^{\prime}}$ and $\ve_{ijk}\,\frac{d\,x^k}{d\,\zeta}\,\frac{d\,y^i}{d\,\tau^{\prime}}$. Therefore, we get that 
\br
\{\mathfrak{e}_{\tau}\(\sigma\)\,\overset{\otimes}{,}\,\mathfrak{b}_{\tau^{\prime}}\(\sigma^{\prime}\)\}_{PB}&=&0
\nonumber\\
\{\mathfrak{e}_{\zeta}\(\sigma\)\,\overset{\otimes}{,}\,\mathfrak{b}_{\zeta^{\prime}}\(\sigma^{\prime}\)\}_{PB}&=&0
\lab{nicepbfefb}\\
\{\mathfrak{e}_{\tau}\(\sigma\)\,\overset{\otimes}{,}\,\mathfrak{b}_{\zeta^{\prime}}\(\sigma^{\prime}\)\}_{PB}&=&
\vartheta\;\IC\; \delta\(\zeta-\zeta^{\prime}\)\,\delta\(\tau-\tau^{\prime}\)\,\theta\(\sigma-\sigma^{\prime}\);\qquad {\rm for}\quad \sigma^{\prime}< \sigma_f^{\prime}
\nonumber\\
\{\mathfrak{e}_{\zeta}\(\sigma\)\,\overset{\otimes}{,}\,\mathfrak{b}_{\tau^{\prime}}\(\sigma^{\prime}\)\}_{PB}&=&
-\vartheta\;\IC\; \delta\(\zeta-\zeta^{\prime}\)\,\delta\(\tau-\tau^{\prime}\)\,\theta\(\sigma-\sigma^{\prime}\);\,\quad{\rm for}\quad \sigma^{\prime}< \sigma_f^{\prime}
\nonumber
\er
with $\IC$ defined in \rf{casimirlike}.  In  \rf{nicepbfefb} we have used the fact the adjoint representation of a compact semi-simple Lie group $G$ is real and unitary, and so it is orthogonal, i.e. for $g\in G$, we have from \rf{adjointrepdef} that 
\be
g\,T_a\,g^{-1}\otimes g\,T_a\,g^{-1}=T_a\otimes T_a
\lab{adjointrepproperty} 
\ee
In addition, in  \rf{nicepbfefb}, we have that $\theta\(\sigma-\sigma^{\prime}\)$ is the step function, i.e. $\theta\(\sigma-\sigma^{\prime}\)=1$ for $\sigma \geq \sigma^{\prime}$ and zero otherwise, and where we have used the fact that (see \rf{jacobiandef})
\be
\ve_{ijk}\,\frac{d\,x^i}{d\,\sigma}\,\frac{d\,x^j}{d\,\tau}\,\frac{d\,x^k}{d\,\zeta}\,\delta^{(3)}\(x-y\)
=\vartheta\; \delta\(\zeta-\zeta^{\prime}\)\,\delta\(\tau-\tau^{\prime}\)\,\delta\(\sigma-\sigma^{\prime}\)\,;
\qquad\qquad\quad \vartheta=\pm 1
\lab{deltatransform}
\ee

Note that from \rf{nicepbfe}, \rf{nicepbfbloop} and \rf{holonomyl2fluxes}, we get that
\be
\{\mathfrak{e}_{\tau/\zeta}\(\sigma\)\,\overset{\otimes}{,}\,{\cal T}_{\tau}\(\tau\)\}_{PB}=0
\lab{pbmathfracketauzetacalt}
\ee
From \rf{nicepbfb} and \rf{nicepbfbloop} we get
\be
\{\mathfrak{b}_{\tau/\zeta}\(\sigma_f\)\,\overset{\otimes}{,}\,{\cal T}_{\tau}\(\tau\)\}_{PB}=0
\lab{pbmathfrackbtauzetacalt}
\ee
Consequently
\be
\{{\cal T}_{\tau}\(\tau_1\)\,\overset{\otimes}{,}\,{\cal T}_{\tau}\(\tau_2\)\}_{PB}=0
\lab{pbcaltcalt}
\ee

Using \rf{pbrel}, \rf{nicepbfew}, \rf{nicepbfe}, \rf{nicepbfbloop} and \rf{nicepbfefb}, we get from \rf{caljdef2static}, \rf{jmatterdefstatic}, \rf{caljgdefstatic} and \rf{caljmagdefstatic}  that
\br
&&\{\mathfrak{e}_{\tau/\zeta}\(\sigma\)\,\overset{\otimes}{,}\, {\cal J}_{\rm spatial}\(\tau^{\prime}\)\}_{PB}=
\beta\,\{\mathfrak{e}_{\tau/\zeta}\(\sigma\)\,\overset{\otimes}{,}\, \rho_G\(\tau^{\prime}\)\}_{PB}=
\nonumber\\
&&=-i\,\beta\,\vartheta\,\delta\(\zeta-\zeta^{\prime}\)\,\delta\(\tau-\tau^{\prime}\)\,\int_{\sigma_i}^{\sigma_f}d\sigma^{\prime}\;\theta\(\sigma-\sigma^{\prime}\)\,\sbr{\IC}{\one\otimes\frac{d\,\mathfrak{e}_{\tau/\zeta}\(\sigma^{\prime}\)}{d\,\sigma^{\prime}}}
\nonumber\\
&&=i\,\beta\,\vartheta\,\delta\(\zeta-\zeta^{\prime}\)\,\delta\(\tau-\tau^{\prime}\)\,\sbr{\IC}{\mathfrak{e}_{\tau/\zeta}\(\sigma\)\otimes\one}
\lab{emathfrackjspatial}
\er
where we have made an integration by parts using the fact that $\frac{d\,\theta\(\sigma-\sigma^{\prime}\)}{d\,\sigma^{\prime}}= -\delta\(\sigma-\sigma^{\prime}\)$, and  we have also used \rf{casimirzero}.

From \rf{pbrel}, \rf{nicepbfbw}, \rf{nicepbfb} and  \rf{nicepbfbloop} we get that 
\be
\{\mathfrak{b}_{\tau/\zeta}\(\sigma_f\)\,\overset{\otimes}{,}\, {\cal J}_{\rm spatial}\(\tau\)\}_{PB}=0
\lab{bmathfrackjspatial}
\ee

Therefore, from \rf{holonomyl2fluxes}, \rf{pbxcala}, \rf{pbcaltcalt}, \rf{emathfrackjspatial} and \rf{bmathfrackjspatial} we get that
\br
&&\{{\cal T}_{\tau}\(\tau_1\,,\,\zeta_1\,,\,\alpha_1\,,\,\beta_1\)\,\overset{\otimes}{,}\,{\cal A}\(\zeta_2\,,\,\alpha_2\,,\,\beta_2\)\}_{PB}=
\nonumber\\
&&=
i\,e^3\,\beta_1\,\beta_2\,\vartheta\,\delta\(\zeta_1-\zeta_2\)\,\one \otimes V_{(2)}\(\tau_1\)\,\sbr{\IC}{\mathfrak{e}_{\tau}\(\sigma_f\,,\,\tau_1\,,\,\zeta_1\)\otimes\one}\,\one \otimes V_{(2)}^{-1}\(\tau_1\)
\lab{pbcaltcala}
\er
where the subscript $(2)$ in $V_{(2)}$ means it depends upon the parameters $\alpha_2$ and $\beta_2$. 
In addition from \rf{pbrel}, \rf{caljdef2static}, \rf{pbxcala}, \rf{nicepbw3}, \rf{nicepbfbw} and \rf{nicepbfew} we have that
\be
\{W\(\sigma_1\,,\,\tau_1\,,\,\zeta_1\)\,\overset{\otimes}{,}\,{\cal A}\(\zeta_2\,,\,\alpha_2\,,\,\beta_2\)\}_{PB}= 0
\lab{pbwcala}
\ee

\section{The Poisson brackets for the charge densities} 
\label{sec:pbchrgedensities}
\setcounter{equation}{0}

Using \rf{pbrel} and \rf{liebasis}, we get that the equal time Poisson brackets of the matter densities \rf{matterdensities} lead to two commuting copies of the Lie algebra of the gauge group
\br
\pbr{\rho_{a}^{\psi}\(x\)}{\rho_{b}^{\psi}\(y\)}&=&f_{abc}\,\rho_{c}^{\psi}\(x\)\,\delta^{(3)}\(x-y\)
\nonumber\\
\pbr{\rho_{a}^{\vp}\(x\)}{\rho_{b}^{\vp}\(y\)}&=&f_{abc}\,\rho_{c}^{\vp}\(x\)\,\delta^{(3)}\(x-y\)
\lab{currpb}\\
\pbr{\rho_{a}^{\psi}\(x\)}{\rho_{b}^{\vp}\(y\)}&=&0
\nonumber
\er
Consequently, we have
\br
\{J_0\(x\)\,\overset{\otimes}{,}\,J_0\(y\)\}_{PB}=i\,\delta^{(3)}\(x-y\)\,\sbr{\IC}{\one\otimes J_0\(x\)}
=-i\,\delta^{(3)}\(x-y\)\,\sbr{\IC}{ J_0\(x\)\otimes\one}
\lab{currpbtimes}
\er
where we have defined the Casimir-like operator
\be
\IC\equiv T_a\otimes T_a
\lab{casimirlike}
\ee
satisfying
\be
\sbr{\IC}{\one \otimes L+L\otimes \one}=0; \qquad \rightarrow \qquad \sbr{T_a}{L}\otimes T_a=-T_a\otimes\sbr{T_a}{L}
\lab{casimirzero}
\ee
with $L$ being any element of the Lie algebra  of the gauge group $G$.

Using the relations \rf{nicepbfb}, \rf{nicepbfe} and \rf{nicepbfefb}, and the fact that $\frac{d\,\theta\(\sigma-\sigma^{\prime}\)}{d\,\sigma}=\delta\(\sigma-\sigma^{\prime}\)$,  we can calculate the Poisson bracket among the components of the quantity $\rho_G\(\tau\,,\,\zeta\)$ given in \rf{caljgdefstatic}. We obtain
\br
&&\{\rho_G\(\tau\,,\,\zeta\)\,\overset{\otimes}{,}\,\rho_G\(\tau^{\prime}\,,\,\zeta^{\prime}\)\}_{PB}=-\vartheta\,\delta\(\zeta-\zeta^{\prime}\)\,\delta\(\tau-\tau^{\prime}\)\,\times
\nonumber\\
&&\times\int_{\sigma_i}^{\sigma_f}d\sigma\left\{\sbr{T_a}{\fbt{\sigma}}\otimes \sbr{T_a}{\frac{d\,\fez{\sigma}}{d\,\sigma}}-
\sbr{T_a}{\frac{d\,\fez{\sigma}}{d\,\sigma}}\otimes \sbr{T_a}{\fbt{\sigma}}
\right. \nonumber\\
&& \left. -\sbr{T_a}{\fbz{\sigma}}\otimes \sbr{T_a}{\frac{d\,\fet{\sigma}}{d\,\sigma}}+
\sbr{T_a}{\frac{d\,\fet{\sigma}}{d\,\sigma}}\otimes \sbr{T_a}{\fbz{\sigma}}
\right\}
\er
We now use the Jacobi identity to show that for any two elements $X$ and $Y$ of the Lie algebra of the gauge group, we get
\be
\sbr{T_a}{X}\otimes\sbr{T_a}{Y}-\sbr{T_a}{Y}\otimes\sbr{T_a}{X}=T_a\otimes\sbr{\sbr{X}{Y}}{T_a}
\ee
Therefore, we get that
\br
\{\rho_G\(\tau\,,\,\zeta\)\,\overset{\otimes}{,}\,\rho_G\(\tau^{\prime}\,,\,\zeta^{\prime}\)\}_{PB}&=&-i\,\vartheta\,\delta\(\zeta-\zeta^{\prime}\)\,\delta\(\tau-\tau^{\prime}\)\;\sbr{\IC}{\one\otimes \rho_G\(\tau\,,\,\zeta\)}
\nonumber\\
&=&i\,\vartheta\,\delta\(\zeta-\zeta^{\prime}\)\,\delta\(\tau-\tau^{\prime}\)\;\sbr{\IC}{\rho_G\(\tau\,,\,\zeta\)\otimes \one}
\lab{calgpbrel}
\er
with $\IC$ defined in \rf{casimirlike}. 

The matter densities, introduced in \rf{jomatterdef}, clearly Poisson commutes with the Wilson line. Therefore, using \rf{currpbtimes}, \rf{nicepbw3} and \rf{deltatransform} we get that the quantities 
$\rho_M\(\tau\,,\,\zeta\)$, defined in \rf{jmatterdefstatic}, satisfy 
\br
\{\rho_M\(\tau\,,\,\zeta\)\,\overset{\otimes}{,}\,\rho_M\(\tau^{\prime}\,,\,\zeta^{\prime}\)\}_{PB}&=&-i\,\vartheta\,\delta\(\zeta-\zeta^{\prime}\)\,\delta\(\tau-\tau^{\prime}\)\;\sbr{\IC}{\one\otimes \rho_M\(\tau\,,\,\zeta\)}
\nonumber\\
&=&i\,\vartheta\,\delta\(\zeta-\zeta^{\prime}\)\,\delta\(\tau-\tau^{\prime}\)\;\sbr{\IC}{\rho_M\(\tau\,,\,\zeta\)\otimes \one}
\lab{calmpbrel}
\er
The quantities $\mathfrak{e}_{\tau/\zeta}\(\sigma\)$ and $\mathfrak{b}_{\tau/\zeta}\(\sigma\)$ Poisson commute with the Wilson line. Therefore, we get that 
\be
\{\rho_G\(\tau\,,\,\zeta\)\,\overset{\otimes}{,}\,\rho_M\(\tau^{\prime}\,,\,\zeta^{\prime}\)\}_{PB}=0
\lab{calgmpbrel}
\ee
From \rf{nicepbfbw} and \rf{nicepbfew}, and the fact that the matter fields Poisson commute with the gauge fields, we get that
\be
\{\rho_M\(\tau\,,\,\zeta\)\,\overset{\otimes}{,}\,\mathfrak{b}_{\tau^{\prime}/\zeta^{\prime}}\(\sigma^{\prime}\)\}_{PB}=\{\rho_M\(\tau\,,\,\zeta\)\,\overset{\otimes}{,}\,\mathfrak{e}_{\tau^{\prime}/\zeta^{\prime}}\(\sigma^{\prime}\)\}_{PB}=0
\ee
and consequently
\be
\{\rho_M\(\tau\,,\,\zeta\)\,\overset{\otimes}{,}\,\rho_{\rm mag.}\(\zeta^{\prime}\,,\,\tau^{\prime}\)\}_{PB}=0
\ee
From \rf{nicepbfb} and \rf{nicepbfbloop} we get that
\be
\{\rho_G\(\tau\,,\,\zeta\)\,\overset{\otimes}{,}\,\mathfrak{b}_{\tau^{\prime}/\zeta^{\prime}}\(\sigma_f^{\prime}\)\}_{PB}=0
\ee
and so
\be
\{\rho_G\(\tau\,,\,\zeta\)\,\overset{\otimes}{,}\,\rho_{\rm mag.}\(\zeta^{\prime}\,,\,\tau^{\prime}\)\}_{PB}=0
\ee
So, the density of magnetic charge $\rho_{\rm mag.}$ Poisson commute with the densities $\rho_M$ and $\rho_G$ of electric charges. From \rf{nicepbfb} we see the magnetic charge densities commute among themselves
\be
\{\rho_{\rm mag.}\(\zeta\,,\,\tau\)\,\overset{\otimes}{,}\,\rho_{\rm mag.}\(\zeta^{\prime}\,,\,\tau^{\prime}\)\}_{PB}=0
\lab{rhomagabelian}
\ee
Using \rf{nicepbfefb} we get that
\br
\{\rho_G\(\tau\,,\,\zeta\)\,\overset{\otimes}{,}\,\mathfrak{e}_{\tau^{\prime}/\zeta^{\prime}}\(\sigma_f^{\prime}\)\}_{PB}&=&i\,\vartheta\,\delta\(\zeta-\zeta^{\prime}\)\,\delta\(\tau-\tau^{\prime}\)\,\sbr{\IC}{\mathfrak{e}_{\tau^{\prime}/\zeta^{\prime}}\(\sigma_f^{\prime}\)\otimes \one}
\nonumber\\
&=&-i\,\vartheta\,\delta\(\zeta-\zeta^{\prime}\)\,\delta\(\tau-\tau^{\prime}\)\,\sbr{\IC}{\one\otimes \mathfrak{e}_{\tau^{\prime}/\zeta^{\prime}}\(\sigma_f^{\prime}\)}
\er
In components, we get
\be
\{\rho_G^a\(\tau\,,\,\zeta\)\,,\,\mathfrak{e}^b_{\tau^{\prime}/\zeta^{\prime}}\(\sigma_f^{\prime}\)\}_{PB}=
-\vartheta\,f_{abc}\, \mathfrak{e}^c_{\tau^{\prime}/\zeta^{\prime}}\(\sigma_f^{\prime}\)\,\delta\(\zeta-\zeta^{\prime}\)\,\delta\(\tau-\tau^{\prime}\)
\ee

\section{The Abelian Gauss theorem in $\IR^3$} 
\label{sec:abeliangauss}
\setcounter{equation}{0}

In the text, we have to deal with boundary terms that in many cases it needs the use of the Abelian Gauss theorem. In order to get the factors correct in terms of the scanning variables, we derive such a theorem here using the notation of Section \ref{subsec:nonabelianstokes}.

Let us consider the case where the space is $\IR^3$, and let the two form $B_{ij}$, $i,j=1,2,3$, in \rf{connectionl2} be abelian, with a vanishing connection $C_{i}=0$. Working with the Hodge dual 
\be
b_i\equiv \ve_{ijk}\,B_{jk}
\ee
we get that from \rf{connectionl2} and \rf{holonomyl2} that
\be
\frac{d\,\ln V}{d\,\tau}={\cal T};\qquad\qquad \qquad  {\cal T}=\frac{1}{2}\,\int_{\sigma_i}^{\sigma_f}d\sigma\, b_i\,\ve_{ijk}\,\frac{d\,x^j}{d\,\sigma}\,\frac{d\,x^k}{d\,\tau}
\ee
Then from \rf{holonomyl3} and  \rf{kdelta}, we get that
\be
\frac{d\,\ln V}{d\,\zeta}={\cal K};\qquad\qquad \qquad  {\cal K}=\frac{1}{2}\,\int_{\tau_i}^{\tau_f}d\tau\,\int_{\sigma_i}^{\sigma_f}d\sigma\,\frac{\partial\, b_l}{\partial\,x^l}\,\ve_{ijk}\frac{d\,x^i}{d\,\sigma}\,\frac{d\,x^j}{d\,\tau}\,\frac{d\,x^k}{d\,\zeta}
\ee
Using the definition of Jacobian \rf{jacobiandef}, we then get that
\be
\vartheta\, \int_{\tau_i}^{\tau_f}d\tau\,\int_{\sigma_i}^{\sigma_f}d\sigma\,b_i\,\ve_{ijk}\,\frac{d\,x^j}{d\,\sigma}\,\frac{d\,x^k}{d\,\tau}\mid_{\zeta=\zeta_f}=\int_{\IR^3} d^3x\, \frac{\partial\, b_l}{\partial\,x^l}
\lab{abeliangauss}
\ee

In dealing with the surface terms, we encounter other types of integrals over spatial surfaces, labelled by the parameter $\zeta$, and having the form
\be
I_{\zeta,i}\(f\)\equiv \int_{\tau_i}^{\tau_f}d\tau\,\int_{\sigma_i}^{\sigma_f}d\sigma\, \Delta\(\sigma\,,\,\tau\,,\,\zeta\)\,\frac{\partial\;}{\partial\,x^i}\(f\(x\)\,\delta^{(3)}\(x-y\)\)
\lab{iifdef}
\ee
where $\Delta$ is the Jacobian, defined in \rf{jacobiandef}, $f\(x\)$ is a c-number function, and the points $x^i$ and $y^i$ are parameterized as $x^i\equiv x^i\(\sigma\,,\, \tau\,,\,\zeta\)$, and $y^i\equiv y^i\(\sigma^{\prime}\,,\, \tau^{\prime}\,,\,\zeta^{\prime}\)$. 

Note that we have
\be
\frac{d\,\sigma}{\d\,x^i}=\frac{\vartheta}{\Delta}\,\ve_{ijk}\,\frac{d\,x^j}{d\,\tau}\,\frac{d\,x^k}{d\,\zeta};\qquad
\frac{d\,\tau}{\d\,x^i}=\frac{\vartheta}{\Delta}\,\ve_{ijk}\,\frac{d\,x^j}{d\,\zeta}\,\frac{d\,x^k}{d\,\sigma};\qquad
\frac{d\,\zeta}{\d\,x^i}=\frac{\vartheta}{\Delta}\,\ve_{ijk}\,\frac{d\,x^j}{d\,\sigma}\,\frac{d\,x^k}{d\,\tau}
\lab{niceinvrel}
\ee
Replacing \rf{niceinvrel} into \rf{iifdef} we get, after integration by parts,
\br
I_{\zeta,i}\(f\)&=&\vartheta\,\int_{\tau_i}^{\tau_f}d\tau\,\int_{\sigma_i}^{\sigma_f}d\sigma\,\ve_{ijk}\,\left[\frac{d\;}{d\,\sigma}\(\frac{d\,x^j}{d\,\tau}\,\frac{d\,x^k}{d\,\zeta}f\(x\)\,\delta^{(3)}\(x-y\)\)
\right.
\nonumber\\
&+& \left. \frac{d\;}{d\,\tau}\(\frac{d\,x^j}{d\,\zeta}\,\frac{d\,x^k}{d\,\sigma}f\(x\)\,\delta^{(3)}\(x-y\)\)
+\frac{d\;}{d\,\zeta}\(\frac{d\,x^j}{d\,\sigma}\,\frac{d\,x^k}{d\,\tau}f\(x\)\,\delta^{(3)}\(x-y\)\)
\right]
\lab{iidefaux}
\er
since the terms containing second derivatives of $x^i$ cancel out. Using the fact that, on the variations of loops, surfaces, and volumes, the reference point $x_R$ is kept fixed, we get that
\be
\frac{d\,x^i}{d\,\tau}=\frac{d\,x^i}{d\,\zeta}=0 \qquad {\rm at}\qquad \sigma=\sigma_i\;,\sigma_f\quad{\rm and}\quad \tau=\tau_i\;,\tau_f
\lab{coniifdef}
\ee
Therefore
\be
I_{\zeta,i}\(f\)=\vartheta\,\frac{d\;}{d\,\zeta}\,\int_{\tau_i}^{\tau_f}d\tau\,\int_{\sigma_i}^{\sigma_f}d\sigma\,\ve_{ijk}\,\frac{d\,x^j}{d\,\sigma}\,\frac{d\,x^k}{d\,\tau}f\(x\)\,\delta^{(3)}\(x-y\)
\lab{iifdef2}
\ee
Note, however, that we have 
\be
\ve_{ijk}\,\frac{d\,y^i}{d\,\sigma^{\prime}}\,\frac{d\,x^j}{d\,\sigma}\,\frac{d\,x^k}{d\,\tau}\,\delta^{(3)}\(x-y\)=0;\qquad
\ve_{ijk}\,\frac{d\,y^i}{d\,\tau^{\prime}}\,\frac{d\,x^j}{d\,\sigma}\,\frac{d\,x^k}{d\,\tau}\,\delta^{(3)}\(x-y\)=0
\ee
since that delta function imposes $\frac{d\,y^i}{d\,\sigma^{\prime}}$ and $\frac{d\,x^j}{d\,\sigma}$ to be parallel, and similar for $\frac{d\,y^i}{d\,\tau^{\prime}}$ and $\frac{d\,x^k}{d\,\tau}$. Therefore, we conclude that
\be
\frac{d\,y^i}{d\,\sigma^{\prime}}\, I_{\zeta,i}\(f\)=0;\qquad\qquad \frac{d\,y^i}{d\,\tau^{\prime}}\, I_{\zeta,i}\(f\)=0
\lab{iifdef3}
\ee

\section{Poisson brackets involving the constraints} 
\label{sec:pbconstraint}
\setcounter{equation}{0}

We now calculate some quantities relevant for the evaluation of the first term on the right-hand side of \rf{prepbacalacal}.  In order to simplify the notation, we introduce the operator, acting on a given quantity $X$, as
\be
{\cal O}_{\zeta}\(X\)\equiv \int_{\tau_i}^{\tau_f} d\tau \int_{\sigma_i}^{\sigma_f}d\sigma\,
\Delta\(\sigma\,,\,\tau\,,\,\zeta\)\, d_{ba}\(x\)\,
T_b\otimes \pbr{{\cal C}_a\(x\)}{X}\equiv T_b\otimes {\cal O}^b_{\zeta}\(X\)
\lab{calozetadef}
\ee
which is integrated on a surface labelled by $\zeta$, with its points parameterized as $x^i=x^i\(\sigma\,,\,\tau\,,\,\zeta\)$, $\Delta$ is the Jacobian defined in \rf{jacobiandef}, and where we have denoted the non-integrable factor $d_{ba}\(V\(\tau\)\,W^{-1}\(\sigma\)\)$, simply by $d_{ba}\(x\)$. 

Note that the operator \rf{calozetadef} is similar to the operator \rf{ocaldef}, and satisfies the same properties, i.e., it is a linear operator, satisfies the Leibniz rule, and commutes with the space derivatives and space integrals. 

We now consider that operator acting on the Wilson line $W$, integrated from the reference point $x_R$, up to a point $y^i=y^i\(\sigma_2\,,\,\tau_2\,,\,\zeta_2\)$. Using \rf{constraintgengauge} and \rf{nicepbw} we get that
\br
&&{\cal O}^b_{\zeta_1}\(W\(\sigma_2\)\)=-i\,e\,W\(\sigma_2\)\,\int_{\tau_i}^{\tau_f} d\tau_1 \int_{\sigma_i}^{\sigma_f}d\sigma_1\,
\Delta\(\sigma_1\,,\,\tau_1\,,\,\zeta_1\)\, d_{ba}^{(1)}\(x\)\,\times 
\\
&&\times\,\int_{\sigma_i}^{\sigma_2}d\sigma^{\prime}_2 W^{-1}\(\sigma^{\prime}_2\)\left\{\sbr{i\,e\,A_i\(z\)\,}{T_a}\,\delta^{(3)}\(x-z\)-\frac{\partial\,\delta^{(3)}\(x-z\)}{\partial\,x^i}\,T_a\right\}\,\frac{d\,z^i}{d\,\sigma^{\prime}_2}\,W\(\sigma^{\prime}_2\)
\nonumber
\er
where we have denoted $x^i=x^i\(\sigma_1\,,\,\tau_1\,,\,\zeta_1\)$ and $z^i=z^i\(\sigma^{\prime}_2\,,\,\tau_2\,,\,\zeta_2\)$, and the supperscript $(1)$ in $d_{ba}^{(1)}\(x\)$, means that it depends upon the parameters $\alpha_1$ and $\beta_1$, through the holonomy defined in \rf{holonomyl2fluxes}, which we now denote as $V_{(1)}$, i.e. $d_{ba}^{(1)}\(V_{(1)}\(\tau_1\)\,W^{-1}\(\sigma_1\)\)$. We now use \rf{wdefa},  \rf{deltatransform}, and the definition of the quantity $I_i$ in \rf{iifdef}, to get
\br
\frac{i}{e}\,W^{-1}\(\sigma_2\)\,{\cal O}^b_{\zeta_1}\(W\(\sigma_2\)\)&=&
-\int_{\sigma_i}^{\sigma_2}d\sigma^{\prime}_2\,W^{-1}\(\sigma^{\prime}_2\)\,T_a\,W\(\sigma^{\prime}_2\)\,\frac{d\,z^i}{d\,\sigma^{\prime}_2}\,I_{\zeta_1,i}\(d_{ba}^{(1)}\)
\lab{calow}\\
&+&\delta\(\zeta_1-\zeta_2\) \int_{\sigma_i}^{\sigma_2}d\sigma^{\prime}_2\,\frac{d\;}{d\,\sigma^{\prime}_2}\(d_{ba}^{(1)}\(z\)\,W^{-1}\(\sigma^{\prime}_2\)\,T_a\,W\(\sigma^{\prime}_2\)\)
\nonumber
\er
Since the adjoint representation matrix of a compact Lie group is orthogonal, we get  (irrespective of the value of the parameters $\alpha$ and $\beta$), that 
\be
d_{ba}\(z\)\,T_a=W\(\sigma^{\prime}_2\)\,V^{-1}\(\tau_2\)\,T_b\,V^{-1}\(\tau_2\)\,W^{-1}\(\sigma^{\prime}_2\)
\ee
Therefore 
\be
\frac{d\;}{d\,\sigma^{\prime}_2}\(d_{ba}\(z\)\,W^{-1}\(\sigma^{\prime}_2\)\,T_a\,W\(\sigma^{\prime}_2\)\)=\frac{d\;}{d\,\sigma^{\prime}_2}\(V^{-1}\(\tau_2\)\,T_b\,V\(\tau_2\)\)=0
\ee
Using \rf{iifdef3}, we get that the first term on the right-hand side of \rf{calow} also vanishes, and so
\be
{\cal O}^b_{\zeta_1}\(W\(\sigma_2\)\)=0
\lab{calowzero}
\ee
It now follows from \rf{bfrakdefsimple2} and \rf{calowzero} that
\be
{\cal O}^b_{\zeta_1}\(\mathfrak{b}_{\tau_2/\zeta_2}\(\sigma_f\)\)=0
\lab{calobfrack}
\ee

From \rf{calozetadef} and \rf{ebfrakdef}, we get, using \rf{constraintgengauge}, \rf{deltatransform} and \rf{calowzero}, that
\be
{\cal O}^b_{\zeta_1}\(\mathfrak{e}_{\tau_2/\zeta_2}\(\sigma_f\)\)=i\,e\,\delta\(\zeta_1-\zeta_2\)\,\sbr{\mathfrak{e}_{\tau_2/\zeta_2}\(\sigma_f\)}{V_{(1)}^{-1}\(\tau_2\)\,T_b\,V_{(1)}\(\tau_2\)}
\lab{caloefrack}
\ee
From \rf{holonomyl2fluxes}, \rf{calobfrack} and \rf{caloefrack}, we get
\be
{\cal O}^b_{\zeta_1}\({\cal T}_{\tau_2}^{(2)}\(\tau_2\)\)=e^2\,\beta_2\,\delta\(\zeta_1-\zeta_2\)\,\sbr{\mathfrak{e}_{\tau_2}\(\sigma_f\)}{V_{(1)}^{-1}\(\tau_2\)\,T_b\,V_{(1)}\(\tau_2\)}
\lab{calocalttau}
\ee
where the supperscript $(2)$ in ${\cal T}_{\tau_2}^{(2)}\(\tau_2\)$ means it depends upon the parameters $\alpha_2$ and $\beta_2$, in the definition \rf{holonomyl2fluxes}, which are different from those parameters appearing in the holonomy $V_{(1)}\(\tau\)$ in \rf{calocalttau}, also defined through  \rf{holonomyl2fluxes}. 

From  \rf{jomatterdef}, \rf{constraint2}, \rf{currpb}, \rf{constraintgengauge} and \rf{calozetadef} we have
\br
{\cal O}^c_{\zeta_1}\({\cal C}_b\(y\)\)&=&-e\,f_{abd}\,\int_{\tau_i}^{\tau_f} d\tau_1 \int_{\sigma_i}^{\sigma_f}d\sigma_1\,
\Delta\(\sigma_1\,,\,\tau_1\,,\,\zeta_1\)\, d_{ca}^{(1)}\(x\)\left[ {\cal C}_d\(x\)\,\delta^{(3)}\(x-y\)
\right. 
\\
&+&\left. E_i^d\(y\)\frac{\partial\,\delta^{(3)}\(x-y\)}{\partial\,x^i}+E_i^d\(x\)\frac{\partial\,\delta^{(3)}\(x-y\)}{\partial\,y^i}-\frac{\partial\,E_i^d\(x\)}{\partial\,x^i}\,\delta^{(3)}\(x-y\)
\right]
\nonumber
\er
where $x^i=x^i\(\sigma_1\,,\,\tau_1\,,\,\zeta_1\)$ and $y^i=y^i\(\sigma_2\,,\,\tau_2\,,\,\zeta_2\)$. Using \rf{deltatransform} and \rf{iifdef}
\br
{\cal O}^c_{\zeta_1}\({\cal C}_b\(y\)\)&=&-e\,f_{abd}\left\{\delta\(\zeta_1-\zeta_2\)\left[ d_{ca}^{(1)}\(y\){\cal C}_d\(y\)-\frac{\partial\,\(d_{ca}^{(1)}\(y\)\,E_i^d\(y\)\)}{\partial\,y^i}\right] + E_i^d\(y\)\, I_{\zeta_1,i}\(d_{ca}^{(1)}\)
\right.
\nonumber\\
&+&\left. \int_{\tau_i}^{\tau_f} d\tau_1 \int_{\sigma_i}^{\sigma_f}d\sigma_1\,
\Delta\(\sigma_1\,,\,\tau_1\,,\,\zeta_1\)\, d_{ca}^{(1)}\(x\)\,E_i^d\(x\)\frac{\partial\,\delta^{(3)}\(x-y\)}{\partial\,y^i}
\right\}
\lab{calozeta1calcb}
\er

Using \rf{calozeta1calcb} we have that
\br
&&{\cal I}_{cd}\equiv \int_{\tau_i}^{\tau_f} d\tau_1\int_{\sigma_i}^{\sigma_f}d\sigma_1
\Delta\(\sigma_1,\tau_1,\zeta_1\) d_{ca}^{(1)}\(x\) 
\int_{\tau_i}^{\tau_f} d\tau_2\int_{\sigma_i}^{\sigma_f}d\sigma_2
\Delta\(\sigma_2,\tau_2,\zeta_2\)\ d_{db}^{(2)}\(y\)\times
\nonumber\\
&&\times\; \pbr{{\cal C}_a\(x\)}{{\cal C}_b\(y\)}
\nonumber\\
&&=\int_{\tau_i}^{\tau_f} d\tau_2\int_{\sigma_i}^{\sigma_f}d\sigma_2\,
\Delta\(\sigma_2\,,\,\tau_2\,,\,\zeta_2\)\, d_{db}^{(2)}\(y\)\,{\cal O}^c_{\zeta_1}\({\cal C}_b\(y\)\)
\nonumber\\
&&=-e\,f_{abe}\,\int_{\tau_i}^{\tau_f} d\tau_2\int_{\sigma_i}^{\sigma_f}d\sigma_2\,
\Delta\(\sigma_2\,,\,\tau_2\,,\,\zeta_2\)\, d_{db}^{(2)}\(y\)\times 
\nonumber\\
&&\times \left\{\delta\(\zeta_1-\zeta_2\)\,
\left[ d_{ca}^{(1)}\(y\){\cal C}_e\(y\)-\frac{\partial\,\(d_{ca}^{(1)}\(y\)\,E_i^e\(y\)\)}{\partial\,y^i}\right]
+E_i^e\(y\)\, I_{\zeta_1,i}\(d_{ca}^{(1)}\) \right\}
\\
&&-e\,f_{abe}\,\int_{\tau_i}^{\tau_f} d\tau_1 \int_{\sigma_i}^{\sigma_f}d\sigma_1\,
\Delta\(\sigma_1\,,\,\tau_1\,,\,\zeta_1\)\, d_{ca}^{(1)}\(x\)\,E_i^e\(x\)\left[ I_{\zeta_2,i}\(d_{db}^{(2)}\)
-\delta\(\zeta_1-\zeta_2\)\,\frac{\partial\,d_{db}^{(2)}\(x\)}{\partial\,x^i}\right]
\nonumber
\er
and so
\br
&&{\cal I}_{cd}
=-e\,f_{abe}\,\delta\(\zeta_1-\zeta_2\)\,\int_{\tau_i}^{\tau_f} d\tau_2\int_{\sigma_i}^{\sigma_f}d\sigma_2\,
\Delta\(\sigma_2\,,\,\tau_2\,,\,\zeta_2\)\times
\nonumber\\
&&\times 
\left[ d_{ca}^{(1)}\(y\)\, d_{db}^{(2)}\(y\)\,{\cal C}_e\(y\)-\frac{\partial\,\(d_{ca}^{(1)}\(y\)\,d_{db}^{(2)}\(y\)\,E_i^e\(y\)\)}{\partial\,y^i}\right]
\nonumber\\
&&-e\,f_{abe}\,\int_{\tau_i}^{\tau_f} d\tau_2\int_{\sigma_i}^{\sigma_f}d\sigma_2\,
\Delta\(\sigma_2\,,\,\tau_2\,,\,\zeta_2\) d_{db}^{(2)}\(y\)\,E_i^e\(y\)\, I_{\zeta_1,i}\(d_{ca}^{(1)}\) 
\nonumber\\
&&-e\,f_{abe}\,\int_{\tau_i}^{\tau_f} d\tau_1 \int_{\sigma_i}^{\sigma_f}d\sigma_1\,
\Delta\(\sigma_1\,,\,\tau_1\,,\,\zeta_1\)\, d_{ca}^{(1)}\(x\)\,E_i^e\(x\) I_{\zeta_2,i}\(d_{db}^{(2)}\)
\lab{cali}
\er
where we have used the fact that the terms multiplied by $\delta\(\zeta_1-\zeta_2\)$, are in fact on the same surface (labelled by $\zeta_1$ or equivalently $\zeta_2$), and so the dummy integrals in $\tau_1$ and $\sigma_1$, can be swapped to the dummy integrals in $\tau_2$ and $\sigma_2$. In addition, the supperscript $(s)$ in $d^{(s)}$ means it depends on the parameters $\alpha_s$ and $\beta_s$, for $s=1,2$. 

We now use \rf{iifdef2} to write
\br
&&\int_{\tau_i}^{\tau_f} d\tau_1 \int_{\sigma_i}^{\sigma_f}d\sigma_1\,
\Delta\(\sigma_1\,,\,\tau_1\,,\,\zeta_1\)\, d_{ca}^{(1)}\(x\)\,E_i^e\(x\) I_{\zeta_2,i}\(d_{db}^{(2)}\)
\nonumber\\
&&=\vartheta\,\frac{d\;}{d\,\zeta_2}\,\int_{\tau_i}^{\tau_f}d\tau_2\,\int_{\sigma_i}^{\sigma_f}d\sigma_2\,\ve_{ijk}\,\frac{d\,y^j}{d\,\sigma_2}\,\frac{d\,y^k}{d\,\tau_2}d_{db}^{(2)}\(y\)\,d_{ca}^{(1)}\(y\)\,E_i^e\(y\)\delta\(\zeta_1-\zeta_2\)
\lab{icdaux1}
\er
since the integrals in $\tau_1$ and $\sigma_1$ commute with the $\zeta_2$-derivative, and where we have used \rf{deltatransform}. A similar manipulation can be performed on the  term containing $I_{\zeta_1,i}\(d_{ca}^{(1)}\)$ on the right hand side of \rf{cali}.

Performing manipulations similar to those done from \rf{iifdef} through \rf{iifdef2} we get that
\br
&&\int_{\tau_i}^{\tau_f} d\tau_2\int_{\sigma_i}^{\sigma_f}d\sigma_2\,
\Delta\(\sigma_2\,,\,\tau_2\,,\,\zeta_2\)\,\frac{\partial\,\(d_{ca}^{(1)}\(y\)\,d_{db}^{(2)}\(y\)\,E_i^e\(y\)\)}{\partial\,y^i}=
\nonumber\\
&&=\vartheta\,\frac{d\;}{d\,\zeta_2}\,\int_{\tau_i}^{\tau_f}d\tau_2\,\int_{\sigma_i}^{\sigma_f}d\sigma_2\,\ve_{ijk}\,\frac{d\,y^j}{d\,\sigma_2}\,\frac{d\,y^k}{d\,\tau_2}d_{db}^{(2)}\(y\)\,d_{ca}^{(1)}\(y\)\,E_i^e\(y\)
\lab{icdaux2}
\er
Therefore, using \rf{icdaux1} and \rf{icdaux2}, we get
\br
&& {\cal I}_{cd}=-e\,f_{abe}\,\delta\(\zeta_1-\zeta_2\)\,\int_{\tau_i}^{\tau_f} d\tau_2\int_{\sigma_i}^{\sigma_f}d\sigma_2\,
\Delta\(\sigma_2\,,\,\tau_2\,,\,\zeta_2\)\,d_{ca}^{(1)}\(y\)\, d_{db}^{(2)}\(y\)\,{\cal C}_e\(y\)
\nonumber\\
&&+e\,\vartheta\, f_{abe}\, \delta\(\zeta_1-\zeta_2\)\,\frac{d\;}{d\,\zeta_2}\,\int_{\tau_i}^{\tau_f}d\tau_2\,\int_{\sigma_i}^{\sigma_f}d\sigma_2\,\ve_{ijk}\frac{d\,y^j}{d\,\sigma_2}\,\frac{d\,y^k}{d\,\tau_2}d_{ca}^{(1)}\(y\)d_{db}^{(2)}\(y\)\,E_i^e\(y\)
\nonumber\\
&&-e\,\vartheta\, f_{abe}\, \frac{d\;}{d\,\zeta_1}\,\int_{\tau_i}^{\tau_f}d\tau_1\,\int_{\sigma_i}^{\sigma_f}d\sigma_1\,\ve_{ijk}\,\frac{d\,x^j}{d\,\sigma_1}\,\frac{d\,x^k}{d\,\tau_1}d_{ca}^{(1)}\(x\)\,d_{db}^{(2)}\(x\)\,E_i^e\(x\)\delta\(\zeta_1-\zeta_2\)
\nonumber\\
&&-e\,\vartheta\, f_{abe}\, \frac{d\;}{d\,\zeta_2}\,\int_{\tau_i}^{\tau_f}d\tau_2\,\int_{\sigma_i}^{\sigma_f}d\sigma_2\,\ve_{ijk}\,\frac{d\,y^j}{d\,\sigma_2}\,\frac{d\,y^k}{d\,\tau_2}d_{ca}^{(1)}\(y\)\,d_{db}^{(2)}\(y\)\,E_i^e\(y\)\delta\(\zeta_1-\zeta_2\)
\er
Consider now
\br
&&f_{abe}\,d_{ca}^{(1)}\(y\)\, d_{db}^{(2)}\(y\)\,T_c\otimes T_d=
\nonumber\\
&&=f_{abe}\, V_{(1)}\(\tau_2\)\,W^{-1}\(\sigma_2\)\,T_a\,W\(\sigma_2\)\,V_{(1)}^{-1}\(\tau_2\)\otimes V_{(2)}\(\tau_2\)\,W^{-1}\(\sigma_2\)\,T_b\,W\(\sigma_2\)\,V_{(2)}^{-1}\(\tau_2\)
\nonumber\\
&&=-i\,V_{(1)}\(\tau_2\)\,W^{-1}\(\sigma_2\)\,\sbr{T_b}{T_e}\,W\(\sigma_2\)\,V_{(1)}^{-1}\(\tau_2\)\otimes V_{(2)}\(\tau_2\)\,W^{-1}\(\sigma_2\)\,T_b\,W\(\sigma_2\)\,V_{(2)}^{-1}\(\tau_2\)
\nonumber\\
&&=-i\,V_{(1)}\(\tau_2\)\,\sbr{T_b}{W^{-1}\(\sigma_2\)\,T_e\,W\(\sigma_2\)}\,V_{(1)}^{-1}\(\tau_2\)\otimes V_{(2)}\(\tau_2\)\,T_b\,V_{(2)}^{-1}\(\tau_2\)
\nonumber\\
&&=-i\,V_{(1)}\(\tau_2\)\otimes V_{(2)}\(\tau_2\)\,\sbr{\IC}{W^{-1}\(\sigma_2\)\,T_e\,W\(\sigma_2\)\otimes \one}\,V_{(1)}^{-1}\(\tau_2\)\otimes V_{(2)}^{-1}\(\tau_2\)
\er
where we have used \rf{adjointrepproperty}. Using \rf{ebfrakdef} and \rf{calrdef} we then have
\br
 && {\cal I}_{cd}\,T_c\otimes T_d=i\,e\,\delta\(\zeta_1-\zeta_2\)\,\int_{\tau_i}^{\tau_f} d\tau_2\int_{\sigma_i}^{\sigma_f}d\sigma_2\,
\Delta\(\sigma_2\,,\,\tau_2\,,\,\zeta_2\)\times
\nonumber\\
&&\times\, V_{(1)}\(\tau_2\)\otimes V_{(2)}\(\tau_2\)\,\sbr{\IC}{W^{-1}\(\sigma_2\)\,{\cal C}\,W\(\sigma_2\)\otimes \one}\,V_{(1)}^{-1}\(\tau_2\)\otimes V_{(2)}^{-1}\(\tau_2\)
\lab{calicdtensor}\\
&&+i\,e\,\vartheta\, \left[ -\delta\(\zeta_1-\zeta_2\)\,\frac{d\, {\cal Y}\(\zeta_2\,,\,\tau_f\)}{d\,\zeta_2}\,
+ \frac{d\, \({\cal Y}\(\zeta_1\,,\,\tau_f\)\,\delta\(\zeta_1-\zeta_2\)\)}{d\,\zeta_1}
+ \frac{d\, \({\cal Y}\(\zeta_2\,,\,\tau_f\)\,\delta\(\zeta_1-\zeta_2\)\)}{d\,\zeta_2}\right]
\nonumber
\er

\section{Invariance of the Hamiltonian under the commutator of two transformations generated by the charges} 
\label{sec:htwocom}
\setcounter{equation}{0}

We now show that the commutator of two transformations generated by the charges, given in  \rf{commutatortransf}, is a symmetry of the total Hamiltonian $H_T$ \rf{completehamiltonian}, i.e. 
\be
\sbr{\delta_{N_1,\alpha_1,\beta_1}}{\delta_{N_2,\alpha_2,\beta_2}}\, H_T\cong  0
\lab{hamiltcommtransf1}
\ee
From \rf{comutatortwotransf}, \rf{upsilondef} and \rf{calmtildedef},   we have that 
\br
\sbr{\delta_{N_1,\alpha_1,\beta_1}}{\delta_{N_2,\alpha_2,\beta_2}}\, X&\cong& \ve_1\,\ve_2 \,\(\alpha_1-\alpha_2\)\,{\rm Tr}_{RL}\left[Q^{N_1}_{(1)}\(\zeta_f\)\otimes Q^{N_2}_{(2)}\(\zeta_f\)\,
\times  \right.
\lab{upsilonadef}\\
&\times&\left. \int_{\zeta_i}^{\zeta_f}d\zeta\,\int_{\tau_i}^{\tau_f}d\tau\,\int_{\tau_i}^{\tau}d\tau^{\prime}\,\int_{\sigma_i}^{\sigma_f}d\sigma\, \Delta(y)\Upsilon_a\(\sigma,\tau,\tau^{\prime},\zeta\)\,\pbr{X}{{\cal C}_a\(y\)}\right]
\nonumber
\er
where we have defined  $\Upsilon_a$ as 
\br
\Upsilon_a(\sigma, \tau, \tau^\prime, \zeta) = ie\vartheta\sbr{Q_{\otimes}^{-1}\(\zeta\)\,\frac{d\,{\cal Z}\(\zeta,\tau\)}{d\,\tau}\,Q_{\otimes}\(\zeta\)}{\beta_1\,T_b\otimes\one \,d_{ba}^{(1)}(y)+ \beta_2\,\one \otimes T_b\,d_{ba}^{(2)}(y)} 
\lab{upsilonexpression}
\er
where $\Upsilon_a$ involves the points $y^i=y^i\(\sigma\,,\,\tau^{\prime}\,,\,\zeta\)$ and  $x^i=(\sigma,\tau,\zeta)$. $\Delta$ is the Jacobian defined in \rf{jacobiandef}, and where we have denoted the non-integrable factor $d_{ba}\(Q^{-1}(\zeta)\,V\(\tau^{\prime}\)\,W^{-1}\(\sigma\)\)$, simply by $d_{ba}\(y\)$. The supperscript $(s)$ in $d^{(s)}_{ba}$ denotes the dependence on the parameters $\(\alpha_s,\,\beta_s\)$. 

We have not considered the Poisson bracket $\pbr{X}{\Upsilon_a(\sigma, \tau, \tau^\prime, \zeta)}$ in \rf{upsilonadef}, because it will be multiplied by ${\cal C}_a\(y\)$, and so it will not contribute when the constraints are imposed. 

In order to simplify the calculations, we evaluate the commutator of transformations for each term of the total Hamiltonian  $H_T$, given in \rf{hamiltonianparts}. 

Taking $X$ as $H_E$ in \rf{upsilonadef}, where $H_E$ is given \rf{hamiltonianparts}, it is straightforward from \rf{vanishingdensityPB2}, that
\be
\sbr{\delta_{N_1,\alpha_1,\beta_1}}{\delta_{N_2,\alpha_2,\beta_2}}\, H_E = 0
\lab{hecommtransf}
\ee

Now, using the relations \rf{constraintgengauge} to obtain how the remaining terms of $H_T$ transforms under the constraints \rf{constraint2}, we get the following
\br
\pbr{H_\psi + H_\vp}{{\cal C}_a(y)} &=& e\,\int d^3z\,\frac{\partial \;}{\partial z^i}\(J^a_i(z)\,\delta^{(3)}(y-z)\)
\lab{hconsttransf}\\
\pbr{H_C}{{\cal C}_a(y)}&=& e\,f_{abc} \left[A^b_0(y)\,{\cal C}_c(y) + \int d^3z\,E^c_i(y)\,\frac{\partial\;}{\partial z^i}\(A_0^b(z)\,\delta^{(3)}(y-z)\)\right]\nonumber\\
\pbr{H_B}{{\cal C}_a(y)} &=& -\vareps_{ijk}\int d^3z\,\frac{\partial\;}{\partial z^i}\left[B_k^a(z)\frac{\partial \delta^{(3)}(y-z)}{\partial y^j} + ef_{abc}A^c_j(y)B^b_k(z)\delta^{(3)}(y-z)\right]\nonumber
\er
where $J_i$ are the spatial components of the matter currents given in \rf{mattercurr}. Notice that we can use the abelian Gauss law \rf{abeliangauss} to rewrite the integrals on the variable $z^i = z^i(\sigma_z,\tau_z,\zeta_z)$ in terms of an integral on the two-sphere $S^2_{\infty}$ of spatial infinity. 

To obtain the transformations of each term of the total Hamiltonian, we need to deal with the integrals involving $\Upsilon_a$ and the Dirac delta $\delta^{(3)}(y-z)$. Hence, we can calculate the integrals involving those Dirac delta and ${\Upsilon}_a$ terms as 
\br
{\cal I}^{\Upsilon}_a
&\equiv&\int_{\zeta_i}^{\zeta_f} d\zeta \int_{\tau_i}^{\tau_f} d\tau \int_{\tau_i}^{\tau} d\tau^\prime\int_{\sigma_i}^{\sigma_f} d\sigma\, {\Upsilon}_a(\sigma,\tau,\tau^\prime,\zeta) \,\Delta(y)\,\delta^{(3)}(y-z(\sigma_z,\tau_z,\zeta_z))\nonumber\\
&=& \int_{\tau_z}^{\tau_f}d\tau\,  {\Upsilon}_a(\sigma_z,\tau_z,\zeta_z, \tau)
\lab{upsilontildedelta}
\er
and, when it involves the derivative of the Dirac delta:
\br
&&{\cal I}^{\Upsilon}_{a,i}\equiv\int_{\zeta_i}^{\zeta_f} d\zeta \int_{\tau_i}^{\tau_f} d\tau \int_{\tau_i}^{\tau} d\tau^\prime\int_{\sigma_i}^{\sigma_f} d\sigma\, {\Upsilon}_a(\sigma,\tau,\tau^\prime,\zeta) \,\Delta(y)\,\frac{\partial\;}{\partial y^i}\delta^{(3)}(y-z(\sigma_z,\tau_z,\zeta_z))\nonumber\\
&&=\int_{\zeta_i}^{\zeta_f} d\zeta \int_{\tau_i}^{\tau_f} d\tau \int_{\tau_i}^{\tau} d\tau^\prime\int_{\sigma_i}^{\sigma_f} d\sigma \,\Delta(y)\,\left[\frac{\partial\;}{\partial y^i}\left( {\Upsilon}_a\delta^{(3)}(y-z)\) - \delta^{(3)}(y-z)\, \frac{\partial  {\Upsilon}_a}{\partial y^i}\right]
\lab{upsilontildederdelta1}
\er
Note that there is a total derivative in the first term on the right-hand side of \rf{upsilontildederdelta1}, involving the integrals on $\sigma$ and $\tau^\prime$. We have the same expression as in \rf{iifdef} despite the integration interval. By similar reasoning that leads to \rf{iidefaux}, and using the conditions \rf{coniifdef}, we  obtain that
\br
&&\int_{\tau_i}^{\tau} d\tau^\prime\int_{\sigma_i}^{\sigma_f} d\sigma \,\Delta(y)\,\frac{\partial\;}{\partial y^i}\left( {\Upsilon}_a\delta^{(3)}(y-z)\) =\lab{upsilontildederdeltafpart}\\
&&=\vareps_{ijk}\vartheta \int_{\tau_i}^{\tau}d\tau^\prime \int_{\sigma_i}^{\sigma_f}d\sigma\left[\frac{d\;}{d\,\tau^{\prime}}\(\frac{d\,y^j}{d\,\zeta}\frac{d\,y^k}{d\,\sigma} {\Upsilon}_a\delta^{(3)}\(y-z\)\)
+\frac{d\;}{d\,\zeta}\(\frac{d\,y^j}{d\,\sigma}\frac{d\,y^k}{d\,\tau^{\prime}} {\Upsilon}_a\delta^{(3)}\(y-z\)\)\right]\nonumber\\
&&=\vareps_{ijk}\vartheta\left\{  \int_{\sigma_i}^{\sigma_f}d\sigma\(\frac{d\,x^j}{d\,\zeta}\frac{d\,x^k}{d\,\sigma} {\Upsilon}_a\delta^{(3)}\(x-z\)\)\right.\nonumber\\
&&\left.+\frac{d\;}{d\,\zeta}\left[\int_{\tau_i}^{\tau} d\tau^\prime\int_{\sigma_i}^{\sigma_f} d\sigma\(\frac{d\,y^j}{d\,\sigma}\frac{d\,y^k}{d\,\tau^{\prime}} {\Upsilon}_a\delta^{(3)}\(y-z\)\)\right]\right\}\nonumber
\er
where we have denoted the points at the loop labelled by $\tau$,  by $x^i = x^i(\sigma, \tau, \zeta)$. Substituting \rf{upsilontildederdeltafpart} into \rf{upsilontildederdelta1} and using \rf{deltatransform} to integrate the last term involving the Dirac delta on the right-hand side of \rf{upsilontildederdelta1}, we obtain that
\br
{\cal I}_{a,i}^{\Upsilon}&=&\vareps_{ijk}\vartheta\left[\int_{\tau_i}^{\tau_f}d\tau\int_{\tau_i}^{\tau} d\tau^\prime\int_{\sigma_i}^{\sigma_f} d\sigma\(\frac{d\,y^j}{d\,\sigma}\frac{d\,y^k}{d\,\tau^{\prime}} {\Upsilon}_a\delta^{(3)}\(y-z\)\)\right]_{\zeta=\zeta_f}
\lab{upsilontildederdelta2}\\
&+&\vareps_{ijk}\vartheta\int_{\zeta_i}^{\zeta_f}d\zeta\int_{\tau_i}^{\tau_f}d\tau\int_{\sigma_i}^{\sigma_f} d\sigma\(\frac{d\,x^j}{d\,\zeta}\frac{d\,x^k}{d\,\sigma} {\Upsilon}_a\delta^{(3)}\(x-z\)\) - 
\int_{\tau_z}^{\tau_f}d\tau\, \frac{\partial \;}{\partial z^i} {\Upsilon}_a(\sigma_z,\tau_z,\zeta_z, \tau).
\nonumber
\er

Now, consider the commutator of transformations acting in the matter sector $H_M = H_\psi + H_\vp$. Taking $H_M$ as $X$ in \rf{upsilonadef}, and using \rf{hconsttransf} and \rf{upsilontildedelta}, we have the following
\br
&&\sbr{\delta_{N_1,\alpha_1,\beta_1}}{\delta_{N_2,\alpha_2,\beta_2}}\, H_M\cong \vartheta\,e\,\ve_1\,\ve_2 \,\(\alpha_1-\alpha_2\)\times
\nonumber\\
&&\times{\rm Tr}_{RL}\left[Q^{N_1}_{(1)}\(\zeta_f\)\otimes Q^{N_2}_{(2)}\(\zeta_f\)\,  \left(\vareps_{imn}\int_{\tau_i}^{\tau_f} d\tau_z \int_{\sigma_i}^{\sigma_f} d\sigma_z\frac{dz^m}{d\sigma_z}\frac{dz^n}{d\tau_z} \,J^a_i(z)\,{\cal I}^{\Upsilon}_{a}(z)\right)_{\zeta_z = \zeta_f}\right]
\lab{hamiltcommtransfM}
\er
The surface, in the scanning of $\mathbb{R}^3$, corresponding to $\zeta_z=\zeta_f$ is the two sphere $S^2_{\infty}$ at spatial infinity. The parameters $\sigma_z$ and $\tau_z$ are angles on such a sphere, and $\zeta$ is associated with the radial direction. The tangent and radial vectors  behave as
\be
\frac{dz^i}{d\sigma_z/\tau_z } \rightarrow r, \qquad \qquad \frac{dz^i}{d\zeta_z} \rightarrow s(\sigma,\tau), \qquad \qquad r \rightarrow \infty
\lab{vecbeh}
\ee
where $s(\sigma, \tau)$ is a function of the angles of the sphere at $\zeta_z = \zeta_f$. Consequently, the Jacobian at spatial infinity behaves as
\be
\Delta(z) \rightarrow r^2, \qquad \qquad \qquad r \rightarrow \infty.
\lab{jacobianbeh}
\ee
If the matter currents satisfy the boundary conditions \rf{boundcond}, then at large distances, it should fall as
\be
J_i^a \rightarrow \frac{1}{r^{2+\delta^{\prime}}}, \qquad \qquad \qquad r \rightarrow \infty
\lab{spatcurr}
\ee
with $\delta^\prime>0$. Hence, the product of $J_i$ with the surface element should fall as $1/r^{\delta^\prime}$ at large distances. In addition, from the expression of $ {\Upsilon}_a$ \rf{upsilonexpression}, we observe that except for the magnetic flux $\mathfrak{b}_\tau$ in the definition of ${\cal Z}$, see \rf{calztauzetadefscaled}, the remaining terms are all phase factors and do not contribute to the magnitude of $ {\Upsilon}_a$ at infinity. So $ {\Upsilon}_a$ should fall as $\mathfrak{b}_\tau$ falls, thus from \rf{ebfrakbehaviour}, we have
\be
 {\Upsilon}_a \rightarrow \frac{1}{r^{\delta - \frac{1}{2}}}, \qquad \qquad \qquad r \rightarrow \infty
\lab{upsfall}
\ee
with $\delta>0$. So, since the integral on $\tau$ in \rf{upsilontildedelta} is an angle integral, using \rf{upsfall} we have that
\be
{\cal I}^{\Upsilon}_a \rightarrow \frac{1}{r^{\delta - \frac{1}{2}}}\qquad\qquad\qquad r \rightarrow \infty
\lab{ilaw}
\ee
Considering the asymptotic behavior of \rf{spatcurr} and \rf{ilaw} in \rf{hamiltcommtransfM}, we get into that 
\be
\sbr{\delta_{N_1,\alpha_1,\beta_1}}{\delta_{N_2,\alpha_2,\beta_2}}\, H_M \rightarrow \frac{1}{r^{\delta+ \delta^\prime -1/2}},\qquad \qquad \qquad r \rightarrow \infty
\ee
and so
\be
\sbr{\delta_{N_1,\alpha_1,\beta_1}}{\delta_{N_2,\alpha_2,\beta_2}}\, H_M  \cong 0;\qquad\qquad \qquad {\rm if}\qquad \delta+\delta^{\prime}>\frac{1}{2}
\lab{hamiltcommtransfM1}
\ee

When the constraints \rf{constraint2} hold true, the first term on the right-hand side of the expression involving $H_C$ in \rf{hconsttransf} vanishes. Then, taking $X$ as $H_C$ in \rf{upsilonadef},  performing the integral over $y^i = y^i(\sigma, \tau^\prime, \zeta)$, and using \rf{upsilontildedelta}, we get that the commutator of transformations acting on $H_C$ becomes  
\br
&&\sbr{\delta_{N_1,\alpha_1,\beta_1}}{\delta_{N_2,\alpha_2,\beta_2}}\, H_C\cong \vartheta\,e\,\ve_1\,\ve_2 \,\(\alpha_1-\alpha_2\)\times
\lab{hamiltcommtransfC}\\
&&\times{\rm Tr}_{RL}\left[Q^{N_1}_{(1)}\(\zeta_f\)\otimes Q^{N_2}_{(2)}\(\zeta_f\)\,  \left(\vareps_{imn}\,f_{abc}\,\int_{\tau_i}^{\tau_f} d\tau_z \int_{\sigma_i}^{\sigma_f} d\sigma_z\frac{dz^m}{d\sigma_z}\frac{dz^n}{d\tau_z} \,A_0^b(z)\,E^c_i(z)\,{\cal I}^{\Upsilon}_{a}(z)\right)_{\zeta_z = \zeta_f}\right]
\nonumber
\er

Since $\sigma_x$ and $\tau_x$ are angle variables on the surface $S^2_{\infty}$, corresponding to $\zeta=\zeta_f$, we get that $\frac{dz^i}{d\sigma_z/\tau_x} \rightarrow r^2$, as $r\rightarrow \infty$.  Then, from \rf{boundcond}, \rf{bcaoai} and \rf{ilaw}, we get that
\be
\sbr{\delta_{N_1,\alpha_1,\beta_1}}{\delta_{N_2,\alpha_2,\beta_2}}\, H_C \cong 0;\qquad\qquad\qquad{\rm if}\qquad \delta > \frac{1}{6}
\lab{hamiltcommtransfC1}
\ee

Finally,  taking $X$ as $H_B$ in \rf{upsilonadef}, the transformation of $H_B$ can be written using the result in \rf{hconsttransf}, \rf{upsilontildedelta}, and \rf{upsilontildederdelta2}. Thus, we have the following
\br
&&\sbr{\delta_{N_1,\alpha_1,\beta_1}}{\delta_{N_2,\alpha_2,\beta_2}}\, H_B\cong -\vartheta\,\ve_1\,\ve_2 \,\(\alpha_1-\alpha_2\)\times
\lab{hamiltcommtransfB}\\
&&\times{\rm Tr}_{RL}\left[Q^{N_1}_{(1)}\(\zeta_f\)\otimes Q^{N_2}_{(2)}\(\zeta_f\)\,  \left[\left(\ve_{ijk}\,\int_{\tau_i}^{\tau_f} d\tau_z \int_{\sigma_i}^{\sigma_f} d\sigma_z\frac{dz^j}{d\sigma_z}\frac{dz^k}{d\tau_z} \,B^a_k(z)\,{\cal I}^{\Upsilon}_{a,i}(z)\right)_{\zeta_z = \zeta_f}\right.\right.\nonumber\\
&&\left.\left.+\left(e\,\ve_{ijk}\,f_{abc}\,\int_{\tau_i}^{\tau_f} d\tau_z \int_{\sigma_i}^{\sigma_f} d\sigma_z\frac{dz^j}{d\sigma_z}\frac{dz^k}{d\tau_z} \,A_j^c(z)\,B^b_k(z)\,{\cal I}^{\Upsilon}_{a}(z)\right)_{\zeta_z = \zeta_f} \right]\right]
\nonumber
\er

Using \rf{ilaw}, \rf{vecbeh}, \rf{jacobianbeh} and the relation \rf{jacobiandeltarel}, the only term that survives when ${\cal I}^{\Upsilon}_{a,i}$ is taken at spatial infinity is the first on the right-hand side in \rf{upsilontildederdelta2}. In addition, if the magnetic field $B_k$ satisfies the condition \rf{boundcond}, then $A_k$ should satisfy \rf{bcaoai}. Considering that \rf{ilaw}, the last term on the right-hand side of \rf{hamiltcommtransfB} vanishes by similar reasoning to the right-hand side of \rf{hamiltcommtransfC}. Hence, \rf{hamiltcommtransfB} becomes
\br
&&\sbr{\delta_{N_1,\alpha_1,\beta_1}}{\delta_{N_2,\alpha_2,\beta_2}}\, H_B\cong -\vartheta\,\ve_1\,\ve_2 \,\(\alpha_1-\alpha_2\){\rm Tr}_{RL}\left[Q^{N_1}_{(1)}\(\zeta_f\)\otimes Q^{N_2}_{(2)}\(\zeta_f\)\, \right.\times \nonumber\\
&&\left.\times
\left[\int_{\tau_i}^{\tau_f} d\tau_z \int_{\sigma_i}^{\sigma_f} d\sigma_z\int_{\tau_i}^{\tau_f}d\tau\int_{\tau_i}^{\tau} d\tau^\prime\int_{\sigma_i}^{\sigma_f} d\sigma\,B^a_k(z) {\Upsilon}_a\delta^{(3)}\(y-z\)\times\right.\right.\nonumber\\
&&\left.\left.\times\(\frac{dz^m}{d\sigma_z}\frac{dz^n}{d\tau_z}\frac{d\,y^m}{d\,\sigma}\frac{d\,y^n}{d\,\tau^{\prime}}-\frac{dz^m}{d\sigma_z}\frac{dz^n}{d\tau_z}\frac{d\,y^n}{d\,\sigma}\frac{d\,y^m}{d\,\tau^{\prime}}\)\right]_{\zeta,\zeta_z = \zeta_f}\right]
\lab{hamiltcommtransfB3}
\er
where the right-hand side does not fall fast enough at spatial infinity to vanish. However, from the expression \rf{upsilonexpression}  of $ {\Upsilon}_a$ and the cyclic property of trace, the terms inside in \rf{hamiltcommtransfB3}, can be rearranged and written as a commutator between the charge operators $Q(\zeta_f, \alpha_s, \beta_s)$ and $B_k^a d_{ba}(Q_{s}^{-1}(\zeta_f)V_{s}(\tau_z)W^{-1}(\sigma_z))T_b$ with $s=1,2$. So, from \rf{fwftildewconstant}, the magnetic field $B_i$ at spatial infinity lies in the direction of the generator $c$, which together with ${\widetilde c}$, belongs to a Cartan subalgebra. From the discussion that follows the expression \rf{upsilondef}, when the constraints \rf{constraint2} and the Bianchi identity $D_i B_i =0 $ are imposed, the charge operator $Q(\zeta_f, \alpha_s, \beta_s)$ can be expressed as a exponentiation of the same Cartan subalgebra generators, $c$ and ${\widetilde c}$ (see \rf{exponentiationcc}). Therefore, such commutators vanish, and we obtain that
\be
\sbr{\delta_{N_1,\alpha_1,\beta_1}}{\delta_{N_2,\alpha_2,\beta_2}}\, H_B\cong 0.
\lab{hamiltcommtransfB4}
\ee
Collecting the results \rf{hecommtransf}, \rf{hamiltcommtransfM1}, \rf{hamiltcommtransfC1}, \rf{hamiltcommtransfB4} we verify that \rf{hamiltcommtransf1} is indeed satisfied.

\end{document}